\documentclass[rmp,aps,twocolumn,nofootinbib]{revtex4}

\usepackage{graphics}
\usepackage{epsfig}
\usepackage{amsmath}

\newcommand{\newc}{\newcommand}

\newcommand {\gapprox}
   {\raisebox{-0.7ex}{$\stackrel {\textstyle>}{\sim}$}}
\newcommand {\lapprox}
   {\raisebox{-0.7ex}{$\stackrel {\textstyle<}{\sim}$}}

\newcommand{\pom}{{\rm I\!P}}
\newcommand{\reg}{{\rm I\!R}}
\newcommand{\alphapom}{\alpha_{_{\rm I\!P}}}

\newcommand{\xpom}{x_{_{\rm I\!P}}}

\newc{\Lumi}{{\cal L}}

\newc{\ra}{\rightarrow}
\newc{\Ra}{\Rightarrow}

\newc{\gev}{{\rm GeV}}

\newc{\rpv}{{\not \!\! R_p}}
\newc{\rpvm}{{\not  R_p}}

\newc{\gsim}{{\stackrel{>}{\sim}}}
\newc{\lsim}{{\stackrel{<}{\sim}}}
\newc{\sleq} {\raisebox{-.6ex}{${\textstyle\stackrel{<}{\sim}}$}}
\newc{\sgeq} {\raisebox{-.6ex}{${\textstyle\stackrel{>}{\sim}}$}}

\def\3{\ss}

\newc{\ETJ}{E^{{\rm jet}}_T}

\def\xgo{x_\gamma^{\rm obs}}

\def\q2{{\rm Q}^2}
\def\p2{{\rm P}^2}
\def\gev2{{\rm GeV}^{2}}
\def\F2g{F_2^\gamma}
\def\f2c{F_2^{\rm charm}}

\def\d0{D^{0}}

\def\begr{\begin{flushright}}
\def\endr{\end{flushright}}

\def\begl{\begin{flushleft}}
\def\endl{\end{flushleft}}

\DeclareMathAlphabet{\mathsc}{OT1}{cmr}{m}{sc}

\newcommand{\bmf}[1]{\mbox{\boldmath $#1$}}

\newlength\dlf

\begin{document}

\title{The Hadronic Final State at HERA}

\author{Paul R. Newman}
\email{p.r.newman@bham.ac.uk}
\affiliation{School of Physics and Astronomy, University of Birmingham, Birmingham B15 2TT, UK }
\author{Matthew Wing}
\email{m.wing@ucl.ac.uk}
\affiliation{Department of Physics and Astronomy, University College London, Gower Street, London WC1E 6BT, UK; \\
DESY, Notkestrasse 85, 22607 Hamburg, Germany}

\begin{abstract}  
The hadronic final state in electron--proton collisions at HERA has 
provided a rich testing ground 
for development of the theory of the strong force, QCD.  In this review, 
over 200\,publications from the H1 and ZEUS 
Collaborations are summarised.  Short distance physics, the measurement of 
processes at high energy 
scales, has provided rigorous tests of perturbative QCD and constrained 
the structure of the proton as well as allowing 
precise determinations of the strong coupling constant to be made.  
Non-perturbative or low energy processes 
have also been investigated and results on hadronisation 
interpreted together with those from other experiments.  
Searches for 
exotic QCD objects, such as pentaquarks, 
glueballs and instantons have been performed.  
The subject of diffraction has been 
re-invigorated through its 
precise measurement, such that it can now be described by 
perturbative QCD.  After discussion of HERA, the H1 and ZEUS detectors 
and the techniques used to reconstruct differing 
hadronic final states, the above 
subject areas are elaborated.  The major achievements are then condensed 
further in a final section 
summarising what has been learned.
\end{abstract}                                                                 

\date{\today}
\maketitle
\tableofcontents

\section{Introduction}
\label{sec:intro}

HERA (the Hadron Elektron Ringanlage) \cite{Voss:1994sr} 
at the DESY laboratory in 
Hamburg, Germany is, to date, 
the only example of a storage
ring devoted to producing collisions between leptons and hadrons. 
It was located in a tunnel of length $6.3\,{\rm km}$, housing
two independent beam-pipes, the first
storing electrons or positrons, here generically referred to as electrons, the second protons. 
In an initial phase of operation (HERA-I, 1992--2000), the electron beam energy
was predominantly $27.5\,{\rm GeV}$ and the protons were at $820\,{\rm GeV}$ (1992--1997) 
or $920\,{\rm GeV}$ (1998--2000). 
In its second, higher-luminosity phase (HERA-II, 2002--2007), the 
electrons remained at the 
same energy, but were longitudinally polarised, and the proton 
energy also remained at $920\,{\rm GeV}$ except for the last 
few months, when lower energies 
(575 and 460\,GeV) were used 
primarily for a measurement of the longitudinal proton structure function.
Experiments were located at up to four points around the ring, of which this
review is concerned with the two,
H1 and ZEUS, where the electrons and the protons were brought to collision, yielding
an $ep$ centre-of-mass energy $\sqrt{s} \leq 318\,{\rm GeV}$.

\begin{figure}[htb]
\begin{center}
~\epsfig{file=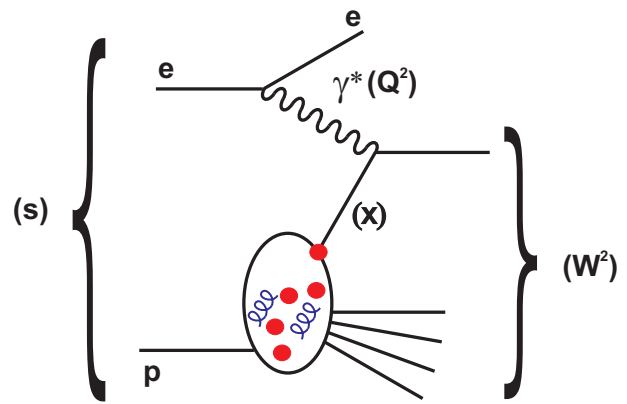,width=0.45\textwidth}
\caption{Schematic illustration of a generic 
electron--proton DIS process,
in which an exchanged photon of virtuality $Q^2$ couples to
a quark carrying a fraction $x$ of the proton's longitudinal
momentum. The electron--proton centre-of-mass energy is denoted $\sqrt{s}$
and the photon--proton centre-of-mass energy (equivalently the 
invariant mass of the hadronic final state) is denoted $W$.}
\label{fig:epbasic}
\end{center}
\end{figure}

For the bulk of the kinematic region accessible at HERA, the 
electron--proton collisions proceed via the exchange of a photon, 
as shown in Fig.~\ref{fig:epbasic}. 
The possible processes 
are conveniently sub-divided into two
categories according to the photon virtuality, $Q^2$. 
The region with  $Q^2 \gg 1\,{\rm GeV^2}$ corresponds to the 
short distance `Deep Inelastic Scattering' (DIS) regime, where the 
photon is usually considered as a structureless
$t$-channel exchange. In contrast, in the 
`photoproduction' regime, $Q^2 \rightarrow 0$, the photon can be thought
of as a distinct object, decoupling from the electron well in advance
of the proton target and interacting as a separate 
entity, often through its own partonic structure (see Section~\ref{pqcd}). 
To a large extent, analyses
at HERA have treated these
two kinematic regions separately, though 
particularly where there are other
hard scales in the problem, this is a largely artificial distinction. 

Conventionally, the term `Hadronic Final State' is taken to mean
the full final state after the removal of the scattered beam lepton
and any electroweak radiation clearly associated with it. 
It therefore includes any 
further leptons or gauge bosons produced in the
photon--proton interaction. As illustrated in Fig.~\ref{fig:epbasic},
the invariant mass of the hadronic final state is denoted $W$, and is 
equivalent to the centre-of-mass energy of the $\gamma^{(*)} p$
collision. 
Expressed in terms of the commonly used `inelasticity' invariant, $y$, 
$W^2 \approx ys - Q^2$. When a hard scale is present, the photon 
couples to a single quark, which carries a fraction $x$ of the 
proton's momentum, where $x \approx Q^2 / (sy)$ is the Bjorken variable.
For a more formal introduction to deep inelastic scattering, 
see for example~\cite{Devenish:2004pb}.

The asymmetric beam energies and near-hermetic instrumentation of
the HERA experiments
were well suited to the study of the hadronic final
state over a wide range of $W$, $Q^2$ and $x$ values. 
These favourable kinematics allowed, for the first time, 
a detailed exploration of the dynamics of the excitations of a single
hadron under a multitude of different circumstances. The vast 
majority of the processes under study are driven by the strong interaction, 
such that where hard scales are present, the theory of
quantum chromodynamics (QCD)~\cite{Gross:1973id,Politzer:1973fx,Gross:1973ju,Politzer:1974fr} and our understanding of 
proton and photon structure were tested with unique precision.
This has led to new insights and a deeper understanding of QCD in general,
as well as providing the stimulus for ever-more sophisticated calculations.  

HERA hadronic final state data provide a very wide range in the energy 
scale of the process. The softest processes are governed by hadronic
mass scales of $\mathcal{O} (1)$\,GeV. Increasing the scale from this
starting point allows the closely-controlled study of the transition from 
a regime which must be described in terms of hadronic objects
to one which can be described perturbatively in terms of partons.
At the other extreme, the highest transverse energy jets provide
scales of $\mathcal{O} (100)$\,GeV, 
resolving the structure of electron--parton interactions at
the $10^{-18} \ {\rm m}$ level. The corresponding 
kinematic range in the Bjorken variable is approximately 
$10^{-4} < x < 0.5$.  
Together with a new window on
fragmentation and hadronisation phenomena, 
these aspects of HERA data have
provided a laboratory of unprecedented precision 
and kinematic range for the 
elucidation of the QCD dynamics of standard DIS processes, which has
led to a vastly improved understanding of the structure 
of the proton and improved measurements of fundamental 
parameters of the Standard Model 
such as the coupling constant of the strong force.

Beyond the best understood domain,  
HERA data have provided access to new kinematic regions and processes
with which to test QCD and extend the range of applicability of existing predictions. 
Many hadronic final state problems have 
contained multiple hard scales, provided
by large transverse momenta and
heavy quark masses as well as $Q^2$. Understanding the delicate interplay
between these different scales has provided exacting challenges to theory.
In photoproduction, HERA data scan the transition from the 
classic $ep$ scattering picture, 
where the photon interacts in a point-like manner, to the situation
where the photon interacts via its own distinct partonic structure, 
more reminiscent of hadron--hadron scattering, facilitating a controlled
study of the differing characteristics of the two extremes. 
Diffractive and related processes in which the proton
stays intact or converts to a neutron via a charge-exchange reaction
imply more complex interactions in which no net colour is exchanged.  
HERA has provided 
an explosion of precise and eclectic measurements of
these hard 
exclusive and semi-inclusive processes with cleanly identified
experimental signatures, leading to a detailed 
understanding in the framework of QCD. 
As discussed and illustrated in considerably more detail in the following,
the multitude of 
Hadronic Final State 
processes studied in $ep$ collisions at 
HERA has thus led to something of a revolution in 
the development and testing of QCD.

\section{Reconstruction of Final States at HERA}
\label{sec:reco}

\subsection{The H1 and ZEUS Detectors}
\label{sec:h1-zeus}

\begin{figure}[htp]
\begin{center}
~\epsfig{file=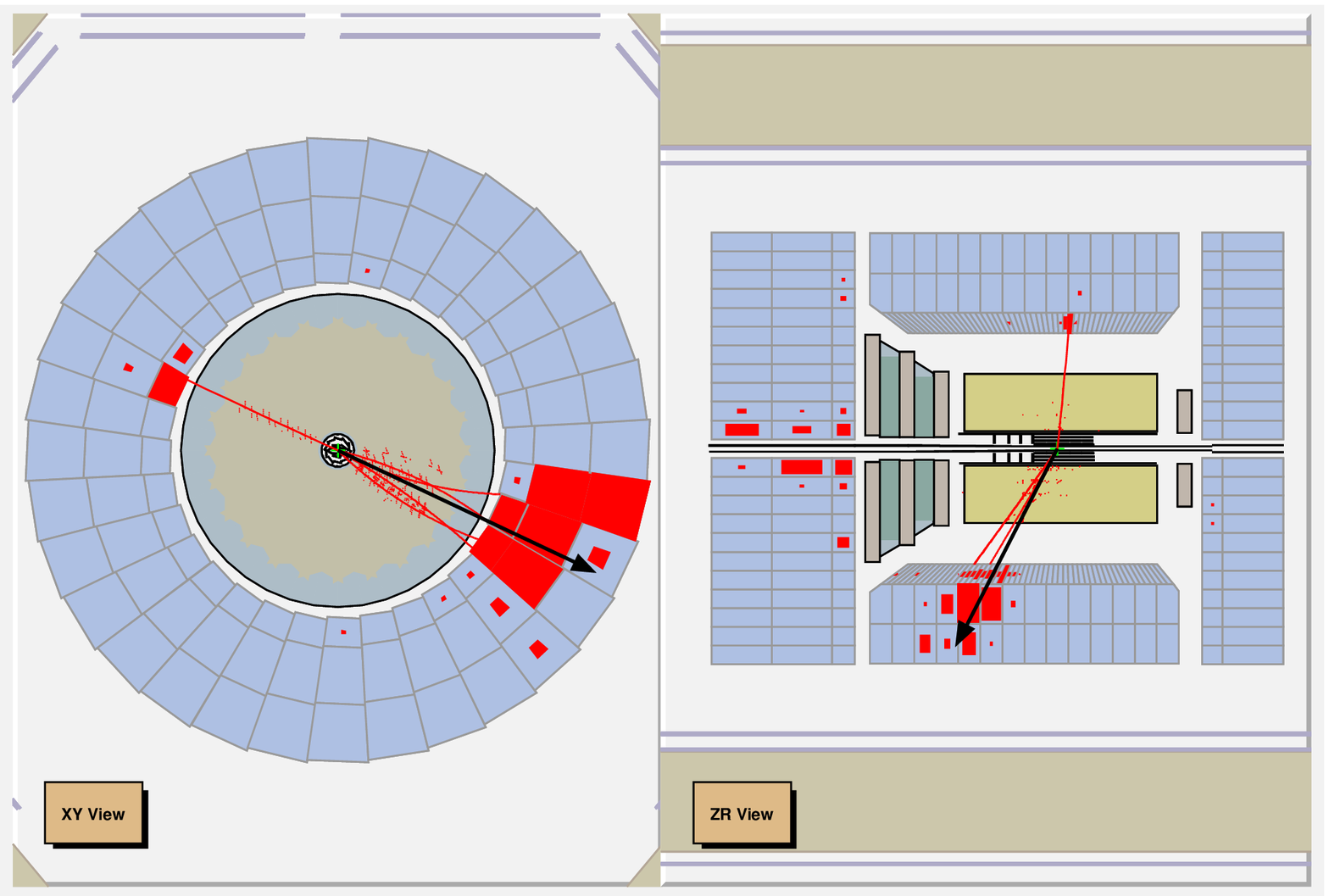,height=5.72cm}
~\epsfig{file=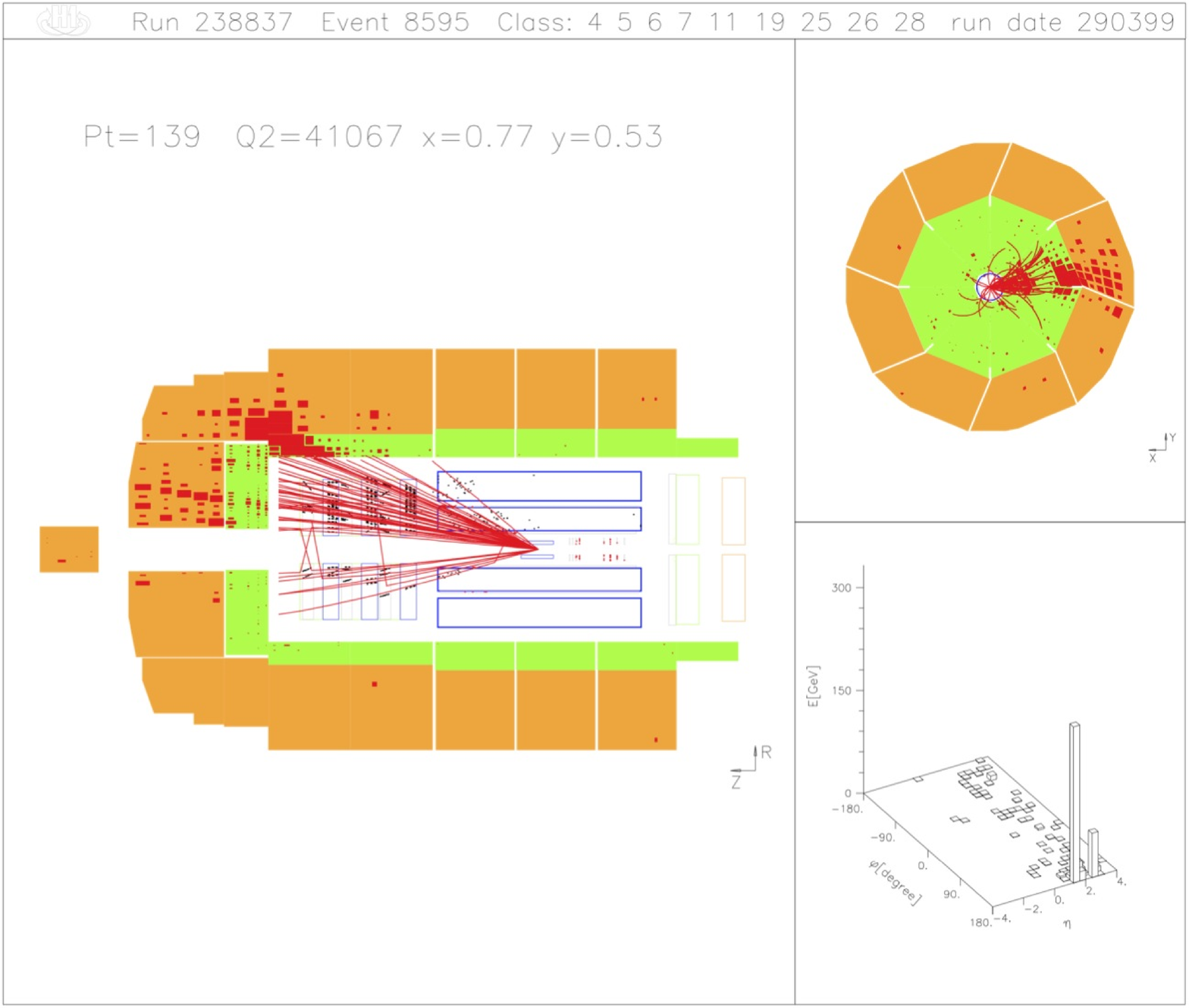,height=7.3cm}
\caption{Event displays of the (top) ZEUS and (bottom) H1 detectors 
showing a neutral 
current and a charged current DIS event, respectively.}
\label{fig:detectors}
\end{center}
\end{figure}

To visualise the physics processes and reconstruction requirements, 
displays of a neutral and 
a charged current DIS event are shown in Fig.~\ref{fig:detectors} for the ZEUS~\cite{zeusdet} and
H1~\cite{Abt:1996hi,Abt:1996xv} detectors, respectively.  In a neutral current event, where 
a photon, or at high $Q^2$ a $Z^0$, is exchanged, a deposit of energy in 
the calorimeter consistent with an electron, 
matched to a track, can be seen 
back-to-back in azimuth with a hadronic jet.  
In a charged current event, where a $W^\pm$ is exchanged, there is 
no activity to balance the hadronic jet, indicating the presence of an undetected 
neutrino.  The figures also show the key parts of the central detectors.  In 
both cases, the electron beam enters the detector from the 
left and the proton beam from the right; due to the significantly larger energy of 
the protons there is more instrumentation in the direction of the outgoing 
proton beam\footnote{A right-handed Cartesian coordinate system 
is used throughout this review, with 
the $Z$ axis pointing in the proton beam direction, referred to as the `forward' direction, and 
the $X$ axis pointing towards the centre of HERA.  The coordinate origin is at the nominal 
interaction point.  The pseudorapidity is defined as $\eta = - \ln \left( \tan \frac{\theta}{2} \right)$, where the polar angle, $\theta$, is measured 
with respect to the proton beam direction.}.  

Other than the forward--backward asymmetry in instrumentation, the 
detectors were 
similar to other general-purpose detectors at high energy colliders such as 
those at LEP, 
the Tevatron and the LHC.  Both H1 and ZEUS detectors had microvertex 
detectors for measuring 
weak decays, surrounded by a central drift chamber for the precise tracking 
of charged particles.  Forward and rear tracking detectors gave additional 
charged-particle information with the rear detectors particularly useful for 
measuring low-angle scattered electrons at small $Q^2$ where the DIS cross section is largest.  
Beyond the tracking detectors 
were calorimeters in which a single technology was used for the ZEUS detector with the H1 
detector also including a dedicated higher-precision electron calorimeter in the rear direction.  
Finally, large muon detectors surrounded the calorimeters.  Further, smaller detectors 
with dedicated purposes, such as for electron or proton tagging, were placed along the beam-pipe 
in both directions.  These detector
components are discussed further in the following 
subsections.

\subsection{Charged Particles}
\label{sec:reco-particles}

Tracks from charged particles were primarily reconstructed by cylindrical drift chambers 
surrounding the interaction point\,:~central jet chambers (CJC1 and 
CJC2)~\cite{Abt:1996hi,Abt:1996xv,Burger:1989eb} in H1 and the 
central 
tracking detector (CTD)~\cite{Harnew:1988ye,Foster:1992mj,Foster:1993ja} 
in ZEUS.  
These were supplemented by silicon vertex 
detectors \cite{Polini:2007sw,Pitzl:2000wz,Eick:1996gv}
placed between the beam-pipe and the drift chambers and in front of forward and rear 
tracking devices.  The H1 and ZEUS drift chambers
covered polar-angle regions of 
$20^\circ < \theta < 160^\circ$ and $15^\circ < \theta < 164^\circ$,
respectively, and were operated in 
uniform axial magnetic fields of 1.16\,T and 1.43\,T.  The high magnetic 
fields 
ensured excellent charge separation up to about 20\,GeV.  The tracking 
resolutions were 
both $\sigma(p_T) / p_T \approx (0.002-0.003)\,p_T$, with $p_T$ in GeV.

Pulse height measurements from the sense wires 
were also used to calculate the specific energy 
loss due to ionisation, 
${\rm d}E/ {\rm d}x$.  The resolutions of the ${\rm d}E/ {\rm d}x$
measurements 
for well-reconstructed tracks were 
under 10\%~\cite{Steinhart:1999xy,Bartsch:2007zza}.  
The ${\rm d}E/ {\rm d}x$
measurements were 
used as a method of particle identification, using likelihood tests 
to distinguish between electrons, pions, kaons, protons and deuterons.

\subsection{Electrons, Photons and Muons}
\label{sec:reco-e-gamma-mu}

Both detectors had large general purpose sampling calorimeters which covered most of the solid 
angle for the measurement of electromagnetic (EM) objects.  The H1 Collaboration had a 
finely segmented liquid 
argon (LAr)~\cite{Andrieu:1993kh} calorimeter 
with lead or stainless steel absorber, complemented by a 
lead--scintillating fibre Spaghetti Calorimeter (SpaCal)~\cite{Appuhn:1996na} 
which covered polar-angle ranges of $4^\circ < \theta < 153^\circ$ and 
$153^\circ < \theta < 177^\circ$, respectively.  
Under test beam conditions, the energy 
resolutions were 
$\sigma(E) / E \approx 11 \% / \sqrt{E}$ and $\approx$\,7\%/$\sqrt{E}$ 
for electrons in the 
LAr calorimeter and SpaCal, respectively, with $E$ in GeV.  
The ZEUS detector had a 
uranium--scintillator 
calorimeter~\cite{Derrick:1991tq,Andresen:1991ph,Caldwell:1992wc,Bernstein:1993kj} which covered a polar-angle range of 
\mbox{$2.5^\circ < \theta < 178.5^\circ$}.  Under test beam conditions, the single-particle energy 
resolution for the ZEUS calorimeter was 18\%/$\sqrt{E}$ for electrons, with $E$ in GeV.

One of the primary design considerations of the experiments was the accurate 
reconstruction of the scattered electrons, 
which should display the signature of isolated high-energy 
deposits in the EM calorimeter.  Various requirements were 
made on the EM clusters and 
the isolation, usually including a
matching of the clusters of energy to a 
charged particle track, such that DIS events 
were selected with high purity and high 
efficiency~\cite{Abramowicz:1995zi,Glazov:1998vv}.  
A typical minimum energy requirement on the scattered electron was 10\,GeV 
for the H1 LAr and 
ZEUS calorimeters and 6.5\,GeV for the SpaCal.  In order to make the 
best measurements 
of the longitudinal structure function at high $y$, these minimum energy 
requirements were reduced in dedicated analyses 
to 3.4\,GeV~\cite{Aaron:2008tx} in the SpaCal and 6\,GeV~\cite{Chekanov:2009na} in the ZEUS calorimeter.

The variables $Q^2$, $x$ and $y$ can be reconstructed from the energy and 
polar angle of the scattered electron only, using the so-called ``electron method'', 
which arises from simple consideration of the DIS kinematics.  Several other reconstruction 
methods exist which make use of combinations of the measurements of the scattered electron 
and the hadronic final state~\cite{Bentvelsen:1992fu,Hoeger:1991wj,Amaldi:1979yh,Bassler:1994uq}.  
The reconstruction method used depends on the analysis and the regions in which particles 
are measured; e.g.\ for charged current events where there is no isolated high-energy electron, 
the Jacquet--Blondel method was used.  Typical relative resolutions in 
$Q^2$ were between about 2\% and 10\%, depending on the reconstruction method.

Photon candidates, typically above 5\,GeV in transverse energy, were 
identified in the 
central calorimeters by reconstructing narrow clusters in the EM calorimeter with 
no track pointing to them.  Backgrounds from $\pi^0$ and $\eta$ particles were removed 
by considering the shape of the cluster and of the calorimeter shower; example quantities 
were the transverse radius of the cluster and the ratio of the highest energy EM cell 
to the total cluster energy~\cite{Aaron:2010uj,Collaboration:2009dqa}.  Photons of lower 
energy could be identified in Compton scattering processes 
($e p \rightarrow e \gamma p$) in which the final state consists 
only of the photon, an electron and a lack of hadronic activity.

Muons were identified by a combination of tracking, calorimetry and the large muon chambers 
surrounding the calorimeters.  Depending on the given analysis and kinematic cuts and whether 
a highly pure or highly efficient sample was required, different algorithms were used to 
combine information from the various sub-detectors.  Wherever possible, a track matched 
with a muon signature in another sub-detector was used for the momentum measurement.  A 
minimum ionising particle (mip) signal in the calorimeters was also used as a high efficiency, but low purity, signal for 
a muon.  Finally, the clearest signal for a muon was provided by the muon chambers.  In the 
case of H1, the return yoke of the magnetic coil was the outermost part of the detector and 
was equipped with streamer tubes forming the 
central muon detector \cite{Abt:1996xv} and tail catcher 
($4^\circ < \theta < 171^\circ$).  
In the forward region ($3^\circ < \theta < 17^\circ$), a set 
of drift chamber layers formed the forward muon 
detector \cite{Biddulph:1993bz},
which allowed a momentum measurement, 
together with an iron toroidal magnet.  
The ZEUS central and rear muon chambers \cite{Abbiendi:1993mi}
consisted of 
limited-streamer tube chambers placed behind the calorimeter, 
inside and outside a magnetised 
iron yoke, covering a polar-angle region of $34^\circ < \theta < 171^\circ$.  
The forward muon detector \cite{Abbiendi:1993mi}
consisted of six trigger planes of limited-streamer tubes and four planes of drift chambers 
covering the angular region $5^\circ < \theta < 32^\circ$.  The efficiency of muon 
reconstruction depended on the method used, with the detection of a mip signal having almost 
100\% efficiency and isolated muons above 2\,GeV being detected 90\% and 55\% of the time 
in the H1 and ZEUS muon systems, respectively~\cite{Aaron:2009sma}.

\subsection{Heavy Flavour Identification}
\label{sec:reco-hfl}

As in other high-energy physics experiments, heavy quarks were identified by several means 
such as reconstructing a given meson mass, requiring a large momentum of a lepton perpendicular 
to the axis of an associated jet, or reconstructing displaced vertices using a silicon detector 
placed close the interaction point.

\begin{figure}[htp]
\begin{center}
~\epsfig{file=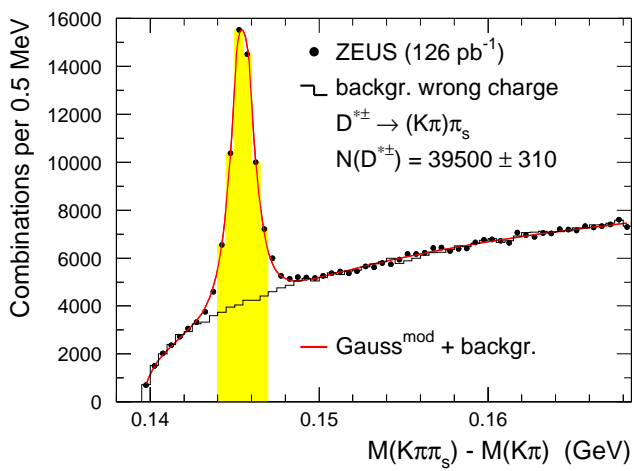,height=6.25cm}
\caption{The distribution of the mass difference 
$m(D^*) - m(D^0)$
for $D^*$ candidates and a background estimate.  From~\cite{Chekanov:2008zt}.}
\label{fig:dstar}
\end{center}
\end{figure}

All ground state charm mesons have been reconstructed by combining tracking information from 
potential decay products.  When only considering this information and no further particle 
identification or measurements of secondary vertices, the decay 
$D^* \to D^0 \pi_s \to K \pi \pi_s$ has the purest signal; an example is shown in 
Fig.~\ref{fig:dstar}.  No beauty mesons have been fully reconstructed due to the significantly 
smaller statistical sample.

Before the use of silicon microvertex detectors, the principal way to identify beauty events 
was to reconstruct leptons and measure their momentum,  $p_T^{\rm rel}$, perpendicular to the axis of 
an associated jet.  This gave a large sample due to a 10\% branching ratio which is much higher 
than any given meson's branching fraction to easily usable hadronic decays.  Due to the quark's larger mass, leptons from 
beauty decays are concentrated at higher values of $p_T^{\rm rel}$ than leptons from charm decays or  
in events initiated by light quarks.  Typical purities achieved for beauty events by this method were about 20\%.

\begin{figure}[bp!]
\begin{center}
~\epsfig{file=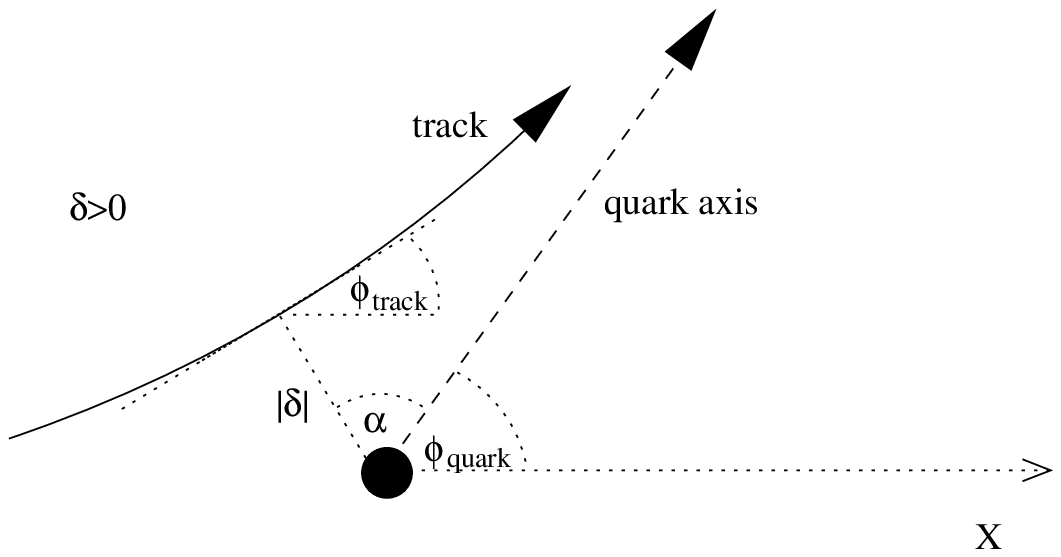,height=4cm}
~\epsfig{file=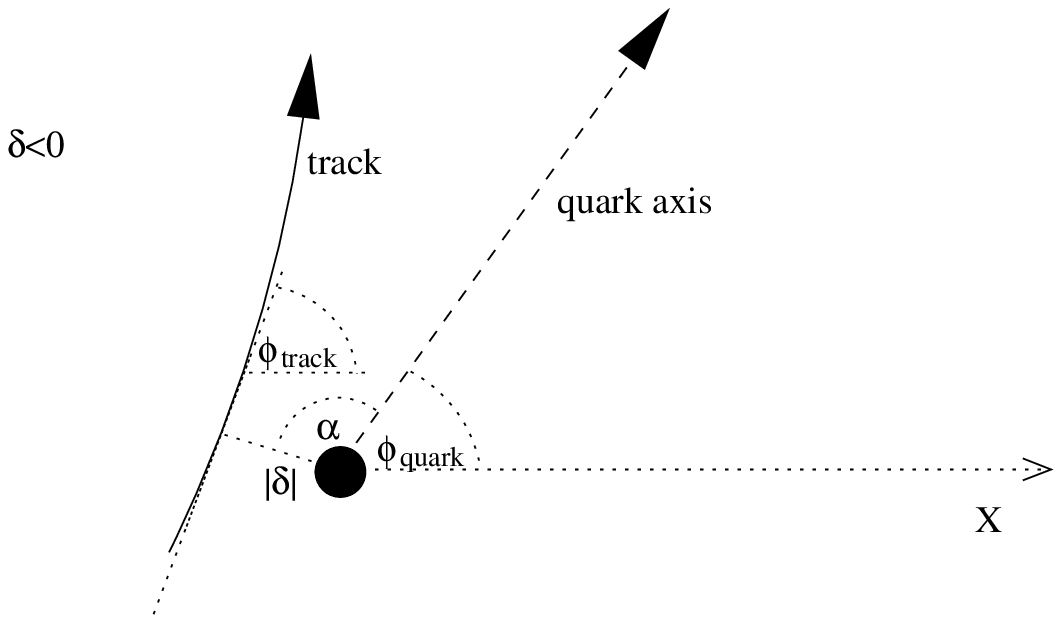,height=4.4cm}
\caption{Diagrams of a track in the $XY$ plane.  The direction of the struck quark, experimentally approximated 
as a jet axis, is used to define the impact parameter, $\delta$.}
\label{fig:2ndvtx}
\end{center}
\end{figure}

\begin{figure}[htp]
\begin{center}
~\epsfig{file=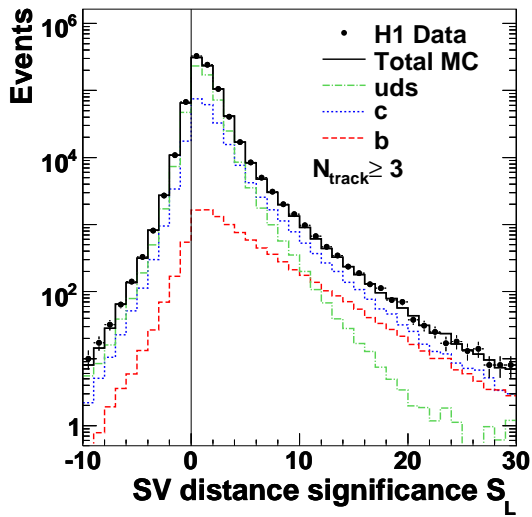,height=7cm}
\caption{Distribution of the significance 
at which the transverse distance between the 
primary and secondary vertices is non-zero.  
Included in the figure is a decomposition of the data by fitting 
Monte Carlo templates for light, $c$ and $b$ quarks.  From~\cite{Aaron:2009ut}.}
\label{fig:sig}
\end{center}
\end{figure}

The reconstruction of secondary vertices, using a microvertex detector, from the weak decay of 
heavy quarks, provides a method for tagging charm and beauty with high efficiency or, by applying 
stringent requirements, with high purity.  
The H1 central silicon tracker (CST) \cite{Pitzl:2000wz}
consisted of two 
layers of double-sided silicon strips, covering an angular 
range of $30^\circ < \theta < 150^\circ$ 
for tracks passing through both layers.  
The ZEUS microvertex detector (MVD) \cite{Polini:2007sw} consisted of 
barrel (BMVD) and forward (FMVD) sections 
with, respectively, three cylindrical layers and four vertical 
planes.  The BMVD provided polar-angle coverage for tracks with three measurements from $30^\circ$ 
to $150^\circ$.  The FMVD extended the coverage to $7^\circ$.  A commonly used variable is the 
impact parameter, $\delta$, of a track which is the transverse distance of closest approach of the 
track to the primary vertex, see Fig.~\ref{fig:2ndvtx}.  If the angle $\alpha$ is less than 
$90^\circ$, $\delta$ is defined as positive, otherwise $\delta$ is defined as negative.  Negative 
values arise only due to imperfect resolution and hence events initiated by light quarks are symmetric 
about $\delta = 0$.  An asymmetry to high positive values occurs due to heavy quark decays.  
A more powerful discriminator is the significance, $S_L$, at which the transverse distance between the 
primary and secondary vertices, $L_{xy}$ is non-zero.  Thus $S_L$ is defined as $L_{xy}/\sigma(L_{xy})$, where 
$\sigma(L_{xy})$ is the uncertainty on $L_{xy}$.  An example distribution is shown in 
Fig.~\ref{fig:sig}, where a strong asymmetry is observed.

To achieve the most precise measurements for beauty and also charm production, the above 
and other discriminating variables were combined in a neural network~\cite{Aaron:2009ut} 
or a discriminating test function~\cite{Chekanov:2008aaa}.

\subsection{Hadronic Jets}
\label{sec:reco-jets}

In the first results on jet production from H1 and ZEUS, jets were reconstructed from 
the calorimeter cells using the JADE clustering~\cite{Bartel:1986ua,Bethke:1988zc} or 
cone algorithms.  Due to the theoretical 
problems with cone algorithms, such as ambiguity in the seed finding and overlapping jets, both 
collaborations switched to the inclusive $k_T$ clustering 
algorithm~\cite{Catani:1993hr,Ellis:1993tq} for the vast majority of their publications.  

Calorimeter cells alone have continued to be used for some publications, due to 
the finer granularity achieved in the forward parts 
of the detector compared with the results of using some pre-clustering before the jet finding.  However, 
energy flow objects (EFOs) have also been in common use by both collaborations.  These 
objects are based on an algorithm which combines tracking and calorimeter information 
in order to optimise the resolution.  In a typical jet at HERA, the amount 
of tracking information used, and hence 
the improvement in the jet energy resolution when including tracks, is 
about 10--20\%.  In certain situations, when several tracks are 
explicitly reconstructed 
in a jet, such as a heavy-flavour meson~\cite{Chekanov:2008ur}, the fraction 
of tracking 
information 
used in the EFO 
input to a jet reaches about 50\% and leads to a significant reduction 
in the systematic uncertainties such as that due to the calorimeter hadronic energy scale.  

The single-hadron energy resolutions, as measured in test beams, are 
$\sigma(E) / E \approx 50\% / \sqrt{E} \oplus 2\%$ 
and 35\%/$\sqrt{E}$, with $E$ in GeV, for the H1 and ZEUS calorimeters, respectively.  A 
detailed parametrisation of the jet energy resolution has not been performed.  However, a  
transverse-energy resolution of 9\% has been achieved at H1 for 
$E_T^{\rm jet} > 25$\,GeV~\cite{Caron:2002yc} and at ZEUS for 
$E_T^{\rm jet} > 14$\,GeV~\cite{Chekanov:2001bw}.

The precision of initial measurements of jet cross sections was limited by the uncertainty 
in the relative jet energy scale between data and Monte Carlo.  Due to the cross section 
rapidly falling 
with increasing energy, a 1\% uncertainty in the determination of the scale led 
to about a 5\% uncertainty in the cross section.  Therefore, a reduction in the energy scale uncertainty 
from initial 
estimates of about 5\% (25\% uncertainty in the cross section) was necessary to be able to fully exploit the 
data for example in precise extractions of parton densities or the strong 
coupling constant (see Section~\ref{sec:alphas}).

\begin{figure}[htb] \unitlength 1mm
  \begin{center}
    \begin{picture}(60,140)
      \put(-5,60){\epsfig{file=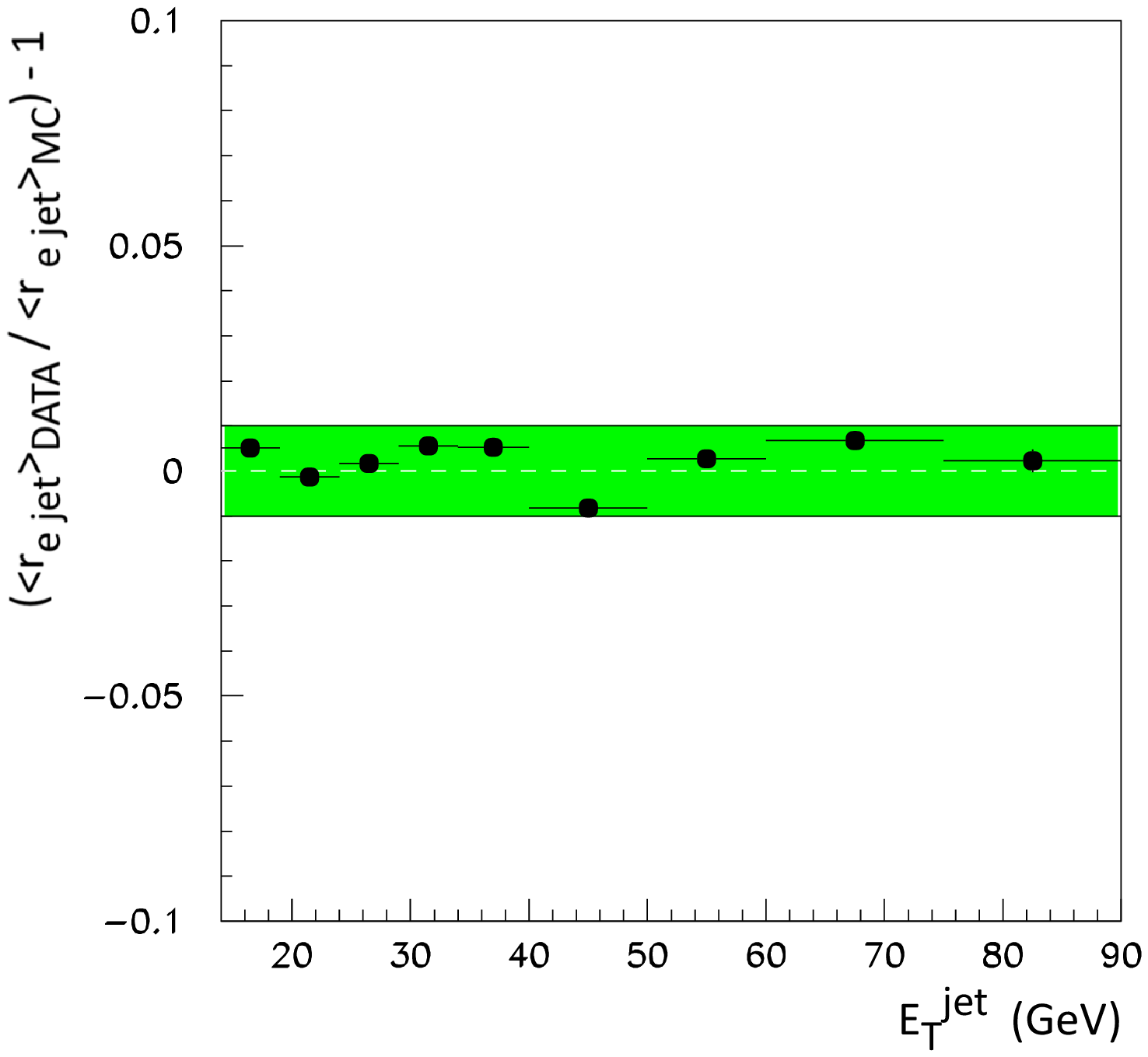,width=6.6cm,angle=270}}
      \put(-5,60){\epsfig{file=Figures/Reconstruction/ptdacellzufo.epsi,width=7cm}}
      \put(14,125){{\large{(a)}}}
      \put(14,52){{\large{(b)}}}
    \end{picture}
  \end{center}
  \caption{ZEUS jet energy scale uncertainty as a function of 
(a) $\eta^{\rm jet}$ and (b) $E_T^{\rm jet}$, 
showing that the scale is known to 1\%. 
In (a), the relative difference between 
the hadronic jet transverse energy, $E_T^{\rm jet}$, and the transverse 
momentum of the 
scattered electron, calculated using the 
double-angle 
method, $p_T^{\rm DA}$, is shown as a function of the jet 
pseudorapidity, $\eta^{\rm jet}$. 
In (b), the relative difference 
between the data and Monte Carlo simulation is shown for the 
quantity $\langle r_{e\, \rm jet} \rangle$ as a function of the jet transverse 
energy, where 
$r_{e\,\rm jet}$ is the ratio of jet to electron transverse energies.}
\label{fig:reco-hadescale}
\end{figure}

In order to 
precisely determine the hadronic jet energy scale, various methods were 
employed, by e.g.\ using tracking information in a jet or cross-calibrating jets reconstructed 
in different parts of the calorimeters.  However, the use of neutral current DIS events provided the most  
powerful calibration tool 
based on the momentum balance between an outgoing jet and the 
scattered electron.  This relied on an accurate 
reconstruction of the electron and a precise determination of the EM energy 
scale uncertainty.  
For the most precise calibrations, closely related techniques
exploiting the double angle kinematic 
reconstruction method \cite{Bentvelsen:1992fu,Hoeger:1991wj}
were employed. 
The difference between the jet and electron transverse energies for 
data and Monte Carlo simulations are shown for the ZEUS data in 
Fig.~\ref{fig:reco-hadescale} versus the pseudorapidity 
and transverse energy of the 
jet.  This demonstrates that the uncertainty on the determination of the jet energy scale 
is 1\%~\cite{Wing:2002fc,Chekanov:2001bw,Chekanov:2001if}.  H1 followed 
similar procedures and achieved an uncertainty of 1.5\%~\cite{Aktas:2006qe}.

\subsection{Photoproduction and DIS and How to Tell Them Apart}
\label{sec:reco-php-dis}

Results from HERA are often (somewhat arbitrarily) classified as 
pertaining to either photoproduction 
or DIS.  In a DIS event, the exchanged photon is virtual, with a squared four-momentum 
typically larger than about 1\,GeV$^2$, and a high-energy scattered electron observed 
in the main calorimeters.  Photoproduction is 
usually defined by the absence of the scattered 
electron in the main calorimeters, implying a virtuality $Q^2<1$\,GeV$^2$ and a median value 
of $10^{-4}-10^{-3}$\,GeV$^2$.  The value of 1\,GeV$^2$ is operational and depends on the 
exact coverage of the calorimeter.  In photoproduction, the photon is quasi-real and the concept 
of a photon structure is introduced.

Small calorimeters were placed along the direction of the outgoing electron beam at distances 
up to 40\,m from the interaction point, see Fig.~\ref{fig:h1-layout}.  These were 
used to tag the electrons in photoproduction events over a narrower range in $Q^2$ and $y$.
For example, 
in the case of the H1 calorimeter at $Z=-33$\,m, values of $Q^2$ were smaller than 
0.01\,GeV$^2$ and the acceptance range was approximately $0.3<y<0.7$.  This 
naturally led to smaller samples compared with 
those obtained via the anti-tag of an electron in the main 
calorimeters, but improved the reconstruction resolution.

\subsection{Very Forward Taggers}
\label{sec:reco-fwd}

\begin{figure*}[tb]
\begin{center}
~\epsfig{file=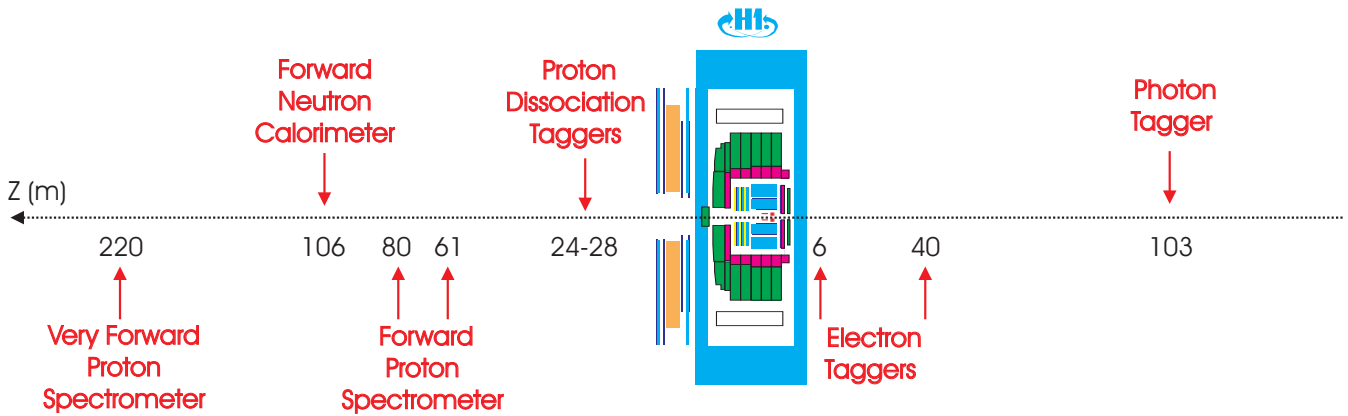,width=\textwidth,bbllx=-600,bblly=188,bburx=620,bbury=558,clip=}
\caption{Schematic illustration of the beamline instrumentation for the
example of the H1 experiment.  The detector components shown are all described 
in the text with the exception of the 103\,m photon tagger which was used for luminosity 
determination via the Bethe--Heitler $e p \to e p \gamma$ process and the proton 
dissociation taggers which were used to identify forward rapidity gaps (see 
Section~\ref{sec:diffraction}).}
\label{fig:h1-layout}
\end{center}
\end{figure*}

Both experiments had very-forward particle taggers to detect low-angle protons and neutrons, 
particularly for measurements of diffractive production (see Section~\ref{sec:diffraction}).

The H1 forward proton spectrometer (FPS)~\cite{Adloff:1998yg,VanEsch:1999pi} and ZEUS leading 
proton spectrometer (LPS)~\cite{Derrick:1996vw} detected positively charged particles carrying 
a substantial fraction of the incoming proton energy.  The particle trajectories were 
measured using a system of detectors that could be inserted very close 
(typically a few mm) to the proton beam, positioned between 64\,m and 80\,m 
and 24\,m and 90\,m from the interaction point for the FPS and LPS, respectively.  
The leading 
proton detectors approached the beam in both the horizontal and the vertical planes, with 
complementary acceptances in scattered proton energy.
Despite 
providing a pure sample of protons, the acceptance was only around 2\%.  
The effective transverse-momentum resolution was dominated by the intrinsic 
transverse-momentum spread of the proton beam at the interaction point, which was about 
40--45\,MeV in the horizontal plane and about 100\,MeV in the vertical plane.  

Both H1 and ZEUS had forward neutron 
calorimeters~\cite{Adloff:1998yg,Bhadra:1994by,Bhadra:1997vi,Bhadra:1997qj} installed in the 
HERA tunnel at \mbox{$\theta=0$} and at \mbox{$Z=106$\,m} from the interaction point in the 
proton-beam direction.  Both devices consisted of a main calorimeter, with hadronic energy 
resolution of 
\mbox{$\sigma(E)/E \approx (60-70) \% / \sqrt{E}$}, with $E$ in GeV, 
supplemented by a pre-shower calorimeter in the case of H1 and a scintillator hodoscope in the 
case of ZEUS.  The forward neutron calorimeters had a limited angular coverage, being sensitive 
to neutrons of less than about 
0.05$^\circ$ or 0.8\,mrad.  Both devices achieved spatial resolutions of about 2\,mm.

Figure~\ref{fig:h1-layout} shows an illustration, for the example of the H1 experiment, of 
the layout of the additional components along the beamline in both directions away from the 
main detector.

\subsection{Triggering}
\label{sec:reco-trig}

Both H1~\cite{Sefkow:1995ca,Nicholls:1998us,Baird:2001sz} and 
ZEUS~\cite{Smith:1992im,Allfrey:2007zz} experiments mainly used a three-level trigger 
system of progressive sophistication in order to select the most interesting events 
on-line and remove background from 
beam-gas interactions and low energy $ep$ collisions in which additional 
statistics would not improve any measurement.  The H1 trigger consisted of two hardware 
levels and a software filter farm, whereas ZEUS had one hardware level and two software 
filters.  The first two levels often considered simple energy 
sums in the calorimeter or the reconstructed vertex position.  At the third level, a 
simplified and fast version of much of the offline reconstruction code, such as jet and 
tracking algorithms, was used.  From a nominal HERA 
bunch crossing rate of 10\,MHz, the trigger 
system reduced the rate such that data were written to storage at a rate of $\approx$\,10\,Hz.

Trigger filters which selected events with a 
high-energy electron candidate were very efficient selectors of DIS events.  Due to 
bandwidth restrictions, these were often pre-scaled or 
were combined with additional requirements, such 
as a jet or a reconstructed meson in the event, depending on the physics motivation.  
Several hundred different combinations of triggers existed to try and cover all 
interesting physics channels, whilst remaining within the limits dictated by data transfer 
rates and processing speeds.  

The principal trigger chains in the H1 and ZEUS experiments, both in terms of bandwidth and 
frequency of use in physics analyses, were designed to select inclusive DIS and 
jet events.  The information to select an electron or jet candidate was often supplemented 
in the trigger with tracking information and, in the 
case of the photoproduction selection at H1, with an 
electron tag in the beam-pipe calorimeters.

\section{Hadronic Final States and Short Distance Physics}
\label{hfs}

In this section, measurements are presented of the hadronic final state  
in the presence of a hard scale, typically the virtuality, $Q^2$, of the exchanged 
boson, the transverse energy of a jet or the mass of a heavy quark in 
the event.  A high or hard scale implies that the reaction occurs at short 
distance and hence allows perturbative calculations to be performed.  
Almost all processes and cross sections at HERA, 
both DIS and photoproduction, involving a jet of high transverse energy 
or a heavy quark have been calculated in the 
DGLAP~\cite{Gribov:1972ri,Lipatov:1974qm,Dokshitzer:1977sg,Altarelli:1977zs} 
approximation to 
next-to-leading order (NLO) in QCD perturbation theory.  In many cases, 
several different approaches to the NLO QCD calculations are available for a 
given process and input parameters such as the strong coupling constant 
can also be varied.  Comparisons between the calculations and with data have led 
to improvements in the predictions.  Various other approximations to QCD are also available 
which emphasise different aspects of parton cascade dynamics and evolution.  
These include the BFKL~\cite{Balitsky:1978ic,Kuraev:1976ge,Kuraev:1977fs} 
and CCFM~\cite{Ciafaloni:1987ur,Catani:1989yc,Catani:1989sg,Marchesini:1994wr} 
approaches. A thorough review of non-DGLAP 
predictions of standard short distance processes 
is not carried out here since NLO corrections are not 
usually available. However, these approaches are considered later in the
context of low $x$ physics (Section~\ref{sec:boundaries}) 
and diffraction (Section~\ref{sec:diffraction}).

Measurements of jet and heavy flavour production are directly sensitive 
to the gluon density in the proton and have hence been used in QCD fits 
to constrain the parton densities in the proton.  Heavy quark data in DIS 
are also used to extract the beauty, $F_2^{b\bar{b}}$, and charm, 
$F_2^{c\bar{c}}$, contribution to the proton structure, $F_2$.  These results 
are reviewed in detail elsewhere~\cite{Klein:2008di,Perez:2012um}.  Measurements of 
photoproduction are also sensitive to the structure of the photon and the 
data can in principle be used in fits to constrain the parton densities in 
the photon.  A more detailed review of photoproduction and its constraints on 
photon structure can also be found elsewhere~\cite{Butterworth:2005aq}; 
however, new results since that review are discussed here.

\subsection{Perturbative QCD Theory of the Hadronic Final State}
\label{pqcd}

A brief description of perturbative QCD related to the hadronic final state is given 
in this section.  Fuller accounts can be found elsewhere~\cite{Ellis:1991qj,Brock:1993sz,Dissertori:2003pj}.

\begin{figure}[htp]
\begin{center}
\vspace{0cm}
\hspace{-1.5cm}~\epsfig{file=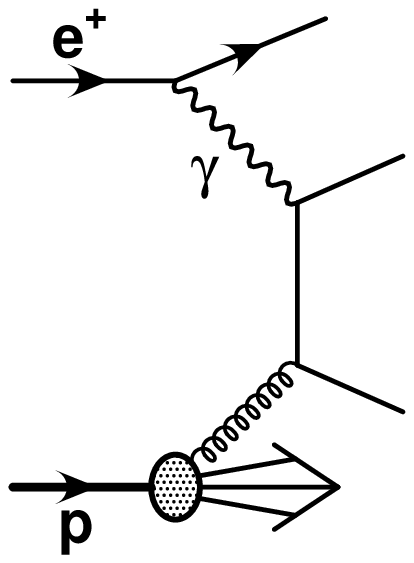,height=4cm}\hspace{1cm}
~\epsfig{file=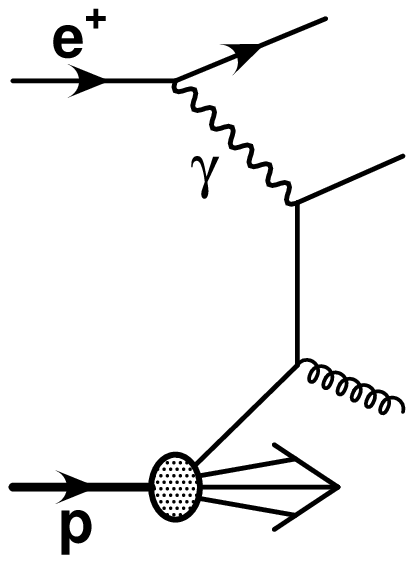,height=4cm}
\put(-165,2){\makebox(0,0)[tl]{(a)}}
\put(- 53,2){\makebox(0,0)[tl]{(b)}}
\caption{Illustrations of the (a) boson--gluon-fusion and (b) $u$-channel QCD 
Compton processes.  Along with $s$-channel QCD Compton scattering, these are 
the LO QCD processes in DIS and direct photoproduction, i.e.\ the lowest-order 
process involving at least one power (or vertex) of $\alpha_s$.}
\label{fig:DIS-feyn}
\end{center}
\end{figure}

Given that the lowest-order DIS process, a quark-parton model (QPM) event 
(see Fig.~\ref{fig:epbasic}), contains a scattered electron recoiling against a 
jet, it may seem trivial to describe jet cross sections in DIS.  However, once 
the sizeable phase space for parton radiation is considered in the context of 
the wide range of possible jet algorithms, the situation becomes far more subtle.  
Jet cross sections are generally presented in the Breit 
frame~\cite{feynman:book,Streng:1979pv} in which the exchanged virtual boson is 
purely space-like, with 3-momentum $\bmf{q} = (0, 0, Q)$, and is collinear with 
the incoming parton, such that QPM events do not contribute at large transverse 
energies.  Therefore leading-order (LO) QCD processes, Fig.~\ref{fig:DIS-feyn}, 
dominate jet cross sections in DIS. 

From the diagrams, it can be seen that the boson--gluon-fusion process 
is related to the gluon density in the proton.  This is dominant at low 
$Q^2$, where low-$x$ partons are most important, whereas the QCD Compton 
process becomes more important with increasing $Q^2$ since it is related 
to the quark density in the proton.  Measurements of jet cross sections are 
therefore sensitive to the strong coupling constant, $\alpha_s$.  
When combined with inclusive DIS cross-section measurements, they allow its 
precise extraction simultaneously with the parton densities in the 
proton, as discussed in Section~\ref{sec:alphas}.  This can be seen from a 
general schematic formula for perturbative QCD calculations of DIS 
jet processes\,:
\begin{align}
 {\rm d}\sigma_{e p \rightarrow e + {\rm jets} + X } = 
 \displaystyle\sum_{a} 
 \displaystyle\int_0^1 & {\rm d}\hat{\sigma}_{e a \rightarrow cd }(x,\alpha_s (\mu_R), \mu_F, \mu_R)\, \nonumber \\
   & f_{a/P}(x,\mu_{F}) \, {\rm d}x \, , 
 \label{eq:DISpert}
\end{align}
where the sum is over the possible partons, $a$, in the proton given by the 
parton density function (PDF), $f_{a/P}$.  The 
factorisation and renormalisation scales are denoted by $\mu_F$ and $\mu_R$ and
may be given by $\surd Q^2$, the jet transverse energy, or a combination of the two.  
The short-distance cross section, ${\rm d}\hat{\sigma}_{e a \rightarrow cd }$, 
depends on $x$, the strong coupling, $\alpha_s$, $\mu_F$ and $\mu_R$.

\begin{figure}[htp]
\begin{center}
~\epsfig{file=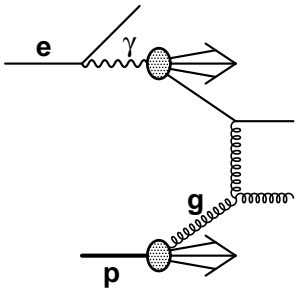,height=4cm}
\caption{An example of a LO resolved jet photoproduction process, containing a hard scattering 
between a quark from the photon and a gluon from the proton.}
\label{fig:res-feyn}
\end{center}
\end{figure}

In photoproduction, where the electron escapes detection and continues down the 
beam-pipe, the virtuality, $Q^2$, is low and the hard scale is given instead 
by the transverse energy of the jets.  The diagrams shown in 
Fig.~\ref{fig:DIS-feyn} also apply to the LO direct jet photoproduction process 
where direct-photon events are classified as those in which all of the photon's 
momentum participates in the hard interaction.  Equation~\ref{eq:DISpert} is modified 
to the general formula\,:
\begin{align}
 {\rm d}\sigma_{e p \rightarrow e + {\rm jets} + X } = 
 \displaystyle\sum_{a} 
 \displaystyle\int_0^1    & {\rm d}\hat{\sigma}_{\gamma a \rightarrow cd }(x,\alpha_s (\mu_R), \mu_F, \mu_R)\,  \nonumber \\
   & \displaystyle f_{\gamma/e} \,  f_{a/P}(x,\mu_{F}) \, {\rm d}x \,,
\label{eq:DIRECTpert}
\end{align}
where the term $f_{\gamma/e}$ represents the probability of the electron radiating a photon 
and is given by the Weizs\"{a}cker-Williams 
formula~\cite{Frixione:1993yw,vonWeizsacker:1934sx,Williams:1934ad}.  
Another class of events, resolved-photon processes, also contribute 
to the photoproduction cross section.  At LO, such processes are 
classified as those in which only a fraction of the photon's momentum participates 
in the hard interaction.  For such events, the photon can be considered as 
developing a structure, the parton densities of which are probed by the hard 
scale of the interaction.  This means that the $ep$ collision can be viewed as a 
hadron--hadron collision in which partons from both the photon and the proton 
participate in the hard process.  Therefore many extra diagrams contribute in LO QCD 
to the photoproduction cross section; an example is shown in 
Fig.~\ref{fig:res-feyn}, in which a quark from the photon collides with a gluon 
from the proton.

A general schematic formula for perturbative QCD calculations of photoproduction 
processes is given by\,:
\begin{widetext}
\begin{equation}
 {\rm d}\sigma_{e p \rightarrow e + {\rm jets} + X } = 
 \displaystyle\sum_{a,b} \int_0^1
 {\rm d}x_{\gamma} 
 \displaystyle\int_0^1 {\rm d}x_{p} \, 
   f_{\gamma/e} \, f_{b/\gamma}(x_\gamma,\mu_{F\gamma}) \, 
   f_{a/p}(x_p,\mu_{Fp}) \,
   {\rm d}\hat{\sigma}_{ab \rightarrow cd }(x_\gamma,x_p,\alpha_s(\mu_R),\mu_{F\gamma}, \mu_{Fp}, \mu_R)\, ,
 \label{eq:pertxsec}
\end{equation}
\end{widetext}
where $x_p$ and $x_\gamma$ are the longitudinal momentum fractions of the parton 
$a$ in the proton and the parton $b$ in the photon, respectively.  The term 
$f_{a/p}$ ($f_{b/\gamma}$) represents the PDFs of partons with flavour $a$ ($b$) 
in the proton (photon). The factorisation scale for the proton (photon) is 
denoted by $\mu_{Fp}$ ($\mu_{F\gamma}$), and $\mu_R$ is the renormalisation 
scale. The factorisation and renormalisation scales are often assumed to have the same 
value in calculations, although this is not necessarily the case and hence for generality, they 
are here treated separately.  The term $d\hat{\sigma}_{ab\rightarrow cd}$ is the hard (partonic) 
cross section.  In the case where parton $b$ is the entire photon, 
$f_{b/\gamma}(x_\gamma,\mu_{F\gamma})$ is $\delta(1-x_\gamma)$ and Eq.~\ref{eq:pertxsec}
describes direct photoproduction and reduces to Eq.~\ref{eq:DIRECTpert}.

The separation between resolved and direct processes has more to do with the 
limitations of our ability to calculate QCD cross sections than with fundamental 
physics.  The separations are not unique beyond LO.  For example the LO resolved-photon 
process in Fig.~\ref{fig:res-feyn} can also be considered as a direct-photon 
process in NLO QCD.  Nevertheless, the labels `direct' and `resolved'
are useful tools for exploring the world of photon physics. 

The NLO corrections for the production of two hard partons at HERA, both in photoproduction 
and DIS, were calculated in the 1990s (see below), thereby allowing comparisons with 
inclusive-jet and dijet measurements, by 
applying a jet algorithm to the partons 
in the final state of an NLO parton-level event generator.  
These NLO QCD calculations 
generally give 
an accurate prediction of the normalisation and the shapes of basic kinematic 
distributions.  However, 
in order to compare with observables measurable from the data,
corrections for hadronisation 
using Monte Carlo models are necessary.  A multiplicative 
hadronisation
correction factor is determined 
from the ratio of the cross sections at the hadron and parton levels 
in the Monte Carlo simulation.  As the simulations 
are only based on LO matrix elements 
with parton showering, their applicability is questionable.  
However, some level of control over the procedure 
can be assured by checking the compatibility   
of the dependences 
on important variables such as jet transverse energy and angle 
between the 
NLO calculation and the parton level Monte Carlo simulation.  For some 
event properties and 
kinematic configurations, the NLO QCD calculations are not very 
reliable, due to the fact that they only allow at most one parton to be 
radiated in 
addition to the primary jet pair.  Calculations at the next order, 
next-to-next-to-leading order (NNLO) in 
QCD~\cite{Alekhin:2009ni,JimenezDelgado:2008hf,Martin:2009iq}, have 
been performed 
for inclusive DIS but not with final-state objects
such as jets or heavy quarks present.

In DIS, NLO QCD calculations are available for the production of jets in neutral 
current~\cite{Mirkes:1995ks,Graudenz:1997gv,Catani:1996vz,Potter:1998jt,Nagy:2001xb} 
and charged 
current~\cite{Mirkes:1995ks} processes.  The NLO corrections have also 
been calculated for $2 \to 3$ scattering (i.e.\ three-jet cross sections) in 
DIS~\cite{Nagy:2001xb} and can in principle be extended to photoproduction.  
Inclusive hadron production has also been calculated to 
NLO~\cite{Kretzer:2000yf,Kniehl:2000cr,Albino:2005me,Albino:2008fy,deFlorian:2007aj,deFlorian:2007hc}, 
whereas prompt photon production has been calculated to 
$\mathcal{O}(\alpha^3)$~\cite{GehrmannDeRidder:2000ce,GehrmannDeRidder:2006vn,GehrmannDeRidder:2006wz}.  

In photoproduction, NLO QCD calculations are available for the production of 
jets~\cite{Aurenche:2000nc,Frixione:1997ks,Frixione:1995ms,Frixione:1997np,Klasen:1996it,Harris:1997hz,Gordon:1992tw}, 
hadrons~\cite{Fontannaz:2002nu,Binnewies:1995pt},  
and prompt photons~\cite{Krawczyk:2001tz,Zembrzuski:2003nu,Fontannaz:2001ek,Fontannaz:2003yn,Gordon:1994sm}.  

The above perturbative calculations all require some choices of input parameters, 
and also need to be corrected for hadronisation, which lead 
to uncertainties in the predictions.  The 
renormalisation and factorisation scales, 
the proton and photon PDFs, the value of $\alpha_s$ 
and, where appropriate, fragmentation functions all need to be chosen.  
The uncertainties are usually dominated by varying the 
renormalisation scale by a factor of two.  However they vary depending on the phase 
space and distribution measured; the precision of the predictions are discussed 
where appropriate in the following sections.  It should be noted that the scale variation 
by a factor of two is merely convention and bears no relation to e.g.\ a one-sigma uncertainty.  This 
should therefore be treated with caution.  In one example
fit to data~\cite{Chekanov:2005nn}, the variation produced 
unacceptable $\chi^2$ values and so a variation of $\sqrt{2}$ was chosen.  
Measurements in which variation of the scale by a factor of two appear to be an 
under-estimation are discussed in Section~\ref{sec:boundaries}.

The above parton-level calculations for jet production 
assume massless partons,  
which is also a possible procedure when 
calculating heavy-quark production
(the so-called ``massless'' 
scheme~\cite{Binnewies:1997gz,Kniehl:1996we,Binnewies:1997xq}).  
These calculations are for photoproduction; 
no heavy flavour calculations exist 
using this scheme for DIS.  Here charm and beauty are regarded 
as active flavours in the PDFs of the proton and photon and are 
fragmented from massless partons into massive hadrons after the hard process.  
This scheme should be applicable at high transverse momenta. 
For momenta of the 
outgoing heavy quark of the order of the quark mass, 
the fixed-order or ``massive'' 
scheme in photoproduction~\cite{Frixione:1994dv,Frixione:1995qc} and in 
DIS~\cite{Harris:1997zq,Harris:1995tu,Harris:1995pr} is 
more appropriate.  In the massive scheme, $u$, $d$ and $s$ are the only active 
flavours in the structure functions of the proton and charm and beauty are 
produced only in the hard subprocess.  Compared with inclusive jet production, 
these 
calculations are subject to significant additional 
uncertainties from the mass of the heavy quark 
and, where appropriate, the transition of the quark to a hadron.  More details 
of the different schemes and in particular their relevance for determination of 
PDFs is given in~\cite{Abramowicz:1900rp} and references therein.

In addition to NLO calculations, 
data are also often compared with predictions from Monte Carlo models which 
incorporate LO matrix elements matched with leading-logarithm 
parton showers.  The Monte Carlo models generally give a more complete and realistic final state, 
but are unreliable 
in normalisation due to the fact that the matrix elements are 
currently only LO.  
The approaches~\cite{mcatnlo1,Nason:2004rx} of matching matrix elements calculated at NLO with parton showers 
is widely used for LHC processes.  This has been 
extended to HERA physics but only for heavy flavour production~\cite{Toll:2010zz}.

\subsection{Comparisons with Data}

\subsubsection{Jet Production}
\label{sec:jet-production}

\begin{figure}[htp]
\begin{center}
~\epsfig{file=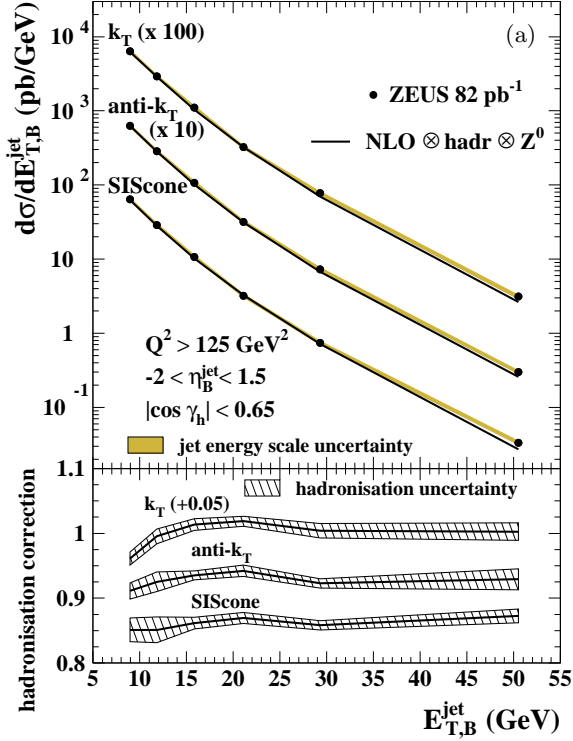,height=10cm}
\put(-28,272){\makebox(0,0)[tl]{(a)}}\\
~\epsfig{file=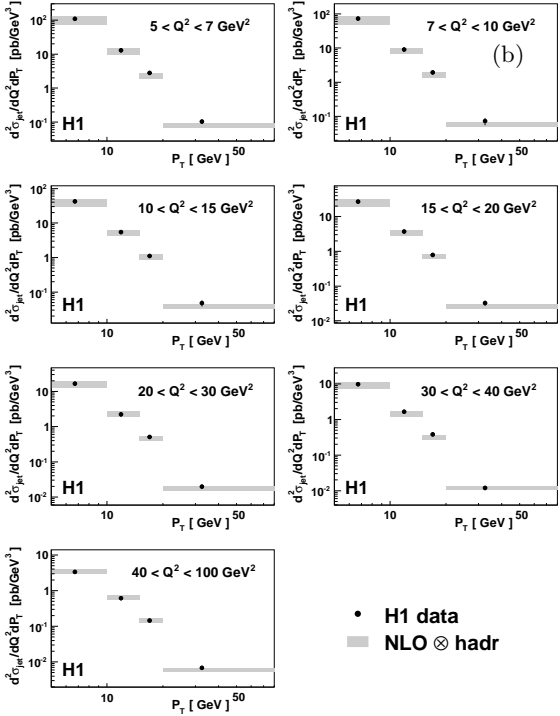,height=10.2cm}
\put(-28,257){\makebox(0,0)[tl]{(b)}}\\
\caption{(a) Measurement of ${\rm d}\sigma/{\rm d}E^{\rm jet}_{T,B}$ for inclusive-jet production 
in the Breit Frame in DIS for different jet algorithms.  The lower part of the figure 
shows the hadronisation corrections applied to the NLO calculations.  
(b) Measurement of ${\rm d}^2\sigma/{\rm d}Q^2\,{\rm d}P_{T}$ 
in different regions of $Q^2$ for inclusive-jet production in the Breit Frame in DIS 
using the $k_T$ jet algorithm.  Note that $P_T$ is equivalent to $E^{\rm jet}_{T,B}$.  
The data in (a) and (b) are compared with NLO QCD 
predictions (corrected for hadronisation and $Z^0$ effects).  (a) 
From~\cite{Abramowicz:2010ke} and (b) from~\cite{Aaron:2010ac}.}
\label{fig:DISJets}
\end{center}
\end{figure}

A measurement of the inclusive jet cross section 
is shown as a function of the 
jet transverse energy in the Breit frame, 
$E_{T, \rm B}^{\rm jet}$, in Fig.~\ref{fig:DISJets}.
The cross section falls 
by three orders of magnitude 
as $E_{T, \rm B}^{\rm jet}$ increases from $9 \ {\rm GeV}$ to 
$50 \ {\rm GeV}$ and the uncertainty on the measurement
remains below 5\% for $E_{T, \rm B}^{\rm jet} < 30$\,GeV, 
dominated by the uncertainty on the jet energy scale.  
The statistical uncertainties only become dominant above 30\,GeV.
The excellent description of these data is a triumph of QCD.
The theoretical uncertainties are of broadly similar size 
to those from experiment, though larger at low 
$E_{T, \rm B}^{\rm jet}$ and smaller at larger $E_{T, \rm B}^{\rm jet}$.

Figure~\ref{fig:DISJets}(a) illustrates the need to specify the choice 
of algorithm when discussing jet cross sections.  The quality of the 
theoretical description is approximately the same for the 
$k_T$~\cite{Catani:1993hr}, anti-$k_T$~\cite{Cacciari:2008gp} and 
SIScone~\cite{Salam:2007xv} algorithms, though the cross sections 
themselves and the necessary hadronisation corrections are 
algorithm-dependent.  Figure~\ref{fig:DISJets}(a) shows that theory 
describes the data for all jet algorithms in this large $Q^2$ range.
In Fig.~\ref{fig:DISJets}(b), H1 data for measurements using the $k_T$ 
algorithm are shown at lower and in different regions of $Q^2$; the NLO QCD 
prediction also describes the data well here.  In general, measurements of 
inclusive-jet~\cite{Chekanov:2006yc,Chekanov:2002sz,Chekanov:2002be,Chekanov:2006xr,Aaron:2010ac,Aaron:2009vs,Aaron:2009he,Adloff:2000tq,Aktas:2007pb,Adloff:2002ew,Abramowicz:2010ke}, dijet~\cite{Abramowicz:2010ih,Chekanov:2006xr,Adloff:2000tq,Aaron:2009vs,Aaron:2009he,Breitweg:2001rq,Chekanov:2001fw} and trijet~\cite{Aaron:2009vs,Aaron:2009he,Chekanov:2005ve,Adloff:2001kg,Chekanov:2008ih} 
production in DIS are all well 
described by NLO QCD, particularly at high $E_T$ or high $Q^2$ and at central 
values of pseudorapidity.  Such a description of the kinematic trends of jet 
production in DIS allows an extraction of the parton densities in the proton and/or 
the value of the strong coupling constant to be made, as discussed in Section~\ref{sec:alphas}.  

Jet photoproduction has the added possibility for the photon to develop 
a structure and so the data can in 
principle be used to extract information on this. 
The structure, or more precisely the parton densities, 
of the photon are generally 
extracted from measurements of DIS $e\gamma$ interactions at $e^+e^-$ 
colliders~\cite{Nisius:1999cv}.  However, 
due to the $Q^{-4}$ dependence of the $ep$ cross section, 
measurements of photoproduction 
at HERA offer larger statistics than are available from 
$e^+e^-$ data, as well as probing 
higher energy scales.  
Jet photoproduction is reviewed 
in more detail elsewhere~\cite{Butterworth:2005aq}, 
with a few of the highlights and new results included here.

\begin{figure}[htp]
\begin{center}
~\epsfig{file=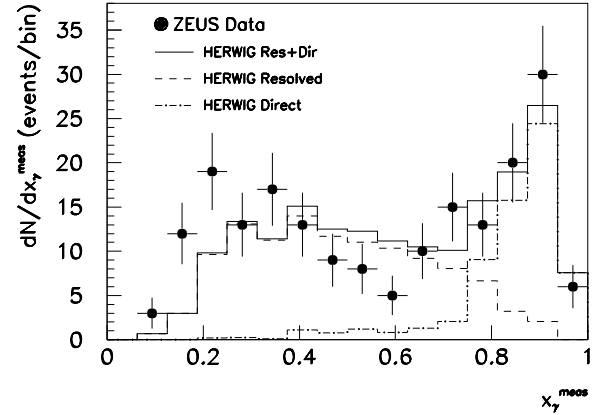,height=5.5cm}
\caption{Raw distribution in 
$x_\gamma^{\rm meas}$ for photoproduction events with two or 
more jets.  The direct- and resolved-photon Monte Carlo predictions are fitted to the data 
with free normalisation. 
From~\cite{Derrick:1993tb}.}
\label{fig:xgamma}
\end{center}
\end{figure}

Some of the first measurements at HERA established the potentially hard scattering nature of 
photoproduction through the observation of two jets with significant transverse 
energy~\cite{Ahmed:1992xj,Derrick:1992zd} in events in which no scattered electron 
was observed in the main detectors.  The need for both direct- and resolved-photon 
interactions in 
describing 
photoproduction at HERA was also shown by comparing data with models of just 
one of the processes or the combined prediction.  A variable particularly 
sensitive to the nature of the photon and the relative fraction of direct- and 
resolved-photon processes is a
hadron- or detector-level estimator of $x_\gamma$, the 
photon's momentum fraction which takes part in the hard scatter.  At LO, $x_\gamma$ is 
identically equal to 1 for direct-photon processes and less than 1 for 
resolved-photon processes.  Experimentally~\cite{Derrick:1993tb} the  
$x_\gamma$ estimator was reconstructed as 

\[ x_\gamma^{\rm meas} = \frac{\sum_{\rm jets}(E-p_z)_{\rm jets}}{\sum_i (E-p_z)_i} \]
where the sums in the numerator and denominator run over all jets and all energy 
deposits in the calorimeter, respectively. The first measurement of a distribution 
in this quantity is shown in Fig.~\ref{fig:xgamma} and is compared with 
a two component fit using direct and resolved photon templates from the 
{\sc Herwig}~\cite{Marchesini:1991ch,herwig2} Monte Carlo programme.  The data 
exhibit a two-peak structure at high and low values of $x_\gamma^{\rm meas}$.  
The direct- and resolved-photon components in the Monte Carlo have very different 
shapes.  The Monte Carlo prediction gives a reasonable representation of the data 
when direct-  and resolved-photon processes are added together.  The resolved-photon 
component describes the low $x_\gamma^{\rm meas}$ region reasonably well but cannot 
describe the data at high $x_\gamma^{\rm meas}$.  These data at high $x_\gamma^{\rm meas}$ 
can only be described with the inclusion of the component from direct-photon 
processes. Hence these data constituted the first observation of direct-photon 
processes in photoproduction.

After these initial findings, a multitude of results from both collaborations 
were 
published, investigating the ability of NLO QCD to describe 
jet photoproduction data
(the reader 
is referred to the H1~\cite{h1pubwww} and ZEUS~\cite{zeuspubwww} paper lists).
These studies addressed
the need for an underlying event due to secondary scatters
and yielded extractions of $\alpha_s$ and measurements of 
quantities sensitive to the structure of the proton and photon.  
The most 
recently published~\cite{Aktas:2006qe,Chekanov:2007qt,Abramowicz:2012jz} and most precise measurements 
of jet photoproduction focus on high transverse energy so as to minimise the effects 
of any underlying event and therefore to provide a clean probe of the structure of 
the proton and photon.

\begin{figure}[htp]
\begin{center}
~\epsfig{file=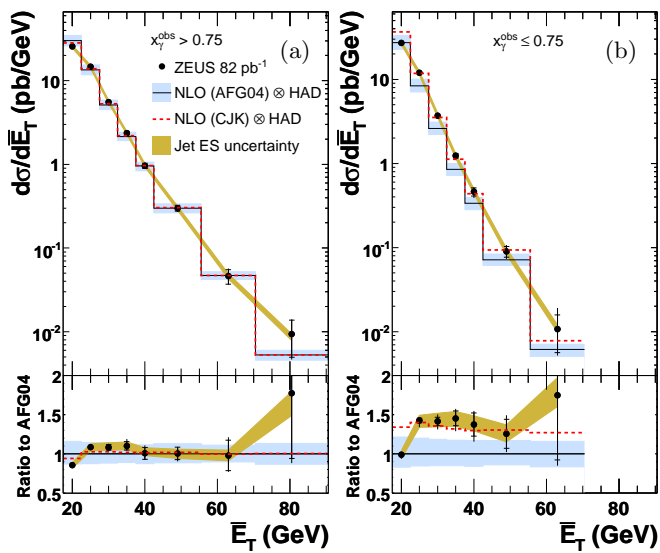,height=7.5cm}
\put(-146,196){\makebox(0,0)[tl]{(a)}}
\put(-23,196){\makebox(0,0)[tl]{(b)}}
\caption{Measured cross-section ${\rm d}\sigma/{\rm d}\bar{E}_T$ for (a) 
$x_\gamma^{\rm obs} > 0.75$ and (b) $x_\gamma^{\rm obs} < 0.75$,  
compared with NLO QCD predictions (corrected for hadronisation) using photon 
PDFs, AFG04 (solid line) and CJK (dashed line).  From~\cite{Chekanov:2007qt}.}
\label{fig:et-gammap}
\end{center}
\end{figure}

Figure~\ref{fig:et-gammap} shows the distribution of the mean transverse energy of the two 
highest-$E_T$ jets, $\bar{E}_T$, in two regions of $\xgo$~\cite{Derrick:1995bg}, defined as
\begin{equation}
\xgo = \frac{\sum_{\rm jet} E_T^{\rm jet}e^{-\eta^{\rm jet}}}
               {2 y E_e},
\label{eq:xgo}
\end{equation}
which is closely related to $x_\gamma^{\rm meas}$, with the high-$\xgo$ region enriched in direct-photon events 
and low $\xgo$ enriched in resolved-photon events.  At the high transverse energies measured 
here the high-$\xgo$ region is well described by NLO QCD, although a difference in shape is 
observed between data and theory.  
As can be seen from the predictions using two rather 
different photon PDFs, AFG04~\cite{Aurenche:2005da} 
and CJK~\cite{Cornet:2004nb}, the sensitivity to the structure of 
the photon is small at high 
$\xgo$, as expected, although the prediction 
using CJK describes the data somewhat 
better.  For $\bar{E}_T < 40$\,GeV, the dominant uncertainty 
on the data 
is due to the jet energy 
scale (see Section~\ref{sec:reco-jets}), which will 
not be improved in future measurements.  
In this region, 
the uncertainties on the data are also 
significantly smaller than those on the NLO QCD predictions.  

At low $\xgo$, the difference in shape between data and NLO QCD in Fig.~\ref{fig:et-gammap}(b) 
is more marked.  For the calculations using AFG04, the data and NLO QCD prediction agree in the 
lowest bin whereas the prediction is significantly below the data at high $\bar{E}_T$.  In contrast, 
the prediction from CJK is too high in the first bin, which dominates the cross section, but 
agrees well at higher $\bar{E}_T$.  
Although the prediction from CJK clearly lies above 
the data, it also gives the best description of the dependence on other variables such as the 
average pseudorapidity of the jets, $\bar{\eta}$~\cite{Chekanov:2007qt}.  All other 
parametrisations of the photon PDFs give a qualitatively similar description of the data to 
that of AFG04.  
The fact that the gluon density in the CJK photon PDF differs 
from the others may hint at the origin of the improved
description.  
These data should thus 
improve our knowledge of the gluon density in the photon PDFs, 
which is 
insufficiently constrained by $e^+e^-$ data.  

\subsubsection{Jet Substructure}

The substructure of a jet gives information on the internal 
pattern of parton radiation, as 
well as details of the hadronisation process.  Given that   
gluons radiate more than quarks,
categorising jets using   
measurements of their substructure 
offers the possibility of obtaining samples which are
enriched in gluon or quark initiators.  
Classifying the jets in an event using this technique
thus allows samples to be obtained which are dominated by
particular parton-level final states, 
in turn giving enhanced control
over the initial state.
Measurements of jet substructure could thus in principle lead to 
improved constraints on the structure of the proton and 
photon, distinguishing for example between 
the $\gamma^* g \rightarrow q \bar{q}$
and $\gamma^* q \rightarrow q g$ processes in 
the DIS case. 

\begin{figure}[htp]
\begin{center}
~\epsfig{file=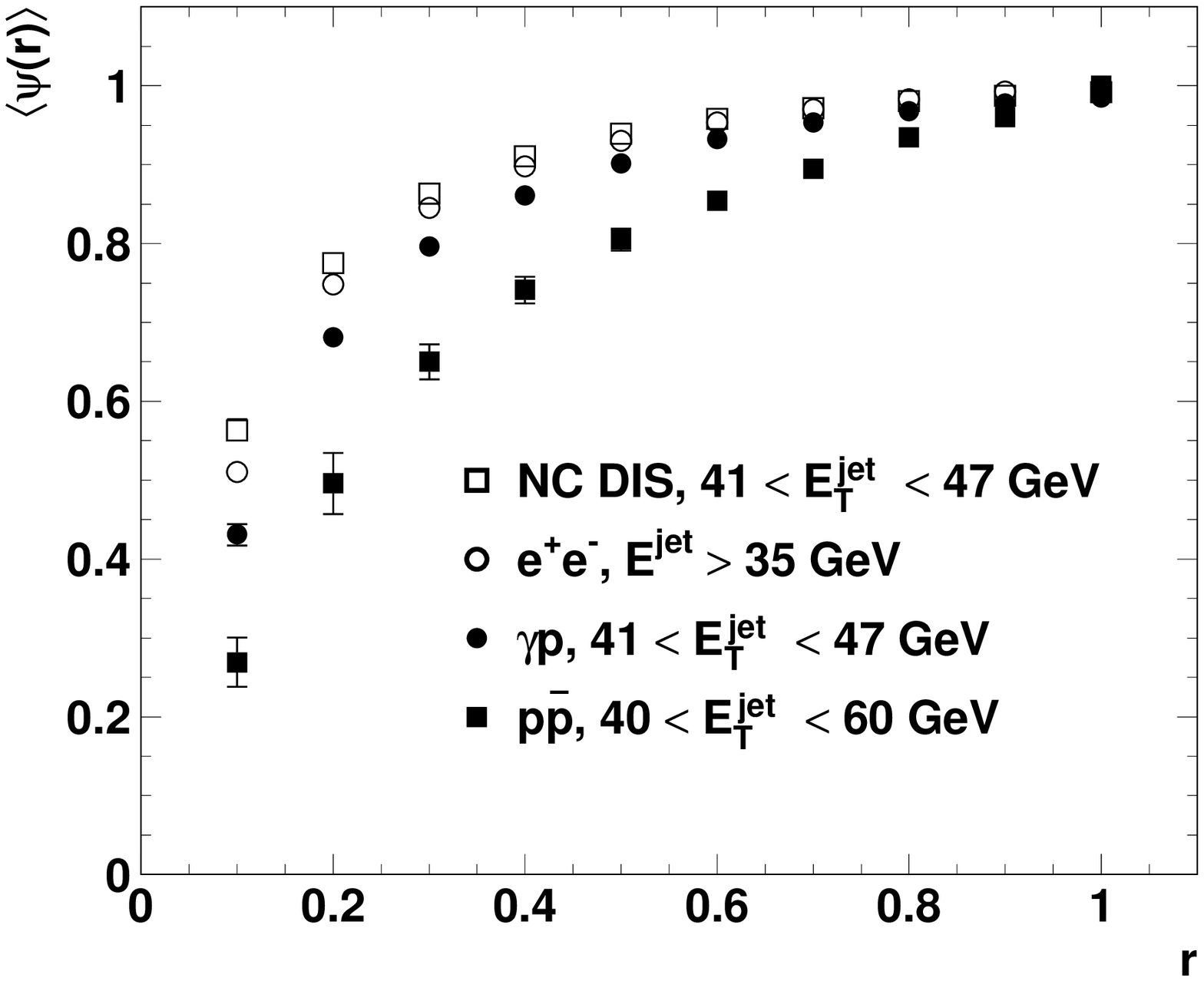,height=7.cm}
\caption{Measured jet shapes 
in $e^+e^-$ scattering~\cite{Akers:1994wj}, from ZEUS in 
DIS~\cite{Breitweg:1998gf} and photoproduction~\cite{Breitweg:1997gg}, and
in $p\bar{p}$ collisions~\cite{Abe:1992wv,Abachi:1995zw}.  
From~\cite{Butterworth:2005aq}.}
\label{fig:jet-sub}
\end{center}
\end{figure}

Jet substructure is generally studied by measuring the jet 
shape~\cite{Ellis:1992qq} and subjet 
multiplicity~\cite{Catani:1992tm,Seymour:1994by,Seymour:1996tf,Forshaw:1999iv}.
The integrated jet shape, 
$\psi(r)$, using only those particles belonging to the jet, is defined as the 
fraction of the jet transverse energy that lies inside a 
cone in the $\eta-\phi$ 
plane of radius, $r$, concentric with the jet axis\,:

\begin{equation}
\psi(r)  = \frac{E_T(r)}{E_T^{\rm jet}},
\label{eq:jet_shape}
\end{equation}
where $E_T(r)$ is the transverse energy within the given cone of radius 
$r$. The mean integrated jet shape, $\langle \psi(r) \rangle$, is defined 
as the averaged fraction of the jet transverse energy inside the cone
$r$\,:

\begin{equation}
\langle \psi(r) \rangle = \frac{1}{N_{\rm jets}} \sum_{\rm jets} \frac{E_T(r)}{E_T^{\rm jet}},
\label{eq:int_jet_shape}
\end{equation}
where $N_{\rm jets}$ is the total number of jets in the sample.  
The substructure of a jet is expected to depend primarily on the 
initiator of the jet and to a lesser extent on the colliding particle.  This is 
supported in Fig.~\ref{fig:jet-sub}, where the mean integrated jet shape 
is shown 
for different processes, \emph{viz.}\ DIS~\cite{Breitweg:1998gf}, $e^+e^-$ 
collisions~\cite{Akers:1994wj}, photoproduction~\cite{Breitweg:1997gg} and 
$p\bar{p}$ collisions~\cite{Abe:1992wv,Abachi:1995zw}.  The results are shown 
for similar jet 
energies for all samples to remove any dependence of the particle 
initiator or the radiation on this quantity.    
In DIS and 
$e^+e^-$ collisions the 
partonic system emerging from the hard interaction 
is expected to consist mainly of quarks.
The similarity between the results from these processes, their
relative narrowness and  
difference from the other samples, is consistent with this.  
The less-collimated and broader jets seen in $p\bar{p}$ 
collisions are indicative 
of gluon-initiated jets dominating the sample.  In photoproduction at these 
energies, direct-photon processes are expected to dominate.  
However, the presence 
of QCD Compton and resolved-photon processes also leads to gluons in the final 
state.  The results in photoproduction 
lie between those from $p \bar{p}$ scattering and those from 
DIS and $e^+ e^-$ collisions, compatible with the 
expected mixture of gluon- and 
quark-initiated jets.

Measurements at HERA of internal jet 
structure~\cite{Adloff:1998ni,Breitweg:1997gg,Breitweg:1998gf,Chekanov:2002ux} 
have been compared with various Monte Carlo models and NLO QCD predictions and shown 
to be well described.  This agreement has also allowed extractions of the strong coupling 
constant~\cite{Chekanov:2002ux,Chekanov:2004kz}.  More recently, measurements have been 
made~\cite{Chekanov:2004kz,Chekanov:2009bc} with the aim of distinguishing between 
gluon- and quark-initiated jets.  By cutting on the jet shape at low $r$, a jet can be 
classified as ``broad'' or ``narrow''~\cite{Chekanov:2004kz} and thereby enriched in 
gluon and quark initiators, respectively.  

\begin{figure}[htp]
\begin{center}
~\epsfig{file=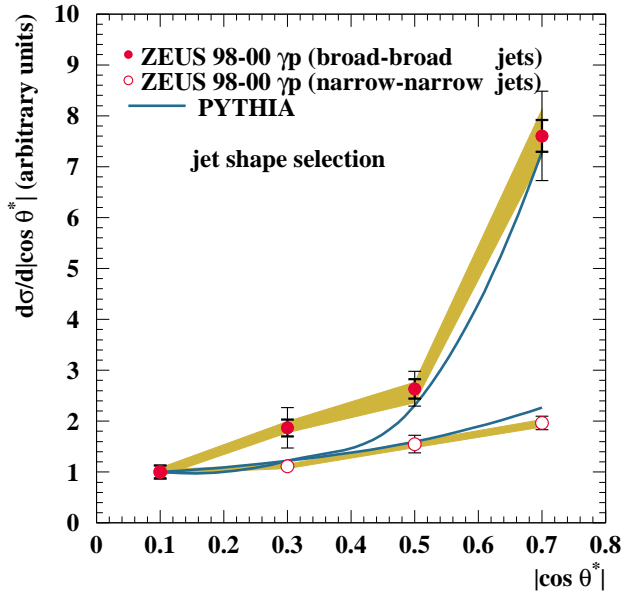,height=8.4cm}
\caption{Measurement of the dijet cross section differential in the cosine of the dijet scattering 
angle, ${\rm d}\sigma/{\rm d}|\cos\theta^*|$, for events with two broad jets or two narrow jets.  
The data are compared with {\sc Pythia} Monte Carlo predictions.  From~\cite{Chekanov:2004kz}.}
\label{fig:jet-sub-gq}
\end{center}
\end{figure}

An example dijet cross section is 
shown in Fig.~\ref{fig:jet-sub-gq}, where both jets are tagged as 
either broad or  
narrow.  
A striking difference is observed between the two samples as a function of the cosine 
of the dijet scattering angle in the dijet centre-of-mass frame, 
with the general trends well described by the {\sc Pythia} 
Monte Carlo~\cite{pythia,Sjostrand:2000wi,Sjostrand:1993yb} predictions.  The dijet scattering 
angle is a revealing quantity as it is sensitive to the propagator, with a 
spin$-\frac{1}{2}$ quark exchange giving a $(1-|\cos\theta^*|)^{-1}$ cross-section dependence 
and a spin$-1$ gluon exchange giving a $(1-|\cos\theta^*|)^{-2}$ cross-section dependence at leading order.  
The shallow rise to high $\cos\theta^*$ for 
the sample with two narrow jets and the steeper rise for the sample with two 
broad jets are indicative of such a difference in propagator.  
The angular distributions 
can be understood in terms of the dominant two-body processes\,: the resolved subprocess 
$q_\gamma g_p \to qg$, mediated by gluon exchange for the broad--broad dijet sample and 
the direct subprocess $\gamma g \to q \bar{q}$, mediated by quark exchange for the 
narrow--narrow dijet sample.
For events with two 
broad jets, {\sc Pythia} predicts the parton final state to consist of 16\% $gg$, 
52\% $qg$ and 32\% $qq$. For events with two narrow jets, {\sc Pythia} predicts the 
parton final state to consist of 71\% $qq$, 28\% $qg$ and 1\% $gg$.  The relatively 
impure sample of gluon-initiated jets is due to the dominance of the boson--gluon fusion 
process (with two quark-initiated jets in the final state) and the 
background from 
$c$ and $b$ quarks, which also yield broad jets due 
to the longer decay chain compared to that for light quarks.  

This detailed understanding of 
jet substructure seeded the development of the new techniques 
to search for the Higgs boson~\cite{Butterworth:2008iy} or
other boosted heavy 
particles~\cite{Abdesselam:2010pt} which are now being used at the 
LHC.

\subsubsection{Prompt Photon Production}

Events containing an isolated `prompt' (or `direct') photon are a 
potentially powerful 
tool to study hard processes.  
Their main attractions lie in the insensitivity of
photons to hadronisation effects and the precise
energy measurements obtainable for isolated electromagnetic objects.  
The number of possible 
processes is also smaller than for the case of jet production; 
to leading order in both QCD and QED, prompt 
photon cross sections in DIS and direct photoproduction
are directly sensitive to the quark content 
of the proton through 
the Compton scattering ($\gamma q \to \gamma q$) process.  
Resolved-photon contributions are dominated by the $gq \to q \gamma$ process,
giving sensitivity to the quark and gluon contents of 
both the proton and photon.
These advantages have to be set against the significantly 
smaller event rates for prompt photon than for jet production and 
the experimental challenge of separating photon
samples from backgrounds due to $\pi^0$ and 
$\eta^0$ meson decays to multi-photon states. 
This separation relies on many discriminating 
variables,
such as the shape of the shower deposited in the calorimeter.  

\begin{figure}[htp]
\begin{center}
~\epsfig{file=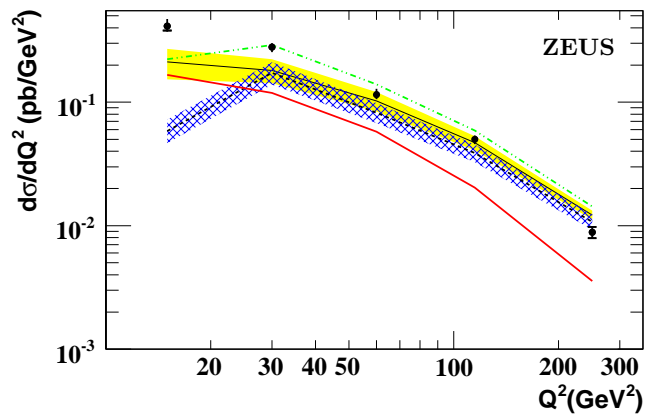,height=6.cm}
\put(-51,143){\makebox(0,0)[tl]{\bf ZEUS}}
\caption{Differential cross section for prompt photon production in DIS with the data 
compared with various theoretical predictions.  An $\mathcal{O}(\alpha^3)$ calculation 
with radiation from the quark line (thick red line) and 
additionally including radiation from the lepton 
line and interference between the two (yellow shaded line) is shown.  The photon is also 
treated by Martin \emph{et al.}\ as a partonic constituent of the proton (blue hatched 
area) and including photon radiation, \emph{i.e.}\ adding the prediction corresponding 
to the thick red line, from the quark line (green dot-dashed line).  
From~\cite{Chekanov:2009dq}.}
\label{fig:photonsDIS}
\end{center}
\end{figure}

Measurements of prompt photon production have been made in both
DIS~\cite{Chekanov:2004wr,Chekanov:2009dq,Aaron:2007eh,Abramowicz:2012qt} and
photoproduction~\cite{Breitweg:1997pa,Breitweg:1999su,Chekanov:2001aq,Chekanov:2006un,Aktas:2004uv,Aaron:2010uj,Abramowicz:2013vfa}.
Both H1 and ZEUS have made measurements in the DIS regime 
using significant fractions of the available data.  
The data have been compared with  
predictions~\cite{GehrmannDeRidder:2000ce,GehrmannDeRidder:2006wz,GehrmannDeRidder:2006vn} 
to order $\alpha^3$, which describe the shapes 
of the measured cross sections, 
${\rm d}\sigma/{\rm d}E_T^\gamma$ and ${\rm d}\sigma/{\rm d}\eta^\gamma$.  
However, the theory is systematically below the data, with the difference 
concentrated at 
low $Q^2$, as shown in Fig.~\ref{fig:photonsDIS}.  
The alternative approach of \cite{Martin:2004dh}
treats photons 
as a partonic 
constituent of the proton, 
introduced by 
including QED corrections to the proton PDFs 
and producing high energy photons in 
the final state. Predictions from this model are also shown in 
Fig.~\ref{fig:photonsDIS}.  
They fall below the data over most of the 
measured range, but are close in the high-$Q^2$ 
region, where lepton emission is expected to be dominant.  
An improved 
description of the data is obtained by appropriately combining the 
two predictions, 
suggesting a need for further calculations 
to exploit the full potential of the 
measurements. 

\begin{figure}[htp]
\begin{center}
~\epsfig{file=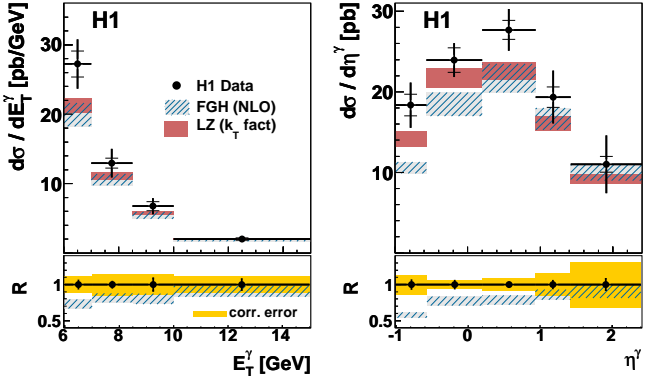,height=5.5cm}
\caption{Differential cross sections for prompt photons in photoproduction.  The data 
are compared with two QCD calculations, one based on collinear factorisation in NLO 
(hatched histogram) and the other based on $k_T$ factorisation.  The lower parts of 
the figures show the ratio of the NLO QCD prediction to the data.  
From~\cite{Aaron:2010uj}.}
\label{fig:photons-php}
\end{center}
\end{figure}

Cross sections for prompt photon photoproduction are shown in 
Fig.~\ref{fig:photons-php}.  The results use almost the full H1 data set and 
have a precision of about 10\%.  
The data 
are compared with an NLO QCD calculation 
based on collinear factorisation and DGLAP evolution~\cite{Fontannaz:2001ek,Fontannaz:2003yn} 
and with a QCD calculation based on 
the $k_T$ factorisation~\cite{Lipatov:2005tz} method which is expected to 
provide a good approximation for a 
significant part of the collinear higher-order QCD corrections. Both 
predictions are below the data with the largest differences at low $E_T^\gamma$ 
and low $\eta^\gamma$.  The prediction based on $k_T$ factorisation is a bit higher 
than that of NLO DGLAP QCD and hence is closer to the data.  The same conclusions as those 
stated here and seen in Fig.~\ref{fig:photons-php} were arrived at in the latest 
ZEUS results~\cite{Chekanov:2006un}.  
The influence of 
high-order QCD terms and hadronisation effects are expected to be 
largest at low transverse energies, 
and further theoretical developments are needed 
in this region in particular. 

Overall, the impact of prompt photon data on the development of
QCD at HERA has been somewhat disappointing compared with the 
successes obtained using jet observables. This is in part due to the 
limited statistics and experimental difficulties encountered in
obtaining cross sections and partly due to the 
need for further improvements in the theoretical understanding   
to fully exploit the data.

\subsubsection{Heavy Quark Production}

Heavy quarks at HERA, as at other colliders, are tagged using many different independent 
techniques as summarised in Section~\ref{sec:reco}.  As with 
inclusive jet cross sections, 
the primary aim of measurements 
is to test QCD 
(see Eq.~\ref{eq:pertxsec}), 
through the sensitivity to 
both the perturbative prediction as discussed in this 
section and the proton and photon PDFs as discussed in 
Section~\ref{sec:alphas}.  As the masses of charm and particularly 
beauty quarks are so 
much larger than $\Lambda_{\rm QCD}$, they can provide a hard scale for perturbative calculations 
that are expected to converge rapidly.  Depending on the process measured, information 
on the fragmentation or even decays can also be extracted  
and searches for excited states performed (see Section~\ref{sec:frag}).

\begin{figure}[htp]
\begin{center}
~\epsfig{file=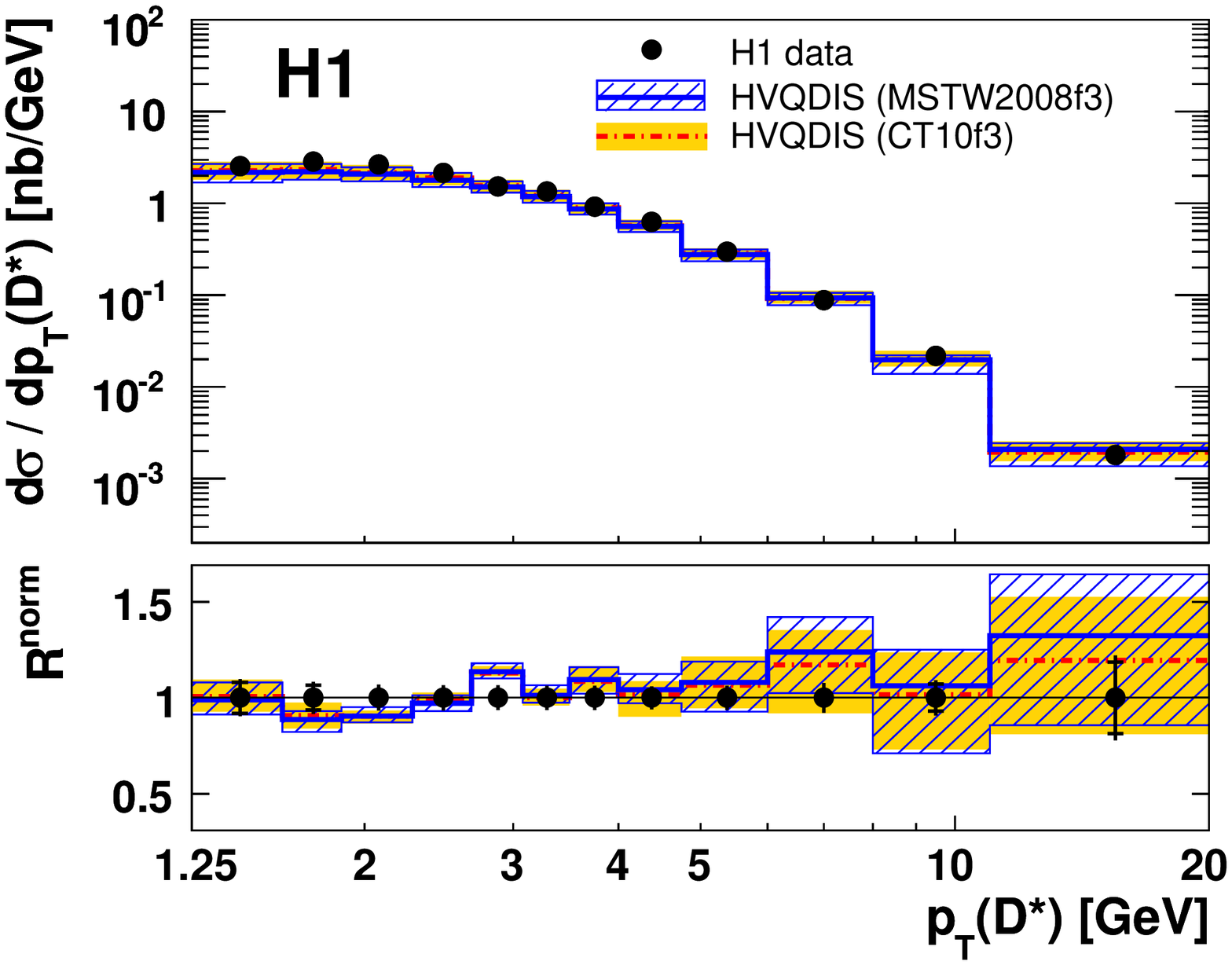,height=6cm}\\
\vspace{0.4cm}
~\epsfig{file=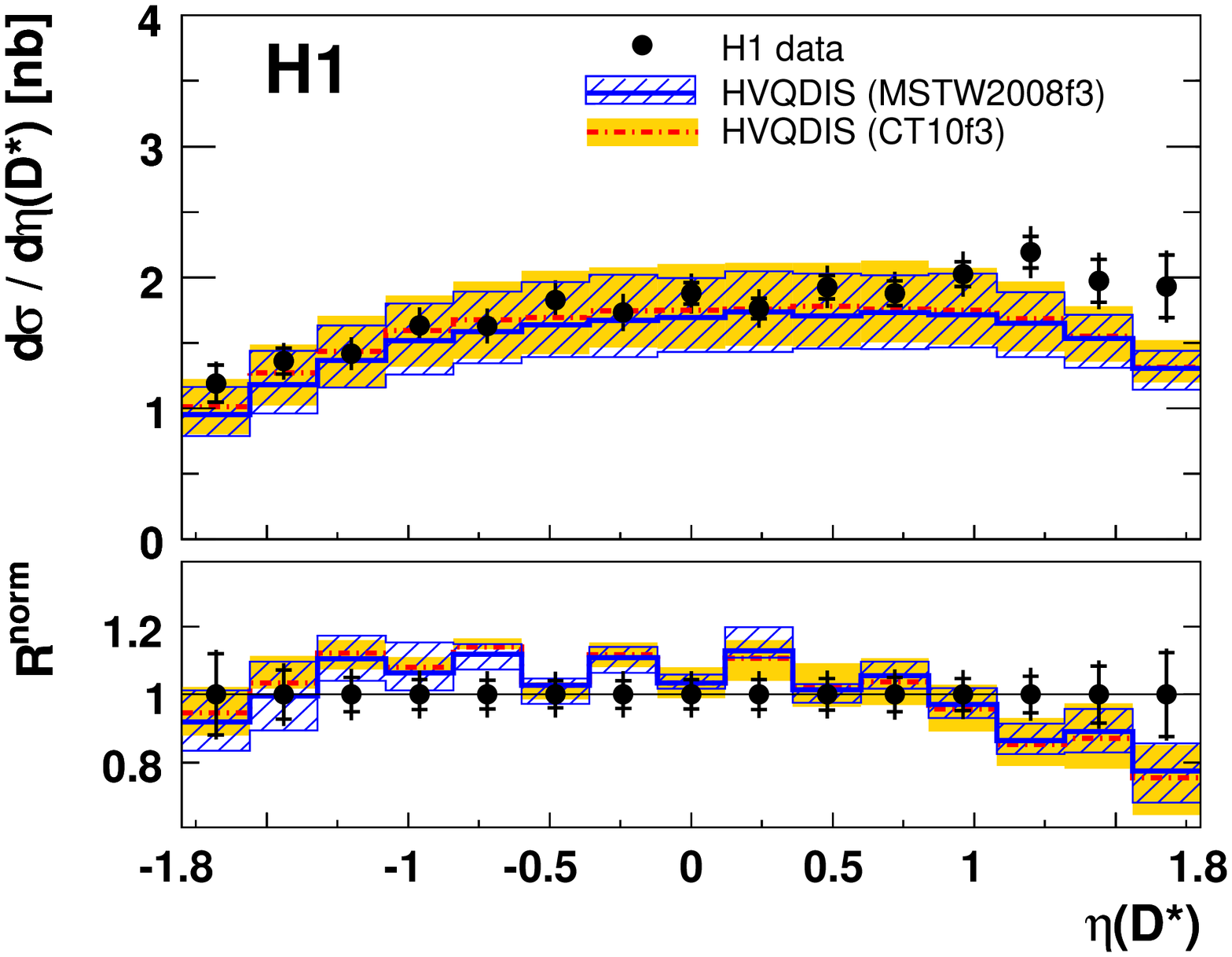,height=6cm}
\caption{Measurements of cross-sections ${\rm d}\sigma/{\rm d}p_T^{D^*}$ and ${\rm d}\sigma/{\rm d}\eta^{D^*}$ for $D^*$ 
production in DIS compared with NLO QCD predictions using two different proton PDFs.  The lower  
plot for each variable shows the ratio of theory to data where each is first normalised 
to its corresponding total cross section.  From~\cite{Aaron:2011gp}.}
\label{fig:dstar-dis}
\end{center}
\end{figure}

The description by NLO QCD of charm production in DIS is generally good, 
as shown in Fig.~\ref{fig:dstar-dis}.  
The QCD prediction, 
``{\sc Hvqdis}''~\cite{Harris:1997zq,Harris:1995tu,Harris:1995pr}, 
is performed in the massive scheme,
which means that it should be most reliable at low 
values of $p_T$.  However, the theory describes the $p_T^{D^*}$ distribution up to 20\,GeV 
and over three orders of magnitude in the cross section.  
Taking the correlated theoretical uncertainties into account, there 
is also a good description of the 
$\eta^{D^*}$ distribution
in both shape and magnitude.  The good description in 
Fig.~\ref{fig:dstar-dis} shows that the 
dynamics of NLO QCD (along with the non-perturbative fragmentation inputs, 
given in Section~\ref{sec:frag}) 
can describe 
charm production over a wide kinematic range.  
Importantly, it also encourages an extrapolation 
using this NLO QCD prediction to the full 
phase-space in $p_T^{D^*}$ and $\eta^{D^*}$ in order 
to give a determination of the charm contribution, $F_2^{c\bar{c}}$, 
to the inclusive proton structure function.  The 
validity of QCD calculations outside the measured region is 
unknown, but at least some confidence is 
given by a good description of the data within the measured region (see 
Section~\ref{sec:alphas}).

Precise measurements of charm photoproduction have been made by reconstructing 
$D^*$ mesons with no explicit 
jet requirements~\cite{Aktas:2006ry,Breitweg:1998yt}.  
The details of the cross sections are not well reproduced by NLO QCD
calculations, 
with notable problems at low values of $p_T^{D^*}$ and in
reproducing the shape of the cross section 
as a function of 
the pseudorapidity of the $D^*$ meson.
However, at the low scales 
measured, $p_T^{D^*} \sim 2$\,GeV, the 
uncertainties from theory are large, typically 
50\%.  More information, 
such as constraints on the proton and photon PDFs, could be 
extracted from the data with improved predictions.  However, 
these and other such 
data are more commonly used to extract information on the 
fragmentation of heavy quarks (see 
Section~\ref{sec:frag}).

\begin{figure}[htp]
\begin{center}
~\epsfig{file=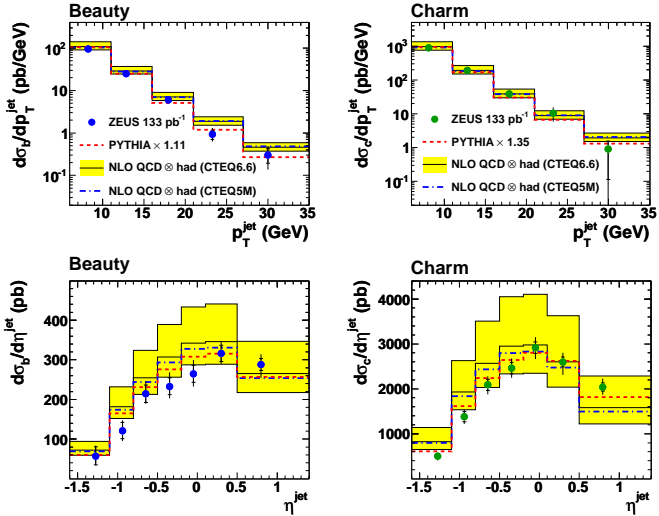,height=6.9cm,bbllx=0,bblly=30,bburx=555,bbury=460,clip=}
\caption{Beauty (left) and charm (right) jet photoproduction cross-sections 
${\rm d}\sigma/{\rm d}p_T^{\rm jet}$ (upper) and ${\rm d}\sigma/{\rm d}\eta^{\rm jet}$ (lower) compared to 
{\sc Pythia} Monte Carlo and NLO QCD predictions.  From~\cite{Abramowicz:2011pz}.}
\label{fig:hqjets}
\end{center}
\end{figure}

Measurements of heavy flavour photoproduction have also 
been made in which the heavy quark 
is part of a jet.  Such measurements are generally 
performed at high transverse energy and, as jets 
are reconstructed, the measurements are usually integrated over the 
fragmentation fraction $z$, with correspondingly reduced 
sensitivity to the details of the hadronisation.
Such measurements thus
potentially offer 
a more precise comparison between data and fixed-order QCD than is the 
case for 
measurements
without jet requirements.  Example results are shown in 
Fig.~\ref{fig:hqjets}. 
The descriptions of both the measured charm and beauty cross sections  
are reasonable, although the uncertainty on the theory is still large.  
Historically this level of agreement 
was not always the case 
for beauty production where at the turn of the millennium, data from the 
Tevatron~\cite{Abe:1993sj,Abe:1993hr,Abe:1995dv,Abe:1995mj,Acosta:2001rz,Abachi:1994kj,Abbott:1999se,Abbott:1999wu,Abbott:2000iv} and the first measurement from HERA~\cite{Adloff:1999nr} were in strong disagreement 
with QCD.  Through improved measurements, presenting results at the hadron rather than quark level 
and better theory (updated fragmentation functions and resummed calculations), this discrepancy 
was resolved and the need for new physics as was postulated at the time, 
rendered unnecessary.
See~\cite{Cacciari:2004ur} for a more complete discussion.

\begin{figure}[htp]
\begin{center}
~\epsfig{file=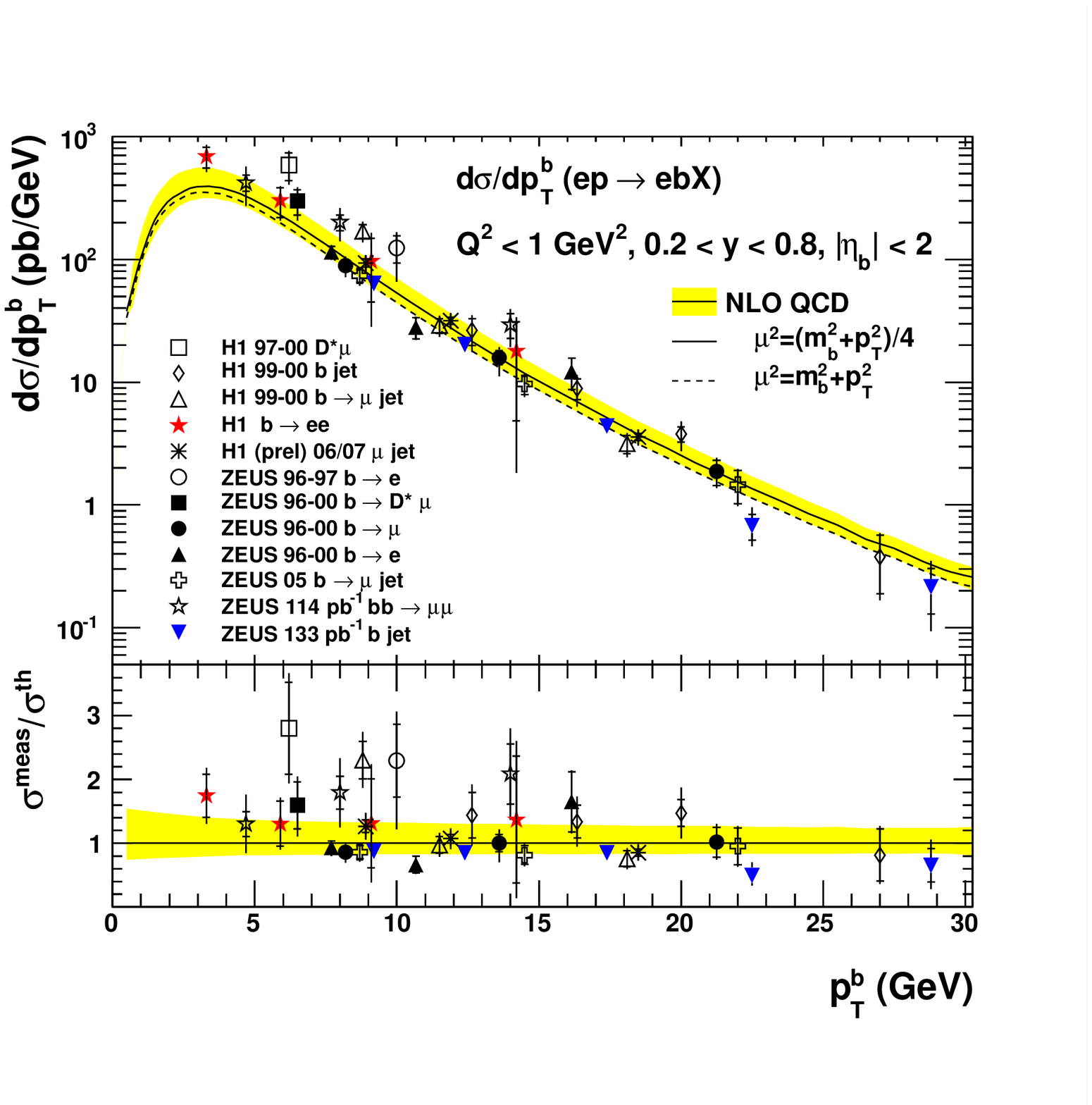,height=8.cm,bbllx=50,bblly=50,bburx=515,bbury=500,clip=}
\caption{Summary of beauty photoproduction measurements at HERA as a function of the 
transverse momentum of the $b$ quark compared with predictions from NLO QCD.  The 
measurements, made using different final states and in different kinematic regions, 
were corrected for branching ratios and extrapolated to the specified kinematic region 
at the quark level for this comparison. }
\label{fig:b-sum}
\end{center}
\end{figure}

Since the first HERA results on beauty production, many channels covering a wide kinematic range 
have been measured in both DIS and photoproduction~\cite{Breitweg:2000nz,Chekanov:2003si,Aktas:2005bt,Aktas:2005zc,Aktas:2006vs,Aaron:2009ut,Aaron:2010ib,Aaron:2012cj,Chekanov:2006sg,Chekanov:2008aaa,Chekanov:2008zz,Chekanov:2008tx,Chekanov:2009kj,Abramowicz:2010zq,Abramowicz:2011kj,Aaron:2012md}.  The  
photoproduction results are summarised in Fig.~\ref{fig:b-sum} and compared with predictions from NLO QCD.  
This figure shows that beauty production at HERA is well described with particularly the most recent and 
most precise data in very good agreement with theory, 
confirming the ability of QCD to describe 
heavy quark production.  

\subsection{Extracting Information From the Data} 
\label{sec:alphas}

As is motivated earlier in this 
article and can be seen from Eqs.~\ref{eq:DISpert}--\ref{eq:pertxsec}, 
the measurements presented in this
Section are sensitive to the structure of the proton and photon 
and to the value of the strong coupling constant.  
Therefore as well as comparing QCD predictions with  
the data, NLO QCD calculations can 
be used to extract the parton densities or $\alpha_s$.  

The measurement of the structure of the proton and 
the extraction of its parton densities is covered in detailed reviews 
elsewhere~\cite{Klein:2008di,Perez:2012um}, 
with the HERA measurements of inclusive DIS having 
provided the strongest constraints  
throughout most of the kinematic range. 
Final-state 
measurements have also been used to constrain the structure of the proton as the processes involved 
are directly sensitive at lowest order to the gluon distribution in the proton (see Fig.~\ref{fig:DIS-feyn}(a)).  
By using photoproduction jet data, high energy scales can 
also be accessed.  Photoproduction jet data 
were included in an NLO QCD fit to ZEUS data only~\cite{Chekanov:2005nn} and found to reduce the 
uncertainty on the gluon distribution by a factor of two at medium to high $x$ ($\ge 0.01$). 

\begin{figure}[htp]
\begin{center}
~\epsfig{file=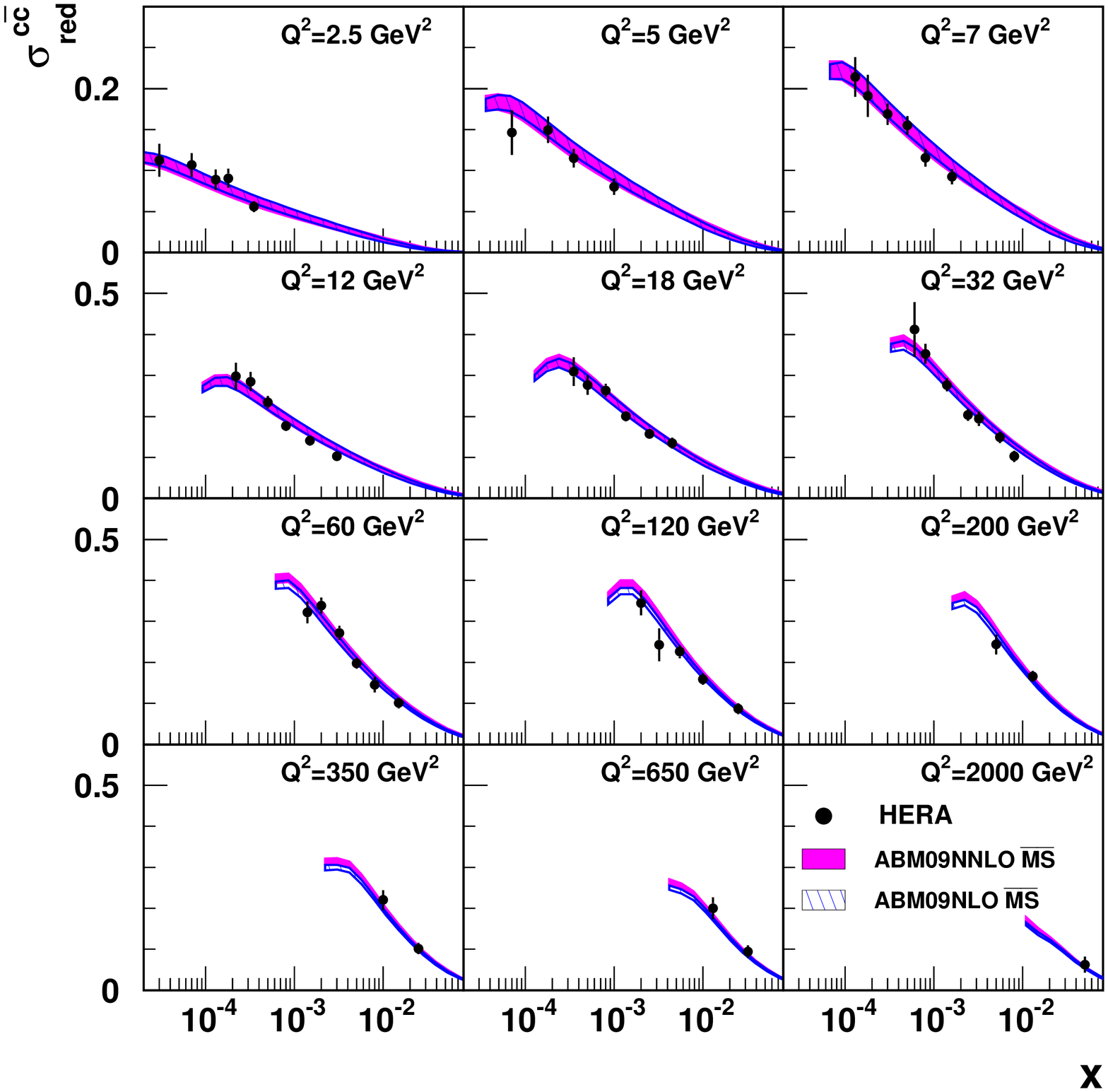,height=8.7cm,bbllx=15,bblly=10,bburx=715,bbury=670,clip=}
\caption{Measurement of the reduced charm cross section at fixed $Q^2$ as a function of $x$,
having combined all available published H1 and ZEUS data.  
The data are compared with a parametrisation of the proton PDF from 
the ABM~\cite{Alekhin:2012ig} group at NLO and NNLO in QCD.  From~\cite{Abramowicz:1900rp}.}
\label{fig:f2cc}
\end{center}
\end{figure}

The H1 and ZEUS measurements of charm production in DIS such as those in Fig.~\ref{fig:dstar-dis} have 
recently been combined~\cite{Abramowicz:1900rp}, accounting for correlations in the uncertainties 
and thereby leading to significantly increased precision.  
In Fig.~\ref{fig:f2cc}, the data are presented as a reduced cross section, 
$\sigma_{\rm red}^{c\bar{c}}$, corresponding for the non-extreme $y$ values considered here to 
the charm contribution to the proton structure function, $F_2^{c\bar{c}}$. 
The data are
compared with predictions based on parametrisations of the parton distribution functions in the 
proton at NLO and NNLO in QCD.  These 
data provide extra constraints on the structure of the proton and the mechanism for charm production.  
In addition, the data are sensitive to the mass of the charm quark~\cite{Alekhin:2010sv,Abramowicz:1900rp} and could 
be used in the context of global fits to extract
this quantity.  
Similar measurements from both collaborations have also been made 
for beauty production, but with much smaller statistics. 

As can be seen in Eq.~\ref{eq:pertxsec} 
and discussed in Section~\ref{sec:jet-production}, 
HERA data on jet photoproduction in particular 
are sensitive to the structure of the photon as well as the proton.  The HERA data complement the measurements 
of $F_2^\gamma$ from $e^+e^-$ colliders in that the HERA data are sensitive at LO to the gluon density 
in the photon, which is poorly constrained 
from $e^+e^-$ data.  Also, the measurements from HERA probe higher 
scales ($\langle \bar{E}_T^2 \rangle \sim 4\,000$\,GeV$^2$) than was accessible at LEP 
(\mbox{$\langle Q^2 \rangle = 780$\,GeV$^2$}~\cite{Abbiendi:2002te}).  There are also far more data 
which are sensitive to the 
heavy flavour structure of the photon from HERA than were 
obtained at LEP.

A number of challenges exist in understanding and using the HERA 
data in fits to the structure of 
the photon.  To achieve as large a data sample as possible, 
the minimum jet transverse energy 
must be as small as possible.  
A further advantage of including the
low transverse energy region is the access to 
correspondingly low values of $x_\gamma$, 
the region where the gluon density is expected to dominate the 
photon structure.  
However, typical cut values are at least 5\,GeV,
dictated by requirements in the 
trigger and the need to maintain a good correlation 
between the directions of the reconstructed jets 
and the partons which seed them. 
Furthermore, at these low values of transverse energy, the 
underlying event and hadronisation uncertainties  
prevent a precise comparison between data 
and QCD predictions. For example,   
in a measurement of inclusive-jet photoproduction 
by H1~\cite{Adloff:2003nr}, the
theoretical uncertainties 
in predicting 
cross sections close to the cut at 5\,GeV on the jet 
transverse energy, were too large to allow discrimination 
between different proton and photon 
PDFs. This measurement and the underlying 
event are discussed in detail in Section~\ref{sec:ue}.  
Jet cross sections 
can thus only be used reliably to constrain photon 
structure at high transverse energy, where the effects of the 
underlying event and hadronisation are expected 
to be minimised, at the expense of reduced statistics and reduced sensitivity to the gluon density at low 
$x_\gamma$.  Figure~\ref{fig:et-gammap} is an example of such a 
measurement where two jets are required 
with transverse energies above 20 and 15\,GeV.

A combined fit of $e^+e^-$ data and HERA dijet data has been performed in order to extract 
the parton densities in the photon~\cite{Slominski:2005bw}.  The dijet data with high 
transverse energies, greater than 14\,GeV, were not well described in the fit.
Part of the 
problem is that the data are more sensitive to the gluon density in the proton than that 
in the photon.  However, as was seen in Fig.~\ref{fig:et-gammap}, there is a significant 
difference between different parametrisations of the photon PDFs, suggesting that there 
is some flexibility which can be explored in order to achieve a better description of the 
data and more stringent constraints on the photon PDFs.  The data also exhibit a strong need 
for higher order calculations (see also Fig.~\ref{fig:DeltaPhi}) and with such programmes, 
more significant constraints on the photon could be made.

\begin{figure}[htp]
\begin{center}
~\epsfig{file=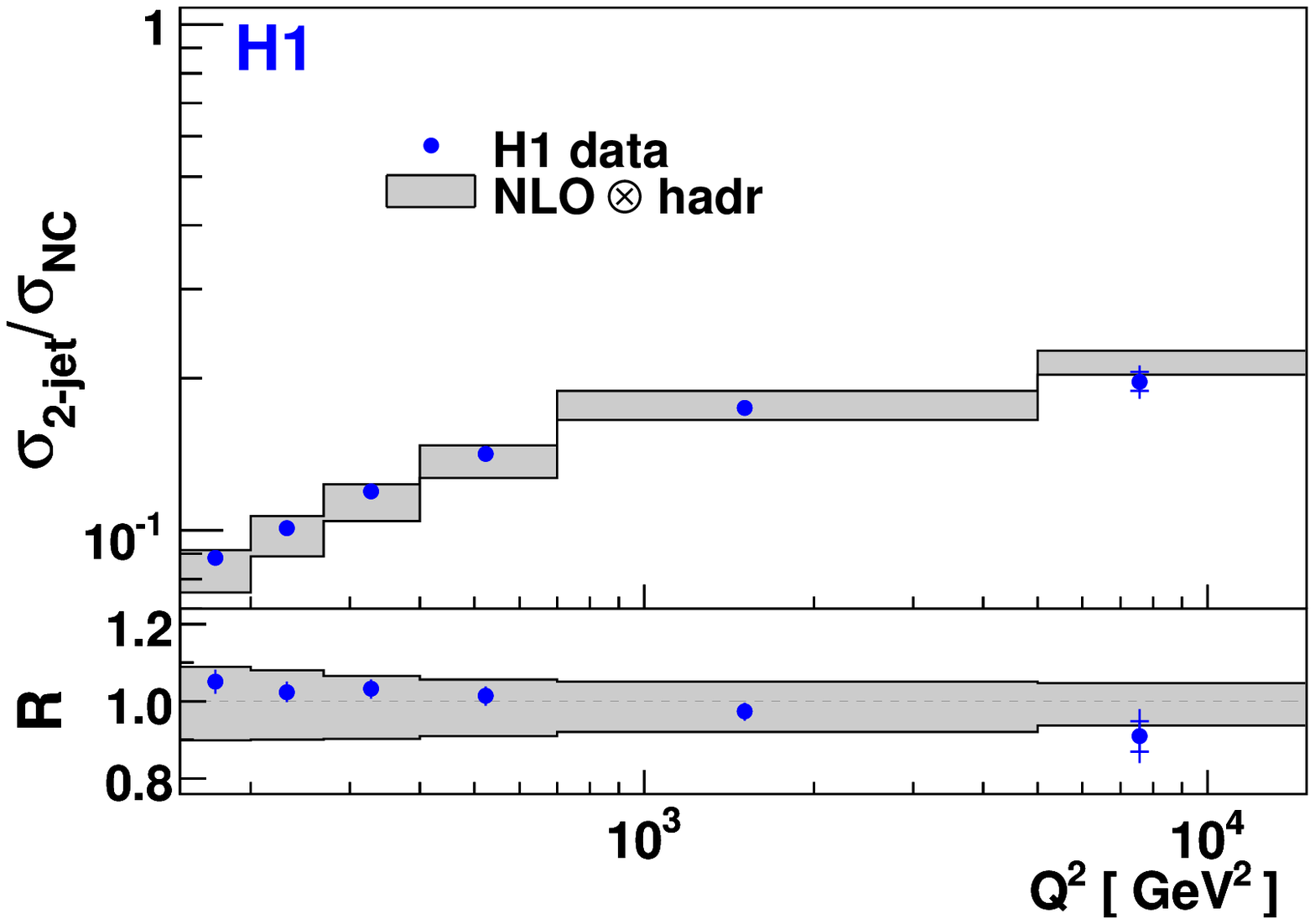,width=8.3cm}
\put(-21,148){\makebox(0,0)[tl]{(a)}}\\
\hspace{-0.3cm}~\epsfig{file=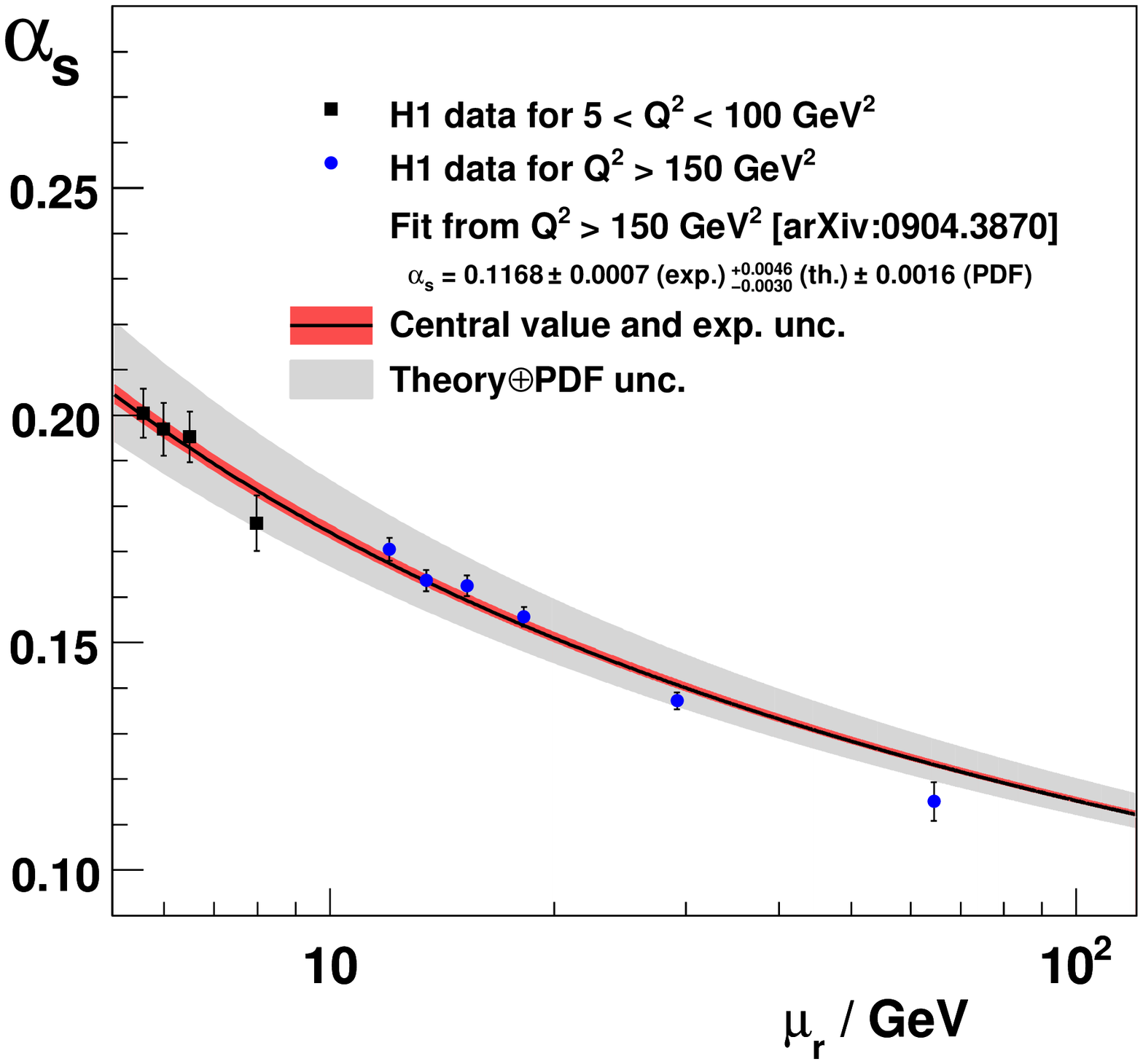,width=8.5cm}
\put(-21,183){\makebox(0,0)[tl]{(b)}}
\caption{(a) Measurement of the ratio of the dijet to the total cross section in DIS,  
compared with NLO QCD predictions (corrected for hadronisation).  (b) Values of 
$\alpha_s$ versus the scale of the process extracted from distributions such as that 
in (a).  The solid line shows the two-loop solution of the renormalisation group 
equation.  
(a) From~\cite{Aaron:2009vs} and (b) from~\cite{Aaron:2009he}.}
\label{fig:alphas}
\end{center}
\end{figure}

The strong coupling constant, $\alpha_s$, has been extracted at HERA in many different processes and 
in a wide kinematic range, thereby providing precise results which clearly display the running 
with the energy scale of the process.  The value of $\alpha_s$ has been extracted using 
inclusive DIS cross sections, event-shape variables (although these receive significant
contributions from non-perturbative processes, see Section~\ref{sec:event-shapes}), jet 
cross sections, and ratios of rates of different processes such as the dijet cross section normalised to  
the total DIS cross section or the three- to two-jet rate.  The ratio of the dijet to the inclusive DIS cross section 
is shown versus the scale, $Q^2$, and compared with an NLO QCD prediction in Fig.~\ref{fig:alphas}(a).  Such a 
ratio reduces some systematic uncertainties which are correlated between the measurements, such as 
the uncertainty on the luminosity measurement.
The resulting precise data are well described by the theoretical prediction.  
By varying $\alpha_s$ in the theory and fitting to the data, a value can be obtained in bins of the scale $Q^2$ 
and the running of the coupling constant with energy scale demonstrated as in Fig.~\ref{fig:alphas}(b).  A clear 
variation of $\alpha_s$ with $Q^2$ is observed and the variation is well described by the two-loop solution 
of the renormalisation group equation.

\begin{figure}[htp]
\begin{center}
~\epsfig{file=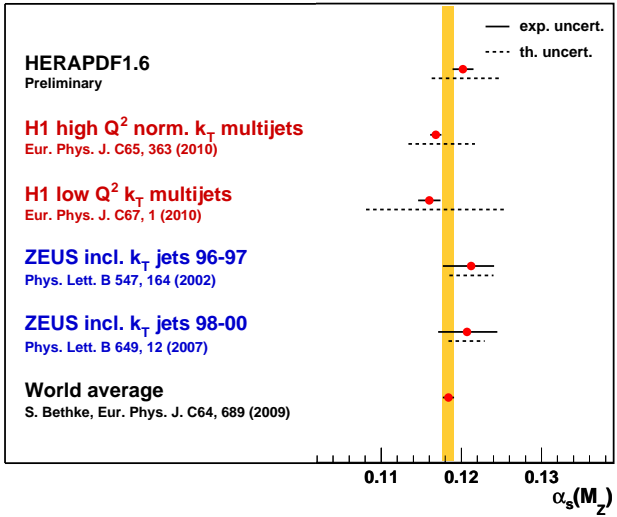,height=7.5cm}\\
\caption{Values of $\alpha_s$ from individual H1 and ZEUS measurements and 
the world average.  Note the extraction 
in the HERAPDF1.6 fit is a `preliminary' result.}
\label{fig:alphas-av}
\end{center}
\end{figure}

A collection of recent determinations of $\alpha_s$, presented 
(as is conventional) at the scale of the $Z$ boson mass, is 
shown in Fig.~\ref{fig:alphas-av}.  The two ZEUS results were extracted 
at high energy scales in order to minimise 
the theoretical uncertainties.  When using the full power 
of the data and including the largest possible 
kinematic region, as for the H1 results, the extracted $\alpha_s$ values have a 
precision comparable to the world average~\cite{Beringer:1900zz}.  
However, 
the theoretical uncertainties, arising mainly from 
missing higher orders in the calculations,
then become large. 
QCD fits to inclusive DIS data yield 
extractions of $\alpha_s$ as well as parton distribution functions.  
Including jet cross section information 
in these fits increases the sensitivity, 
due to the dependence 
on $\alpha_s$ at LO, and yields a precise 
result, as shown in Fig.~\ref{fig:alphas-av} for the 
case of the `HERAPDF1.6' fits. 
Again, the experimental precision of this result is competitive, but it suffers from 
large theoretical uncertainties.  
Development of higher order theoretical calculations and tools
is thus now the limiting factor on the precision of 
strong coupling determinations from $ep$ scattering. 
As $\alpha_s$ is
one of the fundamental parameters of the 
Standard Model and its extrapolation to very large scales
provides constraints on Grand Unification, the case for 
prioritising such calculations is strong.

\subsection{Pushing the Boundaries of Applicability}
\label{sec:boundaries}

Much of this Chapter has been concerned with measurements of jet 
production and 
other hard processes in a region 
where the data are well modelled by
NLO QCD with parton evolution governed by the DGLAP
equations. 
This is generally the case at large 
transverse momenta and in a reasonably 
central region, 
where large scales are present, the strong coupling
$\alpha_s$ is relatively small and momentum fractions $x$ 
are relatively large. 
Going beyond this region of stability requires additional
theoretical tools. 
A successful description at smaller transverse momenta
requires higher orders in the matrix elements 
and an improved understanding of hadronisation and underlying
event phenomena. 
More fundamentally, the kinematics of the non-central region at 
HERA extend into a low-$x$ regime where dynamics beyond the 
DGLAP approximation are often considered
likely to become apparent for the first time. 
These may consist
of novel parton cascade arrangements in which the transverse momentum 
ordering of 
parton emissions between the photon and the proton intrinsic to 
DGLAP is broken and alternative, resummed, schemes, going beyond
fixed order perturbative expansions in $\alpha_s$ -- notably
the BFKL equations \cite{Kuraev:1977fs,Balitsky:1978ic} -- 
are appropriate. The lack of transverse momentum ordering is 
intimately
linked to a breakdown of collinear factorisation and thus to 
the concept of an unintegrated gluon density of the proton, with
finite and variable transverse momentum, as implemented for example in 
the CCFM approach \cite{Ciafaloni:1987ur,Catani:1989yc,Catani:1989sg}. 
For hard scattering processes generating transverse momenta
$p_T > \surd Q^2$ the lack of transverse
momentum ordering has also been modelled
by ascribing a partonic
structure to the virtual photon \cite{Uematsu:1980qy,Gluck:1996ra}. 
At sufficiently low $x$,
phenomena associated with
very high parton densities \cite{Gribov:1968gs}, including 
non-linear evolution and parton `saturation' may 
become important. Whilst such effects are not particularly relevant
to the current discussion of hard processes in the inclusive final
state, they have been considered in detail in the context of 
inclusive and
diffractive DIS, where
the smallest $x$ values accessed at HERA are reached
(see Section~\ref{sec:diffraction}).

In the following,
photoproduction and DIS measurements 
extending beyond the best-understood 
region, for example to higher $\eta^{\rm jet}$, 
lower $E_T^{\rm jet}$ or in higher jet-multiplicity events,
are summarised.  
In some cases, dedicated 
observables have been constructed 
which are most likely to be sensitive to novel effects 
beyond NLO DGLAP QCD.

\subsubsection{Higher Orders}

\begin{figure}[htp]
\begin{center}
~\epsfig{file=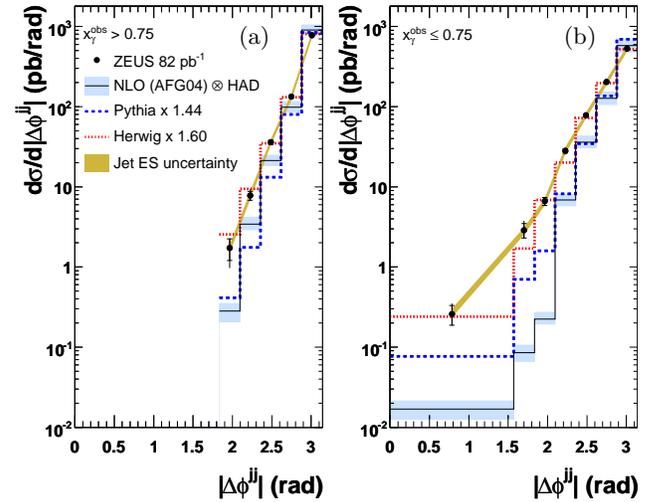,height=6.8cm,bbllx=0,bblly=0,bburx=560,bbury=460,clip=}
\put(-151,181){\makebox(0,0)[tl]{(a)}}
\put(-28,181){\makebox(0,0)[tl]{(b)}}
\caption{Measurement of the difference in azimuth between the two highest transverse energy jets 
for (a) $x_\gamma^{\rm obs} > 0.75$ and  (b) $x_\gamma^{\rm obs} \le 0.75$ compared 
with NLO QCD ($\mathcal{O}(\alpha \alpha_s^2)$, corrected for hadronisation) and Monte Carlo predictions. 
From~\cite{Chekanov:2007qt}.}
\label{fig:DeltaPhi}
\end{center}
\end{figure}

In this section,  
the inadequate description of the high-$E_T^{\rm jet}$ 
photoproduction data in 
Fig.~\ref{fig:et-gammap}(b) by NLO QCD is further investigated.
This discrepancy could be due to a need for refined photon PDFs. However, it 
may also indicate that
NLO QCD is insufficient and higher-order calculations are needed. 
A variable which is particularly 
sensitive to higher orders is the difference in azimuthal 
angle between the two jets of 
highest transverse energy, $\Delta \phi^{\rm jj}$. 
Figure~\ref{fig:DeltaPhi} shows measurements of this quantity
for high and low $\xgo$, 
compared with NLO QCD predictions 
and expectations from the Monte Carlo models, {\sc Herwig} and 
{\sc Pythia}, area-normalised for a comparison of shape.  

For $\xgo > 0.75$, the cross section falls by 
about three orders of magnitude 
over the measured range in $|\Delta \phi^{\rm jj}|$;
more steeply than for $\xgo \leq 0.75$.  At high $\xgo$, NLO QCD agrees with 
the data for the back-to-back configuration
(i.e.\ at the largest $|\Delta \phi^{\rm jj}|$), but 
it has a steeper fall off with increasing decorrelation between 
the jets. 
The prediction from the {\sc Pythia} Monte Carlo 
programme is similar to that for NLO QCD, 
whereas the prediction from the {\sc Herwig} programme 
describes the data well. 
For low $\xgo$, the NLO QCD calculation is much too 
steep and is significantly 
below the data for all values of $|\Delta \phi^{\rm jj}|$ except the highest 
bin. The prediction from the {\sc Pythia} programme is less steep, but still 
gives a poor description. The prediction from the {\sc Herwig} programme is in 
remarkable agreement with the data.

The results here 
illustrate that the parton-shower model in {\sc Herwig} gives a good 
simulation of high-order processes and suggest that matching it 
to NLO QCD (as is done in the programmes {\sc Mc\small{@}nlo}~\cite{mcatnlo1} and 
{\sc Powheg}~\cite{Nason:2004rx}, but is not yet available for these 
processes in $ep$ collisions) would 
give a good description of the data in both shape and normalisation.  
Should such a calculation or other high-order prediction, such as 
a full NNLO QCD treatment, become available, these  
distributions would be ideal tests, as they are inclusive 
quantities of high precision.  

These results 
and conclusions are qualitatively 
similar to those for the azimuthal decorrelation in high $E_T$ dijet 
photoproduction in which at least one of the 
jets is 
tagged as originating from a charm quark~\cite{Chekanov:2005zg} and in the angular 
distribution between a jet and a prompt photon~\cite{Aaron:2010uj}.  
Given that  
$\mathcal{O}(\alpha_s^2)$ calculations are at best LO in the $\Delta\phi$ 
variable, it is perhaps not surprising that the data are not 
always well described 
without NNLO QCD, or higher order, calculations.

\subsubsection{New Low $x$ Phenomena}

\begin{figure}[htp]
\begin{center}
~\epsfig{file=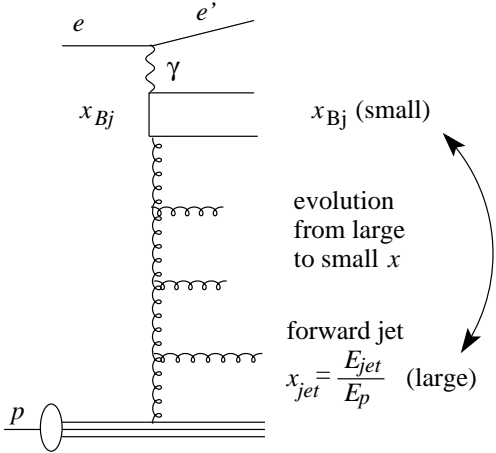,height=6cm}
\caption{Schematic illustration of $ep$ 
scattering with a forward jet taking a fraction $x_{\rm jet}$ 
of the proton momentum.  The evolution in the 
longitudinal momentum fraction, $x$, from large 
$x_{\rm jet}$ to small $x_{\rm Bj}$ is indicated.  
From~\cite{Aktas:2005up}.}
\label{fig:feyn-bfkl}
\end{center}
\end{figure}

In contrast to the case shown in Fig.~\ref{fig:DISJets} and elsewhere where  
high-$Q^2$ events containing jets of high transverse energy in the central part 
of the detector are selected, 
dedicated observables are required to enhance the sensitivity
to low $x$ phenomena. 
An example approach is to measure forward jet production
cross sections in a restricted phase space, as 
illustrated in Fig.~\ref{fig:feyn-bfkl}.
Events containing jets
in the forward direction for which 
the variable $x_{\rm jet} = E^{\rm jet}/E_p$ is much larger than 
Bjorken $x$ (denoted $x_{\rm Bj}$ here) suggest a 
gluon cascade which is strongly ordered in fractional longitudinal momentum,
as expected for BFKL-governed evolution.  
Further requiring that $(E_T^{\rm jet})^2/Q^2$ is of order unity 
restricts evolution in
transverse momentum, thus suppressing standard
DGLAP evolution. 

\begin{figure}[htp]
\begin{center}
~\epsfig{file=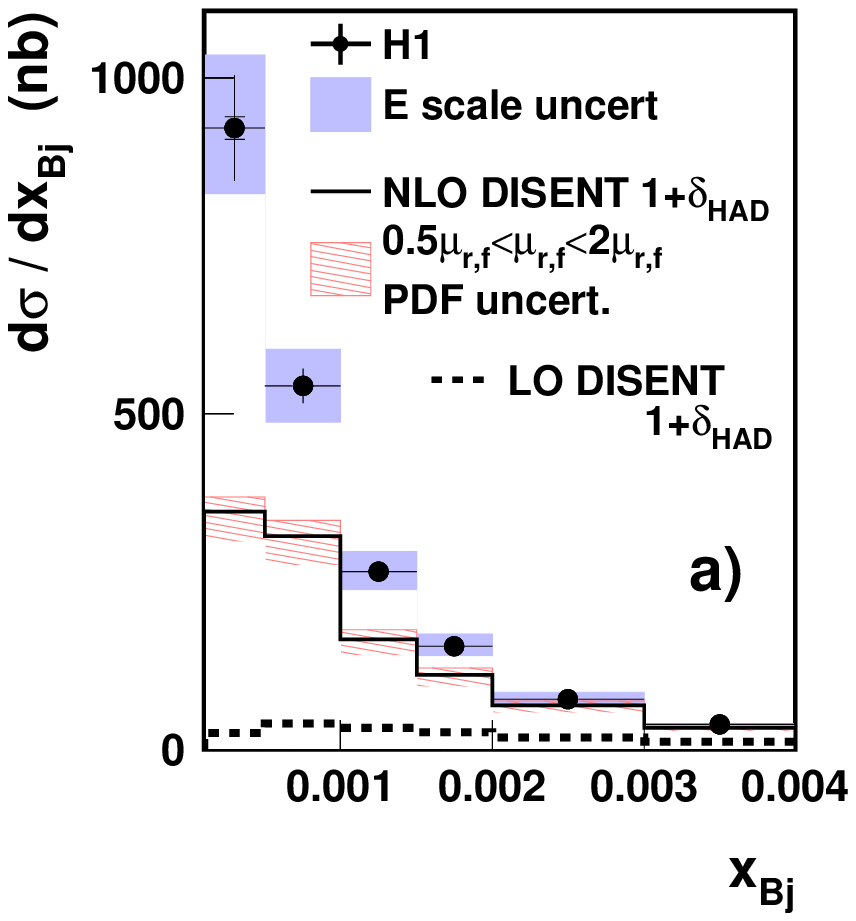,height=6cm}
~\epsfig{file=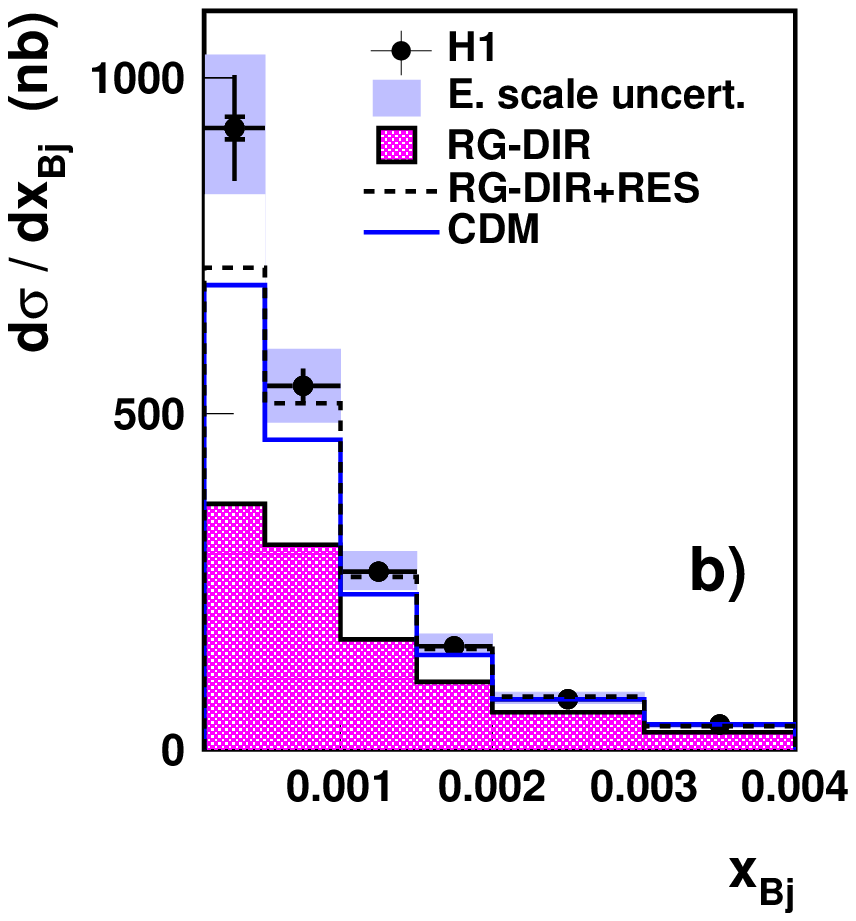,height=6cm}
~\epsfig{file=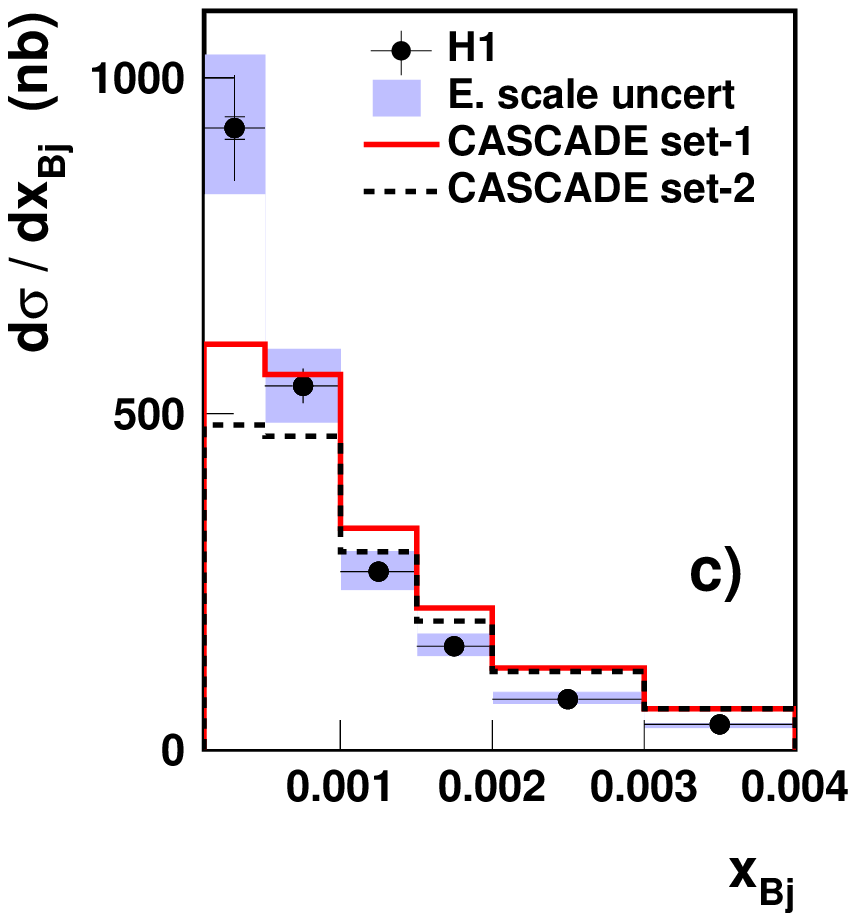,height=6cm}
\caption{Measurement of forward jet production at low $x_{\rm Bj}$ in DIS 
compared with 
(a) NLO DGLAP QCD (corrected for hadronisation) and 
(b) and (c) Monte Carlo models.  
From~\cite{Aktas:2005up}.}
\label{fig:Low-x-Jets}
\end{center}
\end{figure}

An example result using this technique is shown in 
Fig.~\ref{fig:Low-x-Jets} \cite{Aktas:2005up}. 
The events selected here are required to have low 
$Q^2$ and low $x_{\rm Bj}$, specifically $5 < Q^2 <85$\,GeV$^2$ and 
$0.0001 < x_{\rm Bj} < 0.004$, and also to contain jets of  
transverse energy $E_T^{\rm jet} >3.5$\,GeV in the laboratory frame 
in the forward part of the detector, 
$7^\circ < \theta^{\rm jet} < 20^\circ$.
The $x_{\rm jet}$ variable is required to be larger than 
$0.035$ and the scale evolution is restricted via
$0.5 < (E_T^{\rm jet})^2/Q^2 < 5$.
The same data are displayed 
in each of the three sub-figures, 
but they are 
compared with different theoretical calculations and models.  

In Fig.~\ref{fig:Low-x-Jets}(a), the data are compared with LO and 
NLO QCD, based on 
DGLAP evolution, i.e.\ the same 
calculations which described the data well in 
Fig.~\ref{fig:DISJets}.  
The prediction from LO QCD is significantly 
below the data, as expected for the analysis phase space, 
which suppress the 
LO contribution.  
The NLO QCD prediction 
is in agreement with the data at high $x_{\rm Bj}$,
but increasingly deviates from the data with decreasing $x_{\rm Bj}$.  
Notably, at 
low $x_{\rm Bj}$ the NLO QCD prediction is over a factor of 10 higher 
than the LO QCD 
prediction, suggesting that the remaining factor of two difference between 
data and NLO QCD may 
be resolved by the inclusion of NNLO QCD,
assuming that the perturbative series is quickly convergent.
The scale uncertainty band, obtained by the 
conventional method of varying 
the renormalisation scale by a factor of two,
is at the 10\% level. 
However, the order of magnitude
difference between the 
LO and NLO QCD predictions suggests that this may not be fully
representative of the uncertainties due to missing higher orders. 

To ascertain whether predictions based on BFKL 
or otherwise-modified parton evolution provide a better 
description than DGLAP-based models, these results would ideally 
be compared with QCD calculations of 
next-to-leading-logarithmic (NLL) accuracy, summing terms in
either $\ln (Q^2)$ or $\ln (1/x)$.  
Unfortunately, such calculations are not yet available and in their absence, 
Monte Carlo models which incorporate appropriate LO matrix elements 
and leading-logarithmic parton showers are usually used. 
The {\sc Rapgap} 
model~\cite{Jung:1993gf} uses LO matrix elements 
and parton showers based on standard DGLAP evolution.  
As shown in Fig.~\ref{fig:Low-x-Jets}(b)
this prediction (labelled `RG-DIR') lies below the data most 
significantly at low $x_{\rm Bj}$.
The inclusion of a resolved  
virtual photon contribution (RG-DIR+RES, see next section) significantly 
improves the description of the data.  
The colour dipole model (labelled `CDM' in the 
figure)~\cite{Andersson:1988gp,Lonnblad:1994wk}, uses LO matrix elements with parton 
emissions generated by spanning 
colour dipoles between the partons in place of the usual
leading-logarithmic transverse momentum-ordered parton showers.  As the 
dipoles in the CDM radiate independently, there is 
no ordering of the emissions in transverse momentum and hence this 
approach shares a similar characteristic 
with BFKL evolution.  
The description of the data by this model is considerably 
better than that by RG-DIR. 
Given the large point-to-point correlated uncertainties 
in the data and the unquantified
uncertainties in the predictions, the CDM model is in fair agreement 
with the data.  

An alternative model which incorporates BFKL-like characteristics is 
{\sc Cascade}~\cite{Jung:2000hk,Jung:2001hx}.
This model in fact uses the CCFM 
equation~\cite{Ciafaloni:1987ur,Catani:1989yc,Catani:1989sg,Marchesini:1994wr}, which provides a bridge 
between the DGLAP and BFKL descriptions by resumming 
both $\ln(Q^2)$ and $\ln(1/x)$ terms, 
resulting in ordering by emission angle in the parton cascade.
Comparisons of the {\sc Cascade} model 
with the data are shown in Fig.~\ref{fig:Low-x-Jets}(c) for 
two different parametrisations 
of the unintegrated gluon density of the proton.  
The comparison with data is again better than that 
of NLO DGLAP QCD or RG-DIR
at low $x_{\rm Bj}$, though it still falls short of the 
measured cross section at the lowest value and 
there are also substantial discrepancies at high $x_{\rm Bj}$.

\begin{figure}[htp]
\begin{center}
~\epsfig{file=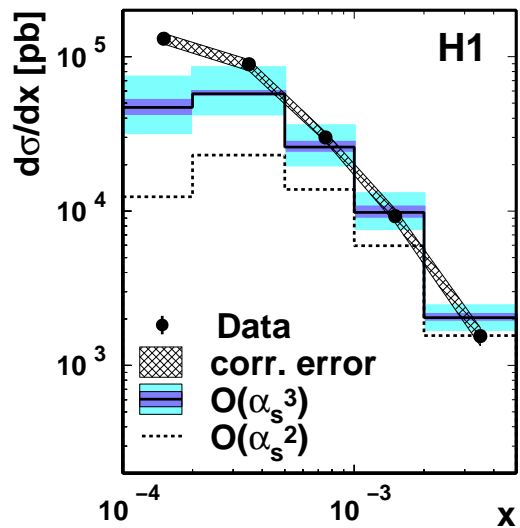,height=7cm}
\caption{Measurement of the cross section as a function of the 
Bjorken scaling variable, $x$, for events in which 
two jets are reconstructed in the forward direction and one central.  
The data are compared with 
second- and third-order QCD calculations.  From~\cite{Aaron:2007xx}.}
\label{fig:Low-x-3Jets}
\end{center}
\end{figure}

Further measurements have been made of related observables
to that shown in Fig.~\ref{fig:Low-x-Jets}.
A similar analysis has been performed~\cite{Aaron:2007xx} 
of events which contain a jet in the central region of the 
detector and two jets in the forward region.  
The data are shown as a function of the Bjorken scaling variable, 
$x$, in Fig.~\ref{fig:Low-x-3Jets} and are compared with 
QCD calculations to different orders in 
$\alpha_s$.  
The 
curves labelled $\mathcal{O}(\alpha_s^2)$ and $\mathcal{O}(\alpha_s^3)$ 
represent LO and 
NLO QCD predictions for this three-jet process, respectively.  
Therefore, again due to kinematic 
restrictions, it is not surprising that 
the LO prediction fails to describe the data.
The NLO prediction is considerably better, but still 
lies below the data at the lowest $x$, indicative of the need for 
higher-order calculations or dynamics beyond the DGLAP approximation.

Other related measurements include
replacing the forward jets by 
forward $\pi^0$ mesons~\cite{Aktas:2004rb,Adloff:1999zx}, which is 
complementary in that it 
probes lower values of $x_{\rm Bj}$, but at the expense of  
reduced signal purity. 
Measurements of azimuthal decorrelations between 
jets at low $x$ \cite{Aktas:2003ja} and between a forward jet and the scattered
lepton \cite{Aaron:2011ef} have also been made.
As in the examples above, 
common themes in these analyses include tendencies for NLO DGLAP QCD
predictions to lie below the data at the lowest $x$ and for the CDM 
model to provide improved descriptions. 

\subsubsection{Virtual Photon Structure}

As discussed in 
Section~\ref{sec:reco-php-dis},
the distinction 
between DIS and photoproduction is somewhat arbitrary.  Although 
no component due to the structure of the 
photon is usually included in calculations of DIS processes, 
such a component may still be relevant where an event contains
a hard process with a scale such as a jet transverse momentum which
is larger than that provided by the virtuality of the photon. 
Under these circumstances, the photon may be considered as being the
object whose structure is being probed. 

\begin{figure}[htp]
\begin{center}
~\epsfig{file=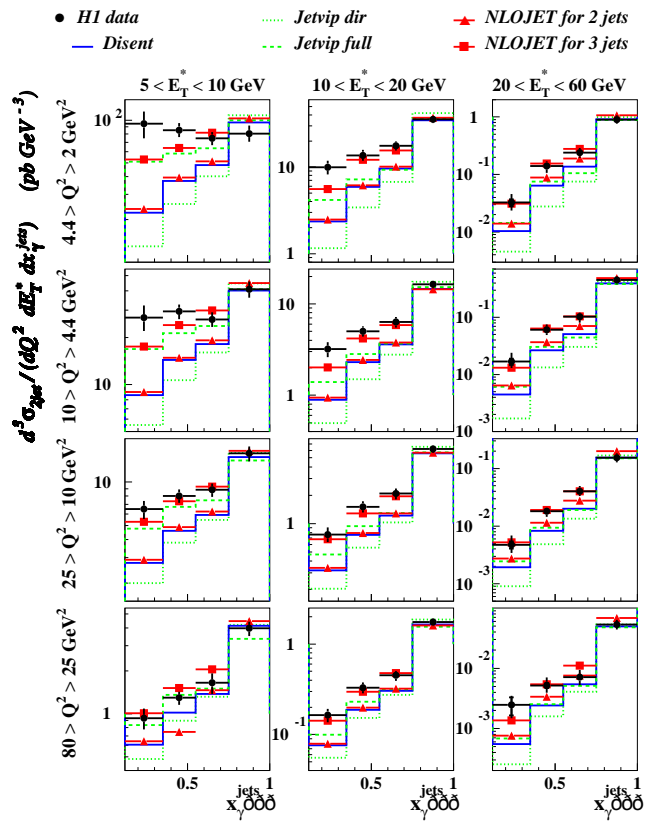,height=11cm}
\caption{Measurement of the dijet cross section in the $\gamma^*p$ frame in DIS as a function 
of $x_\gamma^{\rm jets}$ in different regions of $Q^2$ and $E_T^*$.  The data~\cite{Aktas:2004px} 
are compared with three NLO QCD calculations in DIS for $2 \to 2$ processes, `{\sc Disent}', 
`{\sc Jetvip} dir' and `{\sc Nlojet} for 2 jets', with an NLO QCD calculation for 
$2 \to 3$ processes, `{\sc Nlojet} for 3 jets', and with an NLO QCD calculation in DIS for 
$2 \to 2$ processes, including a component of resolved virtual photon 
structure, `{\sc Jetvip} full'.  All of the theory predictions are corrected for hadronisation 
effects.  From~\cite{Chyla:2005qh}.}
\label{fig:virtual-photon}
\end{center}
\end{figure}

A measurement which is sensitive to the need for a 
resolved virtual photon contribution 
is shown in Fig.~\ref{fig:virtual-photon}.  The cross section for 
dijet production in DIS 
is shown as a function of $\xgo$ (see Eq.~\ref{eq:xgo}), 
here called $x_\gamma^{\rm jets}$.  The 
data~\cite{Aktas:2004px} are shown for different regions of $Q^2$ and 
jet transverse energy $E_T^*$
in the $\gamma^*p$ frame.  
Comparisons are made with various QCD calculations.  
The predictions from {\sc Disent} and `{\sc Nlojet} for 2 
jets' are both NLO QCD predictions 
for dijet production in DIS (i.e.\ $\mathcal{O}(\alpha_s^2)$).  
Both of these calculations successfully describe high $Q^2$ data, but 
for this variable at low $Q^2$ and low $E_T^*$, they lie significantly 
below the data at  
low $x_\gamma^{\rm jets}$.  The other 
calculation shown, `{\sc Jetvip}', should in principle be ideal to 
describe these data as it has 
a component due to the resolved virtual photon which 
may be inferred from the 
difference between the `full' 
and `dir' predictions.  
Although it can be seen that the addition of this component brings 
the NLO QCD calculations significantly closer to the 
data at low $x_\gamma^{\rm jets}$, the 
`dir' component does not 
agree with {\sc Disent} or 
{\sc Nlojet} 
at the level that they agree with each other, though all three are in 
principle calculations of the same contribution.  
Due to this, the {\sc Jetvip} calculations cannot be considered to be
fully reliable. It is unlikely that these calculations will be improved
and hence 
firm conclusions on the influence of 
virtual photon structure on the description of observables such as this
in NLO QCD cannot yet be drawn. 
Alternative predictions have been made~\cite{Chyla:2005qh} using 
the {\sc Nlojet} calculation in its 
mode for calculating three-jet production (`{\sc Nlojet} for 3 jets').  
Although this is not a full NNLO QCD ($\mathcal{O}(\alpha_s^3)$) calculation 
for dijet production, as an NLO QCD calculation 
for $2 \to 3$ processes, it contains a number of the extra diagrams.  
The prediction lies significantly higher than that for 2 jets, 
particularly at low $x_\gamma^{\rm jets}$, and the description of 
the data is significantly 
improved.  

\subsubsection{The Underlying Event}
\label{sec:ue}

\begin{figure}[htp]
\begin{center}
~\epsfig{file=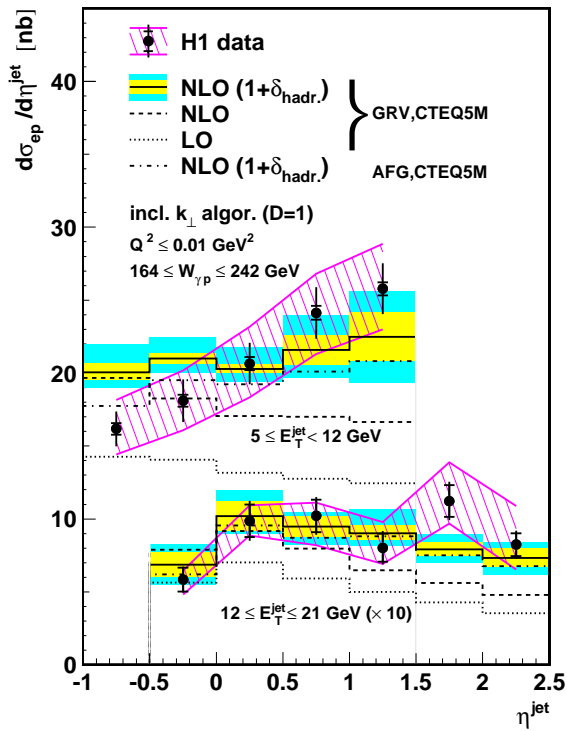,height=10cm}
\caption{Measurements of inclusive jet photoproduction
at the smallest jet transverse energies reached at 
HERA (extending to $E_T^{\rm jet} = 5 \ {\rm GeV}$).
The data are compared with NLO QCD calculations before
and after correcting for underlying event and hadronisation
effects, the combination of which is labelled $\delta_{\rm hadr}$. 
From~\cite{Adloff:2003nr}.}
\label{fig:lowetjets}
\end{center}
\end{figure}

The term `underlying event' is usually used in high transverse momentum
processes
to refer to all hadronic final state particles not originating
from the hard partonic scattering. 
Whilst in principle this includes the influence of the 
proton, and possibly the photon, beam remnants, 
the most interesting component arises from the 
possibility of multiple photon--proton scatterings taking place in 
the same event. 
As with photon structure, 
the underlying event is therefore a concept most naturally
applied to resolved photoproduction, 
but which in principle may 
still be relevant to DIS when the photon virtuality is not significantly
larger than other hard scales in a process.  
The secondary scatters generate additional hadronic energy flow in the 
event, the topology and magnitude of which are poorly understood 
theoretically, but which must be accounted for when measuring jet and 
other related
cross sections. As illustrated in Fig.~\ref{fig:lowetjets}, 
jet measurements are most
sensitive to underlying event issues at relatively low transverse
momentum, where they may 
significantly influence the shapes as well as the normalisations
of distributions. 

\begin{figure}[htp]
\begin{center}
~\epsfig{file=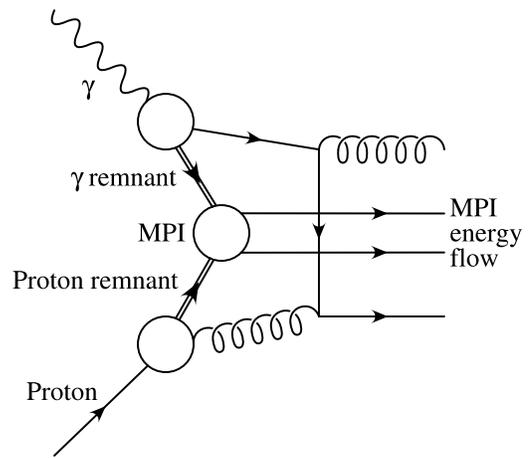,height=6cm}
\caption{A schematic representation of an event with multi-parton interactions.}
\label{fig:mpi}
\end{center}
\end{figure}

At the centre of the development of our understanding of 
the underlying event is 
the possibility of multiple hard interactions~\cite{Landshoff:1978fq}, 
in which more than one pair of partons takes part in separate 
processes that generate large transverse momenta
(see Fig.~\ref{fig:mpi}). Such processes 
are usually modelled~\cite{jimmy,Sjostrand:2000wi}
via an integral over different impact parameters between 
the two hadrons, for each of which secondary partonic scattering is generated
using proton PDFs at a suitably reduced momentum fraction, 
distributed over an appropriate spatial matter distribution.
A consequence of this impact parameter-dependent physical picture is that
the hardest primary interactions are
associated with the most `head-on' of collisions and are thus 
predicted also to exhibit high transverse momentum secondary scatterings. 
This predicted correlation encourages an experimental approach based on
events with high jet multiplicities. 

\begin{figure}[htp]
\begin{center}
~\epsfig{file=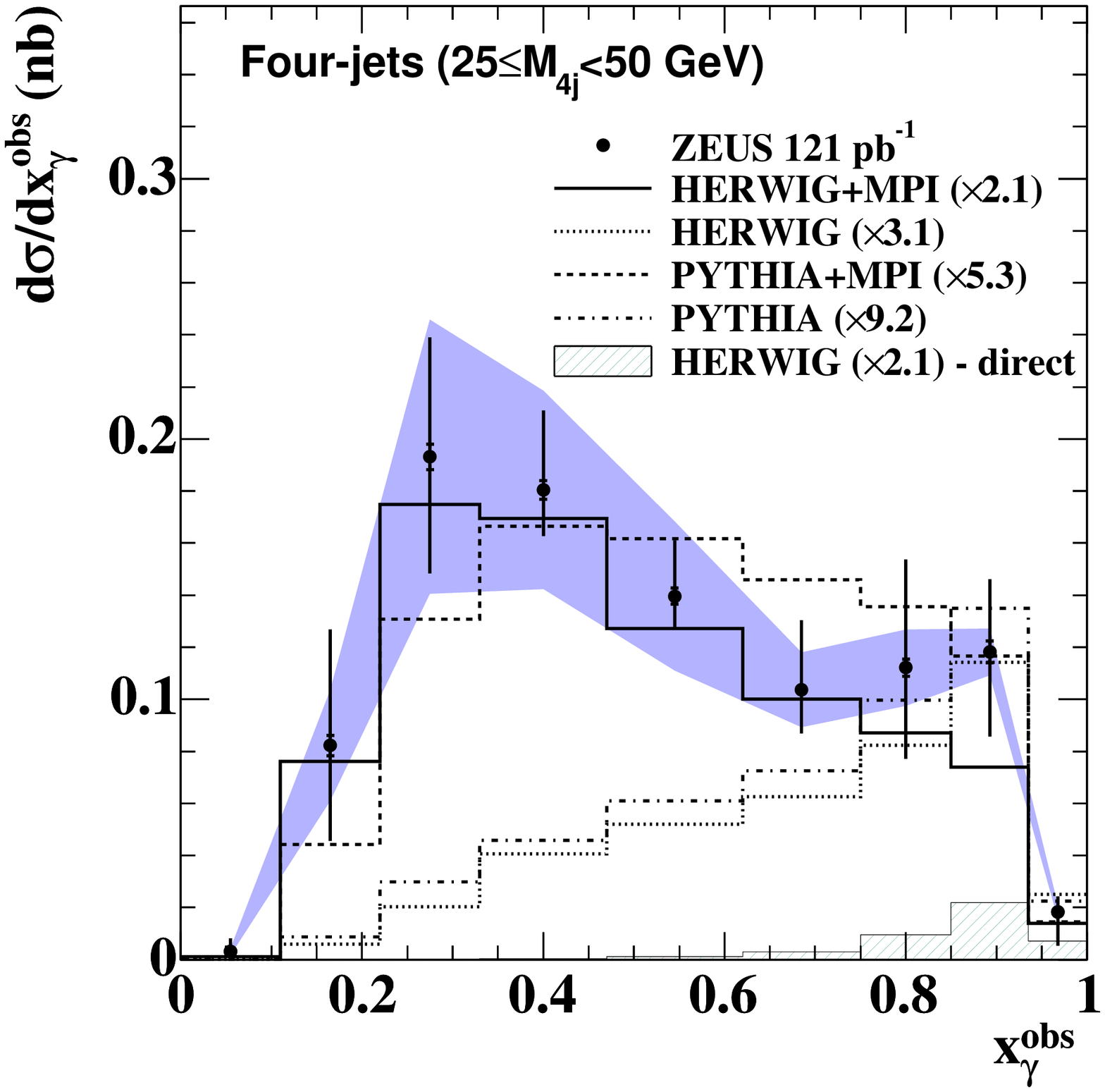,height=8cm}
\caption{Measurement of the four-jet photoproduction 
cross section as a function of $\xgo$ for 
low invariant masses of the four-jet system, 
$25 < M_{\rm 4j} < 50$\,GeV.  The predictions from the 
{\sc Herwig} and {\sc Pythia} MC models are shown both 
with~\cite{jimmy,Sjostrand:2000wi}
and without multi-parton interactions, 
as is the direct-photon component 
of the {\sc Herwig} prediction.  Each prediction is 
area-scaled to the measured high-mass ($M_{\rm 4j} > 50$\,GeV) cross 
section by the factors given in 
the legend.  From~\cite{Chekanov:2007et}.}
\label{fig:mpi-xsec}
\end{center}
\end{figure}

Multi-parton interactions have been studied at the 
Tevatron~\cite{Abe:1993rv,Abe:1997bp,Abe:1997xk,Abazov:2002mr,Acosta:2004wqa}, 
in dijet photoproduction 
events at HERA~\cite{Aid:1995ma,Adloff:2000bs,Derrick:1995bg,Breitweg:1997rx} 
and more recently at the 
LHC~\cite{Aad:2012jba,Aad:2011qe,Aad:2010fh,Khachatryan:2010pv,Chatrchyan:2011id,Chatrchyan:2012tt,Chatrchyan:2013gfi}.  In the case 
of the HERA data, the inclusion of multi-parton interactions in 
Monte Carlo models improves 
the description of the data at low transverse energy for regions 
enriched in resolved-photon 
events, as expected. Supportive of the underlying event models though
this is, it does not in itself constitute 
evidence for the existence of hard secondary scatters.  
To search for direct evidence of such processes, 
photoproduction events containing four jets with $E_T^{\rm jet} > 6$\,GeV  
have been considered.  
Events produced via multi-parton scattering exhibiting this topology 
are in principle distinguishable from four-jet events produced via 
QCD radiation in ordinary $2 \rightarrow 2$ scattering
through the lack of correlation between the angular distributions or momenta 
of the pairs of jets. 
The measured cross section for four-jet photoproduction
is shown 
as a function of 
$\xgo$~\cite{Chekanov:2007et} 
in Fig.~\ref{fig:mpi-xsec}, compared with various 
predictions from Monte Carlo models.  As with the case of 
dijet production, but even more 
significantly, adding models of multi-parton 
interactions to the predictions significantly 
improves the description of the data, with 
models not including this feature being clearly inadequate.  
However, investigation of angular 
correlations in these data did not reveal clear evidence of 
independent secondary scatters.

In summary, although some HERA data show clear 
evidence for some form of underlying event, 
the exact nature of the effect is not yet firmly established.  
Models which are based on multiple 
independent hard secondary scatters significantly improve the 
description of the data, but 
no direct evidence for this process has yet been demonstrated.  
In any case, the HERA data
will help 
to constrain future underlying event models and Monte Carlo tunes.

\subsubsection{Discussion}

The above considerations clearly indicate that an improved description of 
HERA hadronic final state data could be obtained with 
higher-order (either NNLO QCD or NLO QCD with parton 
showering) calculations.  
Tools providing such calculations for comparisons with 
LHC data have generally been successful for similar observables to 
those considered here (see e.g.~\cite{Aad:2012a}). 
All further statements beyond this central conclusion are  
less definite. There is evidence for novel parton cascade dynamics,
similar to those expected for BFKL-dominated evolution, in low $x$ 
HERA hadronic final state data. However, the extent to which such effects 
can be recovered through higher order DGLAP-based calculations
is not yet clear. In the absence of 
such calculations, models based on virtual photon structure 
are often as successful in simulating the effects as those
invoking alternative parton evolution schemes.
The data also show considerable
circumstantial evidence for multi-parton interactions.
However, in the absence of a direct observation of such processes,
it is not yet possible to rule out the possibility that
the apparent need for such effects is a consequence of 
missing higher-order QCD calculations.

\subsection{Event Shapes}
\label{sec:event-shapes}

The hadronic final states of events in DIS and $e^+e^-$ annihilation can be characterised by a number of 
variables that describe the shape of the event.  
These variables are sensitive to perturbative 
and non-perturbative QCD and given these are properties which are measurable for all events, the data 
samples are much larger than can be obtained for example by requiring high-$E_T$ jets in DIS.  
Understanding the hadronisation process is crucial to all measurements in this article and is often modelled 
phenomenologically using Monte Carlo event generators and applying 
the resulting corrections to fixed-order QCD 
calculations.  An alternative approach is to use power corrections, which have an $\mathcal{O}(1/Q)$ 
dependence, and are calculated 
analytically, extending perturbative methods into the 
non-perturbative regime.  Within this 
framework~\cite{Dokshitzer:1995zt,Dokshitzer:1995qm,Dokshitzer:1997ew,Dokshitzer:1997iz,Dokshitzer:1998qp}, 
event-shape variables depend on the strong coupling, $\alpha_s$, and an effective coupling parameter, 
$\alpha_0$, which is expected to be universal for all event shapes.

\begin{figure}[htp]
\begin{center}
~\epsfig{file=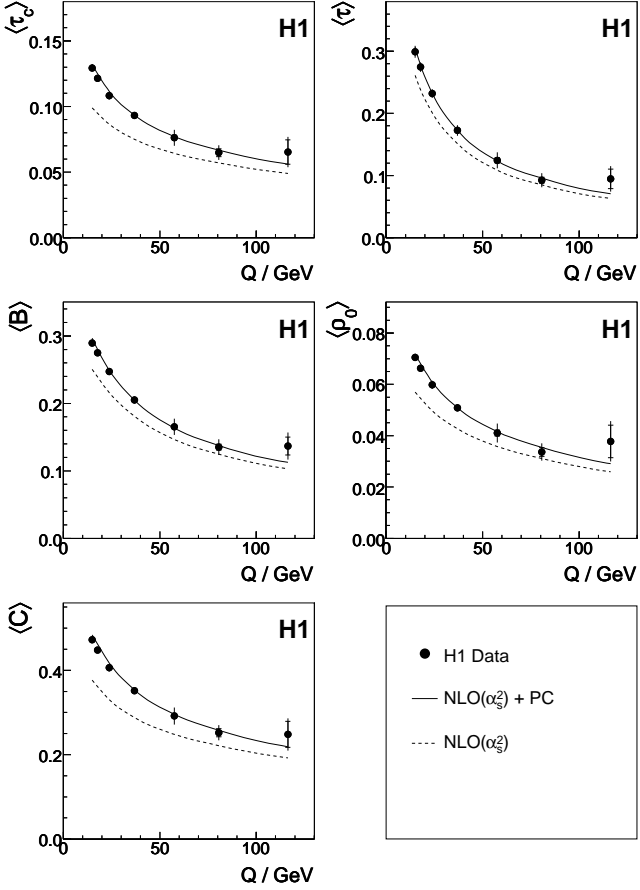,height=12cm}
\caption{Distributions of mean event shapes\,: the thrust variable, $\tau$, which measures the longitudinal momentum 
components projected onto the boson axis; the thrust variable, $\tau_C$, which maximises the sum of the longitudinal 
momenta in the current hemisphere; the broadening variable, $B$, which measures the scalar sum of transverse momenta 
with respect to the boson axis; the squared jet mass, $\rho_0$, normalised to four times the squared scalar momentum sum 
in the current hemisphere; and the $C$-parameter.  
The data, as a function of the scale, $Q$, are compared with the result of a fit based on NLO QCD with power corrections 
(solid line) as well as the contribution from NLO QCD alone (dashed line.)  
From~\cite{Aktas:2005tz}.}
\label{fig:es}
\end{center}
\end{figure}

Event shapes have been measured in $e^+e^-$ 
annihilation (for a recent review and the citations therein, see~\cite{Okorokov:2011mb}) 
and power corrections 
were found to reproduce many aspects of the hadronisation process.  The measurements discussed here 
from H1~\cite{Aktas:2005tz} and ZEUS~\cite{Chekanov:2006hv} use DIS events reconstructed in the Breit frame with a 
minimum $Q^2$ of 196 and 80\,GeV$^2$, respectively.  These most recent measurements have significantly larger 
data samples than previous publications~\cite{Adloff:1999gn,Adloff:1997gq,Chekanov:2002xk,Breitweg:1997ug} and 
therefore allow measurement of differential event-shape distributions rather than just the mean values.  The 
event shapes studied are\,:  the thrust, which measures the longitudinal collimation of the hadronic system relative to 
an appropriate axis; the broadening, which measures the complementary aspect; the jet mass; and a characteristic of 
the event known as the $C$-parameter (for the exact definitions, the reader is referred to the H1 and ZEUS publications 
and their references.)  An example of the measurements is shown in Fig.~\ref{fig:es}, where the mean event shapes 
are compared with a prediction using NLO QCD to describe the perturbative production of partons and power 
corrections to describe the hadronisation process.  The figure also shows the prediction from NLO QCD alone, 
indicating the strong need for power corrections, which, when combined with the NLO QCD prediction, describe the 
data well.  The differential distributions of all variables are similarly well described by the theory if the perturbative 
part is a combination of the NLO QCD prediction matched to a resummed prediction to NLL accuracy.

\begin{figure}[htp]
\begin{center}
~\epsfig{file=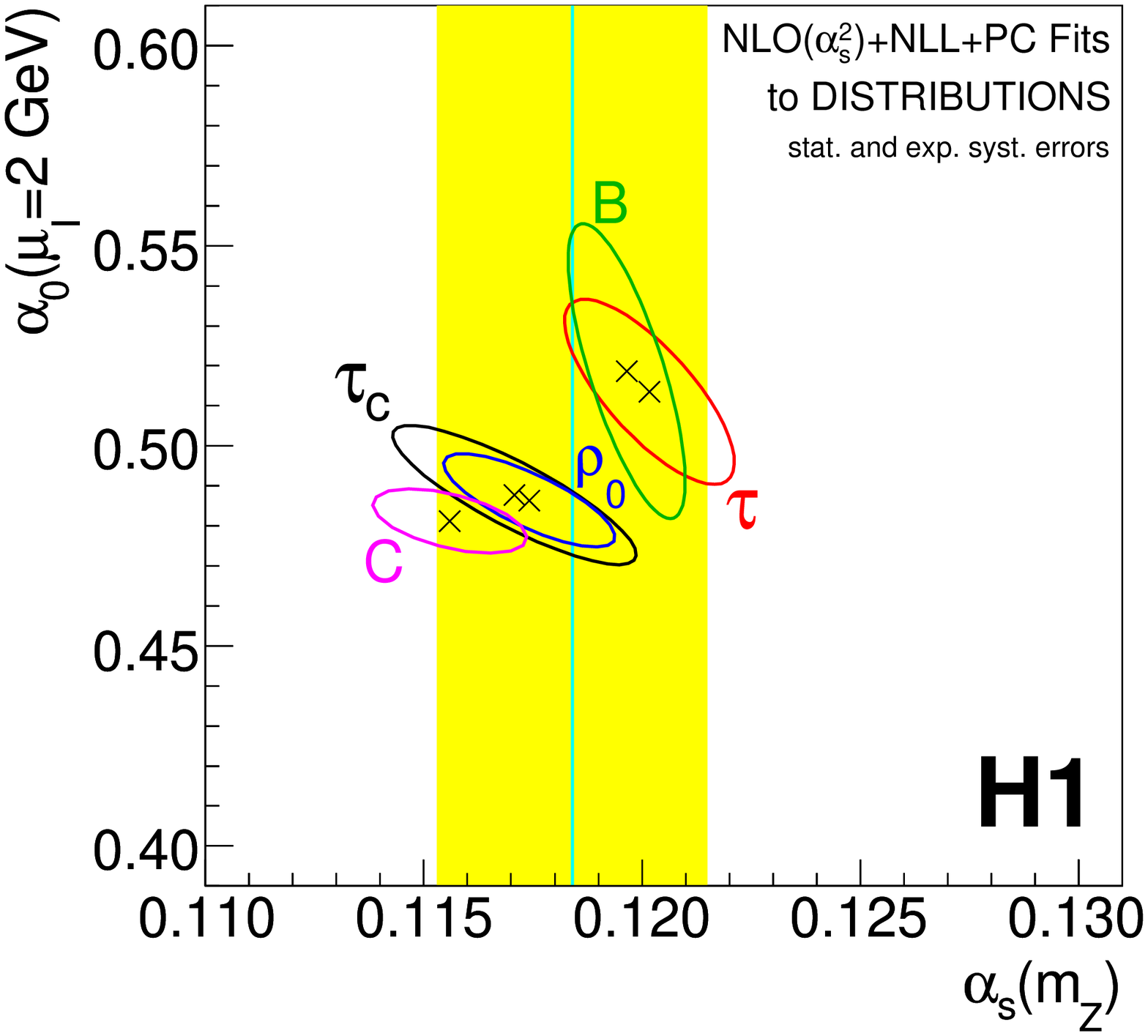,height=8cm}
\put(-40,160){\makebox(0,0)[tl]{\Large{\bf \fontfamily{phv}\selectfont (a)}}}\\
~\epsfig{file=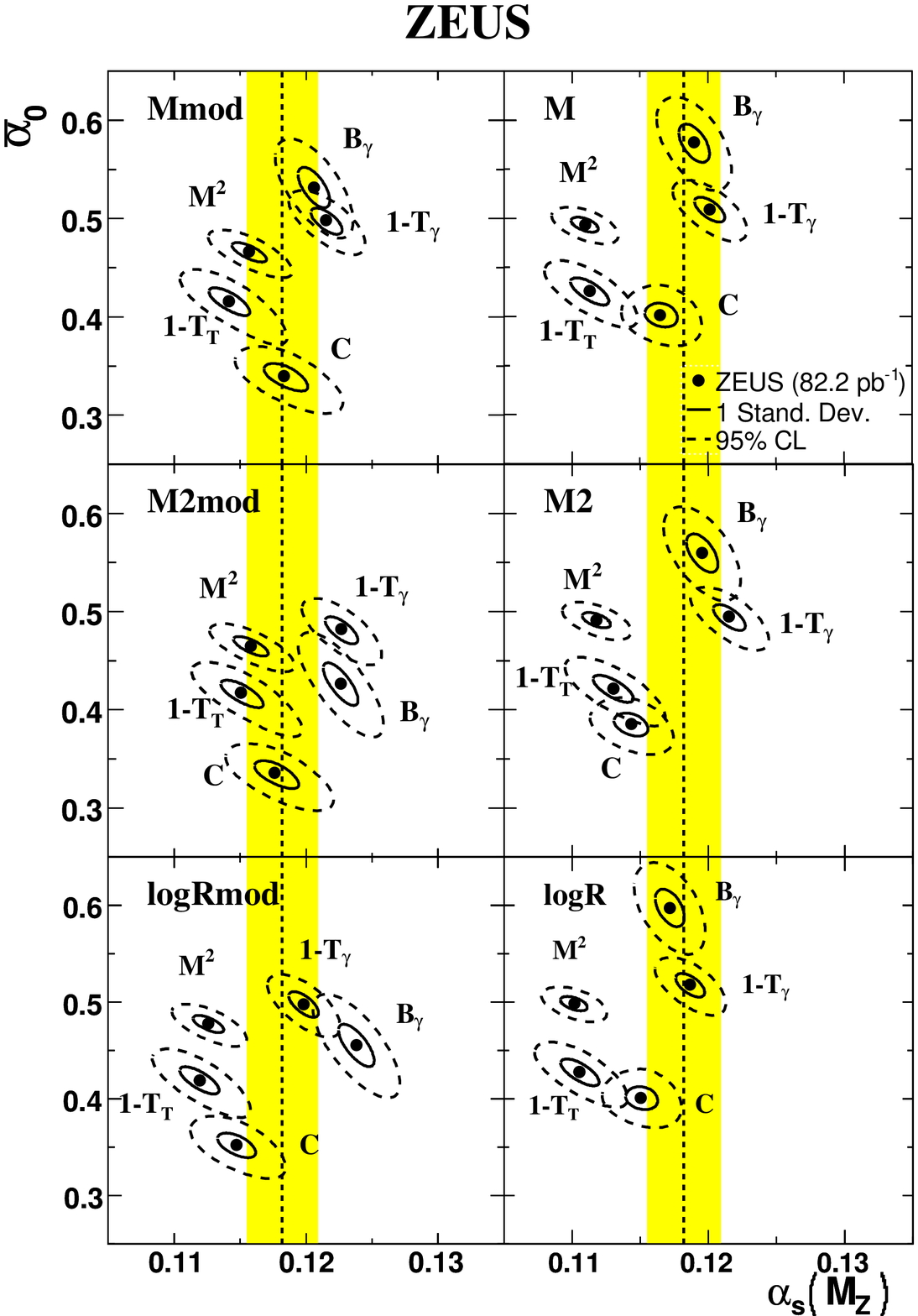,height=6.25cm,bbllx=0,bblly=475,bburx=293,
bbury=700,clip=}
\put(-40,160){\makebox(0,0)[tl]{\Large{\bf \fontfamily{phv}\selectfont (b)}}}
\put(-55,25){\makebox(0,0)[tl]{\Large{\bf \fontfamily{phv}\selectfont ZEUS}}}
\put(-160,-3){\makebox(0,0)[tl]{\Large {\bf \fontfamily{phv}\selectfont 0.11}}}
\put(-100,-3){\makebox(0,0)[tl]{\Large{\bf \fontfamily{phv}\selectfont 0.12}}}
\put(-40,-3){\makebox(0,0)[tl]{\Large{\bf \fontfamily{phv}\selectfont 0.13}}}
\put(-50,-17){\makebox(0,0)[tl]{{\Large{\boldmath${\sf \alpha_s} ({\sf M_Z})$\unboldmath}}}}
\vspace{0.5cm}
\caption{Extracted values from (a) H1 and (b) ZEUS for $\alpha_0$ and $\alpha_s$ for fits of the NLO+NLL+PC predictions 
to all event-shape variables.  The solid lines represent the 1$\sigma$ contours and for (b) the dashed line represents the 
95\% C.L. contour.  The 
world-average value of $\alpha_s$ and its uncertainty are 
shown as the bands, taken from~\cite{Bethke:2004uy}.  
(a) From~\cite{Aktas:2005tz} and (b) from~\cite{Chekanov:2006hv}.}
\label{fig:es-alphas}
\end{center}
\end{figure}

The good description of the differential distributions and their mean values 
allow the parameters of the theory, 
$\alpha_s$ and $\alpha_0$, to be extracted from the data.  The extracted values from fits to the differential distributions 
are shown in Fig.~\ref{fig:es-alphas} in the plane of the two extracted variables for all five event shapes from both 
collaborations. In the case of H1, Fig.~\ref{fig:es-alphas}(a), the values of $\alpha_s$ extracted for the five variables 
agree reasonably well with each other and with the world average.  H1 also investigated the dependence of $\alpha_s$ on the scale, $Q$, 
and observed the expected variation predicted by the renormalisation group equation.  Similarly, the values of $\alpha_0$ 
are all broadly consistent with each other.  Averages of $\alpha_s$ and $\alpha_0$ were then extracted as  
$\alpha_s (M_Z) = 0.1198 \pm 0.0013 ({\rm exp.}) ^{+0.0056}_{-0.0043} ({\rm theo.})$ and 
$\alpha_0 = 0.476 \pm 0.008 ({\rm exp.}) ^{+0.018}_{-0.059} ({\rm theo.})$.  A similar analysis from ZEUS is shown in 
Fig.~\ref{fig:es-alphas}(b), where the values of $\alpha_s$ extracted also agree with each other and the world average. The values 
of $\alpha_0$ are similar but show a larger spread, with the $C$-parameter showing the largest difference from the results of 
H1.  It should be noted that\,: the uncertainty due to variation of the renormalisation scale is significantly larger than the 
experimental uncertainty; the extraction of the parameters is sensitive to the kinematic range chosen for the fits; and the 
parameters are also sensitive to 
the details of the
matching performed to combine the NLO and NLL QCD predictions.  These issues point 
to a need for higher-order calculations and in future a unified approach to fitting the data, both from DIS and $e^+e^-$ 
annihilation.

In summary, the general description of the event-shape distributions and mean values at HERA is good, showing the applicability 
of pQCD, however with indications, as in the rest of this chapter, that orders beyond NLO QCD are needed.  The 
combination of power corrections, to describe the hadronisation process, with pQCD is also supported, although this can 
depend on the precise details of the analysis.  Other aspects and models of hadronisation are discussed in the next chapter.

\section{Non-Perturbative Aspects and Hadronisation}
\label{sec:soft}

In this section, measurements of hadron production and their 
interpretations in QCD, both perturbative and non-perturbative, are discussed.  
Charged particle distributions and Bose--Einstein correlations are 
first discussed and comparisons are made with $e^+e^-$ and other
data to test the universality of fragmentation.  
The parameters of charm fragmentation are then presented and again 
comparisons are made between different reactions.  
Inelastic $J/\psi$ results follow, together with a
discussion of 
HERA's contribution to the understanding of 
charmonium production mechanisms.  Finally, searches for more 
exotic QCD objects, such as deuterons, glueballs, 
instantons and pentaquarks are briefly reviewed. Inclusive energy
flow measurements have also been made at HERA, most recently and 
precisely in \cite{Adloff:1999ws}, but are not discussed further here. 

\subsection{Charged Particle Distributions}

Numerous measurements of charged particles have been made at HERA, investigating both the hard and soft 
aspects of QCD.  Whilst measurements of jet production probe higher scales and cover a wider kinematic range, 
the high statistics  and precise reconstruction when 
measuring individual particles allows some more detailed tests.  
The basic kinematic distributions of all 
charged particles combined
have been measured in both DIS~\cite{Alexa:2013vkv,Aaron:2007ds,Adloff:1996dy,Abt:1994ye,Derrick:1995xg,Derrick:1995jq} 
and photoproduction~\cite{Adloff:1998vt,Abt:1994dn,Breitweg:1997kc} and are in general described by Monte 
Carlo models incorporating leading-order matrix elements and parton showers followed by hadronisation.  Further 
measurements have been made in which specific particle species are tagged, e.g.\ $K_S^0$ mesons, or different 
distributions or properties are investigated, such as angular correlations.  The body of work is too great to cover in 
this short section and the reader is referred to the large number of relevant papers on the H1~\cite{h1pubwww} 
and ZEUS~\cite{zeuspubwww} paper lists.  In this section a few 
results are discussed, focusing on comparisons 
with data from other reactions and 
the general conclusions which can be drawn as a result, 
as well as comparisons with NLO 
QCD predictions using fragmentation functions
(the probability that a parton hadronises into a given hadron
which carries a fraction $z$ of the parton's momentum).

\begin{figure}[htp]
\begin{center}
~\epsfig{file=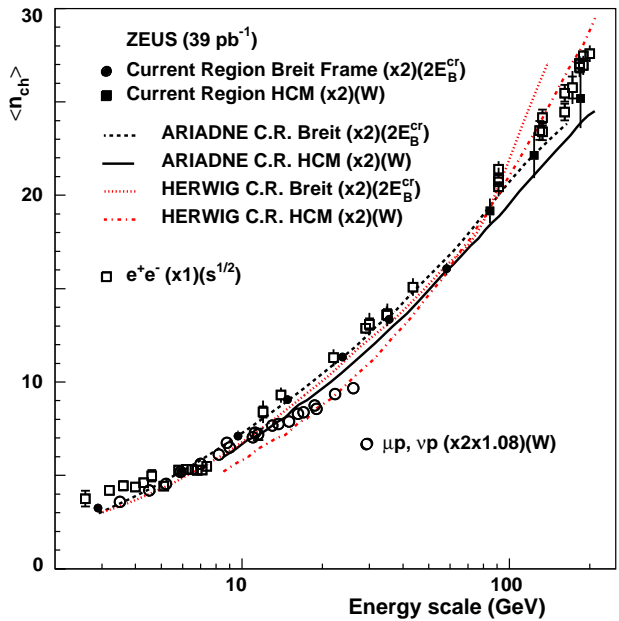,height=8.5cm}
\caption{Mean charged multiplicity, $\langle n_{\rm ch} \rangle$, as a function of the energy scale 
for ZEUS DIS
measurements in the Breit and hadronic centre-of-mass frames, compared with results from 
$e^+e^-$ and fixed-target DIS experiments.  From~\cite{Chekanov:2008ae} and references therein 
for fixed-target and $e^+e^-$ data.}
\label{fig:nch}
\end{center}
\end{figure}

Charged particle multiplicities have been measured in many experiments, at $p\bar{p}$, $pp$, $ep$ and $e^+e^-$ 
colliders~\cite{Beringer:1900zz}, testing models of 
fragmentation and the universality of the process.  In $ep$ 
DIS scattering, there are several combinations of scales and frames that can be 
used~\cite{Aaron:2007ds,Adloff:1998dw,Adloff:1997fr,Aid:1996cb,Chekanov:2008ae,Chekanov:2001sj,Breitweg:1999nt,Derrick:1995ca}.  In the results reviewed here~\cite{Chekanov:2008ae} the final state was reconstructed in the Breit 
and hadronic centre-of-mass frames in which respectively twice the energy of the current region of the Breit frame, 
$2 \cdot E_{\rm B}^{\rm cr}$, and the $\gamma^* p$ centre-of-mass energy, $W$, were used as scales.  The charged particle multiplicities 
in DIS for these different frames and their respective energy scales are shown in Fig.~\ref{fig:nch} in 
comparison with previous DIS experiments and results from $e^+e^-$ experiments.  In general  
charged particle spectra at HERA show the same trend as those from other experiments over about 
two orders of magnitude in the energy scale of the interaction.  In particular, the data from the two $ep$ 
frames with their respective energy scales agree well with the results from $e^+e^-$ experiments, in 
contrast to when the scale $Q$ is used in the Breit frame~\cite{Breitweg:1999nt}.  This results in a lower charged multiplicity at low 
values, $Q<10$\,GeV, than both the $e^+e^-$ data and the DIS data when $2 \cdot E_{\rm B}^{\rm cr}$ is used as the 
scale.  The fixed-target DIS data agree with the HERA and $e^+e^-$ data at low scales, but 
increasingly deviate above about 15\,GeV.

\begin{figure}[htp]
\begin{center}
~\epsfig{file=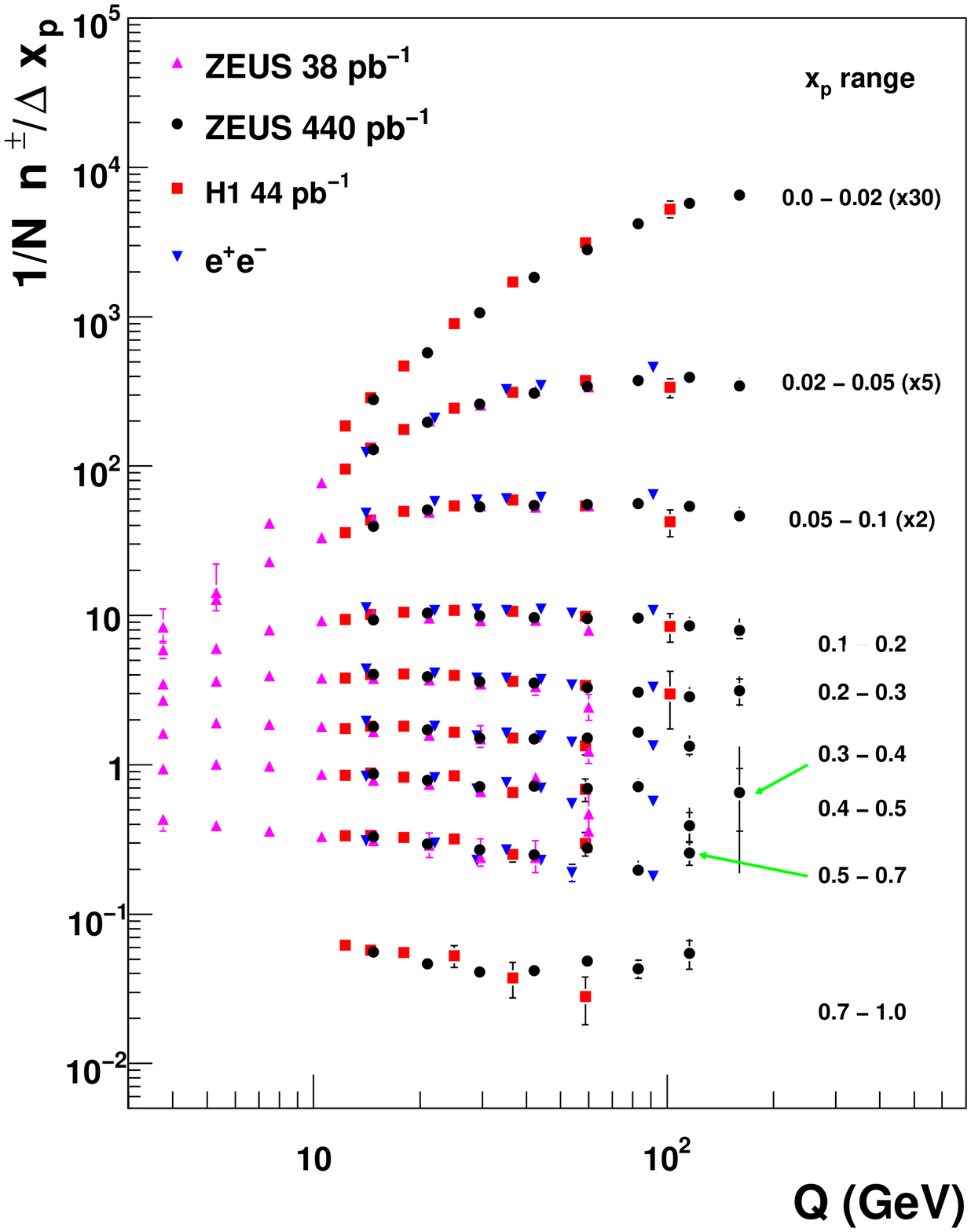,height=10cm}
\caption{The number of charged particles per event per unit $x_p$ 
as a function 
of $Q$ in bins of $x_p$.  Data are shown for H1 and ZEUS and 
are compared with those from $e^+e^-$ 
experiments~\cite{Petersen:1987bq,Braunschweig:1990yd,Li:1989sn,Abreu:1993me}.  
From~\cite{Abramowicz:2010rz}.}
\label{fig:scaled-p}
\end{center}
\end{figure}

A related quantity considered is the number of charged particles per event per unit scaled particle momentum, 
$x_p = 2 P_{\rm Breit}/Q$, where $P_{\rm Breit}$ is the momentum of a hadron in the Breit frame.  This quantity,  
measured in DIS~\cite{Abramowicz:2010rz,Aaron:2007ds,Breitweg:1997ra}, is shown in Fig.~\ref{fig:scaled-p} 
over two orders of magnitude in $Q$ in bins of $x_p$, compared with $e^+e^-$ data.  As $Q$ increases, the 
phase space for soft gluon radiation increases, leading to a rise of the number of soft particles with small $x_p$, 
which along with the decrease with $Q$ at high $x_p$ 
results in clear scaling violations.  The 
comparison between HERA and $e^+e^-$ data in Fig.~\ref{fig:scaled-p} is good for all $Q$ and $x_p$ which, 
along with the comparison in Fig.~\ref{fig:nch}, 
supports the concept of the universality of fragmentation.  
Positively and negatively charged particles 
have also been considered separately and the charge asymmetry 
measured~\cite{Aaron:2009ae} as a function of $Q$ and $x_p$.  
At large $x_p$, the observed charge 
asymmetry is found to increase with $Q$
and correspondingly with Bjorken-$x$, consistent with 
the expectation that this observable is related to the valence quark 
content of the proton.

\bigskip

\begin{figure}[htp]
\begin{center}
~\epsfig{file=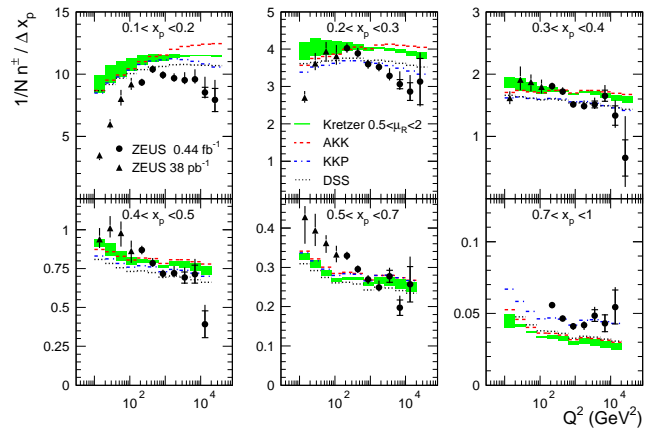,height=5.7cm}
\caption{The number of charged particles per event per unit $x_p$
as a function 
of $Q$ in bins of $x_p$.  The shaded band represents an NLO QCD calculation by Kretzer~\cite{Kretzer:2000yf} 
with its renormalisation uncertainty.  Additional calculations are shown\,:  Kniehl, Kramer and 
P\"{o}tter (KKP)~\cite{Kniehl:2000cr}; Albino, Kniehl and Kramer (AKK)~\cite{Albino:2005me,Albino:2008fy}; 
and De Florian, Sassot and Stratmann (DSS)~\cite{deFlorian:2007aj,deFlorian:2007hc}.  
From~\cite{Abramowicz:2010rz}.}
\label{fig:scaled-nlo}
\end{center}
\end{figure}

The inclusive
data are compared in Fig.~\ref{fig:scaled-nlo} with NLO QCD calculations which use fragmentation functions 
obtained from fits to $e^+e^-$ data.  The predictions from the four different groups shown are relatively similar to each other but do 
not describe the data well.  The scaling violations are poorly 
described, with the theory having a shallower rise 
for low $x_p$ than the data.  
Also, the theory predicts too many particles at low $x_p$, whereas too few are predicted at high 
$x_p$.  The disagreements between data and theory are, in general, in kinematic regions outside those measured in 
$e^+e^-$ collisions, such as low $Q^2$ or high $x_p$.  Hence the theory, 
which is derived from fits to the $e^+e^-$ data, is extrapolated 
to regions in which it may not be applicable.  Future fits should therefore use these HERA data to further 
constrain the fragmentation functions.

\subsection{Bose--Einstein Correlations}

In 1959, Goldhaber \emph{et al.}~\cite{Goldhaber:1959mj,Goldhaber:1960sf} 
observed that pairs 
of like-sign pions 
in $p\bar{p}$ collisions 
had a tendency to have smaller opening angles than pairs of unlike-sign pions.  This is interpreted as being due to the 
symmetrisation of the wave-functions of pairs of identical bosons and 
is hence known as Bose--Einstein correlations 
(BEC).  The effect has since been studied in various 
hadron--hadron collisions, $e^+e^-$ annihilation and 
lepton--nucleon scattering \cite{Alexander:2003ug}.  
The BEC in momentum space are related to the spatial 
dimensions of the production source, with measurements usually being
characterised in terms of the 
effective source size, $r$.  

Whilst measurements of $r$ clearly reflect the 
size of the interacting particles in heavy ion 
collisions \cite{Aamodt:2011mr}, the situation from proton--proton
collisions is less clear, 
with dependences on the 
event multiplicity and transverse momentum of the bosons 
observed in high energy 
experiments \cite{Aamodt:2010jj,Khachatryan:2011hi,Alexopoulos:1992iv}.
Conclusions from BEC studies are to some extent obscured 
by the fact that the
extracted values of $r$ are often strongly 
dependent on the choice of control sample relative to which 
the signal is measured. Values obtained using
unlike-sign control samples are up to a factor of two larger than those
obtained relative to event-mixed control samples. 

In $ep$ scattering, 
the source size might be expected to reflect the size of the 
proton in low energy-scale processes and to evolve to smaller sizes
more similar to those from $e^+ e^-$ data
as the energy scale increases and the electron scatters
from a single parton rather than from the proton as a whole.
It is therefore interesting  
to study whether there is any dependence of the 
effective source size on the kinematic variables, in particular $Q^2$, since 
this reflects the effective 
transverse size of the exchanged virtual photon.  
The HERA case is therefore ideally suited to investigating
whether
$r$ depends on the primary 
interaction or whether it is determined solely by the 
fragmentation stage of the process.

Measurements of BEC at HERA have been performed 
for various different samples and kinematic regions in DIS, 
spanning $0.1 < Q^2 < 8\,000$\,GeV$^2$.  H1 and ZEUS 
have both measured the effect using an inclusive sample 
of hadrons~\cite{Adloff:1997ea,Chekanov:2003gf}.
H1 additionally studied a diffractive DIS event sample~\cite{Adloff:1997ea},
in order to investigate whether a difference between 
the production mechanisms of diffractive and
non-diffractive processes could be established. 
Motivated by a reported dependence of $r$ on the mass of the 
interfering bosons in $e^+e^-$ annihilation 
($r_{\pi \pi} > r_{KK} > r_{\Lambda \Lambda}$~\cite{Alexander:1999rm}),
ZEUS also measured the effect for samples of charged and neutral 
kaons~\cite{Chekanov:2007ad}.  

Selected effective source size measurements from DIS and $e^+ e^-$ collisions
are collected in 
Table~\ref{table-bec}.
No dependence of the source size on the kinematic 
variables, including $Q^2$, was found at HERA. 
There is also no significant variation between the results 
obtained in diffractive and inclusive DIS.
The results for charged and neutral kaons are consistent with one
another and also with those from inclusive charged particle production.  
Furthermore, the results from HERA are 
consistent with lower-energy DIS 
experiments~\cite{Arneodo:1986bza,Korotkov:1993my}.
HERA and the other DIS data 
are also compatible with measurements from $e^+e^-$ 
annihilation~\cite{Choi:1995xb,Althoff:1986wn,Juricic:1988zd,Alexander:1999rm}. 
From these observations, it can be concluded that,
within the DIS regime, Bose--Einstein 
interference in $ep$ scattering does not depend significantly on the 
details of the hard process. The similarity with data from 
$e^+e^-$ annihilation further indicates that the presence of an
incoming hadron is not important if a short distance hard interaction
takes place. The effective source size is thus driven 
primarily by the fragmentation process in DIS. 

\begin{table}[hbt]
  \begin{center}
  \begin{tabular}{|c|c|c|} \hline 
   Process   & Experiment & $r$ (fm)  \\ \hline \hline
             & AMY       &  $0.73 \pm 0.05 \pm 0.20$ \\ \cline{2-3}
   $e^+ e^-$  & TASSO  &  $0.82 \pm 0.06 \pm 0.04$ \\ \cline{2-3}
              & MARK II &  $0.75 \pm 0.03 \pm 0.04$ \\ \cline{2-3}
             & LEP  &  $0.78 \pm 0.01 \pm 0.16$ \\ \hline \hline
   Previous  & EMC       &  $0.84 \pm 0.03$ \\ \cline{2-3}
   DIS       & BBCNC  &  $0.80 \pm 0.04$ \\ \hline \hline
             & H1 &  $0.68 \pm 0.04 ^{+0.02}_{-0.05}$ \\ \cline{2-3}
             & ZEUS &  $0.666 \pm 0.009 ^{+0.022}_{-0.036} $ \\ \cline{2-3} 
   HERA     & H1 (diffractive) & $0.59 \pm 0.13 ^{+0.05}_{-0.05}$ \\ \cline{2-3}
      & ZEUS ($K^\pm K^\pm$)     &  $0.57 \pm 0.09 ^{+0.15}_{-0.08} $ \\ \cline{2-3} 
            & ZEUS ($K^0_S K^0_S$)   &  $0.63 \pm 0.09 ^{+0.11}_{-0.08} $ \\ \hline 
  \end{tabular}
\end{center}
\caption[]
{Table of source sizes, $r$, extracted from hadron production in $e^+e^-$ annihilation 
and in lepton--nucleon scattering.  The uncertainties given for EMC and BBCNC are 
statistical only; the systematic uncertainties are expected to be of a similar size.  
All other results have the statistical and systematic uncertainties shown separately.  The 
control samples used are the unlike-sign samples except for the LEP average and ZEUS 
kaon measurements.  The LEP value is the average of using the unlike-sign and event-mixing 
samples as the control with the second error reflecting half the difference between the two 
and the value when using the unlike-sign sample is the higher of the two $r$ values.  For the 
kaon pairs, an unlike-sign sample is not usable due to the strong signal from 
$\phi \to K^+K^-$. 
}
\label{table-bec}
\end{table}

\subsection{Charm Fragmentation}
\label{sec:frag}

Predictions of observable heavy quark cross sections rely on a knowledge of  
fragmentation parameters.  This is 
particularly true when measuring the production rate of a given e.g.\ charm meson where both the fraction of charm 
quarks hadronising to the given meson and the charm fragmentation function are required inputs in a QCD 
calculation.  Due to the large charm cross section at HERA, precise measurements of charm fragmentation 
parameters have been performed. 
The parameters can be compared with results from $e^+e^-$ data, thereby testing the 
universality of the charm fragmentation process.  As well as the ground state charm mesons, excited charm mesons 
(e.g.\ $D_1$ and $D_2^*$) have also been reconstructed and fragmentation parameters extracted.  These are not 
discussed further here and the reader is referred to the relevant publications~\cite{Chekanov:2008ac,Abramowicz:2012ys}.

\begin{figure}[htp]
\begin{center}
~\epsfig{file=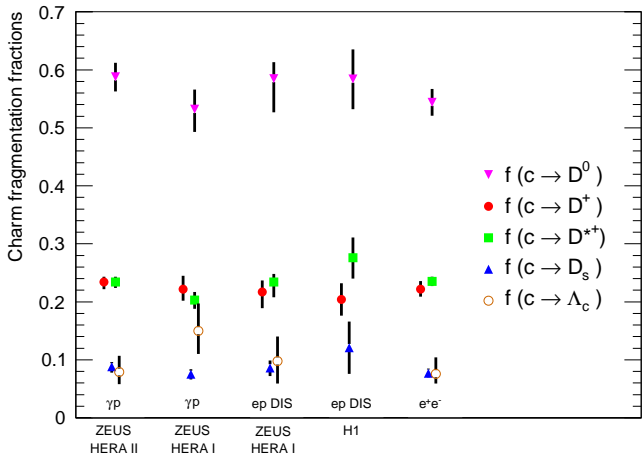,height=6.25cm}
\caption{Measurements of fragmentation 
fractions for various charm hadrons in DIS and photoproduction 
at HERA and also in $e^+e^-$ annihilation at LEP.  From~\cite{Abramowicz:2013eja}}
\label{fig:frag-frac}
\end{center}
\end{figure}

The probability that a charm quark hadronises into a given hadron, the fragmentation fraction, has been determined at HERA 
for all ground state charm hadrons, both in DIS~\cite{Aktas:2004ka,Chekanov:2007ch,Abramowicz:2010aa} and in 
photoproduction~\cite{Chekanov:2005mm,Abramowicz:2013eja}.  The results are shown in Fig.~\ref{fig:frag-frac} along 
with a combined result 
from $e^+e^-$ annihilation~\cite{Gladilin:1999pj,Lohrmann:2011np}.  The results for a given hadron are consistent between the three processes 
and, particularly in the photoproduction regime, the HERA results are of competitive precision to the results from $e^+e^-$ data.

\vspace{0.8cm}

As the values of the fragmentation fractions are consistent with being independent of the process, all measurements have been 
combined~\cite{Lohrmann:2011np}.  The results of this analysis represent our best knowledge of these values and should 
be used in future calculations of charm production.

\begin{figure}[htp]
\begin{center}
~\epsfig{file=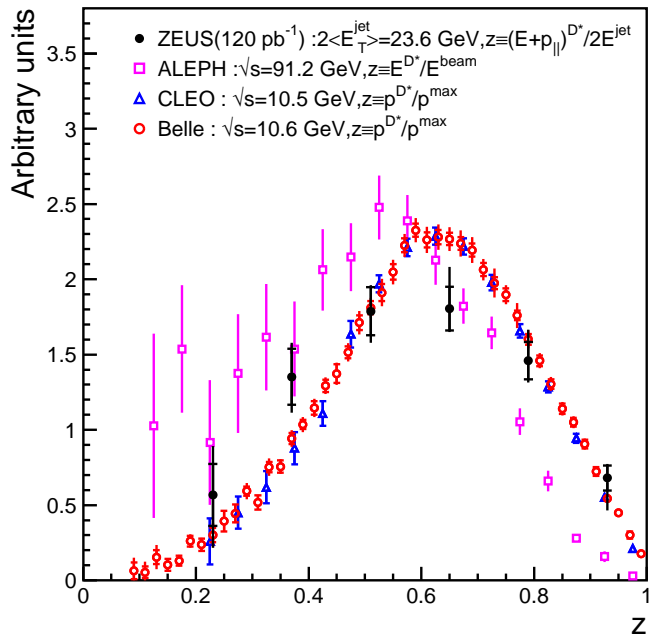,height=8.5cm}
\caption{$D^*$ fragmentation function 
$f(z)$
for ZEUS data compared to measurements in $e^+e^-$ collisions.  To compare 
the shapes, the data sets were normalised to $1/({\rm bin\ width})$ for $z>0.3$.  From~\cite{Chekanov:2008ur}.}
\label{fig:frag-fun-ee}
\end{center}
\end{figure}

The so-called fragmentation function
of a given hadron is the 
distribution in the 
fractional transfer $z$ of a quark's energy to the 
hadron.  This has been measured 
for the $D^*$ meson in DIS by H1~\cite{Aaron:2008ac} and in 
photoproduction by ZEUS~\cite{Chekanov:2008ur}.  Both sets of results showed a strong sensitivity to models of 
fragmentation with parameters in a given model varied and best-fit values extracted.  The measurement from ZEUS tagged 
$D^*$ mesons associated to 
a jet; the relatively high scale of the process, $\sqrt{\hat{s}}$, given by $2 \cdot \langle E_T^{\rm jet} \rangle = 23.6$\,GeV, allowed a 
measurement of the fragmentation function which was
relatively unbiased 
by the kinematic cuts~\cite{Chekanov:2008ur}.  The ZEUS data are shown 
in Fig.~\ref{fig:frag-fun-ee} compared to results from $e^+e^-$ collisions.  
The scales of the process for Belle and CLEO 
are similar, about 10.5\,GeV, whereas for ALEPH 
the scale is 91.2\,GeV.  
It can be seen that the ZEUS data, which are intermediate 
in scale, lie between the 
Belle/CLEO and ALEPH results, which is qualitatively consistent with expectations from scaling violations in QCD in which 
$\langle z \rangle$ decreases with increasing energy~\cite{Cacciari:2005uk}.  Similarly, the $z$ distributions from both 
collaborations were used to extract the parameter $\epsilon$ in the Peterson fragmentation function~\cite{Peterson:1982ak} 
as implemented in the {\sc Pythia} Monte Carlo model and the outcomes were found to agree with the results when fitting the $e^+e^-$ data.  These 
results are consistent with the hypothesis of fragmentation universality 
between $ep$ and $e^+e^-$ processes.

\begin{figure}[htp]
\begin{center}
~\epsfig{file=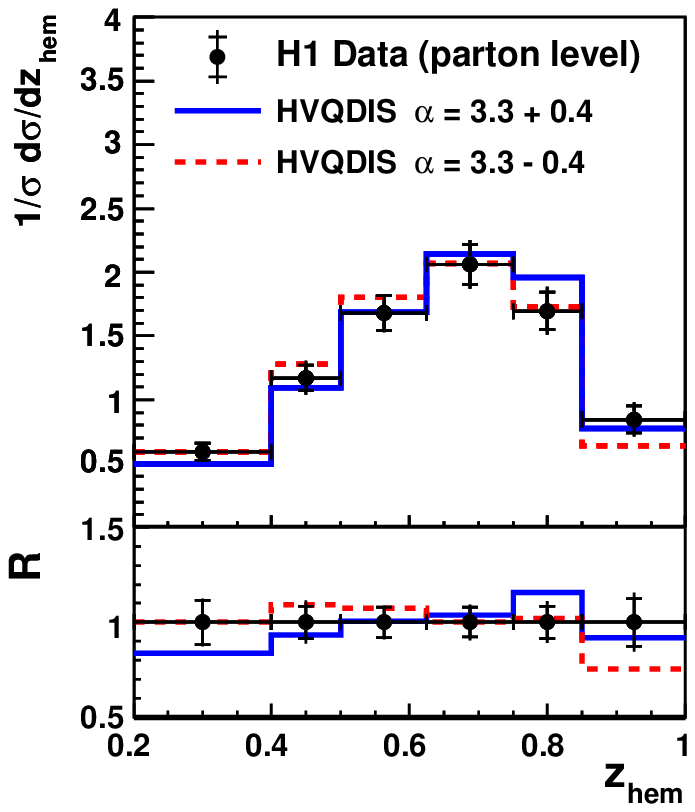,height=4.5cm}
~\epsfig{file=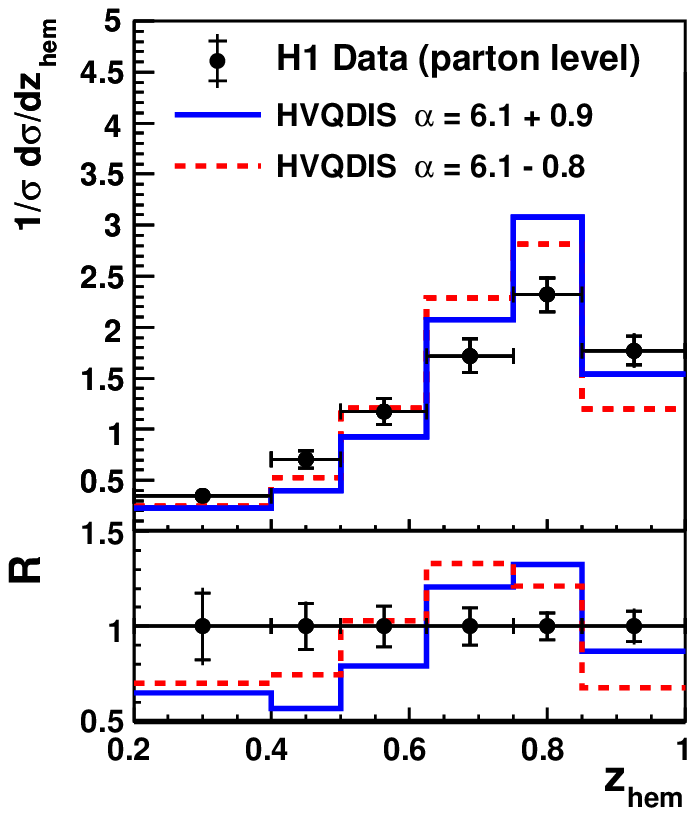,height=4.5cm}
\put(-215,97){\makebox(0,0)[tl]{(a)}}
\put(- 91,97){\makebox(0,0)[tl]{(b)}}
\caption{Normalised $D^*$ cross sections as a function of 
an estimator, $z_{\rm hem}$, of the fragmentation function, $z$, 
showing the best fits of an NLO QCD prediction using the Kartvelishvili fragmentation 
function with the uncertainty range on the
$\alpha$ parameter shown.  The scales of the processes are (a) $8.4 < \sqrt{\hat{s}} < 18$\,GeV and 
(b) $\sqrt{\hat{s}} < 8.4$\,GeV.  From~\cite{Aaron:2008ac}.}
\label{fig:frag-fun-fit}
\end{center}
\end{figure}

Both collaborations fit NLO QCD predictions to the data using the Kartvelishvili \emph{et al.}\ model~\cite{Kartvelishvili:1977pi} 
for fragmentation, 

\begin{equation}
f(z) \propto z^\alpha (1-z),
\end{equation}
where $\alpha$ is a free parameter.  The result from ZEUS of $\alpha = 2.67 ^{+0.25}_{-0.31}$ can be compared with that from 
H1 for lower scales, given in Fig.~\ref{fig:frag-fun-fit}.  A higher value of $\alpha$ implies a high $\langle z \rangle$ and so 
the observation of $\alpha$ increasing with decreasing scale, $\sqrt{\hat{s}}$, is also consistent with the expectations from 
scaling violations in QCD.  However, it should be noted that the quality of the fit for the lowest scale, in Fig.~\ref{fig:frag-fun-fit}(b), 
is poor indicating an incomplete description of the full phase space down to the production threshold.  Future fits of 
fragmentation functions should use these HERA data as part of their input and the uncertainties in the determination of the 
fragmentation parameter should 
ideally be propagated 
as a systematic uncertainty in NLO QCD calculations of a given 
charm cross section.

\subsection{Inelastic $J/\psi$ Production}

Even though the discovery of the $J/\psi$ meson, and hence the charm quark, was made in 1974, its hadroproduction mechanism 
is still uncertain (for a review, see~\cite{Brambilla:2010cs}).  At HERA, $J/\psi$ production is dominated by boson--gluon fusion in which a photon emitted from the 
incoming electron interacts with a gluon from the proton to produce a $c\bar{c}$ pair which subsequently forms a $J/\psi$ meson.  
In the colour-singlet (CS) model~\cite{Chang:1979nn,Berger:1980ni,Baier:1981zz,Baier:1981uk,Baier:1983va}, the   
$c\bar{c}$ pair produced has the same quantum numbers as 
the physical $J/\psi$ bound state, achieved by radiating 
a hard gluon in the perturbative process.  In the colour-octet (CO) 
model~\cite{Caswell:1985ui,Thacker:1990bm,Bodwin:1994jh}, the $c\bar{c}$ pair emerges from the hard process with 
quantum numbers different from those of the $J/\psi$ and evolves into the physical $J/\psi$ state by emitting one or more  soft 
gluons.  The probability for CO processes occurring is governed by long-distance matrix elements that can be obtained from fits to 
experiment.  Predictions within the CS framework have also been made using the $k_T$ factorisation 
approach~\cite{Baranov:1998af,Baranov:2002cf,Lipatov:2002tc} in which the effects of non-zero incoming parton transverse 
momentum are taken into account.  These three approaches have been compared extensively to HERA measurements. 

\begin{figure}[htp]
\begin{center}
~\epsfig{file=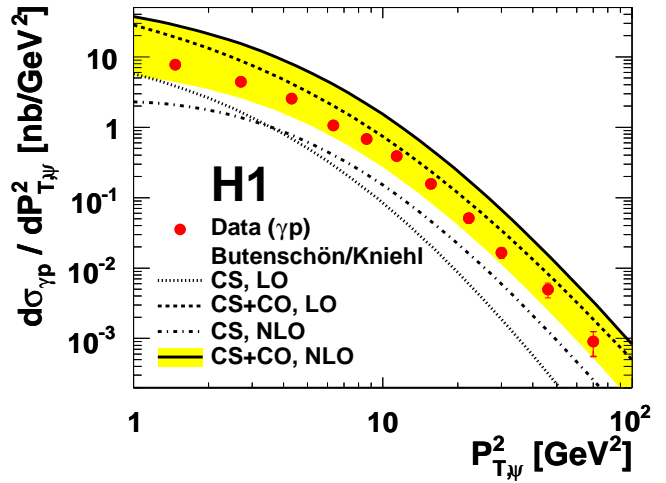,height=6.5cm}
\caption{Photoproduction cross section 
differential in $P_T^2$ of the $J/\psi$ meson for H1 data, 
compared with LO and NLO QCD 
predictions from the CS model and the CS and CO models combined.  The  
uncertainty on the theory (yellow band) arises from the difference between results using an LO 
or a higher-order colour octet contribution~\cite{Butenschoen:2009zy}.
From~\cite{Aaron:2010gz}.}
\label{fig:jpsi-pt}
\end{center}
\end{figure}

Both H1 and ZEUS have measured inelastic $J/\psi$ production in DIS~\cite{Aaron:2010gz,Adloff:2002ey,Chekanov:2005cf} and in 
photoproduction~\cite{Aaron:2010gz,Adloff:2002ex,Aid:1996dn,Ahmed:1994ef,Abramowicz:2012dh,Chekanov:2009ad,Chekanov:2002at,Breitweg:1997we}.
Inelastic $J/\psi$ production has also 
been measured extensively at the Tevatron and 
the resulting data have been used to constrain the 
CS and CO contributions, usually by first applying a CS calculation
and then adding a CO contribution, fitted to 
best describe the data. By combining the CS and fitted CO contributions in 
this manner, the theory is able to describe the $J/\psi$ cross section in 
$p\bar{p}$ collisions~\cite{Cho:1995vh,Cho:1995ce}.  However neither this, nor 
the $k_T$ factorisation approach are able to describe the 
polarisation measurements for 
$J/\psi$ and $\psi(2S)$ mesons (see~\cite{Abulencia:2007us} and 
references therein).  The CO contribution fitted to the 
Tevatron data is expected to be applicable to $ep$ collisions and is used 
to predict H1 and ZEUS data.  
A measurement of $J/\psi$ photoproduction is shown in Fig.~\ref{fig:jpsi-pt} in comparison to LO and NLO QCD 
predictions~\cite{Kramer:1995nb,Artoisenet:2009xh,Butenschoen:2009zy,Butenschoen:2010rq,Butenschoen:2011yh} from the 
CS model and the CS and CO models combined.  
The predictions from the CS model 
lie below the data although the shapes of the $P_T^2$ and inelasticity, $z$  (not shown), distributions are well reproduced.  As 
the uncertainties on the theoretical predictions are large, the next order in QCD may be required to describe the 
data.  Inclusion of a CO 
contribution gives a higher cross section and describes the data well, although again with large uncertainties.  
Predictions based on $k_T$ factorisation in the CS model (using the {\sc Cascade} Monte Carlo 
programme~\cite{Jung:2000hk,Jung:2001hx}, not shown) also gives a good description of the data.

\begin{figure}[htp]
\begin{center}
~\epsfig{file=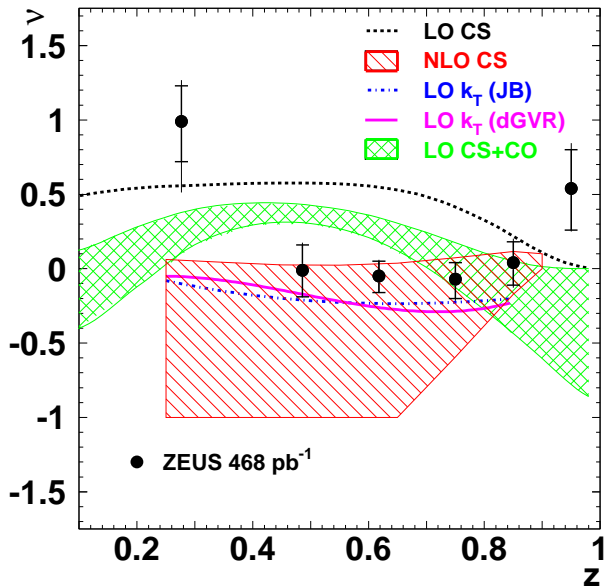,height=8cm}
\caption{Helicty parameter, $\nu$, for the decay azimuthal angle in the frame where the polarisation axis in the $J/\psi$ meson 
rest frame is defined by the flight direction of the $J/\psi$ meson in the $\gamma p$ rest frame.  The value can vary between $-1$ 
and $+1$ and is shown here for different inelasticity values.  From~\cite{Chekanov:2009ad}.}
\label{fig:jpsi-z}
\end{center}
\end{figure}

Measurements of the polarisation of the $J/\psi$ mesons provided 
strong distinguishing power for the Tevatron results and 
comparable analyses were performed at HERA.  An example is shown in Fig.~\ref{fig:jpsi-z}, where the helicity parameter is 
plotted versus the inelasticity for the full ZEUS data sample.  
The theoretical and experimental uncertainties 
are large.  None of the predictions gives a good description over the full phase space and unfortunately discrimination between 
different models is not possible from the HERA data alone.

Unfortunately, the production mechanism for $J/\psi$ mesons 
remains largely unresolved.  
Although HERA has produced precise measurements 
of differential cross sections in DIS and photoproduction and 
measurements of the polarisation, 
progress is limited by the large theoretical uncertainties. 
Further measurements from the Tevatron or 
LHC~\cite{Aad:2011sp,Chatrchyan:2011kc,Abelev:2012gx,Aaij:2011jh,Abelev:2011md} extending to higher 
$P_T$ along with global analyses (see e.g.~\cite{Butenschoen:2012qh}) including the HERA data presented here may lead to a 
deeper insight.

\subsection{Exotic and Unusual QCD} 
\label{sec:exo-qcd}

Many of the searches for new physics at HERA have focused on the hadronic final state.  The situation is perhaps well 
summarised by generic searches, which identify deviations between data and predictions for arbitrary combinations of 
high transverse momentum objects (leptons, photons, jets and missing energy)~\cite{Aaron:2008aa}.  Both here and in 
dedicated studies~\cite{Adloff:1998aw}, H1 data consistently exhibited an excess of events with a single isolated lepton, 
a high transverse momentum jet and large missing energy, corresponding to the topology expected from a $W$ boson 
produced at high $p_T$, recoiling against a quark or gluon. However, no comparable signal was observed by ZEUS 
and excitement about this channel diminished when the full H1 data became available. The final situation from HERA 
is summarised in a combined H1 and ZEUS result~\cite{Aaron:2009ab}. 
Ultimately HERA data across a wide range of signatures and channels are
remarkably consistent with the Standard Model.

Within the Standard Model, a variety of previously unobserved quark and gluon bound states and other novel strong 
interaction phenomena are predicted. Due to its rich, complex and non-Abelian nature, the status of these states and 
phenomena within QCD is not always obvious. As discussed in the following sections, in many such cases observing 
new effects for the first time would constitute a major discovery in its own right. 

\subsubsection{Pentaquark States}

The possibility to have particles consisting of five 
quarks and antiquarks, i.e.\ pentaquarks, was proposed as early as the 
1970s~\cite{Jaffe:1976ih,Strottman:1979qu,Lipkin:1987sk} and is predicted within QCD~\cite{Diakonov:1997mm}.  
In the early part of the 2000s, many collaborations searched for pentaquark states, with several seemingly clear signals, 
but also contradictory results.  The world-wide status 
is summarised in dedicated reviews of the subject,  see for 
example~\cite{Hicks:2012zz,Danilov:2008zza,Hicks:2005gp,Hicks:2004ge,Dzierba:2004db}. Here we concentrate 
on the H1 and ZEUS contributions.

Most pentaquark searches have focussed on the 
lightest irreducible quark combination, termed 
$\Theta^+$, which has a quark content $uudd\bar{s}$
and a mass expected to be around 1530\,MeV.   
The ZEUS Collaboration searched~\cite{Chekanov:2004kn} 
for the $\Theta^+$ in the $K_S^0 p$ decay channel 
and reported a signal of around $4-4.5\,\sigma$ significance with a 
mass of $1521 \pm 1.5$\,MeV in DIS events.  
Superficially this was consistent with other `observations' of the 
$\Theta^+$ pentaquark.  However the comparatively low mass~\cite{Hicks:2005gp} pointed to problems.  H1 
subsequently performed the same analysis~\cite{Aktas:2006ic} and saw no signal, which, along with other null 
results~\cite{Hicks:2012zz,Danilov:2008zza,Hicks:2005gp,Hicks:2004ge,Dzierba:2004db} and the fact that ZEUS only 
saw this at high $Q^2$ (above 20\,GeV$^2$, but not above 1\,GeV$^2$) indicates that the ZEUS result had some flaw or 
must have been a statistical fluctuation.  ZEUS have yet to confirm or refute the result even though a data sample several 
times larger is available.

Both collaborations searched in the $\Xi\pi$ channel~\cite{Chekanov:2005at,Aktas:2007dd} for the pentaquark resonance 
reported by the NA49 Collaboration~\cite{Alt:2003vb}.  The search 
was performed for pentaquarks decaying to the doubly charged $\Xi^- \pi^-$ and the neutral $\Xi^- \pi^+$ final states.  The HERA
collaborations observed a clean resonance 
for the known baryon $\Xi(1530)^0$ decaying to $\Xi^-\pi^+$, but observed no 
other higher-mass resonances and so set production limits in the mass range, 1600--2350\,MeV.  These null results 
were in keeping with the many others 
for this channel (again see pentaquark review 
articles~\cite{Hicks:2012zz,Danilov:2008zza,Hicks:2005gp,Hicks:2004ge,Dzierba:2004db} and references therein).

Assuming the existence of a light pentaquark such as the $\Theta^+$, then a heavier analogue with a charm quark might  
also be expected to exist.  The H1 Collaboration searched for a $D^*p$ resonance and found a clear signal~\cite{Aktas:2004qf} at about 
3.1\,GeV, consistent with a pentaquark with quark content $uudd\bar{c}$.  A similar search was performed by ZEUS with a 
larger statistical sample~\cite{Chekanov:2004qm}, but no signal was observed.  Given the cleanliness of the H1 signal and 
its significance of at least $5-6\,\sigma$, the situation was puzzling.  Further null results were published by other high energy physics 
experiments~\cite{Danilov:2008zza} and a higher-statistics analysis presented by H1~\cite{Krueger:2009zz}.  Although 
unpublished, the newer H1 measurement shows no 
resonance structure around 3.1\,GeV, using a data sample four times larger 
than previously, 
suggesting that the original results 
arose from a statistical fluctuation.

In summary, after initial excitement and `discoveries' of new particles, the results from HERA have gone the way of the 
overall status for pentaquark searches in the many experiments world-wide~\cite{Hicks:2012zz,Amsler:2008zzb}, with no convincing evidence remaining.

\subsubsection{Instanton Searches}

In both electroweak and QCD interactions, the ground state has a rich topological structure, associated with non-perturbative 
fluctuations of the gauge fields, called instantons~\cite{Belavin:1975fg,'tHooft:1976up,'tHooft:1976fv}.  In electroweak interactions, 
instantons are not expected to be observable at current colliders.  In QCD, the effects of instantons are expected to be manifest 
at lower energies~\cite{Ringwald:1994kr,Balitsky:1993jd,Moch:1996bs,Ringwald:1999ze}, although they have yet to be seen.  
The rate of instanton production in DIS at HERA was expected to be 
sizeable~\cite{Ringwald:1994kr,Ringwald:1999ze,Moch:1996bs,Ringwald:1998ek,Ringwald:1999jb,Ringwald:2000gt,Gibbs:1995bz} 
and so presented 
the HERA experiments with an opportunity to discover the effect.
The major challenge for H1 and ZEUS~\cite{Adloff:2002ph,Chekanov:2003ww} 
was that the instanton cross section was nonetheless small in comparison 
with the inclusive, which was hard to suppress sufficiently.

\begin{figure}[htp]
\begin{center}
~\epsfig{file=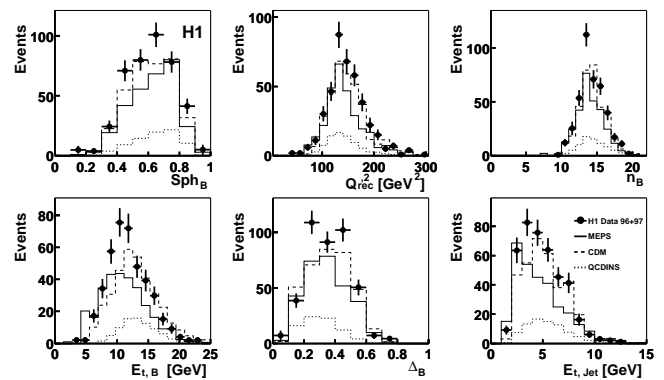,height=5cm}
\caption{Raw distributions in sphericity, $Q^2$, 
charged particle multiplicity,
total transverse energy, isotropy and transverse 
jet energy, compared with predictions for DIS events 
and for instanton-induced events (QCDINS), after
applying a cut on a multivariate discriminant
designed to enhance instanton-like signatures. From~\cite{Adloff:2002ph}.}
\label{fig:instantons}
\end{center}
\end{figure}

The expected 
characteristics of instanton events are 
large transverse energy, high-multiplicity 
and the production of different quark flavours democratically and  
isotropically in their centre-of-mass frame (a `fireball' configuration).  
A search based on these criteria is presented in 
Fig.~\ref{fig:instantons}, which compares the 
predicted normalisations and the
shapes for particular variables between data, the
QCDINS instanton 
model \cite{Ringwald:1999jb}
and inclusive DIS models after applying a cut based on a multivariate 
discriminant. 
The application of cuts such as this suppresses the inclusive
DIS background by factors of typically 1\,000.
Although events exhibiting instanton-like 
characteristics are observed,
their isolation is hampered by the uncertainty on the remaining 
background, 
mainly due to the sometimes 
poor modelling of the hadronic final state by the inclusive 
DIS Monte Carlo simulations. It has therefore not been
possible to isolate regions in which 
a clear signature for instanton production is expected.  
However, the analysis allows upper limits 
to be placed on instanton 
production cross sections which are within a factor of 3--5 of 
the predicted values, depending on the kinematic region.  
Tuning and improving the simulation of the hadronic 
final state using the measurements 
contained in this article may allow a signal for 
instanton production to be extracted in other experiments.

\subsubsection{Search for Glueball States}

The existence of glueballs (bound states of two or more gluons)
is predicted by QCD.  The lightest 
state is expected to have quantum numbers $J^{PC} = 0^{++}$ 
and a mass in the range 1550--1750\,MeV~\cite{Yao:2006px}.  The state 
$f_0(1710)$ is frequently considered to be a 
glueball or tetraquark candidate~\cite{Klempt:2007cp,Albaladejo:2008qa},
since it does not fit into existing multiplet structures, 
but its constituent parton content is not 
established.  As the $K_S^0 K_S^0$ system is expected to couple to glueballs, 
resonances in its invariant-mass 
spectrum have been searched for.

\begin{figure}[htp]
\begin{center}
~\epsfig{file=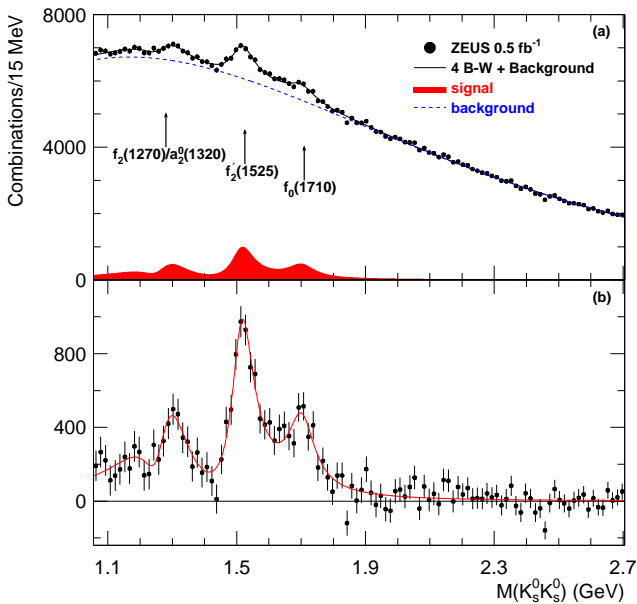,height=8.3cm}
\caption{(a) The measured $K_S^0 K_S^0$ invariant-mass spectrum 
compared with a fit including signal and background
contributions.  (b) The 
background-subtracted $K_S^0 K_S^0$ invariant-mass spectrum with the fit for the signals also shown.  
From~\cite{Chekanov:2008ad}.}
\label{fig:ks-ks}
\end{center}
\end{figure}

In an initial study from ZEUS~\cite{Chekanov:2003wc} of the $K_S^0 K_S^0$ system in DIS, indications of 
two states, $f_2^\prime(1525)$ and $f_0(1710)$, were seen with statistical significances of about three standard 
deviations.   A subsequent measurement~\cite{Chekanov:2008ad} used all events and all data (0.5\,fb$^{-1}$), 
dominated by photoproduction; given the gluon-rich environment, it is expected that photoproduction 
should be more sensitive than DIS to the production of glueballs.  By reconstructing a secondary vertex and 
removing contamination from photon conversions and $\Lambda$ baryons, a clean sample of $K_S^0$ 
candidates were reconstructed
via their decay to $\pi^+\pi^-$~\cite{Chekanov:2008ad}.  
Using over 
$10^6$ $K_S^0$ mesons, pairs were combined to give the invariant-mass spectrum shown in Fig.~\ref{fig:ks-ks}.  
Three peaks around 1300, 1500 and 1700\,MeV are observed with no heavier state seen.

The states were fitted accounting for the interference pattern predicted by SU(3) symmetry arguments~\cite{Faiman:1975ne}.
The first peak at about 1300\,MeV is consistent with the combination $f_2(1270)/a_2^0(1320)$.  The second and third peaks 
are consistent with the states $f_2^\prime(1525)$ and $f_0(1710)$; their masses and widths are consistent with values from the 
PDG~\cite{Yao:2006px} and 
are of similar precision.  The fit yields $4058 \pm 820$ events
for the $f_0(1710)$ resonance, constituting 
a statistical significance of five standard deviations.  The $f_0(1710)$ has a mass consistent with a $J^{PC} = 0^{++}$ glueball candidate, 
although if it is the same state as 
is seen in 
$\gamma \gamma \to K_S^0 K_S^0$~\cite{Althoff:1982df,Acciarri:2000ex}, it is 
unlikely to be a pure glueball state.

\subsubsection{Deuteron and Anti-Deuteron Production}

The production of light stable nuclei, such as deuterons ($d$), in high-energy collisions is poorly 
understood.  Anti-deuterons ($\bar{d}$)
were first observed in 1965~\cite{Massam:1965mw,Dorfan:1965uf} and subsequently in
 $e^+e^-$~\cite{Albrecht:1985zv,Albrecht:1989ag,Akers:1995az,Schael:2006fd,Asner:2006pw}, proton--nucleus 
($pA$)~\cite{Binon:1969qz,Antipov:1971zs,Cronin:1974zm}, proton--proton 
($pp$)~\cite{Alper:1973my,Henning:1977mt,Abramov:1986ti} and 
nucleus--nucleus~\cite{Aoki:1992mb,Appelquist:1996qy,Adler:2001uy,Ahle:1998jv,Bearden:1999iq,Bearden:2002ta,Adler:2004uy,Arsene:2010px,Anticic:2011ny} collisions.  The coalescence model~\cite{Butler:1963pp} was developed to 
describe the production of deuterons and anti-deuterons in heavy-ion collisions.   This approach has also been 
used to describe $d$ or $\bar{d}$ production in $pp$ and $e^+e^-$ interactions.  The invariant differential 
cross section for deuteron production can be parametrised as 

\begin{equation}
\frac{E_d}{\sigma_{\rm tot}} \frac{{\rm d}^3\sigma_d}{{\rm d}p_d^3} = B_2 \left( \frac{E_p}{\sigma_{\rm tot}} \frac{{\rm d}^3\sigma_p}{{\rm d}p_p^3} \right)^2
\end{equation}
where $E_{d(p)}$ and $\sigma_{d(p)}$ are the energy and the production cross section of the $d(p)$, $p_d(p_p)$ is the 
momentum of the $d(p)$ and $\sigma_{\rm tot}$ is the total cross section.  The coalescence parameter, $B_2$, is inversely 
proportional to the volume of the fragmentation region emitting the particle.  If $B_2$ is the same for particles and anti-particles, 
then the production ratio for $\bar{d}/d$ is 
expected to be equal to the square of that for $\bar{p}/p$.

The production of anti-deuterons was first measured at HERA by H1 in photoproduction~\cite{Aktas:2004pq}; ZEUS then 
measured both deuteron and anti-deuteron production in DIS~\cite{Chekanov:2007mv}.  Both collaborations exploited 
the rate of ionisation energy loss, 
${\rm d} E/{\rm d}x$, measured in the inner tracking chambers and the distance of closest approach 
of the track to the event vertex.  Clear signals of $d$ and $\bar{d}$ were seen along with small signals for tritons; no 
heavier nuclei or anti-triton candidates were observed.  After measurement of the invariant cross sections, the ratios 
of the different production rates were measured in order to compare with other colliders and with the coalescence model.

\begin{figure}[htp]
\begin{center}
~\epsfig{file=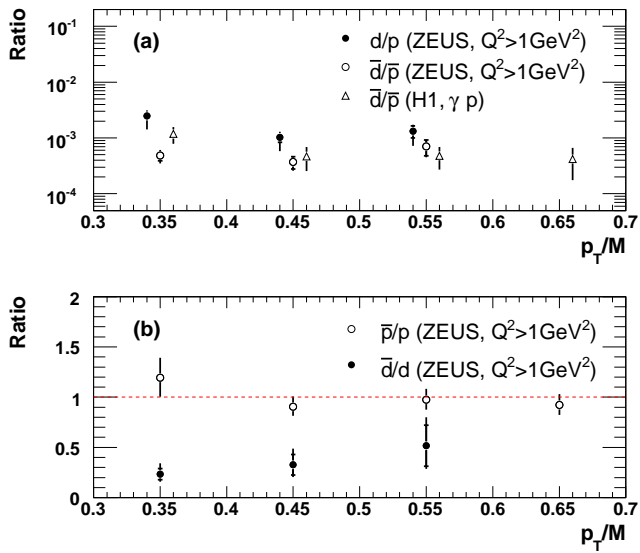,height=7.8cm,bbllx=0,bblly=0,bburx=550,bbury=500,clip=}
\caption{(a) $d/p$ and $\bar{d}/\bar{p}$ production rates in DIS (ZEUS) and in photoproduction (H1) as a function of $p_T/M$.  
(b) $\bar{d}/d$ and $\bar{p}/p$ production rates in DIS as a function of $p_T/M$.  From~\cite{Chekanov:2007mv}.}
\label{fig:deuteron}
\end{center}
\end{figure}

The ratio of production rates is shown in Fig.~\ref{fig:deuteron}.  The production rate of deuterons is three orders of magnitude 
smaller than the rate for protons.  The ratios $d/p$ and $\bar{d}/\bar{p}$ and also the results in DIS and photoproduction in 
Fig.~\ref{fig:deuteron}(a) are consistent with one another and with 
independence of the ratio, particle transverse momentum to mass, $p_T/M$.  
These ratios are also in agreement with those from 
$\Upsilon$ decays and $pp$ and heavy-ion data.  Similarly when analysed within the context of the coalescence model, 
the value of $B_2$ extracted is consistent between $d$ and $\bar{d}$ and DIS and photoproduction.  The results from 
HERA for $B_2$ agree with those from $pp$ data, 
but lie above the results from $e^+e^-$ collisions at the $Z$ pole and from 
heavy-ion data.  However, heavy-ion data at lower centre-of-mass energy tend towards the values from HERA and $pp$ 
collisions.  In Fig.~\ref{fig:deuteron}(b), the ratios $\bar{d}/d$ and $\bar{p}/p$ are shown versus $p_T/M$. The $\bar{p}/p$ 
ratio is consistent with unity as expected from hadronisation of quark and gluon jets.  The ratio $\bar{d}/d$ is about $0.3 \pm 0.1$ 
and is inconsistent with the coalescence model, which predicts that
this ratio should be equal to that for $(\bar{p}/p)^2 \approx 1$, a 
relationship which 
is consistent with data from $pp$~\cite{Alper:1973my,Henning:1977mt,Abramov:1986ti}, $pA$~\cite{Cronin:1974zm} 
and heavy-ion collisions~\cite{Adler:2004uy}. 
To summarise, the 
coalescence model can describe a number of features of $d$ and $\bar{d}$ 
production in high-energy collisions, but given the results here can not provide a complete picture.

\section{Exclusive and Semi-Inclusive Processes}
\label{sec:diffraction}

\subsection{Introduction}
\label{diff:intro}

Diffractive processes in proton-(anti)proton scattering, and to a lesser
extent, other projectiles scattering diffractively on proton targets,
have been the subject of sustained and intense study since well 
before HERA.
Early experimental results and their theoretical
description are extensively covered in a number of review
articles \cite{Kaidalov:1979jz,Alberi:1981af,Goulianos:1982vk,zotov:1988}.
The first results have also now started to appear from the 
LHC~\cite{Aad:2012pw,Chatrchyan:2012vc,:2012sja,Aaij:2013jxj}.
Although it was not widely expected prior to data taking, the study of 
quasi-elastic and
diffractive processes has been one of the most successful areas of
study at HERA, and certainly one of the most prolific in terms of 
publications. Complementary reviews can be found 
in \cite{Wolf:2009jm,Ivanov:2004ax}.

The kinematics of HERA, with a strong forward boost of the hadronic
centre of mass relative to the lab frame for non-extreme $y$ values,
were particularly favourable for the study of diffractive excitations
of the real and virtual photon. 
Two separate cases are usually distinguished. In the (quasi-)elastic case, 
$\gamma^{(*)} p \rightarrow V p$,
the photon coupling to the beam lepton
either remains intact (Deeply 
Virtual Compton Scattering, DVCS) or converts to a vector 
meson $V$.
In the diffractive dissociation case, 
$\gamma^{(*)} p \rightarrow X p$
the system $X$ produced at the
proton vertex is a multi-particle state covering a continuum of 
invariant masses, $M_X$. 

Most of the interest has centred around the case where the 
proton remains intact and is scattered through a small angle (implying
a small absolute value of the 
squared four-momentum transfer at the proton vertex, $|t|$
(Figs.~\ref{VM:feynman}(a) and~\ref{feynman}(a)). 
Measurements of processes in which 
the proton dissociates to large mass systems $Y$, i.e.
$\gamma^{(*)} p \rightarrow XY$ and $\gamma^{(*)} p \rightarrow VY$
were more limited, but have been considered for example in the 
context of the decomposition of the total photoproduction cross
section into diffractive and non-diffractive 
channels \cite{Aid:1995bz}. The only detailed measurement of the $M_X$ 
dependence 
in proton dissociation processes \cite{Adloff:1997mi}
revealed a clear difference compared with 
the elastic-proton case, which is presumably a consequence of the
sub-leading exchanges discussed in 
Section~\ref{sec:sublead}, but has yet to be
interpreted in detail.

Prior to HERA, substantial data were already available
on exclusive vector meson
production in both DIS and 
photoproduction \cite{Bauer:1977iq,Shambroom:1982qj,Binkley:1981kv,Denby:1983az}, with
fixed target data continuing to emerge during and beyond the HERA
era \cite{Arneodo:1994id,Adams:1997bh,Airapetian:2000ni,Airapetian:2009ad,Alexakhin:2007mw,Adolph:2012ht}.
DVCS has also recently been studied in detail in fixed target
experiments at 
HERA \cite{Airapetian:2012pg,Airapetian:2012mq,Airapetian:2006zr} 
and Jefferson 
Laboratory \cite{Camacho:2006hx,Girod:2007aa,Gavalian:2008aa}.
In contrast, the study of the diffractive dissociation of real and virtual
photons is almost entirely the preserve of HERA, with only very limited
fixed target data \cite{Chapin:1985mf}.

Before the first precise measurements emerged from HERA, 
a QCD-based treatment of the exclusive production of vector
mesons was already fairly well 
advanced \cite{Ryskin:1992ui,Brodsky:1994kf,Frankfurt:1995jw}. 
Cross sections were expected to be $Q^2$-suppressed relative to 
inclusive DIS and to be related 
to the
parton densities of the proton, the lowest 
order partonic exchange 
being a pair of gluons with compensating colour 
charges \cite{Low:1975sv,Nussinov:1975mw}. 
There was less consensus on the appropriate 
QCD treatment of single diffractive
photon dissociation. On the one hand, 
`hard' diffractive processes, similar to exclusive vector
meson production were predicted, in which the colour singlet exchange 
couples fully into the hard interaction with the photon.
Such an exchange
may again be perturbatively calculable starting from a 
knowledge of the proton structure.   
In contrast, in `soft' diffraction, the 
colour singlet exchange is a composite virtual object, 
similar to the pomeron
of peripheral hadronic scattering,
from which a single parton participates in the
hard scattering, the remainder forming a low transverse momentum remnant 
system, cleanly separated in rapidity from the oppositely-travelling
intact proton.
This type of configuration \cite{Ingelman:1984ns,Donnachie:1987xh}
produces a leading twist contribution
with similar $Q^2$ dependence to the total cross section, 
limited only by kinematic 
constraints at large $Q^2$. 
In the DIS regime, it can be related
to a concept of diffractive parton densities. 
A major issue in inclusive diffraction at HERA has been determining
where each of these `soft' and `hard' processes are dominant and 
understanding the transition between the two. 

The detailed dependence on centre-of-mass energy of elastic and, 
via the optical theorem, total 
hadron--hadron cross sections has historically been remarkably well
described in a large kinematic domain by Regge
phenomenology~\cite{Regge:1959mz,Regge:1960zc}. In this framework, 
interactions take place via the $t$-channel exchange of
reggeons related to mesons~\cite{Chew:1961yz}
and of the leading vacuum singularity, the
pomeron ($\pom$)~\cite{Chew:1961ev}. The pomeron is
the mediator of diffractive scattering. 
Although the pomeron is essentially a phenomenological object,
associated mainly with soft processes, it can also be generated 
perturbatively. For example, connecting the two exchanged gluons in
the lowest order partonic process by adding
further gluon `rungs' in a leading-logarithmic $1/x$ approach 
leads to the so-called `BFKL pomeron'.

At asymptotically large energies, 
pomeron exchange dominates the elastic channel, such that both elastic and 
total cross sections display a slow increase with centre-of-mass
energy.  Interactions in which one
or both of the hadrons dissociates to higher 
mass states~\cite{Fienberg:1956,Good:1960ba}
also
occur naturally in this approach. 
Such processes are characterised by the presence of
large regions of rapidity space in which no hadrons are produced
and are dominated by diffractive exchange at large
centre-of-mass energy $\sqrt{s}$ and small dissociation 
masses $M_X$. The
inclusive dissociation mass distribution may be treated via
Mueller's generalisation of the optical theorem~\cite{Mueller:1970fa}, 
such that an appropriate Regge description involves diagrams that contain 
three-reggeon couplings.

Pomeron language has been adopted to a large extent at HERA, 
as a matter of convenience. This should not be taken to imply
the existence of a universal
$t$-channel pomeron exchange, since the detailed properties of the
exchange and its Regge trajectory 
clearly vary between processes and with kinematic variables at HERA,
as discussed in detail below.

First evidence for diffractive dissociation
processes involving sufficiently large scales for perturbative QCD
to be applied
came from diffractive dijet production in $p \bar{p}$ collisions
at the SPS as observed by 
UA8 \cite{Brandt:1992zu}. Later data from the Tevatron 
followed -- see e.g.~\cite{Affolder:2000vb}.
These observations
lent experimental support to the idea
that the pomeron might be considered as an 
object with its own 
structure, which might be probed in 
DIS \cite{Ingelman:1984ns,Donnachie:1987xh}.
First ideas on the structure of the pomeron
were in place as early as the 
mid-1980s \cite{Gribov:1984tu,Berger:1986iu}.

\begin{figure}[htb]
\begin{center}
~\epsfig{file=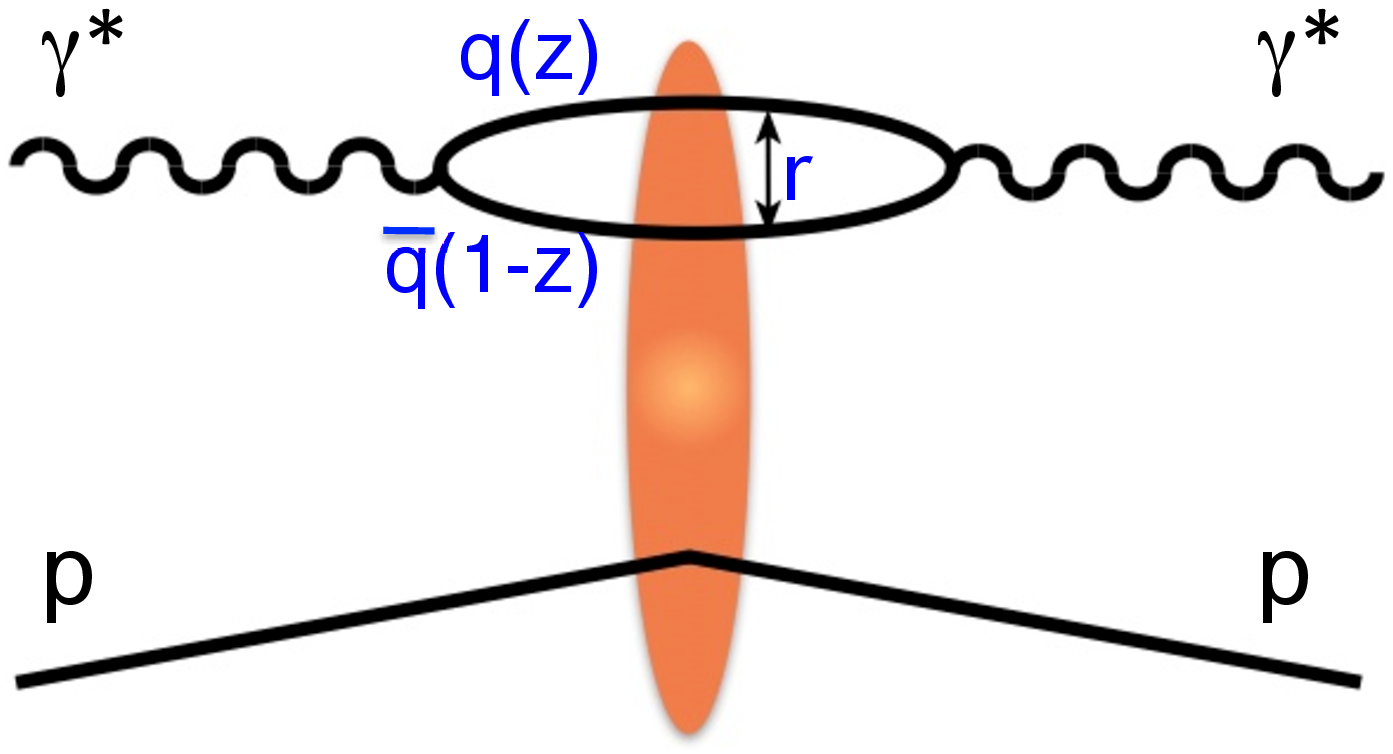,width=0.18\textwidth,angle=270}
\caption{Schematic illustration of the dipole model concept,
in which a virtual photon fluctuates to a $q \bar{q}$ pair with 
relative momentum fractions $z$ and $1-z$, forming a colour dipole
of transverse size $r \sim 1 / \surd Q^2$. The dipole then scatters elastically from the proton
according to a dipole cross section. The example shown corresponds
to the amplitude for the total inclusive $\gamma^* p$ cross section.
However, only small modifications are required to adapt this to the cases of
exclusive or inclusive diffraction.}
\label{Dipolefig}
\end{center}
\end{figure}

Alternatively to Regge theory, an $s$-channel
picture \cite{Miettinen:1978jb,Dederichs:1989gk}
developed around the original Good \& Walker
approach \cite{Good:1960ba} in which 
different projectile Foch states are
absorbed differently on the target. These ideas have enjoyed
a recent revival as part of the quest to describe diffraction at
the LHC \cite{Ryskin:2011qe}. 
The application of this $s$-channel approach to 
HERA data involves the elastic and quasi-elastic scattering
of colour dipoles, corresponding to $q \bar{q}$ (or 
higher multiplicity)
fluctuations of real and virtual photons on the 
proton \cite{Mueller:1989st,Nikolaev:1990ja}. 
An attractive feature of this approach is that 
the possible presence of `saturation' phenomena,
whereby the low $x$ growth of parton densities is tamed as 
ultimately required
to satisfy unitarity,
are easily incorporated at the parametric
level \cite{GolecBiernat:1998js}. Whilst such models
were originally devised to describe the total $\gamma^* p$ cross
section (see Fig.~\ref{Dipolefig}),\footnote{The 
results from the inclusive cross section
have been inconclusive to date, with strong evidence for
saturation only at low scales $\ll 1 \ {\rm GeV^2}$, which preclude a
partonic interpretation \cite{Forshaw:1999uf}.} vector 
meson production and other exclusive processes are easily incorporated
with no further free parameters except those that describe the 
$t$ dependence (see e.g.~\cite{Kowalski:2003hm,Kowalski:2006hc} 
for a modern treatment). 
Diffractive dissociation can also be 
included \cite{Nikolaev:1991et}, 
though has proved
more difficult to describe, due to the need to incorporate higher
terms, corresponding for example to a dipole formed from a
$q \bar{q}$ pair and a gluon \cite{Bartels:1998ea,GolecBiernat:1999qd}.
The dipole approach to diffraction at HERA is discussed further in 
Sections~\ref{vm:dipole} and~\ref{sec:dipoles}.

\subsection{Exclusive Production of $1^{--}$ States}
\label{vm}

\subsubsection{Kinematics and Experimental Selection}
\label{vm:intro}

\begin{figure}[htb] \unitlength 1mm
  \begin{center}
    \begin{picture}(60,36)
      \put(-11,2){\epsfig{file=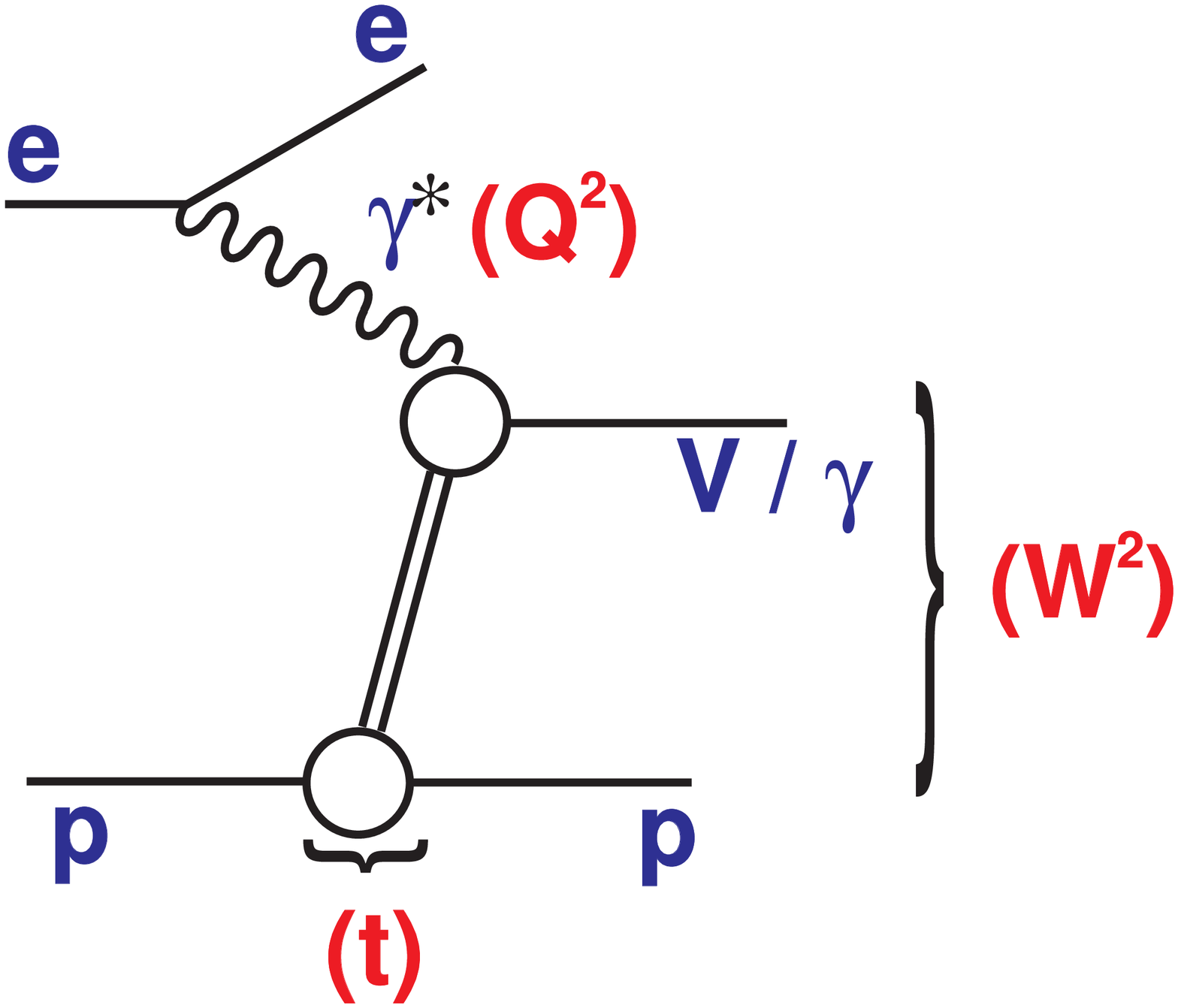,height=0.19\textwidth,clip}}
      \put(33,8){\epsfig{file=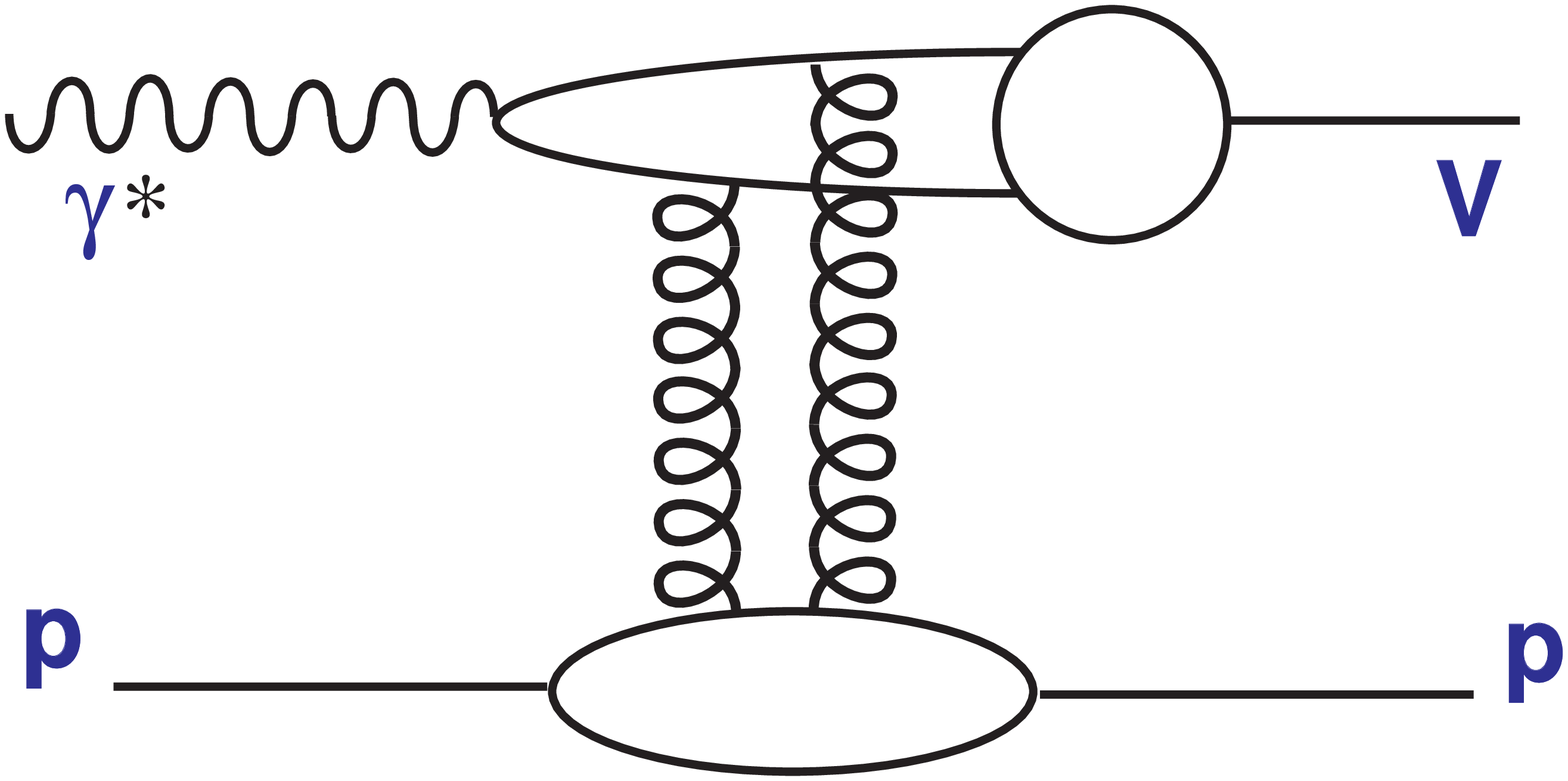,height=0.11\textwidth,clip}}
      \put(8,-3){{\footnotesize{{\bf (a)}}}}
      \put(48,-3){{\footnotesize{{\bf (b)}}}}
    \end{picture}
  \end{center}
  \caption{(a) Generic representation of
exclusive vector meson production via the quasi-elastic
scattering of a real or virtual photon. The
commonly used kinematic variables
discussed in the text are indicated.
(b) Lowest order perturbative diagram for hard vector
meson production, 
involving the exchange of a
pair of gluons between the proton and the real or virtual photon.}
\label{VM:feynman}
\end{figure}

Quasi-elastic vector meson production and DVCS (Fig.~\ref{VM:feynman}(a))
are the simplest diffractive processes that can be studied at HERA.
For a fixed final state vector meson or photon, 
they are usually described in terms of the kinematic variables,
$Q^2$, $W$ and $t$. Distributions in all three of these variables have been 
measured in analyses covering the vector meson species,
$\rho$, $\omega$, $\phi$, $\rho^\prime$, $J/\psi$, 
$\psi^\prime$ and $\Upsilon$ as well as in DVCS. 
No evidence has been found
for the exclusive production 
at the photon vertex
of particles with non-$1^{--}$ quantum numbers 
such as the $\pi^0$ \cite{Adloff:2002dw}, 
as would be expected
for the exchange of the postulated negative $C$-parity partner 
of the pomeron, the odderon \cite{Lukaszuk:1973nt}.

The most precise vector meson data are obtained
by reconstructing two-prong decays via charged decay products 
(notably $\rho^0 \rightarrow \pi^+ \pi^-$, $\phi \rightarrow K^+ K^-$
and $J/\psi \rightarrow e^+ e^-$ or $\mu^+ \mu^-$) and requiring no
further activity beyond the noise levels in the detector, except that
associated with the scattered beam electron. DVCS selections require a
similar lack of activity in the detector beyond the final state electron
and photon, the main 
experimental complication being separating the signal from
the competing purely electromagnetic Bethe--Heitler process, which generates
identical final states.   
In contrast to inclusive diffractive studies, the simplicity of these 
final states and the high precision of the tracking
detectors have allowed precision $t$ 
measurements  
without the need to 
tag the intact outgoing proton. 
This reconstruction method uses 
$|t - t_{\rm min}| = p_T^2 (V)$, where the vector meson transverse
momentum $p_T (V)$ is obtained from its decay products and $|t_{\rm min}|$,
the minimum kinematically accessible value of $|t|$, is usually negligible.

\subsubsection{General Characteristics of Vector Meson Production}
\label{vm:data}

Vector meson production has emerged as a sensitive probe of the transition
from the soft diffractive dynamics which are familiar from hadronic scattering
experiments to a harder regime which may be calculated perturbatively. 
The former regime is encountered wherever no hard scale is present, the 
classic example being $\rho^0$ 
photoproduction \cite{Aid:1996bs,Breitweg:1999jy,Breitweg:1997ed}. Under
such circumstances, the energy dependence of the photon--proton process
is in good agreement with the form predicted by Regge asymptotics: 
\begin{equation}
\sigma^{ \gamma p \rightarrow Vp} \propto 
(W^2)^{2 \alphapom(t) - 2} \ ,
\label{vm:energy}
\end{equation}
where the pomeron trajectory,
$\alphapom(t) = \alphapom(0) + \alphapom^\prime t$, is
assumed to be linear and its intercept 
$\alphapom(0) \simeq 1.085$ \cite{Donnachie:1992ny,Cudell:1996sh}. 
This has been 
found to work well for $\rho^0$ photoproduction data, though 
interestingly, the slope
of the pomeron trajectory has been found \cite{Breitweg:1999jy}
to be significantly smaller than the value of
$\alphapom^\prime \sim 0.25$ \cite{Donnachie:1984xq,Abe:1993xx} 
obtained from soft $pp$ and $p \bar{p}$ 
scattering. 
A possible explanation for this
may be found in process-dependent absorptive
corrections, 
which are absent in DIS, present to some extent in photoproduction and to 
a larger
extent in fully hadronic scattering. Detailed models of these effects 
can be found for example 
in \cite{Kaidalov:2003xf,Kaidalov:2009fp,Gotsman:2007pn}.
Further characteristics of this soft regime 
are \cite{Aid:1996bs,Breitweg:1997ed}
a skewed
lineshape for the $\rho$ meson due to its interference with non-resonant
$\pi^+ \pi^-$ production and a large value,
$B \sim 10 \ {\rm GeV^{-2}}$,
of the slope parameter describing the $t$ dependence according to
\begin{equation}
\frac{{\rm d} \sigma^{\gamma p \rightarrow Vp}}{{\rm d} t} = 
\left( \frac{{\rm d} \sigma^{\gamma p \rightarrow Vp}}{{\rm d} t} \right)_{t=0} \
e^{Bt} \ .
\label{eq:tslope}
\end{equation}
Interpreted in a simple optical model 
as a measurement of the mean impact parameter between the incoming
photon and proton, these $B$ values suggest that the interaction 
takes place over
a transverse distance of typically $1 - 1.5 \ {\rm fm}$. 

\begin{figure}[htp]
\begin{center}
~\epsfig{file=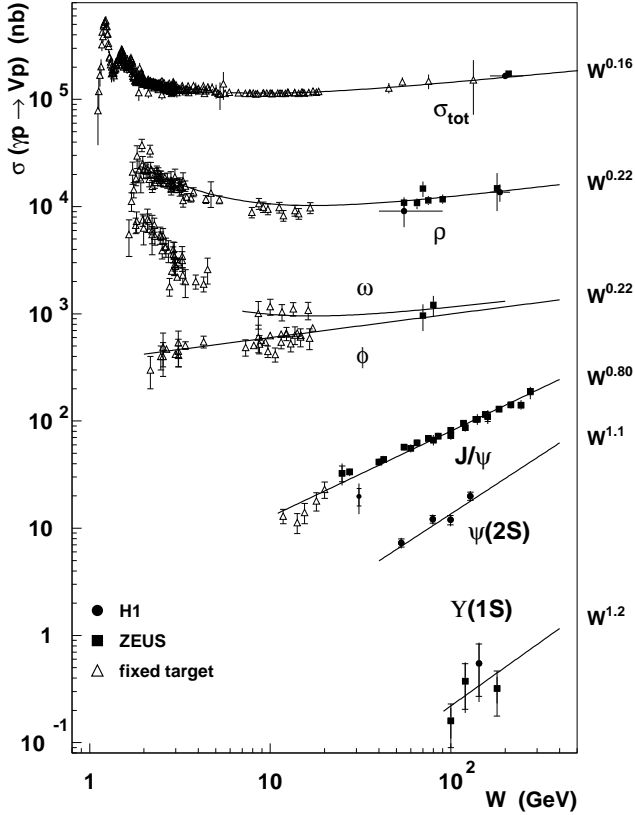,width=8.5cm}
\caption{Compilation of photoproduction cross section measurements as a function
of the $\gamma p$ centre-of-mass energy, $W$. The total cross section and 
various vector meson production cross sections are 
included, with the approximate power law dependences 
$\sigma \propto W^\delta$ indicated for each process.}
\label{vm:wdependence}
\end{center}
\end{figure}

Wherever hard scales are present, usually provided either
by heavy quarks in the vector meson or by large $Q^2$, but sometimes also
by large $|t|$, the qualitative picture changes. The energy dependence 
becomes progressively steeper, such that the $W$ dependence,
parametrised similarly to Eq.~\ref{vm:energy}, yields an increased
effective value of $\alphapom(0)$. This effect is illustrated in 
Fig.~\ref{vm:wdependence}, 
where a compilation of photoproduction data on the total
cross section and different vector meson production cross sections is 
shown as a function of $W$. 
The steepening of the dependence on $W$ for the heaviest vector mesons
can be interpreted in terms of the scale dependence of the proton
gluon density at low $x$, as discussed further below. 

\begin{figure}[htp]
\begin{center}
~\epsfig{file=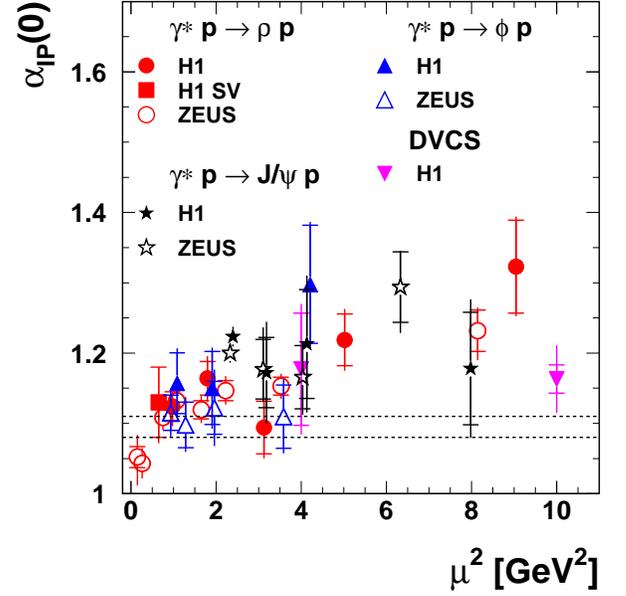,width=8cm}
\caption{Development of the effective 
pomeron intercept (see Eq.~\ref{vm:energy} and surrounding text)
with a characteristic scale, $\mu^2 = (Q^2 + M_V^2)/4$,  
derived from fits to the $W$ dependences 
of various vector meson production data,
as well as a DVCS measurement 
(shown at $\mu^2 = Q^2$). The dashed lines indicate
the range of values which are typical in soft diffraction.
From~\cite{Aaron:2009xp}.}
\label{vm:delta}
\end{center}
\end{figure}

The transition from soft hadronic to perturbative behaviour is 
neatly mapped out in a single process in $\rho^0$ electroproduction
data. As $Q^2$ increases, the $t$ slope parameter $B$ decreases, the 
$W$ dependence becomes steeper and the lineshape skewing disappears,
all in a manner which has been measured with good 
precision \cite{Adloff:1999kg,Aaron:2009xp,Chekanov:2007zr,Breitweg:1998nh}.
An example, also including comparisons with other 
exclusive vector meson production
processes, is shown in Fig.~\ref{vm:delta}. Here, the effective 
pomeron intercept (Eq.~\ref{vm:energy})
is shown as a function of a scale, which is chosen to be
$(Q^2 + M_V^2)/4$ (see Section~\ref{vm:dipole}).
As the scale increases, the effective pomeron
intercept
shifts from values typical
of soft hadronic scattering to values which are compatible with results
for the equivalent quantity 
$\alphapom(0) = 1 + \lambda$ in fits of inclusive low-$x$ 
HERA data to the form 
$F_2(x,Q^2) \propto x^{- \lambda(Q^2)}$ \cite{Adloff:2001rw,Chekanov:2008cw}. 

\begin{figure}[htp]
\begin{center}
~\epsfig{file=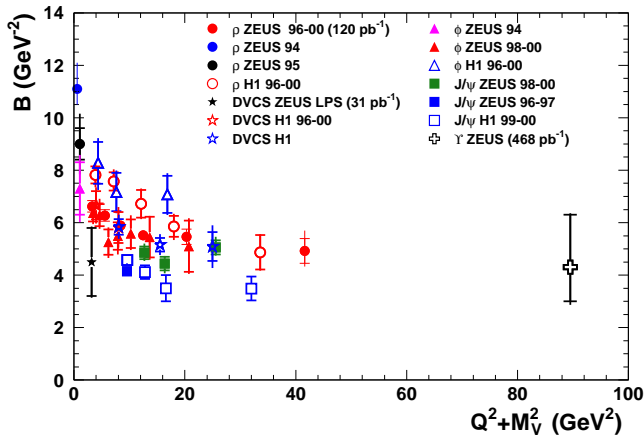,width=8.7cm}
\caption{Exponential $t$ slopes for vector meson electroproduction and 
photoproduction as a function of the characteristic scale
$Q^2 + M_V^2$. From~\cite{Abramowicz:2011fa}.}
\label{vm:tslopes}
\end{center}
\end{figure}

The exponential $t$ slopes of vector meson production processes are also 
found to vary systematically with scale 
and are approximately invariant in
$Q^2 + M_V^2$. A compendium of results is shown in 
Fig.~\ref{vm:tslopes}. Although the uncertainties are often large and
there is some scatter, the data suggest a convergence towards an 
asymptotic value of $B \sim 5 \ {\rm GeV^{-2}}$. 
In optical models, this
can be interpreted as
the point at which the physics is 
entirely short-distance in nature, 
the size of the probe becomes negligible and the slope parameter measures
the size of the proton. 
Quantitatively, this indicates an effective
proton size of around $0.6 \ {\rm fm}$, which is interestingly
smaller than 
the value of $\sim 0.8 \ {\rm fm}$ which is well measured
using electromagnetic probes. 
Interpreting vector meson production in terms of 
gluon exchange, this suggests that the
gluon radius of the proton may be smaller than its quark radius. 

\subsubsection{Vector Meson Production in QCD}
\label{vm:qcd}

The observed relationship between the $W$ dependences 
of inclusive DIS and vector meson production in the presence of a 
sufficiently large 
scale encourages a perturbative approach.
This has evolved considerably in the HERA era, to the point
where most observables can be successfully predicted. Although the 
basic quark charge counting SU(4) prediction for the ratio of cross sections
$\rho : \omega : \phi : J/\psi = 9:1:2:8$ holds approximately
when viewed as a function of the 
scale $(Q^2 + M_V^2)/4$ \cite{Aaron:2009xp}, 
vector meson wavefunction effects remain a significant source of
uncertainty. 

A QCD collinear factorisation 
theorem \cite{Collins:1996fb}, valid for the leading power
of $Q^2$ where $|t| \ll \Lambda_{\rm QCD}^2$ relates 
cross sections for
heavy vector meson production from longitudinally polarised photons
to the
generalised parton densities (GPDs) of the proton \cite{Diehl:2003ny}
(see Section~\ref{sec:dvcs}). 
Neglecting skewing effects, in which
the two exchanged partons carry different fractions of the proton
longitudinal momentum, the GPDs reduce to the 
square of the gluon density of the proton. The 
process is then driven to first approximation by the 
exchange of a pair of gluons from the proton structure, as illustrated in 
Fig.~\ref{VM:feynman}(b). 
Later approaches \cite{Martin:1996bp}
incorporated transversely
polarised photon cross sections at sufficiently large $Q^2$
and light vector mesons in a similar framework, with some degree of
success.

\begin{figure}[htp]
\begin{center}
~\epsfig{file=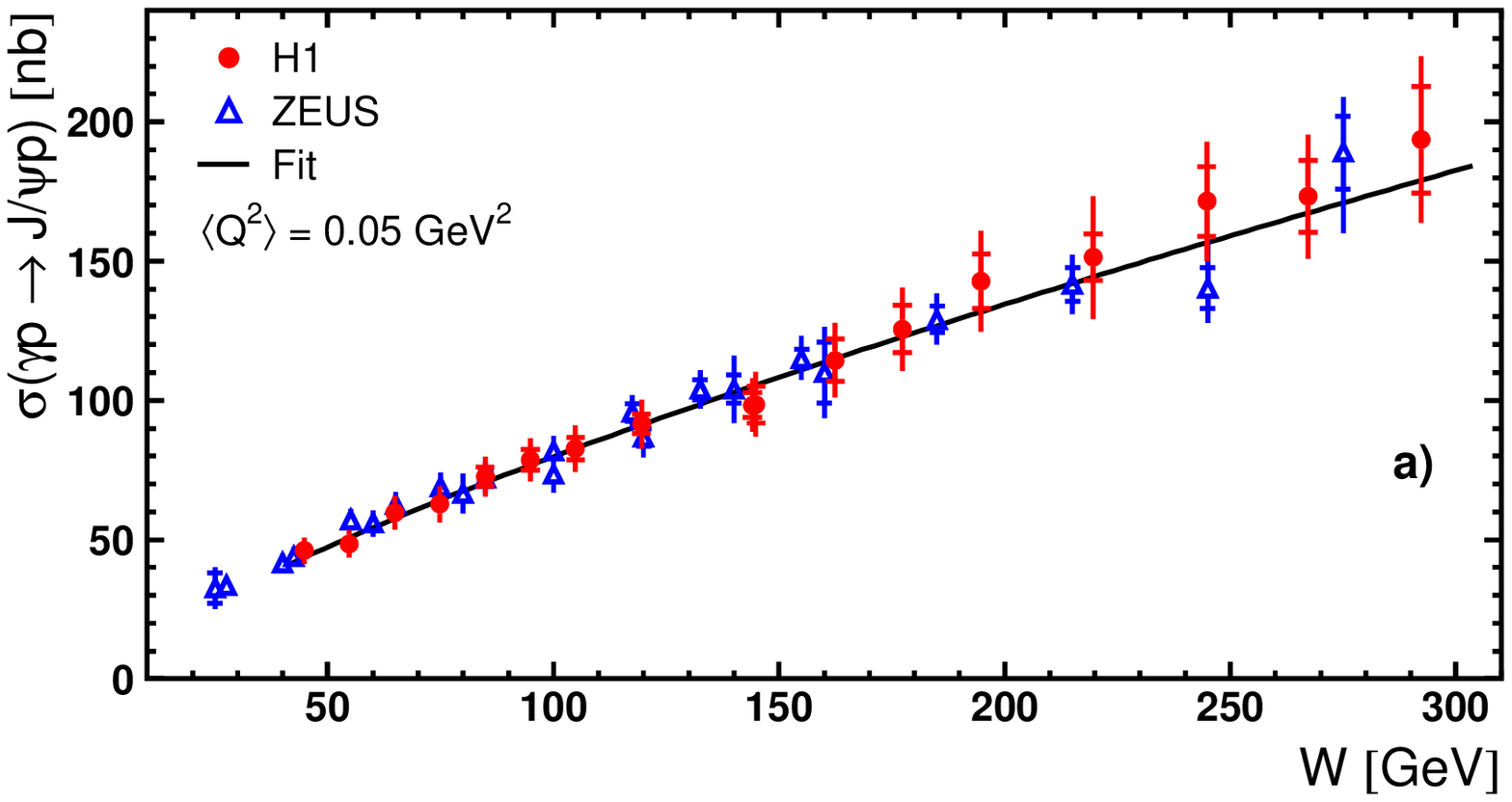,width=0.475\textwidth}
~\epsfig{file=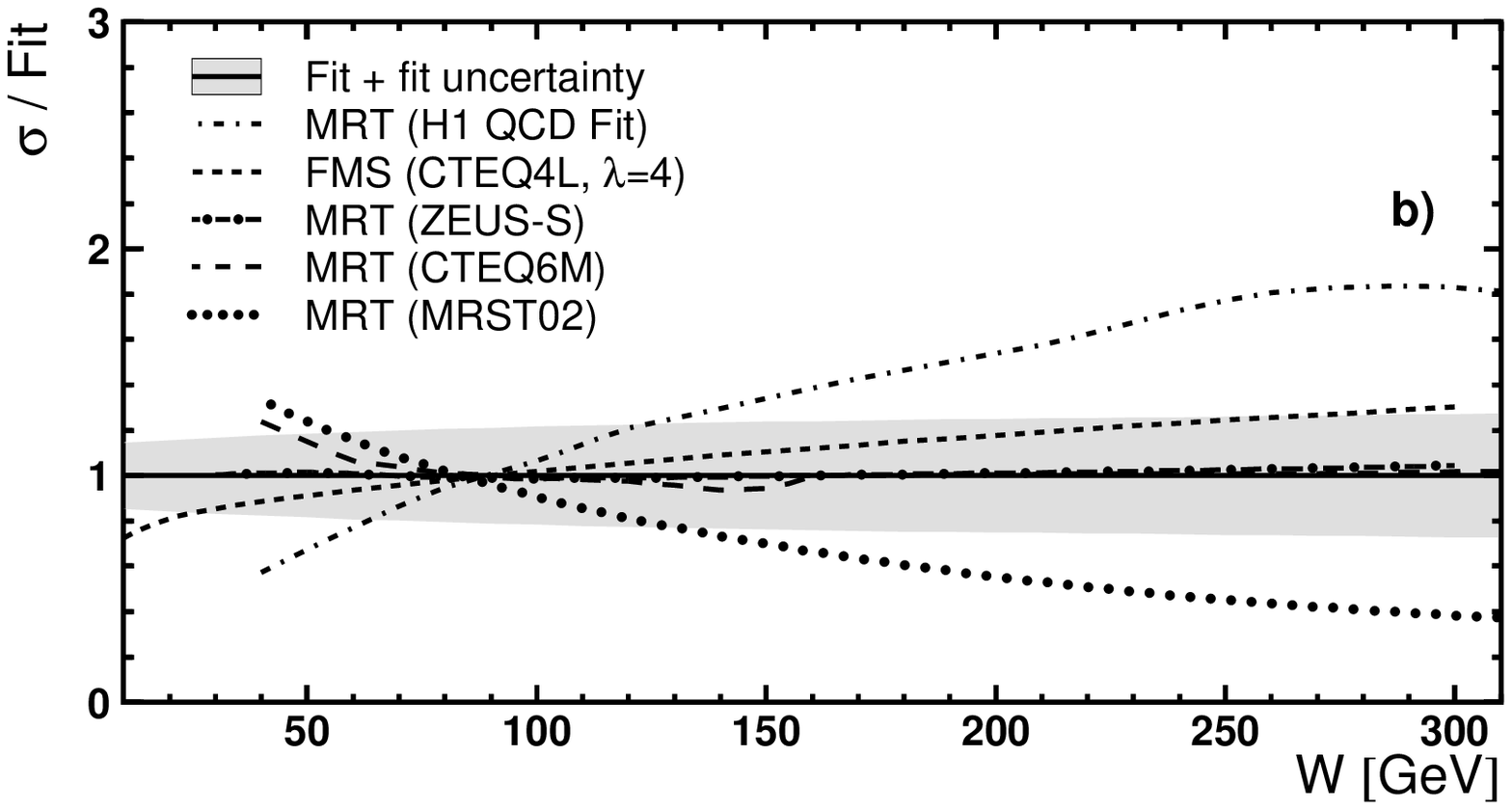,width=0.475\textwidth}
\caption{$J/\psi$ photoproduction cross section as a function of $W$,
compared with a QCD calculation, MRT \cite{Martin:1999wb}, based on 
four different parametrisations of the gluon density of the 
proton and a dipole model, FMS \cite{Frankfurt:2000ez}. 
The predictions are normalised to the data.
From~\cite{Aktas:2005xu}.}
\label{jpsi:wdep}
\end{center}
\end{figure}

Since it is relatively uncomplicated
theoretically, has a reasonably large scale, 
probes small $x$ values and is experimentally clean, 
$J/\psi$ photoproduction is an ideal testing ground for these ideas and,
with sufficiently strong theoretical understanding, a potentially
competitive means of extracting the gluon density of the proton. 
The many measurements at 
HERA \cite{Aktas:2005xu,Chekanov:2002xi,Alexa:2013xxa} have recently
been supplemented by the first measurements from ultra-peripheral
processes at the LHC, extending the $W$ range to beyond $1 \ {\rm TeV}$
for the forward kinematics of LHCb \cite{Aaij:2013jxj}. 
The $W$ dependence of the 
cross section for $J/\psi$ photoproduction has been calculated 
using the proton gluon density by a number of 
groups \cite{Martin:1999wb,Martin:2007sb,Frankfurt:2000ez}. An example 
comparison
with data is shown in 
Fig.~\ref{jpsi:wdep}.
From comparisons between the
predictions with different gluon densities, it is clear that there is
high sensitivity. However, 
the normalisation is not well predicted and 
there remains considerable model dependence.

More detailed comparisons with QCD models require an understanding of the 
helicity structure of vector meson production. The five independent
helicity amplitudes, describing transitions from either longitudinally
or transversely polarised photons to either 
longitudinally or transversely polarised vector mesons, can be extracted
from the production and decay angular distributions via spin density
matrix elements as parametrised in \cite{Schilling:1973ag}. 
Whilst helicity amplitudes for $J/\psi$ production 
at modest $|t|$ are consistent with 
$s$-channel helicity 
conservation \cite{Aktas:2005xu}, significant `helicity flip'
amplitudes, for which the vector meson helicity differs from that of the 
photon, are observed for the case of 
$\rho$ 
production \cite{Adloff:2002tb,Breitweg:1999fm,Aaron:2009xp,Chekanov:2007zr}
and for $J/\psi$ production at large 
$|t|$ \cite{Aktas:2003zi,Chekanov:2009ab}. 
This reflects the often unequal 
sharing of the longitudinal momentum between the two quarks,
corresponding to values of $z$ close to 0 or 1 in Fig.~\ref{Dipolefig}, 
particularly for the transverse polarisation case with $\rho$ mesons, 
This allows the helicity to flip at sufficiently 
large $t$, an effect which 
can be reasonably well predicted in QCD-based models. 

Probably the most important feature of angular distribution analyses is that
they allow an unfolding of the 
cross sections $\sigma_L$ and $\sigma_T$,
corresponding to vector meson production from 
longitudinally and transversely 
polarised vector mesons, respectively.
The $Q^2$ dependences of $\sigma_L$ and $\sigma_T$ are separately
predicted by QCD models. The basic leading order dependences
at $t = 0$ of $\sigma_L \propto 1/Q^6$ and 
$\sigma_T \propto 1/Q^8$ \cite{Brodsky:1994kf} are 
expected to be strongly
violated, in particular due to the $Q^2$ dependence of the
proton gluon density, but also due to the quark virtuality and the running of
$\alpha_s$ \cite{Frankfurt:1995jw}. The HERA data for
all vector meson species exhibit approximate scaling with 
power law dependences of approximately $\sigma_L \propto (Q^2 + M_V^2)^{2.1}$ 
and $\sigma_T \propto (Q^2 + M_V^2)^{2.9}$, which are reasonably
well reproduced by both collinear factorisation and dipole-based 
models \cite{Martin:1996bp,Kowalski:2006hc}. 

As discussed in Section~\ref{vm:data}
the variation of the slope parameter $B$ with centre-of-mass 
energy $W$ is usually parametrised in terms of the slope of
the pomeron trajectory $\alphapom^\prime$. Values close to zero
are measured with small uncertainties in some
hard exclusive processes, notably $J/\psi$ production at
large $|t|$ \cite{Aktas:2005xu,Chekanov:2002xi,Chekanov:2004mw}. 
This is consistent
with expectations in the BFKL 
approach \cite{Nikolaev:1995ed,Brodsky:1998kn}, 
where $\alphapom^\prime$
is related to the average transverse momentum of partons along the
gluon ladder. However, it is hard to draw quantitative conclusions in light
of the smaller-than-expected $\alphapom^\prime$
values for soft processes such as $\rho^0$ photoproduction 
(Section~\ref{vm:data}). 

\subsubsection{Deeply Virtual Compton Scattering and Generalised
Parton Densities}
\label{sec:dvcs}

Due to the underlying exchange of a pair of gluons, which in general
differ in both longitudinal and transverse momentum, 
hard exclusive DIS processes have emerged in recent years as
candidates with which to understand correlations between partons
in the proton and thus the transverse spatial, momentum 
and spin distributions
of partons. This information
is encoded in GPDs \cite{Diehl:2003ny}, which were introduced in
Section~\ref{vm:qcd}.
HERA data offer unique sensitivity to GPDs at low $x$. 
This topic is most commonly associated with the DVCS process, though
it is also relevant to vector meson production.

The dominant underlying parton level process for DVCS at low $x$
is similar to that shown in
Fig.~\ref{VM:feynman}(b), though there is also a 
`handbag' contribution in which the incoming (virtual) and 
outgoing (real) photons couple directly to quarks from the
GPDs. DVCS has the advantage of
avoiding the complication of the vector meson wavefunction, but 
cross sections are suppressed 
due to the final state photon coupling. 
Due to the smaller cross sections compared with their vector meson
counterparts, DVCS studies emerged relatively late in the lifetime
of HERA. 
For the
HERA kinematic region, integrated over azimuthal degrees of freedom, 
interference between DVCS and the competing Bethe--Heitler process
is small and 
DVCS cross sections can be extracted by statistically subtracting the
Bethe--Heitler component using Monte Carlo models tuned in regions in
which the DVCS contribution can be neglected. 

\begin{figure}[htp]
\begin{center}
~\epsfig{file=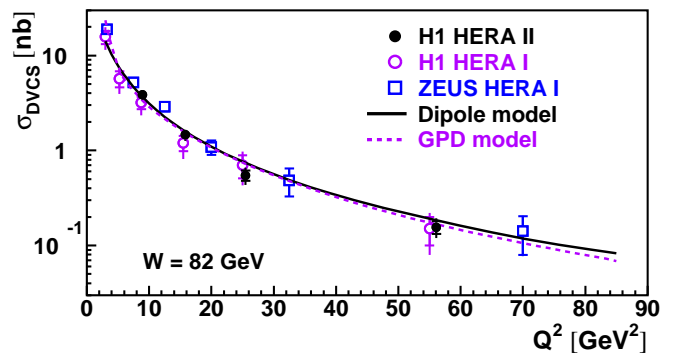,width=8.7cm}
\caption{Dependence of HERA DVCS cross section 
measurements on $Q^2$. 
The data are compared with a GPD-based model \cite{Kumericki:2009uq}
and also with a dipole-based 
model \cite{Marquet:2007qa}. See Section~\ref{vm:qcd} for 
an explanation of the latter. 
From~\cite{Aaron:2009ac}.}
\label{dvcs:fig}
\end{center}
\end{figure}

By the end of HERA operation, $Q^2$, $W$ and $t$ 
distributions had all been measured with good 
precision \cite{Chekanov:2008vy,Aaron:2009ac,Aaron:2007ab}.
As shown in Figs.~\ref{vm:delta} and~\ref{vm:tslopes},  
the exponential $t$ slope is $B \simeq 5 \ {\rm GeV^{-2}}$ and 
parametrisations of the cross section in the usual $W^\delta$
form yield $\delta \approx 0.5 - 0.6$. 
Example $Q^2$ dependence data are shown in Fig.~\ref{dvcs:fig}.
The DVCS cross section falls more 
slowly (roughly as $(Q^2)^{-1.5}$)
with photon virtuality than is the case for vector meson
production.  
This observation is in line with theoretical 
predictions \cite{Frankfurt:1997at} and is partially explained
by the presence of the quark-driven handbag contribution, which
is not relevant to vector meson production.  
Together, these observations 
indicate that DVCS can be considered as a hard process, even at 
relatively small $Q^2$.

Asymmetries in the azimuthal degree of freedom between the lepton and
the (virtual-to-real) photon scattering
planes in DVCS give access to the real part of the amplitude, and 
are most sensitive to the non-diagonal GPDs. 
The first H1 data have appeared on the DVCS 
beam charge asymmetry \cite{Aaron:2009ac}. It is clearly non-zero
and its azimuthal dependence 
has been 
used to confirm that the DVCS amplitude is dominantly imaginary for
HERA kinematics. 
It has not yet been possible to determine 
beam polarisation asymmetries in the manner successfully pursued
in the fixed target configuration by 
HERMES \cite{Airapetian:2012pg,Airapetian:2012mq}.
Nonetheless, the H1 and ZEUS 
data will provide essential low $x$ ingredients in models of GPDs
for a long time to come \cite{Kumericki:2009uq}, the principal
sensitivity being to the GPD, $H$.

\subsubsection{Exclusive Processes in Dipole Models and 
Low {\boldmath $x$} Saturation}
\label{vm:dipole}

Viewed in the proton rest frame, vector meson production 
naturally
factorises into three separate processes, occurring on 
distinct time-scales (\emph{c.f.}\ Figs.~\ref{Dipolefig} and~\ref{VM:feynman}(b)). 
The
incoming real or virtual photon first fluctuates into a $q \bar{q}$ pair,
which creates a colour dipole of transverse size 
$r_\perp \sim 1 / \surd Q^2$ \cite{Wusthoff:1997fz}. 
The dipole then scatters elastically from the proton, 
a process which is generically described in terms of a dipole cross section. 
Finally, the outgoing colour dipole combines back to a photon in the
DVCS case, or else 
hadronises to a final state vector meson
in a manner which depends on the non-perturbative vector meson wavefunction. 
The scale choice of $(Q^2 + M_V^2) / 4$, already used in the previous
sections, corresponds to the inverse of the 
scanning radius of the dipole--proton interaction, for the case where the 
incoming photon longitudinal momentum 
is shared equally between the quark and antiquark
forming the dipole \cite{Frankfurt:1995jw,Ivanov:2004ax}.
It is appropriate for heavy vector mesons and for light vector mesons
produced by longitudinally polarised photons. 

At small enough $r_\perp$, the dipole cross section
can be described perturbatively in the manner discussed 
in Section~\ref{vm:qcd}. At larger $r_\perp$, 
the dipole cross section is 
not known \emph{a priori}, but is
sensitive to any low $x$ 
saturation of the 
proton parton densities.
There are various parametrisations of the dipole cross section
which include saturation phenomena, 
some of which are
based on particular approximations to 
QCD with non-linear evolution such as the Colour Glass 
Condensate \cite{Iancu:2003ge}
and some of which are purely 
phenomenological \cite{GolecBiernat:1998js,GolecBiernat:1999qd,Forshaw:1999uf,Forshaw:2003ki}. 

\begin{figure}[htb]
\begin{center}
~\epsfig{file=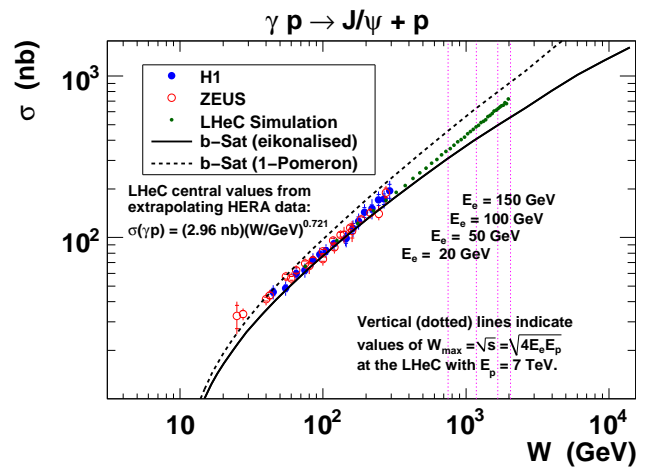,width=0.47\textwidth,bbllx=0,bblly=0,bburx=510,bbury=360,clip=}
\caption{Exclusive $J/\psi$ photoproduction data compared with versions
of the b-Sat dipole model \cite{Kowalski:2006hc} both 
with (eikonalised) 
and without (1-Pomeron) 
saturation effects included. An extrapolation into the
kinematic regime of a possible future $ep$ collider is also included. 
Figure from~\cite{AbelleiraFernandez:2012cc}.}
\label{LHeC:jpsi}
\end{center}
\end{figure}

In the most sophisticated dipole models of vector meson production,
the target matter density, and hence any saturation effects, 
depend on the distance from the centre
of the proton, 
such that the dipole cross section varies with
impact 
parameter \cite{Kowalski:2003hm,Kowalski:2006hc,Rezaeian:2012ji}, 
which in turn is 
closely related to $1/t$ (see also \cite{Marquet:2007qa} for a 
complementary approach). 
When viewed differentially in $t$, the influence of the saturation 
effects thus becomes stronger as $t$ 
increases \cite{Kowalski:2006hc,AbelleiraFernandez:2012cc}.
Such models adequately describe almost all
aspects of vector meson production and DVCS at HERA. An example
comparison with the inclusive $J/\psi$ photoproduction cross section
as a function of $W$, 
also illustrating the predicted size of the saturation effects, is
shown in Fig.~\ref{LHeC:jpsi}. 

\subsection{Inclusive Diffractive Dissociation}

\subsubsection{Kinematics and Experimental Selection of Diffraction}

\begin{figure*}[tb]
\centerline{\hspace*{0.1cm}
            {\Large{\bf{(a)}}}
            \includegraphics[height=0.3\textwidth]{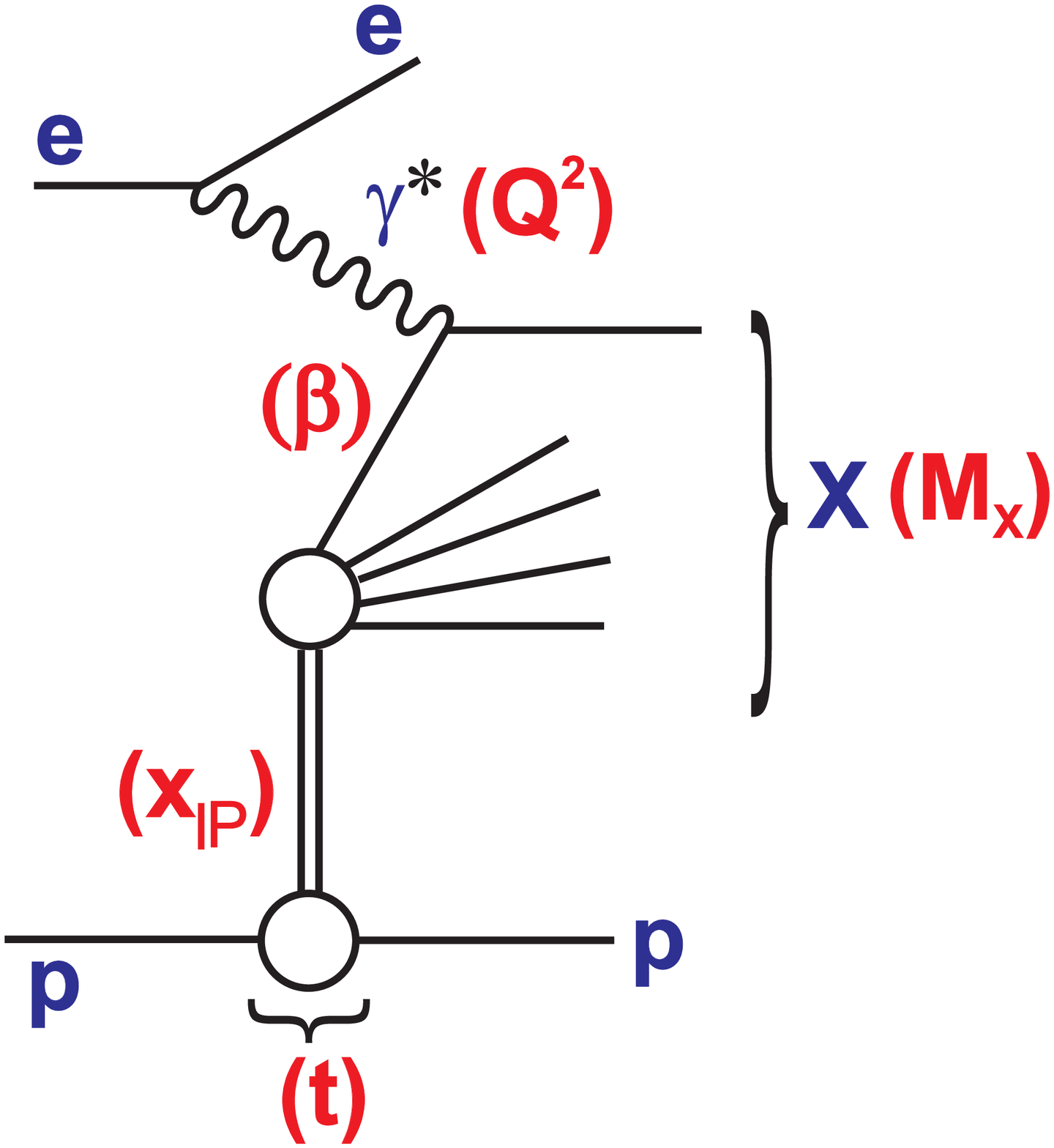}
            \hspace*{0.1cm}
            {\Large{\bf{(b)}}}
            \includegraphics[height=0.3\textwidth]{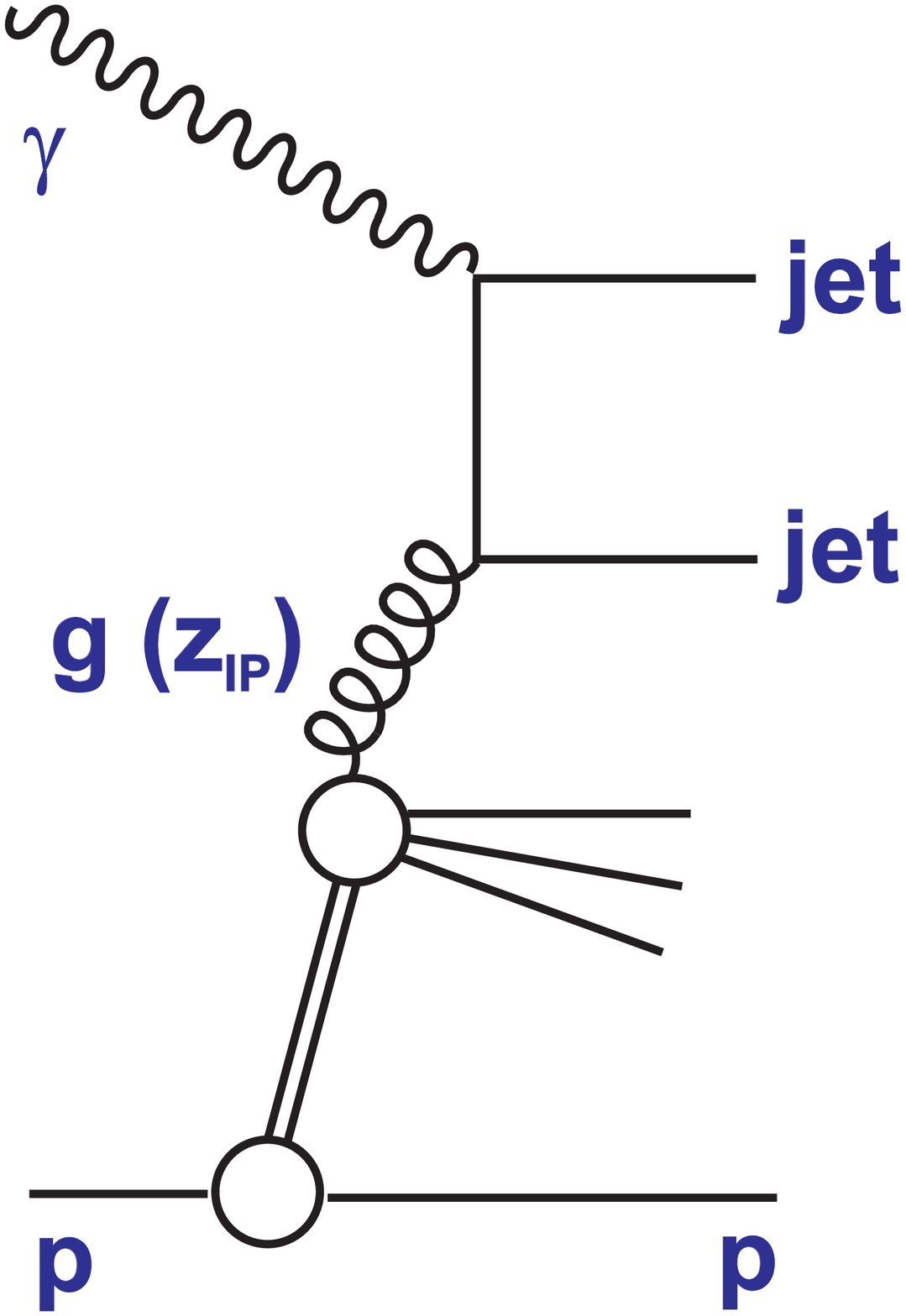}
            \hspace*{0.6cm}
            {\Large{\bf{(c)}}}
            \includegraphics[height=0.3\textwidth]{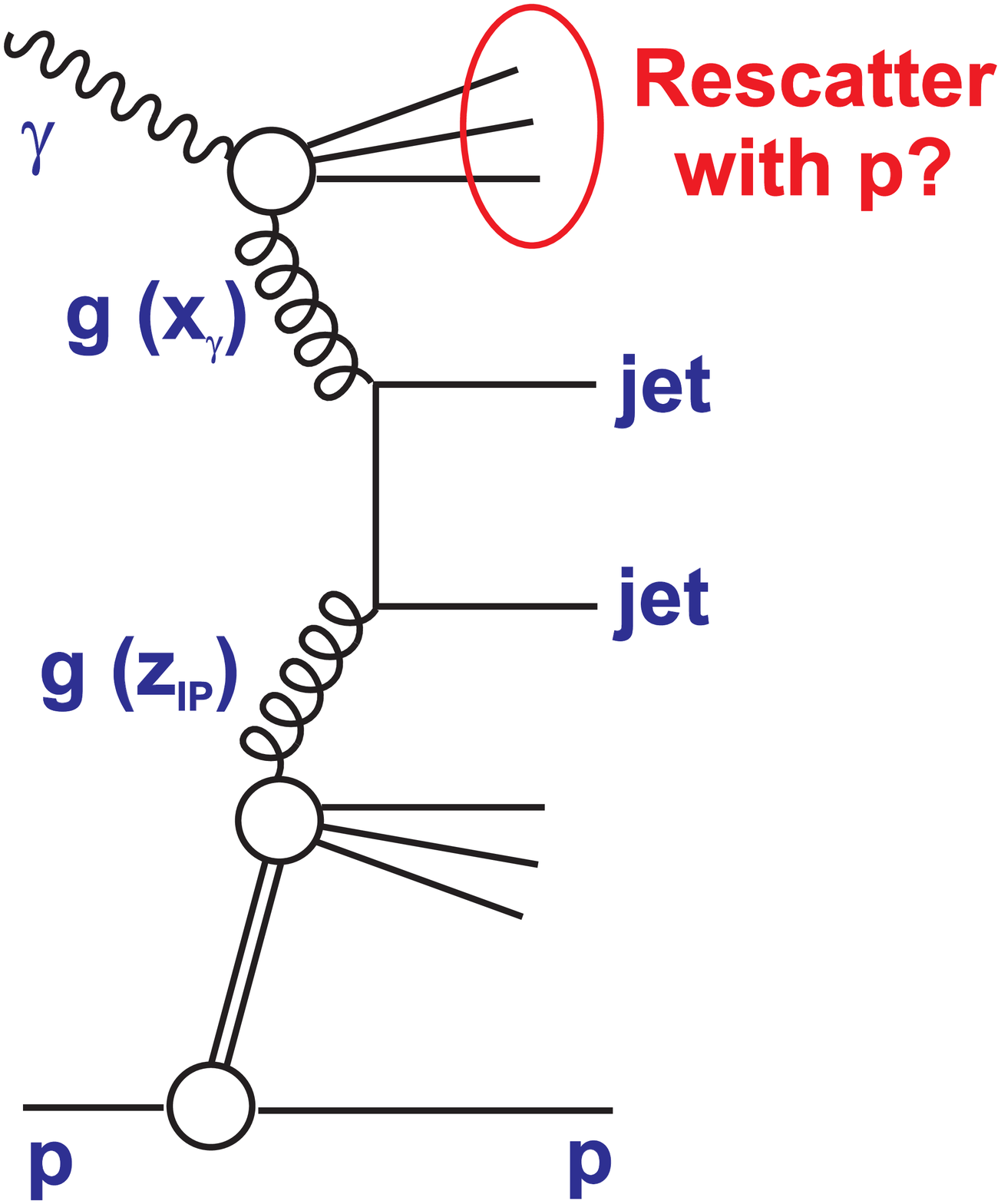}}
\caption{Sketches of diffractive $ep$ processes.
(a) Inclusive DDIS at the level of the quark parton model,
illustrating the kinematic variables 
discussed in the text. (b) Dominant leading order QCD diagram for 
hard scattering in DDIS or direct photoproduction, in
which a parton of momentum fraction $z_{I\!\!P}$ from the DPDFs
enters the hard scattering. (c) A leading
order process in resolved photoproduction involving a parton 
of momentum fraction $x_\gamma$ relative to the photon.}
\label{feynman}
\end{figure*}

The kinematic variables describing the inclusive
diffractive DIS (DDIS) process $ep \rightarrow eXp$ are 
illustrated in Fig.~\ref{feynman}(a).
In addition to $x$, $Q^2$ and $t \, (<0)$, there is one further 
non-trivial invariant, 
corresponding to the mass $M_X$ 
of the diffractive final state. 
In practice, the variable $M_X$ is usually expressed in terms of
\begin{equation}
\beta \;  = \; \frac{Q^2}{Q^2+M_X^2-t}\ .
\end{equation}
Small values of $\beta$ 
thus refer to events with diffractive masses much bigger 
than the photon virtuality, while exclusive processes have 
values of $\beta$ close to unity.
In models based on a factorisable 
pomeron (Section~\ref{factorisation}), $\beta$ may be interpreted as
the fraction of the pomeron longitudinal momentum 
which is carried by the struck parton.
The longitudinal momentum fraction
of the colourless
exchange with respect to the incoming
proton is denoted $x_{_{I\!\!P}}$, such that $\beta \, x_{_{I\!\!P}} = x$.

Experimentally, diffractive $ep$ scattering 
is characterised by the presence of a leading proton in the 
final state, retaining most of the initial state proton energy, and
by a lack of any other hadronic activity in the 
forward (outgoing proton) direction, such that the
system $X$ is cleanly separated and $M_X$ may be measured in the central
detector components.  
These signatures have been widely exploited at HERA to select 
diffractive events by tagging the outgoing proton
in the H1 Forward Proton Spectrometer or 
the ZEUS Leading Proton Spectrometer
(proton-tagging method, see Fig.~\ref{fig:h1-layout}) or
by requiring the presence of a large gap in the rapidity 
distribution of hadronic final state particles 
in the forward region (LRG method). 
In a third approach 
($M_X$ method~\cite{Breitweg:1998gc,Chekanov:2005vv,Chekanov:2008cw}), the inclusive
DIS sample is
decomposed into diffractive and non-diffractive contributions based
on their characteristic dependences on $M_X$.
Whilst the LRG and $M_X$-based techniques yield better
statistics than the LPS method, they suffer from systematic uncertainties 
associated with an admixture 
of proton dissociation to baryon states with small masses, 
typically $M_Y\, \lapprox\, 1.6 \ {\rm GeV}$, which 
is irreducible due to the limited forward detector acceptance.
Detailed comparisons between 
cross sections obtained by the LRG and proton 
tagging methods 
can be found in \cite{Chekanov:2008fh,Aaron:2010aa}.
Neither collaboration observes 
any evidence for deviations from a constant 
ratio of cross sections measured by the two methods, 
indicating that the ratio of probabilities for
the proton to scatter elastically and to undergo a low mass excitation 
is independent of the 
inclusive kinematic quantities. 

\subsubsection{Inclusive Diffraction Data}
\label{f2d}

Observations of diffractive
dissociation in DIS \cite{Derrick:1993xh,Ahmed:1994nw} 
and of jet production in diffractive 
photoproduction \cite{Ahmed:1994ui,Derrick:1994yg} and 
DIS \cite{Derrick:1994ze} 
were among the earliest HERA publications, 
based on rapidity gaps which were much larger than could conceivably
occur at significant rates in standard fragmentation 
models \cite{Andersson:1983ia}.
As increasingly larger datasets became available, 
the precision on the inclusive diffraction
cross sections improved correspondingly, until the final measurements covered
three-fold differential cross sections with a precision of a few 
percent \cite{Aaron:2012ad,Chekanov:2008fh} over large phase space regions. 

\begin{figure}[htp]
\begin{center}
~\epsfig{file=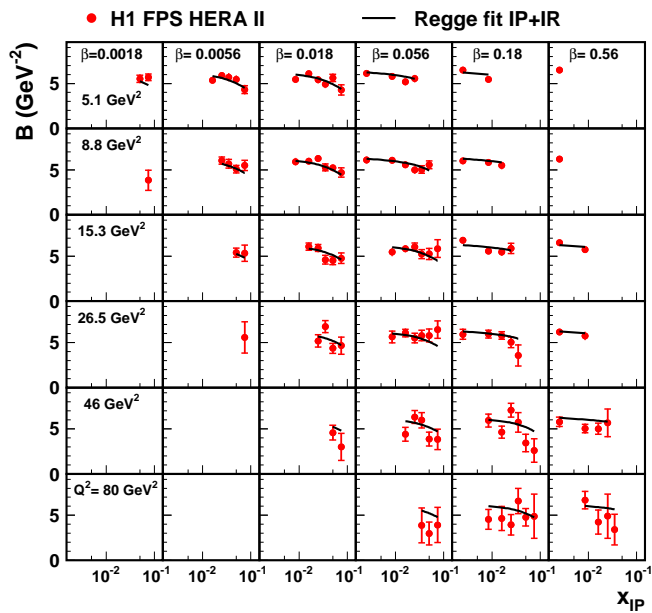,width=8.5cm}
\caption{Example data on the dynamics of the slope parameter, $B$,
describing the $t$ dependence of DDIS,
as measured using the H1 FPS.  From~\cite{Aaron:2010aa}.}
\label{fig:ddistslope}
\end{center}
\end{figure}

In analyses of data with protons tagged in 
the LPS and FPS, the $t$ dependence of DDIS has been measured and,
as in the case of exclusive vector meson production, found to
be compatible with an exponential parametrisation of the form of
Eq~\ref{eq:tslope}.
Example H1 data on the 
slope parameter $B$
are shown in Fig.~\ref{fig:ddistslope} \cite{Aaron:2010aa}.
Similar results, with globally slightly larger $B$ values, are
obtained by ZEUS \cite{Chekanov:2008fh}.
Beyond the low $M_X$ resonance region,
$B$ is typically around $6-7 \ {\rm GeV^{-2}}$,
independently of $\beta$ and $Q^2$, but with some dependence 
on $x_{_{I\!\!P}}$, which is further discussed in Section~\ref{sec:regge}. This is 
indicative of an 
approximate factorisation of the proton vertex from the virtual photon
vertex in DDIS (Section~\ref{factorisation}). 

The semi-inclusive DDIS cross section is usually presented in 
the form of a diffractive reduced cross section $\sigma_r^{D(3)}$,
related to the experimentally measured
differential cross section by
\begin{equation}
\frac{{\rm d}^3\sigma^{ep \rightarrow e X p}}{\mathrm{d} x_{_{I\!\!P}} \ \mathrm{d} x \ \mathrm{d} Q^2} = \frac{2\pi
  \alpha^2}{x Q^4} \cdot Y_+ \cdot \sigma_{r}^{D(3)}(x_{_{I\!\!P}},x,Q^2) \ ,
\label{sigmar}
\end{equation}
where $Y_+ = 1 + (1-y)^2$ and $y$ is the usual Bjorken variable.
The reduced cross section depends 
at moderate scales, $Q^2$, on two
diffractive structure functions
$F_2^{D(3)}$ and $F_L^{D(3)}$ according to
\begin{equation}
\sigma_r^{D(3)} =
F_2^{D(3)} - \frac{y^2}{Y_+} F_L^{D(3)}.
\label{sfdef}
\end{equation}
For $y$ not too close to unity,
$\sigma_r^{D(3)} = F_2^{D(3)}$ holds to very good approximation.
Measurements have also been made of the 
full four dimensional reduced cross section, dependent also 
on $t$ \cite{Chekanov:2008fh,Aaron:2010aa}. 
However, since the 
$t$ dependence factorises to good approximation, this degree of 
freedom is usually integrated out.
A more detailed exposition of the formalism 
of diffractive DIS can be found for example in \cite{Aktas:2006hy}.

\begin{figure*}[tb]
\begin{center}
~\epsfig{file=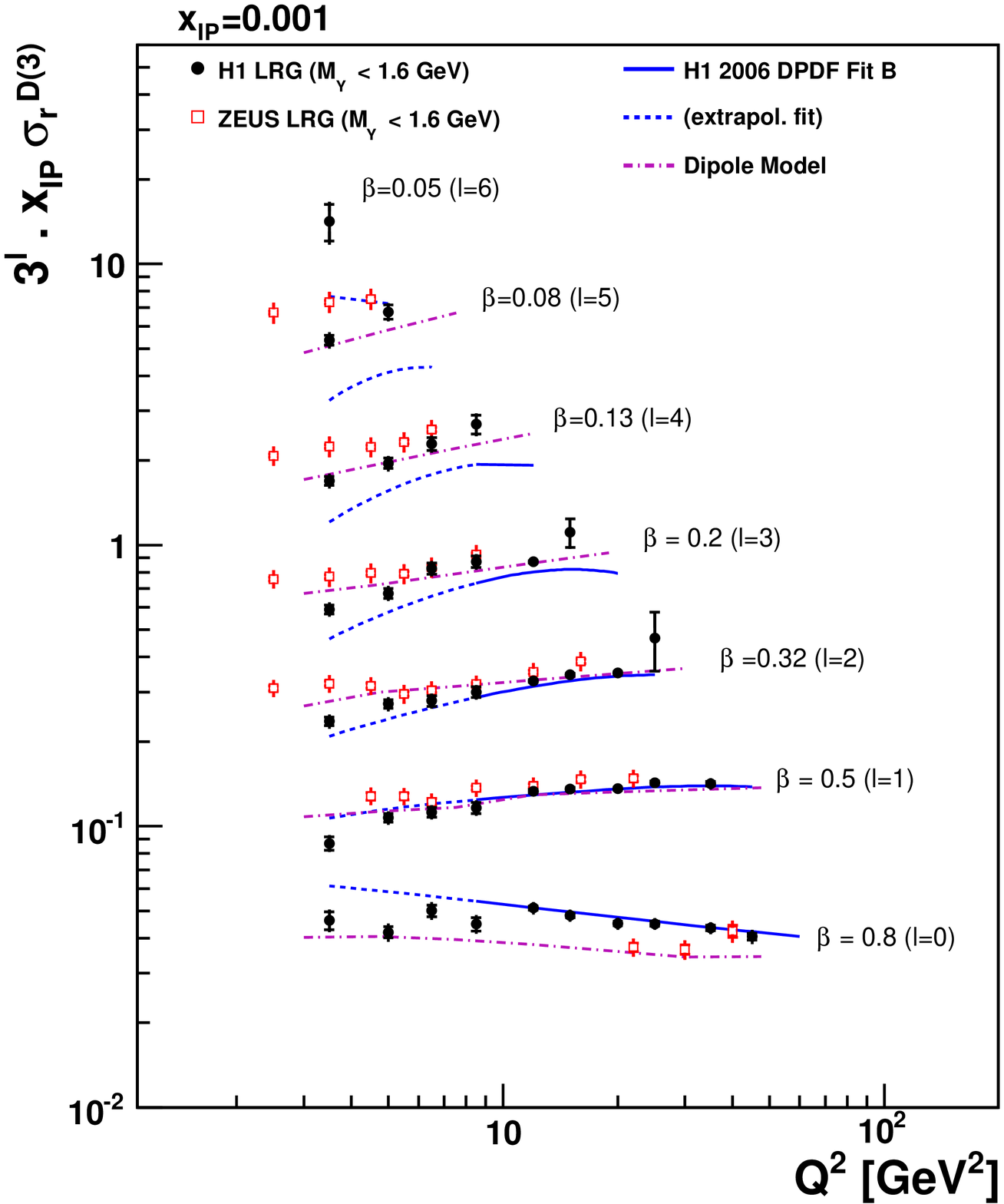,height=6.8cm}
~\epsfig{file=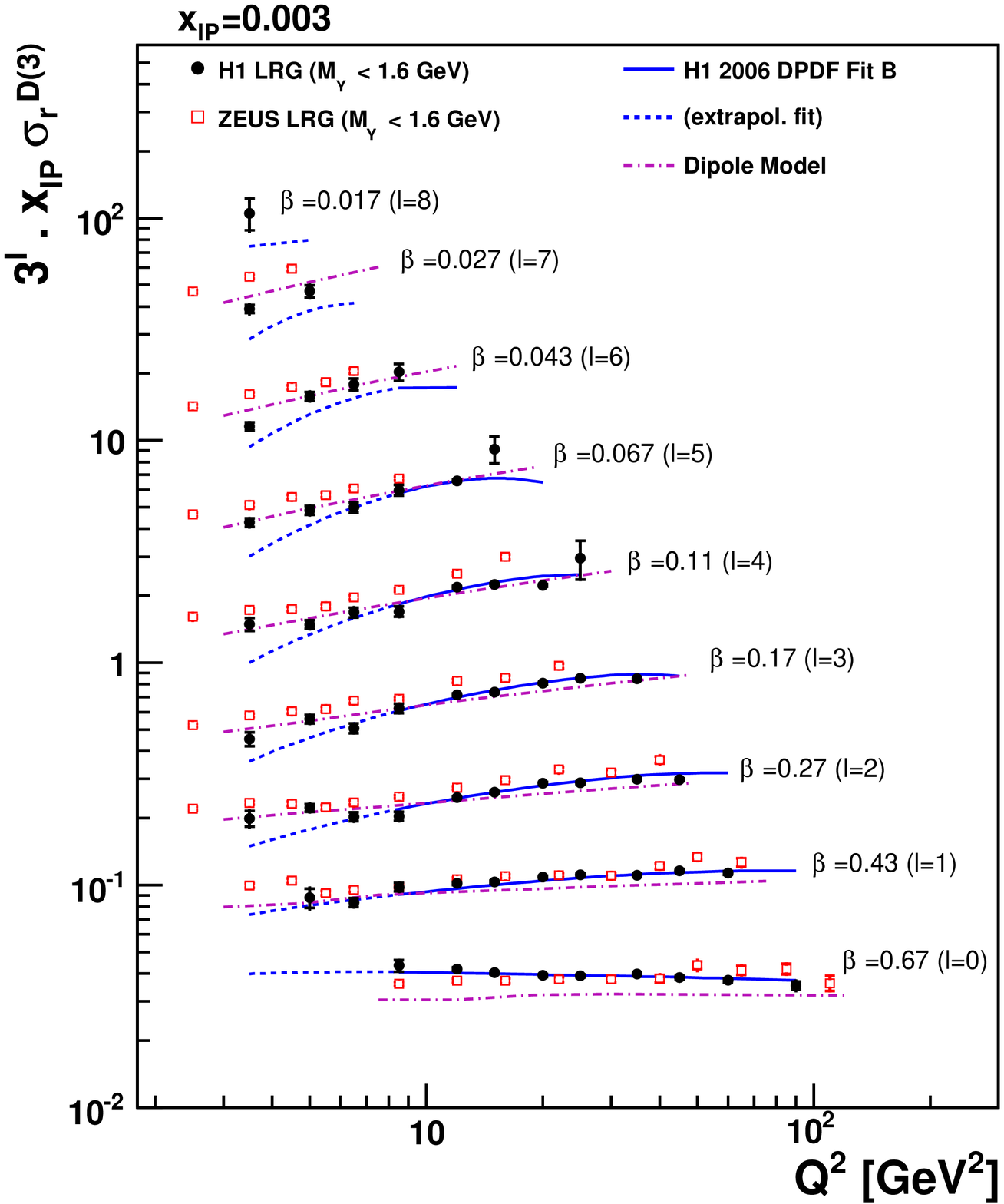,height=6.8cm}
~\epsfig{file=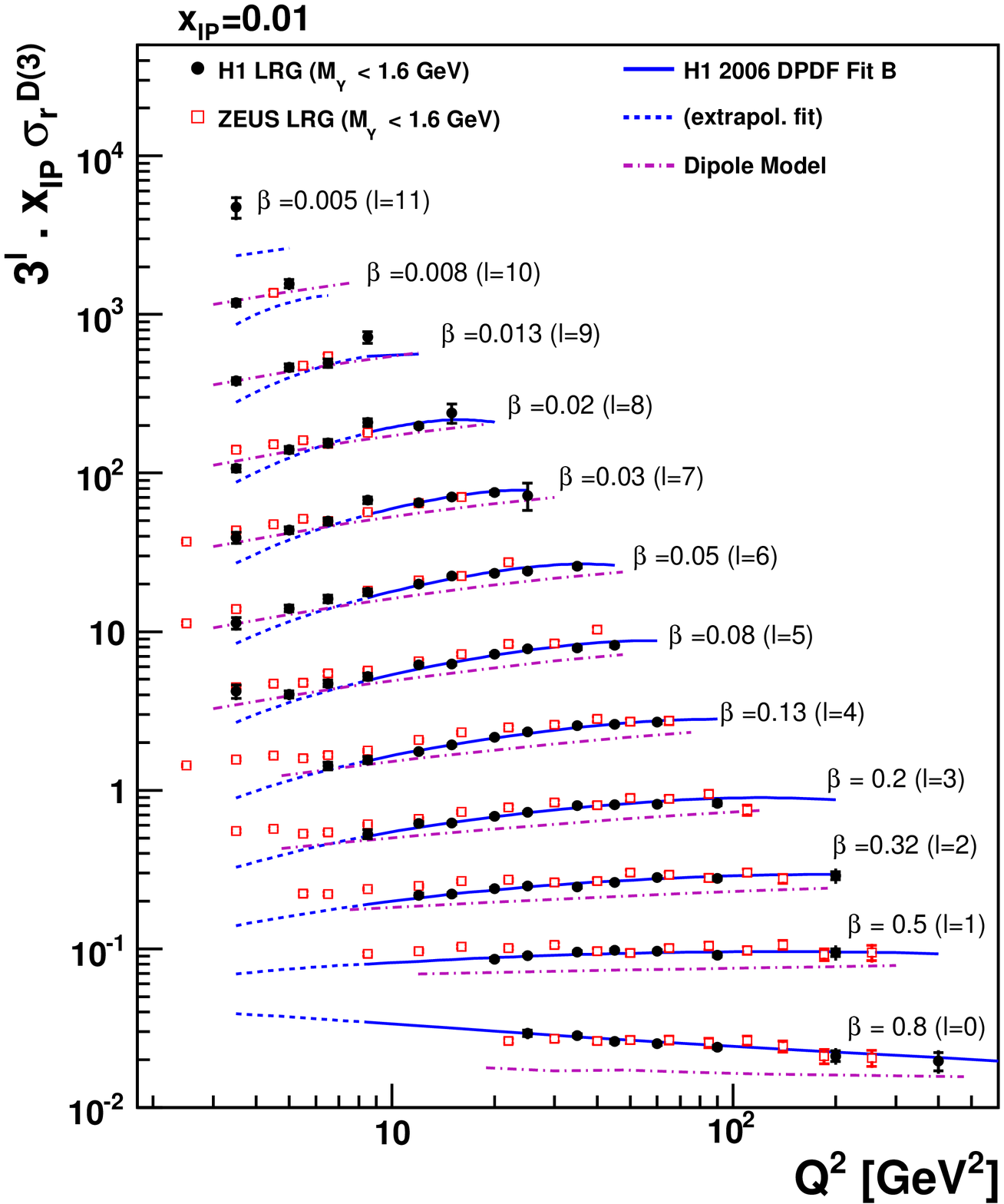,height=6.8cm}
\caption{Comparisons between H1 and ZEUS DDIS data measured using the
LRG method at
$\xpom = 0.001$, $\xpom = 0.003$ and $\xpom = 0.01$.
From~\cite{Aaron:2012ad}.}
\label{LRGcomp}
\end{center}
\end{figure*}

With the most recent $\sigma_r^D$ data, close 
agreement has developed between 
the H1 and ZEUS measurements. 
A first combination
of data from the two experiments, using the proton tagging method, 
has recently been published \cite{:2012vx}. 
The cross-calibration of the systematic 
uncertainties between the two experiments implicit in the averaging
procedure leads to an improvement in precision well beyond that
expected from statistical considerations alone.
Full combinations of data from the two collaborations for the LRG method
have yet to be performed, mainly due to some non-trivial discrepancies,
usually at the edges of the accessible phase space.
An investigation of the residual differences and crude combination may be 
found in \cite{Newman:2009km}. The comparison of final data in the
most precise region 
is illustrated in Fig.~\ref{LRGcomp}. 
Whilst there is good agreement in the bulk of the phase space, 
residual disagreements are apparent at small $\beta$ and
small $x_{_{I\!\!P}}$.
These discrepancies have thus far
prevented a combination of
the H1 and ZEUS data, though they have only a small influence on the
interpretation of the data discussed in the following. 

Various models have been used to interpret DDIS data. Since it has become 
standard, Sections~\ref{factorisation}\,--\ref{gap:survival} 
focus mainly on the `proton vertex
factorisation' approach, in which the diffractive exchange (`pomeron') has
a resolved structure. The 
relevance of DDIS to dipole models is
discussed separately in Section~\ref{sec:dipoles}.
It is worth noting that a third approach, 
in which rapidity
gaps are randomly produced from final state colour interactions in otherwise
standard DIS events, has been at least partially successful. This model
is attractive in its simplicity. Indeed an early 
incarnation \cite{Buchmuller:1995qa}
in which diffractive final states simply emerge randomly 
due to colour rotations in one ninth of all low $x$ DIS events works remarkably
well for inclusive data. Closely related ideas are encoded in 
the `Semi-classical model' \cite{Buchmuller:1997eb,Buchmuller:1998jv}.
The `Soft Colour Interaction' model \cite{Edin:1995gi}
offers a rather different, but again related, approach, 
which later developed into the `Generalised Area Law' \cite{Rathsman:1998tp}
model. All of these approaches have enjoyed success,
particularly in comparisons with $F_2^D$ and $\sigma_r^D$ data,
but none provides a comprehensive description, 
breaking down for example when confronted with more exclusive 
processes such as diffractive
dijet production \cite{Adloff:2000qi}.

\subsubsection{Factorisation Studies}
\label{factorisation}

Cleanly interpreting a three (or even four)-fold differential cross section is 
far from easy. It is simplified considerably if the 
dependences on the relevant kinematic variables can be 
factorised. QCD hard scattering collinear factorisation, when applied to
diffractive DIS \cite{Collins:1997sr}, implies that the cross section for the process 
$ep \rightarrow eXY$ can be written in terms of convolutions of 
partonic cross sections $\hat{\sigma}^{e i} (x, Q^2)$ with 
diffractive parton density functions (DPDFs) $f_i^D$ as
\begin{align}
{\rm d} \sigma^{ep \rightarrow eXY} (x, Q^2, \xpom, t) = 
\sum_i \ 
& f_i^D(x, Q^2, \xpom, t) \nonumber \\
 \otimes \
& {\rm d} \hat{\sigma}^{ei}(x,Q^2) \ ,
\label{equ:diffpdf}
\end{align}
with additional implicit dependences on $\mu_F$, $\mu_R$ and $\alpha_S$
as in Eq~\ref{eq:DISpert}. 
The partonic cross sections
are the same as those for inclusive DIS.
This factorisation formula is valid for 
sufficiently large $Q^2$ and fixed
$\xpom$ and $t$.

The experimental data over most of the
accessible phase space are compatible with 
the deeper `proton vertex' factorisation \cite{Ingelman:1984ns}, in which 
the DPDFs may be
factorised into a term containing only 
variables associated with
the virtual photon vertex ($\beta$, $Q^2$) and a term containing 
variables associated with the 
proton vertex ($\xpom$, $t$):
\begin{equation}
f_i^D(x,Q^2,\xpom,t) = f_{\pom/p}(\xpom,t) \cdot
f_i (\beta=x/\xpom,Q^2) \, .
\label{reggefac}
\end{equation}
This is equivalent to considering
the diffractive exchange as a `pomeron'
with a partonic structure given by the parton
distributions $f_i (\beta,Q^2)$, the variable $\beta$ 
corresponding to the fraction of the pomeron longitudinal momentum
carried by the struck quark. The 
`pomeron flux factor' $f_{\pom/p}(\xpom,t)$ represents the
probability that a pomeron with particular values of $\xpom$ and $t$
couples to the proton. 
Whilst this approach cannot be valid to ultimate precision (e.g.\ it
is clearly violated in the high $\beta$ region where higher twist processes
such as exclusive vector meson production play an important
role), it is justified theoretically for the leading
twist case \cite{Blumlein:2001xf}.

As was first shown in \cite{Adloff:1997sc}, a good description of the data
at relatively large $\xpom$ cannot be obtained by considering pomeron exchange alone. 
As in the case of total hadronic cross sections \cite{Donnachie:1992ny}, 
sub-leading
exchanges become important away from the asymptotic ($\xpom \rightarrow 0$ 
in this case) limit. 
An additional sub-leading
exchange ($\reg$) is therefore usually considered in addition, 
contributing at low $\xpom$ and $\beta$ 
and exhibiting a similar factorisation to the pomeron term, such that 
Eq.~(\ref{reggefac}) is modified to
\begin{align}
f_i^D(x,Q^2,\xpom,t) & =  f_{\pom/p}(\xpom,t) \cdot f_i (\beta,Q^2) \nonumber \\
& + n_\reg \cdot f_{\reg/p}(\xpom,t) \cdot f_i^\reg (\beta,Q^2) \, ,
\label{reggefac2}
\end{align} 
where $n_\reg$ sets the relative normalisation of the sub-leading term.
Further investigations of the sub-leading trajectory contribution can be found in 
Section~\ref{sec:sublead}. 

With the above ansatz, a rather complete description is obtained of
all HERA inclusive diffractive data. 
The energy dependence of DDIS is encoded in the
pomeron flux factor, which is parametrised based on Regge phenomenology,
as discussed further in Section~\ref{sec:regge}. The DPDFs are treated in
a similar way to the case of inclusive DIS, as described in 
Section~\ref{dpdfs}. 
Full details of the standard fitting scheme adopted for the most recent data
by both collaborations can be found 
in \cite{Aktas:2006hy,Aktas:2006hx}. An improved heavy flavour
treatment is described in \cite{Chekanov:2009aa}.

\subsubsection{Energy Dependences and Soft Phenomenology}
\label{sec:regge}

In the fits to DDIS data described in Section~\ref{factorisation}, 
the $\xpom$ and $t$ dependences are
parametrised using a flux factor 
motivated by Regge theory,
\begin{eqnarray}
f_{\pom/p}(\xpom, t) = A_\pom \cdot 
\frac{e^{B_\pom t}}{\xpom^{2\alphapom (t)-1}} \ .
\label{eq:fluxfac}
\end{eqnarray}
The parameters
$B_\pom$ and $\alphapom^\prime$ and their uncertainties are obtained from
fits to FPS or LPS data \cite{Aktas:2006hx}, which also take sub-leading contributions
into account. The $\xpom$ dependence of the data principally determines the
effective pomeron intercept $\alphapom (0)$, appropriate to the DDIS data. The 
most precise determinations are obtained from LRG data, with remarkable 
consistency
between statistically independent data sets: 
\begin{align}
\alphapom(0) = 1.118 \ \pm 0.008  \ {\rm{(exp.)}}  
& ^{+ 0.029}_{- 0.010} \ {\rm{(model)}} \nonumber \\ 
& \text{\cite{Aktas:2006hy}} \, ; \nonumber  \\
\alphapom(0) = 1.117 \ \pm 0.006 \ {\rm{(exp.)}}   
& ^{+ 0.022}_{- 0.007} \ {\rm{(model)}} \nonumber \\
& \text{\cite{Chekanov:2008fh}} \, ; \nonumber \\
\alphapom(0) = 1.113 \ \pm 0.002 \ {\rm{(exp.)}}   
& ^{+ 0.029}_{- 0.015} \ {\rm{(model)}} \nonumber \\
& \text{\cite{Aaron:2012ad}} \, , \nonumber 
\end{align}
the model dependence uncertainties being highly correlated between
the different data sets. 

These $\alphapom(0)$ 
values are only slightly larger than their counterparts obtained
from fits to 
total \cite{Donnachie:1992ny,Cudell:1996sh}
and diffractive \cite{Goulianos:1982vk,Aad:2012pw}
cross sections in $pp$ and $p \bar{p}$ scattering 
and are compatible with results from soft photoproduction at 
HERA \cite{Adloff:1997mi,Breitweg:1997za}. They are 
significantly smaller than the results obtained from the
energy dependences of hard exclusive processes at HERA 
(Section~\ref{vm:data}) and are 
much smaller than values predicted
based on leading-logarithmic 
BFKL \cite{Balitsky:1978ic,Kuraev:1976ge,Kuraev:1977fs} 
or other `hard pomeron' \cite{Donnachie:1998gm} approaches. 
This supports the picture of the dominating process
in DDIS being the application of a hard virtual photon probe 
to an approximately factorisable
exchange object, closely related to the pomeron of 
soft hadronic scattering.

The slope 
of the effective DDIS
pomeron trajectory can be determined from the $\xpom$ dependence of
the slope parameter $B$, as obtained using the LPS and FPS detectors. In this case, the
best values obtained to date are:
\begin{align}
\alphapom^\prime  = -0.01 \ \pm 0.06 \ {\rm{(stat.)}}  
& ^{+ 0.04}_{- 0.08} \ {\rm{(syst.)}} \ 
\pm 0.04 \ {\rm{(model)}} \nonumber \\
& \text{\cite{Chekanov:2008fh}} \, ; \nonumber \\
\alphapom^\prime  = \hspace{0.27cm} 0.04 \ \pm 0.02 \ {\rm{(exp.)}}   
& ^{+ 0.08}_{- 0.06} \ {\rm{(model)}} \nonumber \\
& \text{\cite{Aaron:2010aa}} \nonumber .
\end{align}
Whilst the precision on these results is limited, the trajectory slope is
clearly smaller than the canonical value from 
soft $pp$ and $p \bar{p}$ scattering of
$\alphapom^\prime \sim 0.25$ \cite{Donnachie:1984xq,Abe:1993xx}.
As discussed in Section~\ref{vm:data},
soft vector meson photoproduction data also show lower
$\alphapom^\prime$ values than those usually taken from $pp$ and $p \bar{p}$ 
scattering \cite{Breitweg:1999jy},
suggesting that this is a heavily process dependent parameter,
which is highly sensitive to absorptive corrections. 

\begin{figure}[htp]
\begin{center}
~\epsfig{file=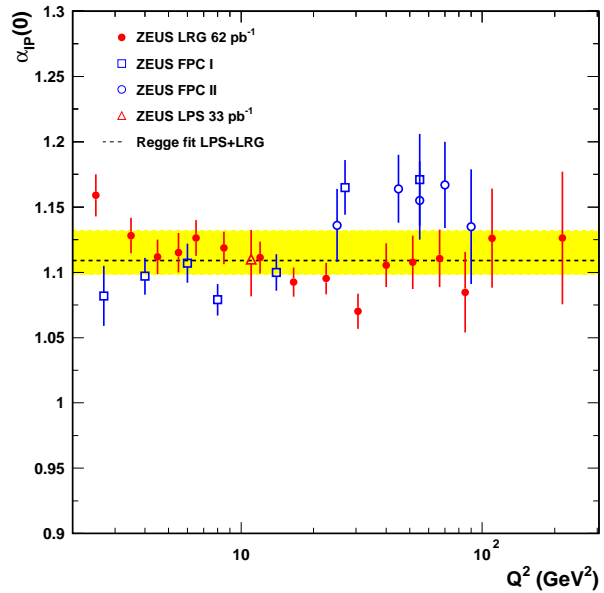,width=9cm}
\caption{Summary 
of ZEUS data on tests of proton vertex factorisation in the form 
of possible $Q^2$ variations of the pomeron intercept in 
fits to LRG and LPS data, 
as well as `FPC' data obtained by the $M_X$ decomposition method. From~\cite{Chekanov:2008fh}.}
\label{alphapom:fac}
\end{center}
\end{figure}

The validity of the factorisation assumption implicit in the fits to 
inclusive diffractive
data is tested by adding further free parameters, which allow the 
pomeron intercept and the
slope parameter $B$ to vary freely between different 
$Q^2$ or $\beta$ bins. A summary of ZEUS
results on possible variations of $\alphapom(0)$ with $Q^2$ is shown in 
Fig.~\ref{alphapom:fac}. Both here and in corresponding H1 studies 
(see e.g.~\cite{Aaron:2012ad}), there is no evidence for any variation of the 
effective pomeron intercept with $Q^2$ or $\beta$ when considering 
data taken using the 
LRG or LPS/FPS methods. 
Similar searches for a $Q^2$ or $\beta$ 
dependence of $\alphapom^\prime$
or $B$ have also yielded null results \cite{Aaron:2010aa}, 
albeit with lesser precision than for the $\alphapom(0)$ case. 

\begin{figure}[htp]
\begin{center}
~\epsfig{file=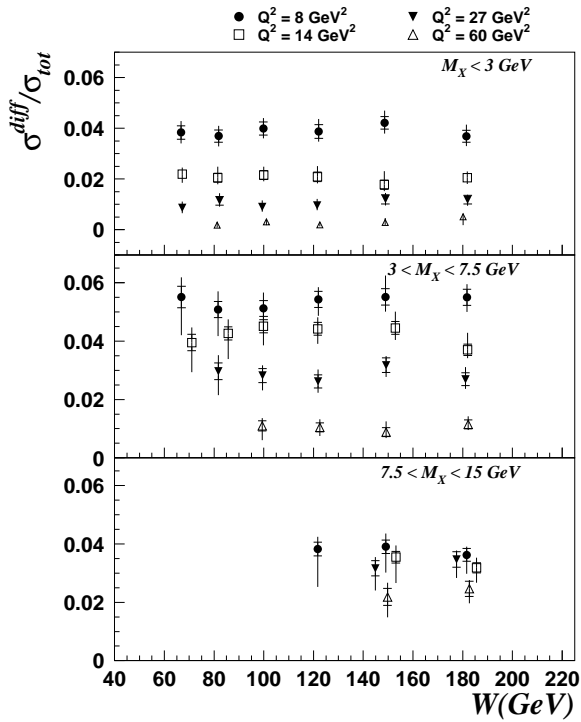,width=9cm}
\caption{The ratio of diffractive to inclusive cross sections 
as a function 
of $W$, integrated over different $M_X$ regions at 
fixed $Q^2$. Obtained using ZEUS 1994 data.  From~\cite{Breitweg:1998gc}.}
\label{zeus:difratio}
\end{center}
\end{figure}

It is informative to compare the photon--proton centre-of-mass energy 
($W$ or equivalently
$1/x$ at fixed $Q^2$) dependences of the diffractive and the inclusive cross
sections. Basic Regge pole phenomenology predicts that the growth 
with centre-of-mass energy of the
diffractive cross section
($\sim (W^2)^{2 \alphapom(t) - 2}$ at fixed $Q^2$ and $\beta$, with 
$\langle t \rangle \sim 1/B \sim 0.2\,{\rm GeV^2}$) 
should be 
faster than that of the total cross section ($\sim (W^2)^{\alphapom(0) - 1}$). 
However, in numerous
studies of HERA data, the ratio 
of diffractive to inclusive cross sections
has been found
to be relatively flat as a function of $W$. 
The first example \cite{Breitweg:1998gc} is shown in Fig.~\ref{zeus:difratio}.  See also
\cite{Aktas:2006hy,Aaron:2012ad} for more recent results. 
This flatness represents a clear breakdown of the 
simple Regge approach, showing 
conclusively that there is no universal pomeron 
in virtual photon--proton scattering, as already
discussed in Section~\ref{vm:data}. The flatness of the ratio has been 
interpreted \cite{GolecBiernat:1998js,GolecBiernat:1999qd}
as evidence
for saturation of low $x$ parton densities, which 
is expected to be
visible in diffraction at
higher $x$ than in the inclusive case, due to the exchange of two, 
rather than one, gluon in 
the simplest interpretation. 
Any saturation effects thus tend to reduce the diffractive cross section
relative to the inclusive in a manner which 
becomes more important as $x$ falls or $W$ grows.
Whilst this is a highly suggestive observation, 
it has also been explained without invoking saturation, 
for example in models which explain rapidity gap formation as
a random process, due to  
soft colour rearrangements \cite{Buchmuller:1998jv,Edin:1995gi}.

\subsubsection{QCD Phenomenology and Diffractive Parton Densities}
\label{dpdfs}

A cursory glance at Fig.~\ref{LRGcomp} immediately
leads to the conclusion that the scaling violations in diffraction remain of
positive sign ($\partial \sigma_r^D / \partial Q^2 > 0$) up to large values of  
$\beta\, \sim\, 0.5$, which may be compared with $x \sim 0.1$
for the
inclusive cross section.\footnote{See \cite{Aktas:2006hy}
for an alternative comparison between scaling violations in 
diffractive and inclusive DIS considered at the same $x$ values in
each case.} 
This is indicative of a 
large role of gluons in the diffractive exchange \cite{Ahmed:1995ns}. 
This qualitative conclusion
is formalised through the extraction of DPDFs, as described in more detail
in this section. 

The DPDFs in Eq.~\ref{equ:diffpdf}
represent probability distributions for partons $i$ in the proton
under the constraint that the proton is scattered to a 
particular system $Y$ with a specified four-momentum. 
They are essentially equivalent to the fracture functions developed
in an earlier approach \cite{Trentadue:1993ka}.
They are not known
from first principles, but 
can be determined from fits to the data using
the DGLAP evolution
equations~\cite{Blumlein:2001xf}.
Due to kinematic constraints,
it is not possible to access the full range of $\beta$ and $Q^2$
using data from only one value of $\xpom$ at HERA.
The parametrisations of the $\xpom$ 
dependence of the DPDFs described in Section~\ref{sec:regge}, together with the 
proton vertex factorisation assumption introduced in 
Section~\ref{factorisation},
is therefore adopted, such that data from multiple $\xpom$ values can 
be used simultaneously in extracting DPDFs which then depend on
$\beta$ and $Q^2$ only.

DPDFs have been extracted using standard NLO QCD procedures, similar to those 
employed to extract inclusive proton PDFs, but with smaller numbers of free parameters
for the momentum fraction dependence of the parton densities at the starting
scale for evolution. This momentum fraction is usually denoted $z$ or 
$z_\pom$ (it is equal to $\beta$ when used to describe the quark coupling to the
exchange boson). In the fits made to date, only two distinct parton densities
are considered: a gluon density and a singlet quark density, which is assumed
to be flavour symmetric 
between up, down and strange quarks and their antiquarks. 
This latter assumption, particularly on the size of the strange quark density, 
is rather ad hoc and remains to be tested in detail.

\begin{figure}[htp]
\begin{center}
~\epsfig{file=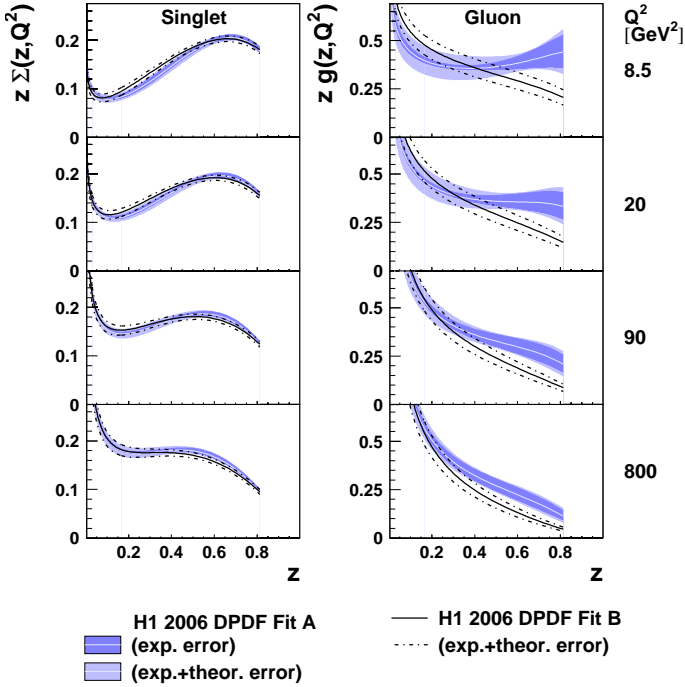,width=9cm}
\caption{DPDFs extracted from an H1 fit to inclusive LRG data 
only. The two different sets of DPDFs shown 
(Fit A and Fit B) have very similar fit qualities, illustrating the
lack of sensitivity to the gluon density at large momentum fractions. 
From~\cite{Aktas:2006hy}.}
\label{H1:DPDFs}
\end{center}
\end{figure}

\begin{figure}[htb] \unitlength 1mm
  \begin{center}
    \begin{picture}(60,30)
      \put(-15,-3){\epsfig{file=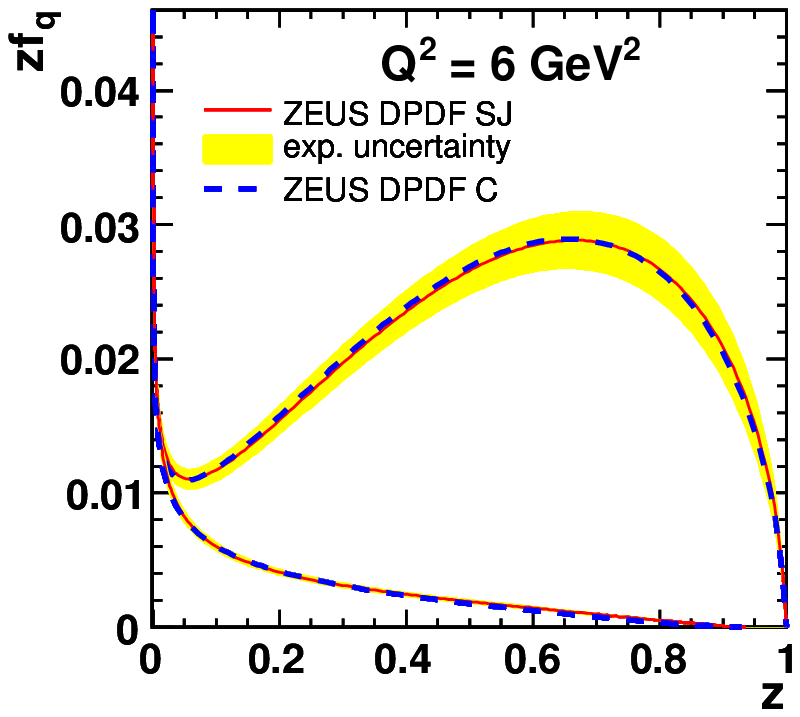,height=0.22\textwidth,clip}}
      \put(30,-3){\epsfig{file=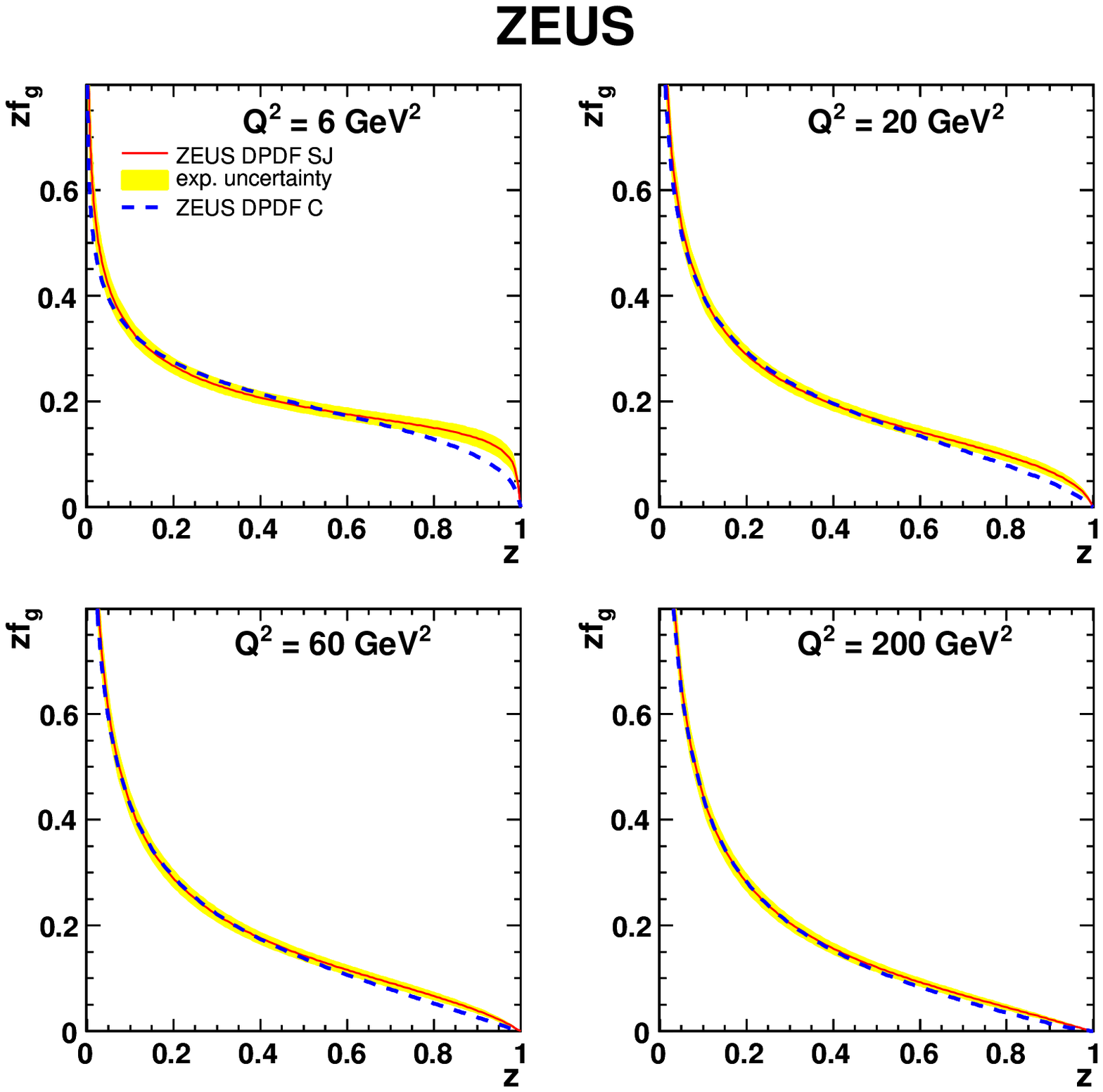,height=0.22\textwidth,clip}}
      \put(8,6){{\scriptsize{charm}}}
      \put(14,18){{\scriptsize{light}}}
    \end{picture}
  \end{center}
  \caption{Diffractive quark (left) and 
gluon (right) distributions 
extracted from a ZEUS fit to inclusive LRG and LPS 
data, with diffractive dijet data also included.  From~\cite{Chekanov:2009aa}.}
\label{ZEUS:DPDFs}
\end{figure}

The most recent
results obtained when fitting inclusive $\sigma_r^D$ data alone can be found
in \cite{Aktas:2006hy,Chekanov:2009aa}, see Fig.~\ref{H1:DPDFs}. 
Since $\sigma_r^D$ measures essentially the
charge-squared-weighted sum of quarks, these fits result in tight constraints
(to around 5\%) on the singlet quark density. From the scaling violations, the
gluon distribution is also rather well constrained at moderate $z$ values (to around $10 - 15 \%$).
However, the sensitivity to the gluon density with $z \, \gapprox \, 0.5$ is poor, due to
the inevitable dominance of the evolution by the
$q \rightarrow qg$ splitting as $z \rightarrow 1$. 
If diffractive jet production (see Section~\ref{apply:dpdf})
is used as an additional constraint, the 
high $z$ gluon precision is
improved considerably, at the expense of additional theoretical
assumptions on the large $\beta$ dynamics. Results of fits of this type can be
found in \cite{Aktas:2007bv,Chekanov:2009aa}, see Fig.~\ref{ZEUS:DPDFs}.

In all fits performed, the DPDFs are dominated by the gluon density, 
which extends
to large values of $z$ and accounts for typically 
60 \% (ZEUS \cite{Chekanov:2009aa}) to
70\% (H1 \cite{Aktas:2006hy}) of the total 
longitudinal momentum of the diffractive 
exchange. 

It is interesting to note that the
DPDFs, particularly the quark 
distribution, resemble the parton densities of the 
photon \cite{Nisius:1999cv}. 
The hadronic structure of the photon is generated from an initial 
$\gamma \rightarrow q \bar{q}$ splitting, with lower $x$ structure emerging
from further splittings and evolution. This superficial similarity is
as might be expected if diffraction emerges from a single gluon exchange,
dressed in a manner which neutralises the colour. The structure then develops
from $g \rightarrow q \bar{q}$ similarly to the photon case, but with an
admixture of $g \rightarrow gg$ initial splittings.

More sophisticated approaches to DPDFs exist, notably 
in which a direct, hard, pomeron contribution is also
included \cite{Martin:2006td} or which include a higher
twist contribution at 
large $\beta$ 
(see Section~\ref{sec:dipoles}) \cite{Royon:2000wg,GolecBiernat:2007kv}.
Whilst these contributions arise naturally
in QCD and must be present at some level, the evidence both from these 
fits and from hadronic final state comparisons 
(Section~\ref{apply:dpdf}) is that they
are numerically small, compared with the standard 
resolved structure contribution. 
The DPDF approach does, however, appear to undergo an infrared
breakdown at larger $Q^2$ 
scales than is the case in inclusive QCD 
fits \cite{Aktas:2006hy,Chekanov:2009aa}. 
Whilst this may provide further evidence for saturation 
effects in DDIS \cite{Frankfurt:2001av}, 
it may also be a consequence of enhanced higher twist
contributions or a less quickly convergent QCD order expansion in the 
diffractive than the
inclusive case.

\subsubsection{Applications of Diffractive Parton Densities}
\label{apply:dpdf}

The DPDFs described in Section~\ref{dpdfs} and their predecessors have been 
used to predict a wide variety of observables in DIS and photoproduction
at HERA, as well as at the Tevatron and the LHC. At one level, this is
done through implementations 
(usually of the H1 Fit B DPDFs \cite{Aktas:2006hy}) in Monte Carlo 
generators \cite{Jung:1993gf,Cox:2000jt,Navin:2010kk}. 
In particular,
the {\sc Rapgap} model has been used extensively as an experimental tool for modelling
both inclusive and hadronic final state diffractive data at HERA. 
The DPDFs have also been 
interfaced to various NLO QCD
calculations in order to compare their predictions with measurements. 

There is a wealth
of literature on this topic, which is summarised
with a few examples below. 
These tests are very similar in design and scope to those discussed in 
Section~\ref{hfs}. However, the level of precision is poorer in the diffractive
case, due to smaller data samples, added experimental complications and 
kinematic restrictions placed for example on jet transverse 
momenta.\footnote{The total available
invariant mass $M_X$ for final state particle production in diffraction
is typically a factor of 10 smaller than the analogous variable, $W$, 
in the inclusive case.} 

Early measurements of
inclusive final state observables in DDIS
were sufficient to rule out a diffractive exchange
with a quark-dominated structure, by showing that the basic event
topology is consistent with the boson--gluon fusion process, 
$\gamma^* g \rightarrow q \bar{q}$ (Fig.~\ref{feynman}(b)), 
yielding copious high $p_T$
particle production and leaving behind a colour-octet,
single gluon-like, remnant of the
diffractive exchange. The full list of inclusive final state observables measured in DDIS now comprises
charged particle spectra, multiplicities and their rapidity 
correlations \cite{Adloff:1998ds,Adloff:1998dw,Chekanov:2001ph},
energy flows \cite{Adloff:1998ds,Chekanov:2001ph} and event
shapes (thrust and sphericity) \cite{Breitweg:1997ug,Adloff:1997nn,Chekanov:2001ph}. 
These measurements are universally in agreement with predictions based on DPDFs
extracted from $\sigma_r^D$ data, reinforcing the 
picture of a diffractive exchange dominated by gluons extending to large momentum fractions. 

\begin{figure}[htb]
\begin{center}
~\epsfig{file=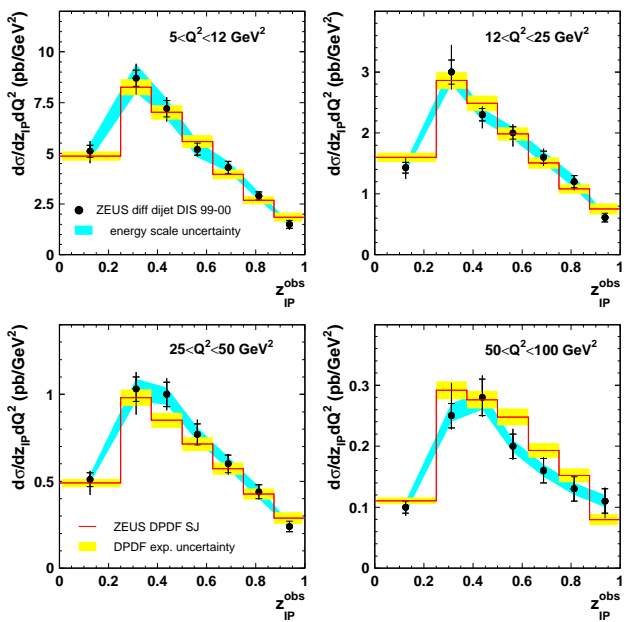, width=0.52\textwidth}
\caption{Comparison of diffractive DIS dijet cross sections, measured
double differentially in $z_\pom$ and $Q^2$,  
with NLO QCD predictions based on DPDFs. In this case, the data shown were used together 
with inclusive diffraction measurements in the fit, to improve the high $\beta$
sensitivity. 
From~\cite{Chekanov:2009aa}.}
\label{diffdis:jets}
\end{center}
\end{figure}

The most precise tests of DPDFs in diffractive DIS  have been obtained from exclusive final
state observables formed from relatively high transverse momentum jet
production, the rate for which is
closely related to the diffractive gluon density. Dijet measurements have been 
made as a function of many different 
variables \cite{Aktas:2007bv,Aktas:2007hn,Chekanov:2007aa,Aaron:2011mp}
and have been extensively used in 
comparisons with DPDF-based NLO 
predictions \cite{Aktas:2007bv,Chekanov:2009aa}. 
An example is shown in Fig.~\ref{diffdis:jets}.
This sort of comparison has 
again been universally successful.
In fact,
the factorisable (`resolved pomeron') model in diffractive DIS
works in a wider range of contexts than might be expected.
For example, DPDF-based predictions also describe 3-jet diffractive
final states fairly 
well \cite{Adloff:2000qi,Chekanov:2001cm}, despite being at the LO, rather 
than the NLO, level in the 3-jet case.  A further intriguing example is 
the case where one of the reconstructed jets is close 
in rapidity to the edge of the rapidity 
gap \cite{Aaron:2011mp}, a topology which can be measured by
exploiting the proton tagging method. This latter case 
rules out a dominant contribution from hard (pomeron remnant-free) 
diffractive 
production of the type discussed in \cite{Hebecker:2000xs,Martin:2006td}, 
which ought not to be 
describable using the DPDF approach. 
The strong experimental evidence for proton vertex factorisation 
and the applicability
of DPDFs to diffractive dijets in DIS, together with the relatively 
high precision 
with which the momentum fraction $z$ can be 
reconstructed
from the kinematics of the 
diffractive final state and the jet pair, have led to 
the argument more
recently being reversed, with jet data being 
used as an input to DPDF extractions \cite{Aktas:2007bv,Chekanov:2009aa},
as discussed in Section~\ref{dpdfs}.

\begin{figure*}[tb]
\begin{center}
~\epsfig{file=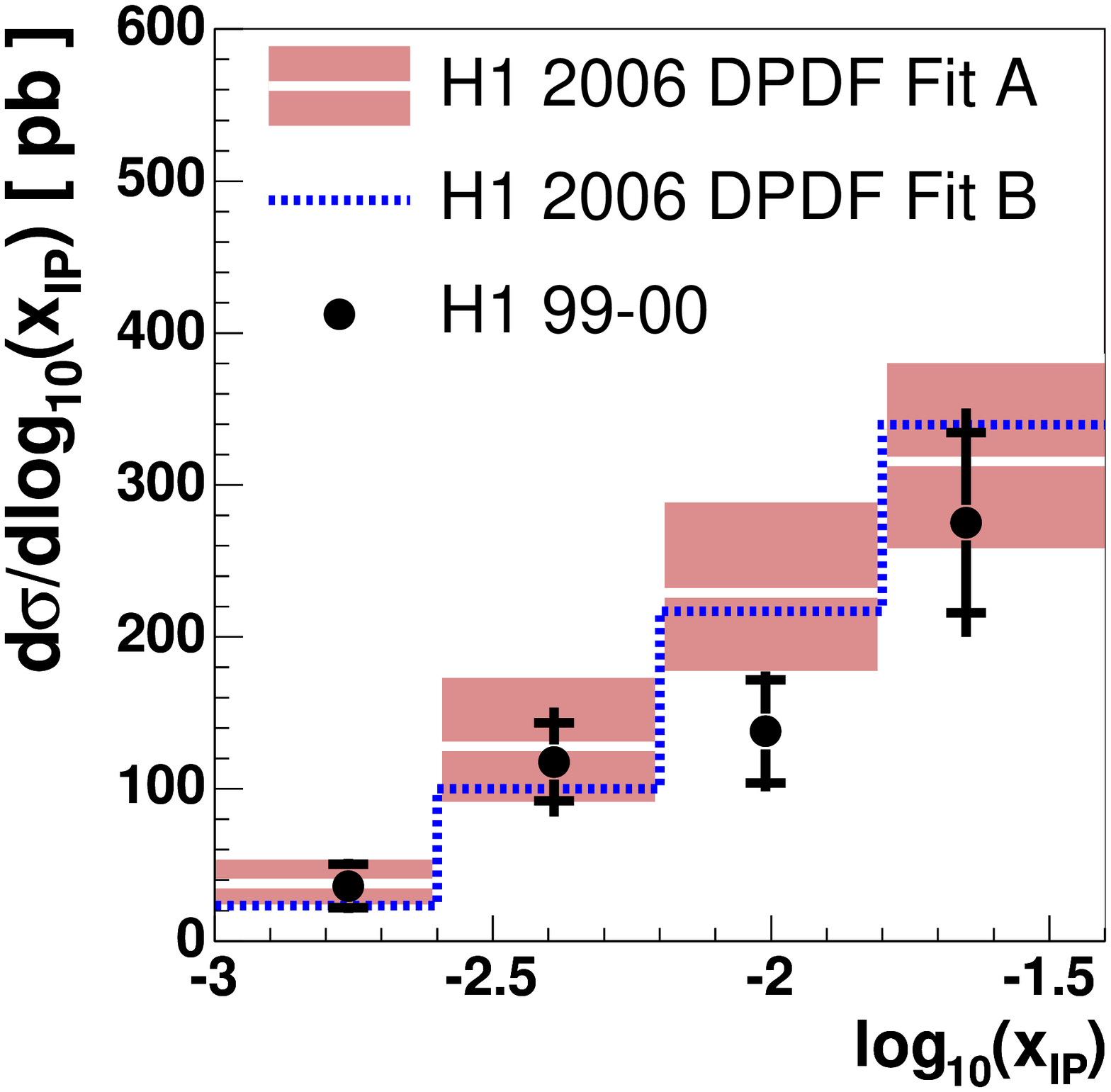,height=5.5cm}
~\epsfig{file=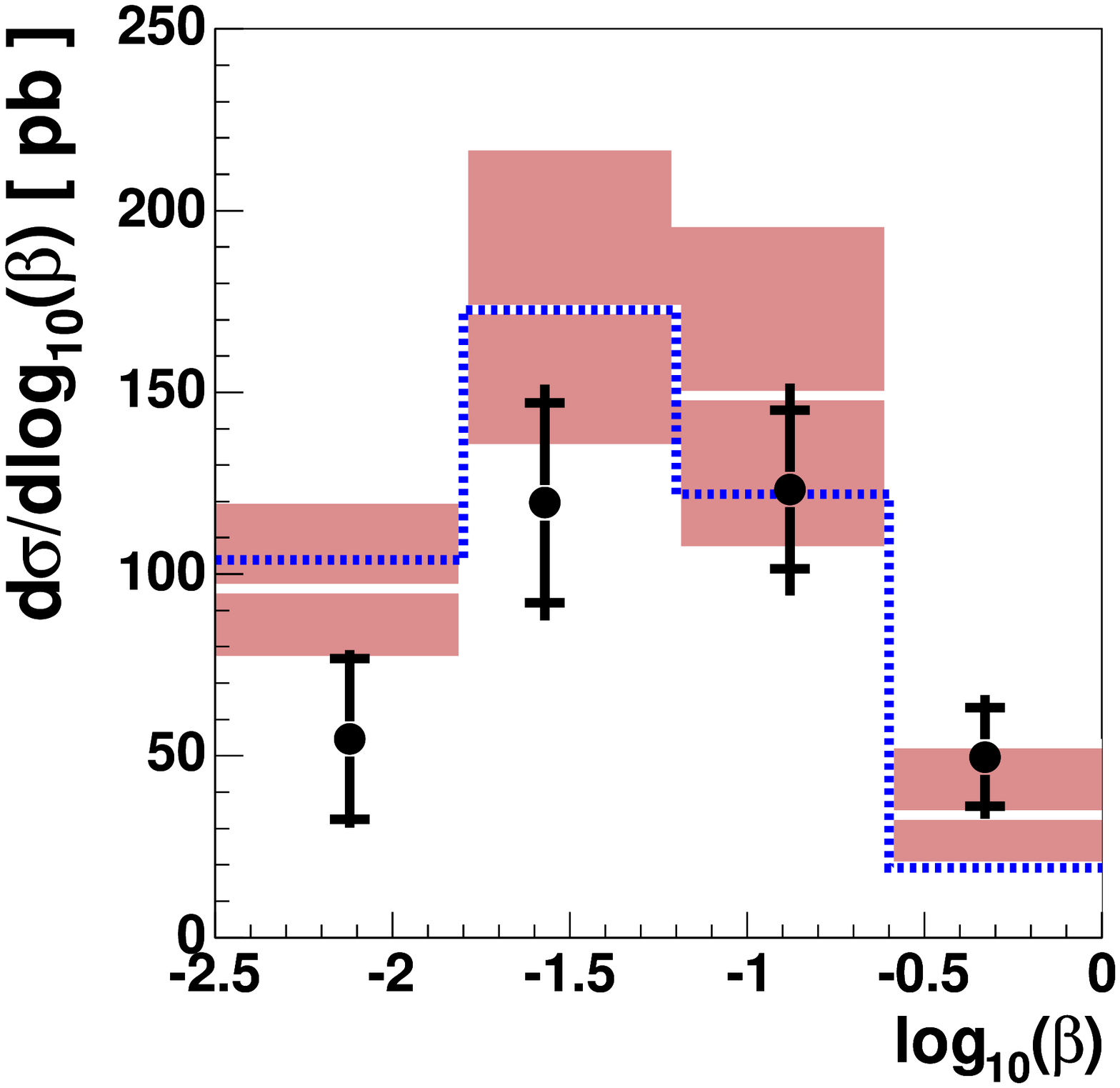,height=5.5cm}
~\epsfig{file=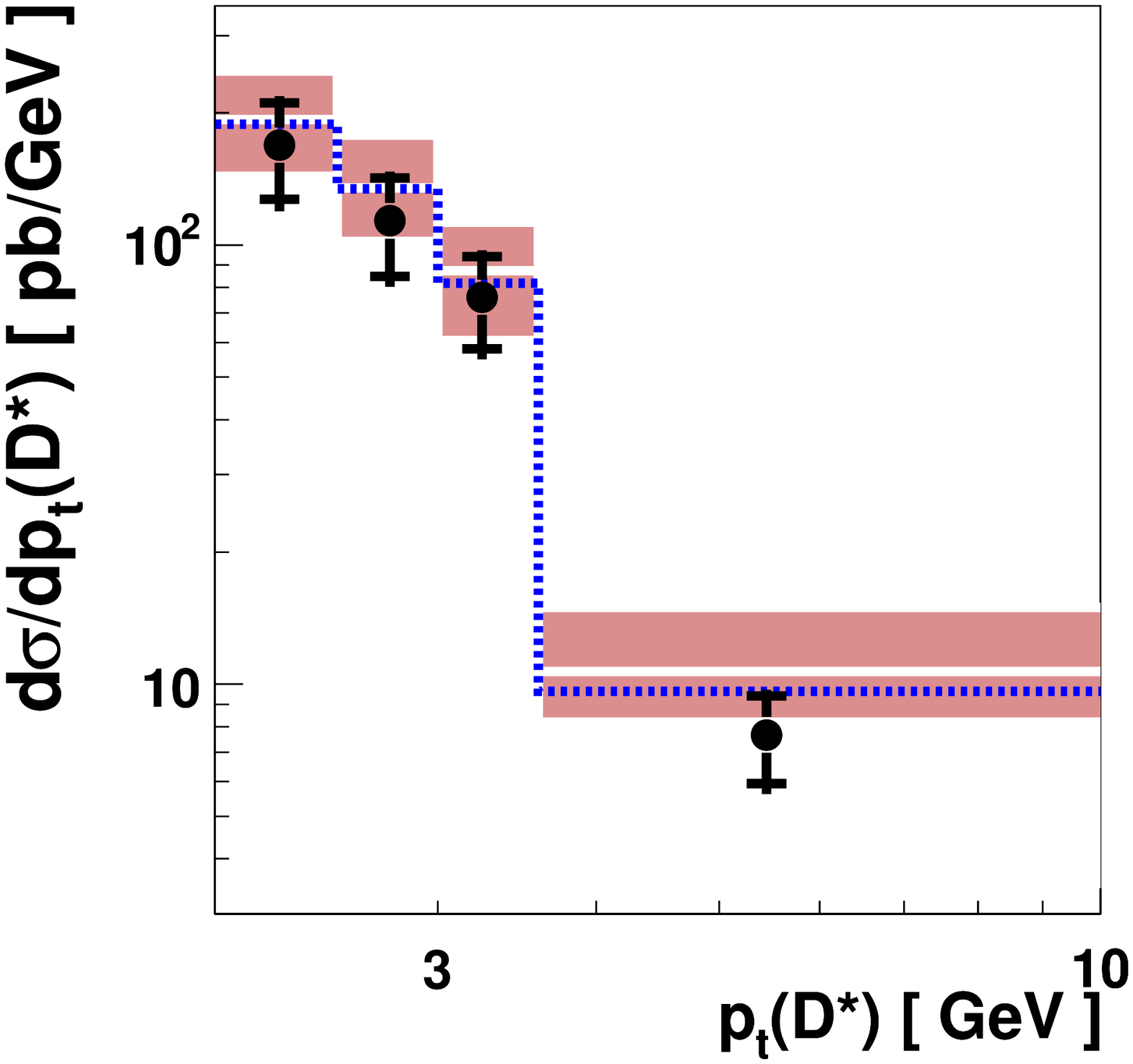,height=5.5cm}
\caption{Comparisons between diffractive open charm ($D^*$) production measurements
and NLO QCD predictions based on DPDFs.
From~\cite{Aktas:2006up}.}
\label{diffdis:charm}
\end{center}
\end{figure*}

Diffractive open charm production in DIS 
\cite{Aktas:2006up,Chekanov:2003gt}
provides a further exclusive final
state with high sensitivity to the diffractive gluon density via the lowest
order contributing process, $\gamma^* g \rightarrow c \bar{c}$. In all measurements
made to date, the charm quarks are tagged through the reconstruction of a $D^*$
meson using the usual $D^* - D^0$ mass difference 
technique (Section~\ref{sec:reco-hfl}).
Due to larger backgrounds
and smaller statistics, charm observables have not yielded the precision 
tests of diffractive factorisation and DPDFs that have been achieved with jets. In
principle, due to the weaker kinematic constraints on $M_X$, 
it should be possible to probe smaller $z$ values by using charm as a
tag. However, the problems of forward-going track reconstruction experienced by
both collaborations has limited progress in this direction. Nonetheless, as can be seen
from the example comparisons in Fig.~\ref{diffdis:charm}, open charm production is
well described by DPDF-based models, providing complementary support for this approach
and a constraint on the heavy flavour treatment in diffractive QCD fits.

A final class 
of DPDF tests in DIS may be performed using only inclusive diffractive
cross sections. 
Whilst these tests require measurements at the extremes of the accessible 
kinematic range, resulting in only limited precision, they are complementary
to hadronic final state constraints. An H1 measurement of the diffractive charged current cross
section is shown to be consistent
with DPDF-based predictions in \cite{Aktas:2006hy}. 
This is the only comparison made to date which is sensitive to the 
light quark flavour
decomposition of the diffractive quark density. The rather ad hoc 
flavour-democratic
assumptions 
made in the standard QCD fitting procedures (Section~\ref{dpdfs}) are
consistent with the data within the large statistical uncertainties. 

\begin{figure}[htb]
\begin{center}
~\epsfig{file=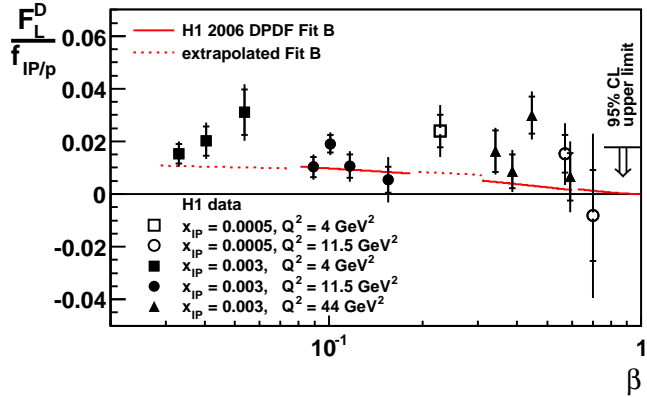,width=0.49\textwidth,bbllx=6,bblly=185,bburx=260,bbury=348,clip=}
\caption{Summary of H1 measurements of the diffractive longitudinal structure 
function $F_L^D$, exploiting data taken with variations in the proton beam energy. 
To allow comparisons between measurements at different $\xpom$ values,
the data are normalised by a diffractive flux factor $f_{\pom/p}$ 
(see Section~\ref{sec:regge}).  The data are compared with an NLO QCD prediction 
based on DPDFs.  
From~\cite{Aaron:2012zz}.}
\label{FLDfig}
\end{center}
\end{figure}

The diffractive
longitudinal structure function, $F_L^D$ \cite{Newman:2005mm} is a further independent observable
with sensitivity to both the quark and the gluon densities in novel ways. 
In particular, it provides a
unique test of the low-$x$ gluon.
$F_L^D$ can in principle be measured from variations in the DDIS cross section
with the azimuthal angle between the lepton and jet scattering planes, though attempts
to observe this effect using the FPS/LPS detectors have yielded results consistent
with zero \cite{Chekanov:2008fh}. Significantly non-zero results have, however, been 
obtained \cite{Aaron:2012zz}
using the Rosenbluth technique, comparing diffractive reduced cross section data at
fixed $\xpom$, $\beta$ and $Q^2$, but different $ep$ centre-of-mass energy, 
exploiting the
data taken at $\sqrt{s} = 460 \ {\rm GeV}$ and $\sqrt{s} = 525 \ {\rm GeV}$ at the
very end of the HERA running. The sensitivity to $F_L^D$ requires highly challenging
measurements at large $y$ (small scattered electron energy). Nonetheless, this structure
function has been measured over a fairly large kinematic region. As summarised in 
Fig.~\ref{FLDfig}, the data are again supportive of the DPDF approach. 

\subsubsection{Limitations of Diffractive Parton Densities}
\label{gap:survival}

Whilst models based on proton vertex factorisation and DPDFs work well to
describe all diffractive processes beyond the lowest $M_X$ resonance region
in DIS, they fail spectacularly when DPDFs extracted from HERA $\sigma_r^D$
data are applied to diffractive $p \bar{p}$ scattering at the Tevatron. 
For example, predictions for diffractive dijet production at the 
Tevatron exceed the data by  
a factor of around 10 \cite{Affolder:2000vb,Klasen:2009bi}. This 
limitation
is predicted as part of the QCD hard scattering factorisation
theorem for diffraction \cite{Collins:1997sr} and is
usually interpreted in terms of
multiple scattering, or `absorptive' effects, which occur in the 
presence of beam remnants.
These effects can be parametrised in terms of a `rapidity gap survival 
probability' \cite{Dokshitzer:1991he,Bjorken:1992er}. 
Similar effects are emerging at the LHC \cite{Chatrchyan:2012vc}. 

\begin{figure}[htb]
\begin{center}
~\epsfig{file=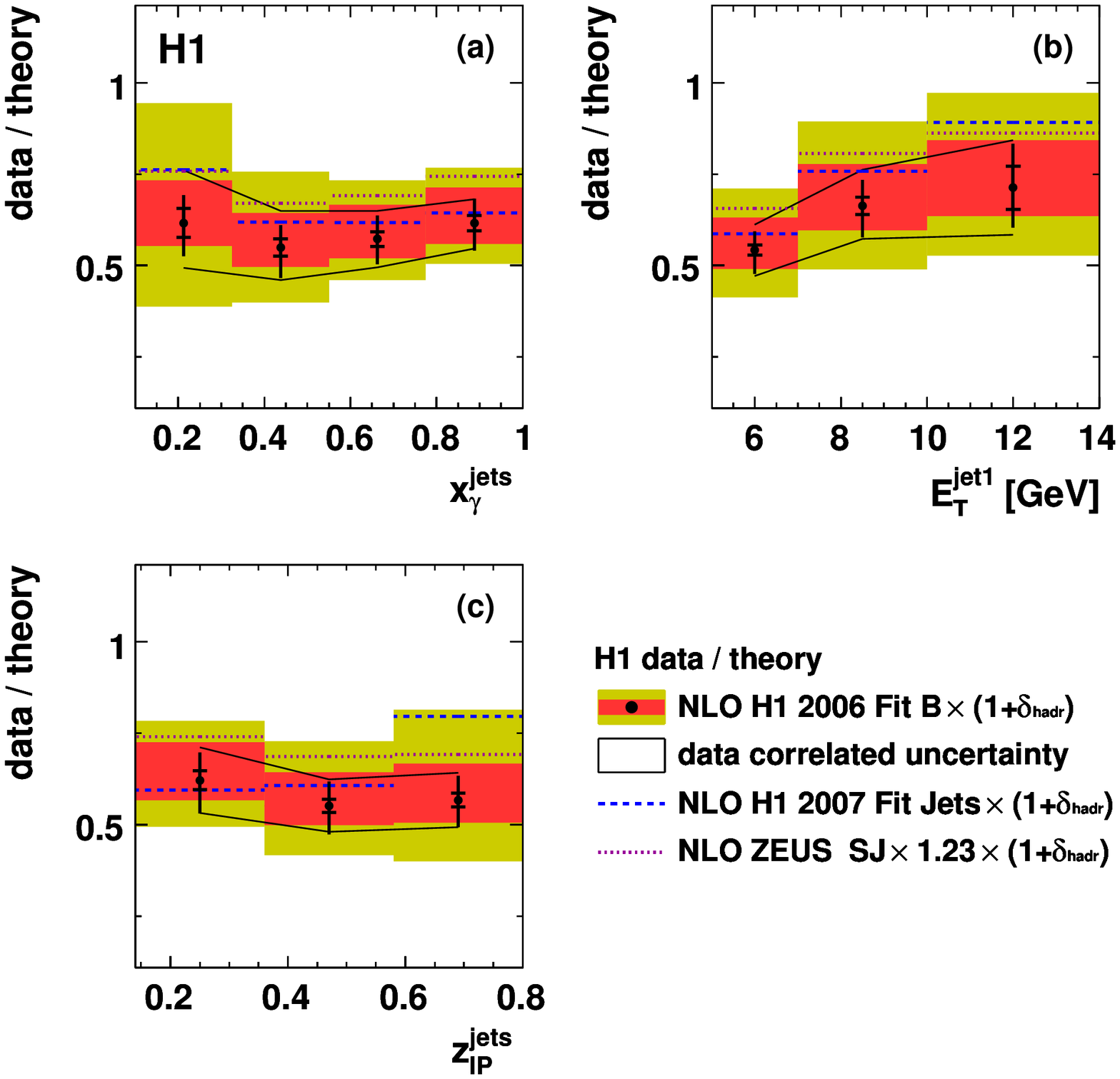,width=0.47\textwidth}
~\epsfig{file=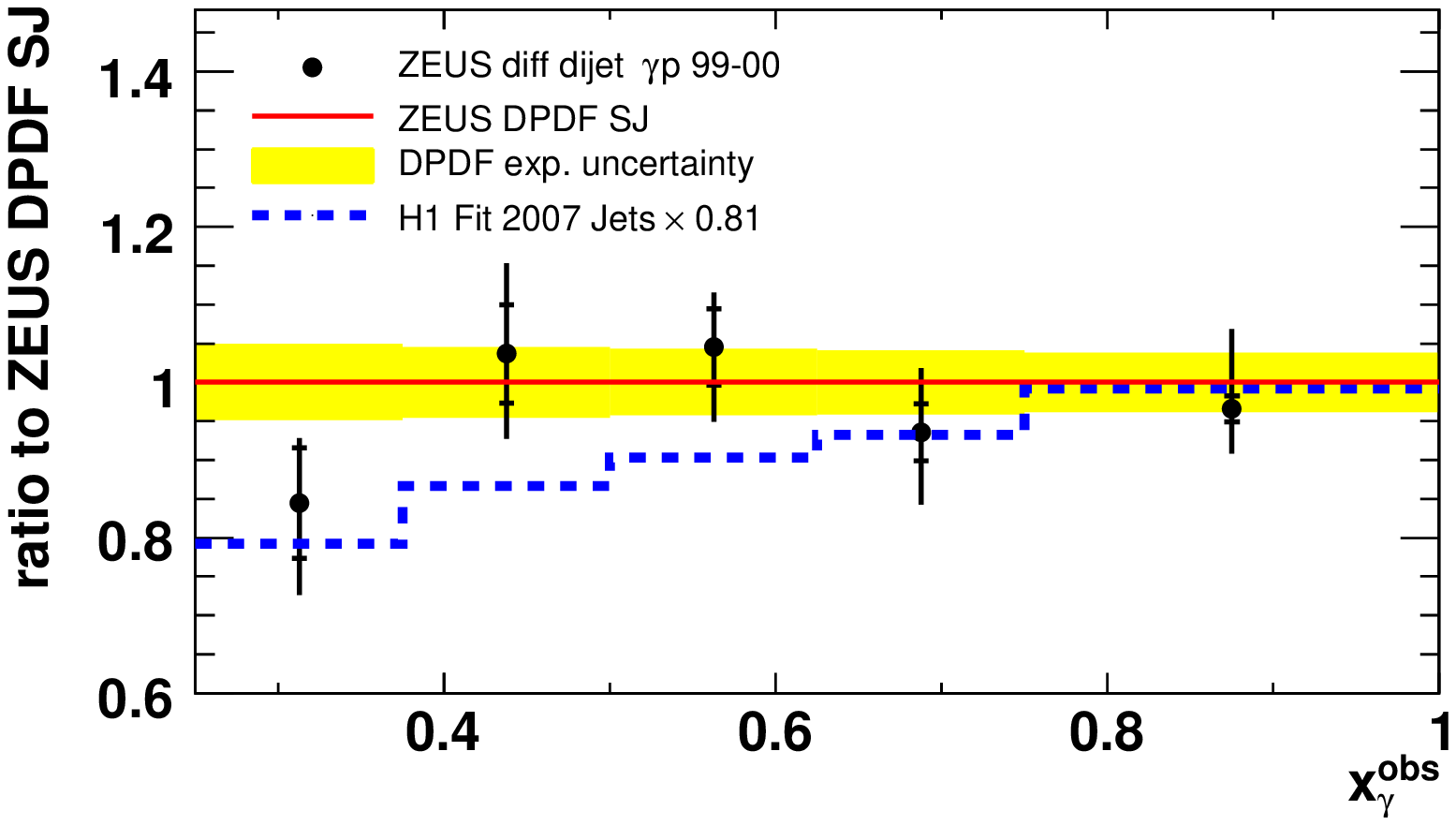,width=0.47\textwidth}
\caption{Diffractive dijet photoproduction data shown as ratios of measured
cross sections to DPDF-based NLO predictions (i.e. as measurements of the
rapidity gap survival probability). Top: H1 data, from \cite{Aaron:2010su} 
Bottom: ZEUS data, from \cite{Chekanov:2009aa}.}
\label{gammap:difjet}
\end{center}
\end{figure}

Measurements of diffractive dijet 
photoproduction have been pursued as a control experiment
for gap destruction models.
In a lowest order interpretation, direct photon processes 
ought to be unaffected by such effects,
whereas they should be present in
resolved photon processes, where the photon interacts through its hadronic
structure (Fig.~\ref{feynman}(c)). 
For example, a theoretical model which successfully describes the Tevatron result
predicted a suppression of the resolved contribution by a factor 
of 0.34 \cite{Kaidalov:2003xf}.
Multiple measurements of diffractive dijet photoproduction have been made by H1 and 
ZEUS, and have been compared with NLO 
calculations based on DPDFs extracted in DDIS.
Whilst the experimental data are just about compatible
between the two collaborations, the conclusions from
H1 and ZEUS differ slightly in this area. 
Notably, neither collaboration sees any evidence for the expected $x_\gamma$
dependence of the survival probability. 
H1 data \cite{Aktas:2007hn,Aaron:2010su} suggest a suppression 
of the data by a factor of around 0.6 relative to NLO QCD predictions, 
independently of $x_\gamma$. ZEUS results, which correspond to larger jet transverse 
energies \cite{Chekanov:2007rh,Chekanov:2009aa}
suggest a smaller suppression and are, in fact, 
consistent with no suppression
whatsoever. Selected data from both collaborations are shown in 
Fig.~\ref{gammap:difjet}.
These  
apparent problems are at least partially resolved by a recent
model in which a more careful treatment of point-like, as distinct from fully resolved, photon 
structure is introduced \cite{Kaidalov:2009fp}, leading to a much larger, and $E_T$ dependent,
survival probability for $0.1 \, \lapprox \, x_\gamma < 1$.

Diffractive charm production has also
been studied in photoproduction \cite{Aktas:2006up,Chekanov:2007pm}. 
The measurements are kinematically
restricted to the large $x_\gamma$ 
direct photon region and are described, within large
experimental and theoretical (scale) uncertainties, 
by DPDF-based predictions without recourse to gap destruction effects.

Rapidity gap survival probabilities and their kinematic dependences
remain an area which is not yet fully resolved. This is unfortunate, since
in addition to the interest this topic generates in its own right, it is an
essential ingredient in predicting hard diffractive cross sections at the LHC. 

\subsubsection{Diffractive DIS in Dipole Models}
\label{sec:dipoles}

\begin{figure}[htb]
\begin{center}
~\epsfig{file=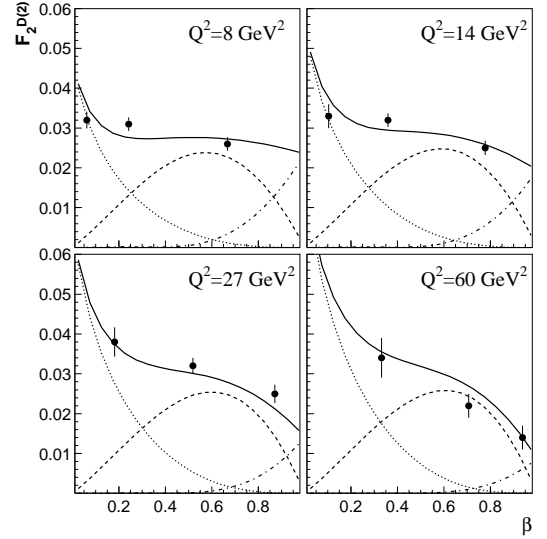,width=0.45\textwidth}
\caption{An example decomposition of the $\beta$ dependence of ZEUS
DDIS data \cite{Breitweg:1998gc} into different dipole terms, 
according to the parametrisation in \cite{GolecBiernat:1999qd}. 
The dashed, dotted and dot-dashed
curves correspond to leading twist transverse $q \bar{q}$, 
leading twist transverse $q \bar{q} g$ and
higher twist longitudinal $q \bar{q}$ contributions, respectively.}
\label{kgbfig}
\end{center}
\end{figure}

The application of the dipole picture to inclusive DDIS has proved 
to be problematic,
mainly due to the need for higher multiplicity fluctuations 
($q \bar{q} g$ and perhaps others) in order to describe the large $M_X$,
small $\beta$ region. This need was first shown in a quantified
manner through the `BEKW' fits to 
$F_2^D$ data in \cite{Bartels:1998ea}.
In this parametrisation, the data at low and moderate $\beta$ are 
described in terms of $q \bar{q}$ and $q \bar{q} g$
dipole fluctuations of transversely polarised photons,
whilst the high $\beta$ region contains a $Q^2$-suppressed non-leading twist 
contribution from $q \bar{q}$ fluctuations of longitudinally polarised
photons. This approach was further developed for comparisons with
inclusive DDIS data for example 
in the saturation model \cite{GolecBiernat:1999qd,GolecBiernat:2007kv}.
An example decomposition of the $\beta$ dependence of DDIS data is shown
in Fig.~\ref{kgbfig}. 
Predictions have also been made for 
hadronic final state observables, based on a two-gluon exchange model
of exclusive $q \bar{q}$ \cite{Bartels:1996ne,Lotter:1996zu} and 
$q \bar{q} g$ \cite{Bartels:1999tn,Bartels:2002ri} production. 
However, progress has been limited by significant theoretical uncertainties 
associated with the low $\beta$ $q \bar{q} g$ contribution and a lack of 
direct evidence for the high $\beta$ higher twist $q \bar{q}$ term.

\subsection{Leading Protons and Neutrons Beyond the Pomeron Region}
\label{sec:sublead}

Both H1 and ZEUS had proton spectrometers (see Fig.~\ref{fig:h1-layout}) with 
acceptance in the range $0.1\, \lapprox\, \xpom\, \lapprox\, 0.3$, well beyond 
the classic diffractive region where pomeron exchange is expected to dominate.  
In this region, the $ep \to eXp$ cross section may be understood in terms of the 
exchange of `sub-leading' exchanges of neutral meson 
states~\cite{Adloff:1998yg,Chekanov:2008tn}.  In the case of ZEUS, the scattered
proton sensitivity extended as far as $\xpom \sim 0.7$.  The acceptance of the 
zero degree forward neutron calorimeters in both experiments extended over a very 
wide range in neutron energies, from low values, where neutron production is 
describable by standard proton fragmentation, to large values, where charged 
colour singlet exchange $ep \to eXn$ becomes dominant.
Interpreted in terms of meson trajectory exchanges, this latter
reaction proceeds only via isospin-1 exchanges, for which the relative
rates of leading neutron and leading proton production are simply
related via a Clebsch--Gordan coefficient of 2. Studying the
leading proton and leading neutron data together 
in a Regge pole model (e.g.~\cite{Szczurek:1997cw})
leads to the 
tentative conclusion 
that the sub-leading trajectory with
$\alpha_\reg (0) \sim 0.5$ in Eq.~\ref{reggefac2}, which becomes important for 
$\xpom \simeq 0.05$, 
may be that associated with the $f_2$ meson.

At larger $\xpom$ values,
leading proton and neutron production can only be described simultaneously
if the exchanged trajectory is dominantly 
the pion, with $\alpha_\pi (0) \simeq 0$ \cite{Bishari:1972tx}, 
though other contributions also 
appear to be present \cite{Kaidalov:2006cw,Khoze:2006hw}. 
The study of leading neutron data at relatively large neutron energies
therefore raises the interesting, but not uncontroversial~\cite{Frankfurt:1997ij}, 
opportunity of measuring the partonic
structure of the pion using a similar factorisation between pion flux and 
structure function to that described for the pomeron
in Section~\ref{factorisation}.     

\begin{figure}[htb]
\begin{center}
~\epsfig{file=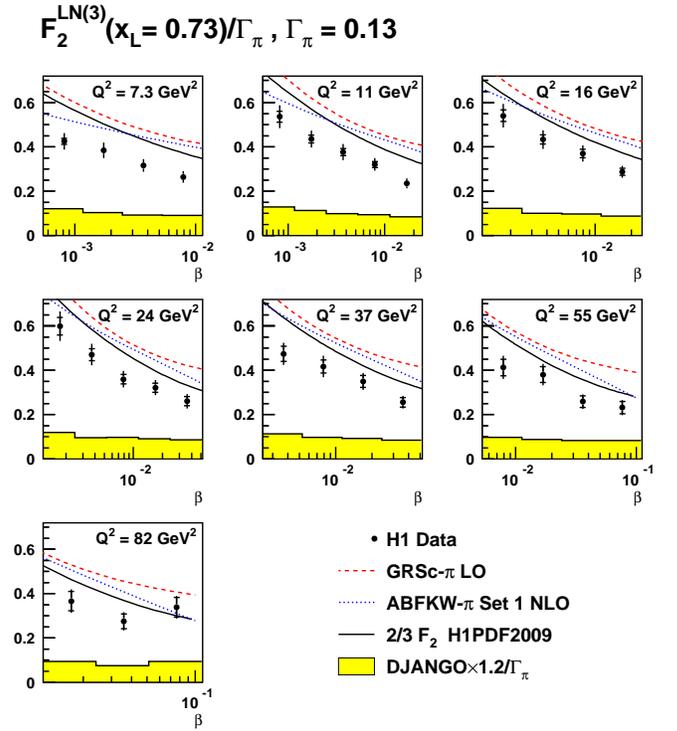,width=0.47\textwidth}
\caption{Leading neutron structure function data
(integrated over neutron transverse momenta up to $200 \ {\rm MeV}$),
divided by a parametrisation \cite{Holtmann:1994rs} of the 
pion flux $\Gamma_\pi$ at
$x_L \equiv 1 - \xpom = 0.73$. The estimated contribution from standard 
fragmentation processes is shown as a yellow band. 
After accounting for this, the data are
compared with two parametrisations of the pion 
structure function \cite{Aurenche:1989sx,Gluck:1999xe}
and with a parametrisation of the proton structure 
function \cite{Aaron:2009kv}
multiplied by $2/3$.
From~\cite{Aaron:2010ab}.}
\label{Neutronfig}
\end{center}
\end{figure}

For $\xpom \, \lapprox \, 0.3$, reconstructed
using the neutron energy $E_n$ and assuming
exclusive production at the proton vertex via  
$\xpom \equiv 1 - E_n / E_p$, inclusive 
leading neutron data in 
DIS \cite{Aaron:2010ab,Chekanov:2007tv,Chekanov:2004wn}
are broadly consistent with the pion
exchange hypothesis.
An example analysis is shown in Fig.~\ref{Neutronfig}.
Leading neutron data at $\xpom = 0.27$ are shown
after dividing by a parametrisation of the pion 
flux. Up to 
residual contributions from standard fragmentation processes and 
possibly other isovector exchanges, the data
then correspond to the pion structure function, $F_2^\pi (\beta, Q^2)$.
Existing pion structure function 
parametrisations are 
broadly in-line with the data, but
clearly overshoot when considered in detail. 
A simple model based on valence quark
counting, such that $F_2^\pi = 2 F_2 / 3$, is slightly closer to the 
data. Considering the large uncertainties in the pion flux factor and the
lack of previous data sensitive to the measured range, 
which extends to 
$\beta < 10^{-3}$, the level of agreement is reasonable. 
An analysis in which HERA leading neutron data are used 
as an input to a pion parton density extraction is, however, yet to
be performed.

In addition to the inclusive
neutron production 
process, measurements have also
been made of dijet \cite{Aktas:2004gi,Chekanov:2009ac} and charm 
quark \cite{Chekanov:2004dk} 
photoproduction in
association with leading neutrons.
Considered together with the inclusive 
leading neutron data, these final
state measurements show evidence for considerable absorptive
corrections. This complication, together with the uncertainties inherent
in factoring out the pion flux, have limited the information finally
extracted on the pion structure function.  

\subsection{Very Hard Diffraction and the BFKL Pomeron}
\label{bfkl}

For the diffractive processes discussed in the previous 
sections, $|t|$ is generally smaller than typical hadronic mass
scales. However, diffractive processes have also been studied at
HERA at $|t|$ values which are large enough to provide a hard scale
in their own right. Under such circumstances, the $t$ channel pomeron
exchange ought to be genuinely hard in the sense that it 
couples as a whole to individual partons.
Cases where the hard sub-process satisfies
$\hat{s} \gg -\hat{t} \gg \Lambda_{\rm QCD}$ correspond to a perturbatively
calculable limit of Regge theory. Where there is no particular ordering
in transverse momentum within the colour singlet exchange, the 
dynamics may be driven primarily by BFKL 
evolution (see Section~\ref{sec:boundaries}). Diffractive processes
at large $|t|$ therefore represent a promising area in which to search
for low $x$ parton dynamics driven by 
the BFKL pomeron. 

One striking signature for perturbative colour singlet exchange
is the photoproduction of pairs of jets separated by a large rapidity
gap. For such `gap between jets'
configurations, $|t| \approx p_{\rm T, jet}^2$ is very large,
typically $|t| > 25 \ {\rm GeV^2}$ in HERA studies, and
calculations using
the leading-logarithmic BFKL approximation \cite{Mueller:1992pe} 
with appropriate modifications \cite{Forshaw:1997wn,Enberg:2001ev}
can be applied to the
scattering between a parton from the proton and a parton from the 
resolved structure of the photon.
However, the situation is complicated by the possibility of secondary
scattering and a rapidity gap survival probability significantly
smaller than unity, similar to that discussed for the diffractive
dijet photoproduction process in Section~\ref{gap:survival}. 
The HERA data \cite{Derrick:1995pb,Adloff:2002em,Chekanov:2006pw}
have shown clear evidence for events with little 
energy flow between jets, which  
occur more frequently than 
expected from hadronisation
fluctuations in standard Monte Carlo models of jet photoproduction. 
The order of magnitude of the signal is also in agreement with that
expected from the BFKL calculations. Whilst this is highly 
suggestive, and is in fact perhaps the best evidence for BFKL dynamics
at HERA, more quantitative conclusions
have been precluded by the uncertainties in the
perturbative calculations, the gap survival probability and the 
residual non-diffractive contributions.

\begin{figure}[htb]
\begin{center}
~\epsfig{file=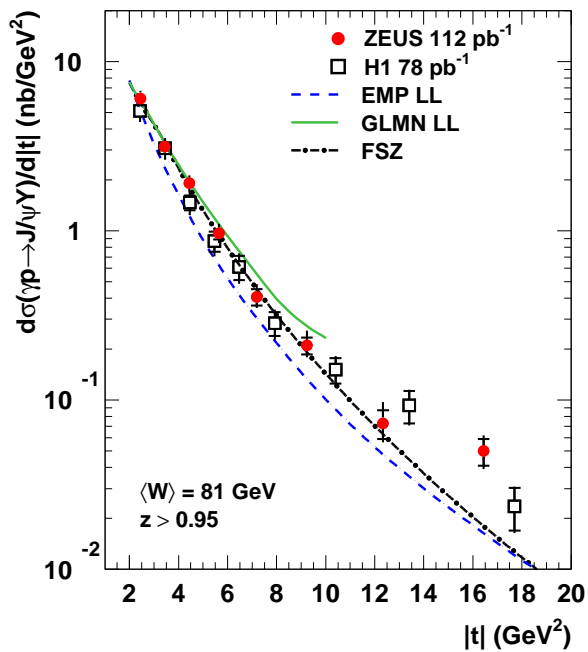,width=0.5\textwidth}
\caption{Dependence on $|t|$ of proton dissociative $J/\psi$ 
production integrated over proton dissociation masses satisfying
$M_Y^2 / W^2 < 0.05$. The data are compared with a prediction 
(`GLMN LL' \cite{Gotsman:2001ne}) which
is based on DGLAP evolution and is expected to be valid
for $|t| < m_\psi^2$. They are also compared
with a BFKL-based prediction (`EMP LL' \cite{Enberg:2002zy})
and a more phenomenological approach (`FSZ' \cite{Frankfurt:2008er}.
From~\cite{Chekanov:2009ab}.}
\label{Hight:Jpsi}
\end{center}
\end{figure}

Another candidate process in which to observe perturbative 
colour singlet exchange is quasi-elastic
vector meson production at large
$|t|$. Although $|t|$ values are smaller here than in the gaps-between-jets
case and the proton essentially always dissociates,
the exclusive production mechanism at the photon vertex
(Fig.~\ref{VM:feynman}(b)) implies significantly reduced complications 
from 
gap destruction effects, even in photoproduction. 
The possibility of perturbative colour singlet
exchange has long been considered for this 
process \cite{Ginzburg:1985tp} and detailed calculations have been performed 
in the leading-logarithmic BFKL 
framework~\cite{Forshaw:1995ax,Bartels:1996fs,Forshaw:2001pf,Enberg:2003jw,Poludniowski:2003yk}.
BFKL evolution is 
expected to be relevant for $|t| > m_V^2$, where
$m_V$ is the vector meson mass. 
For the case of heavy vector meson production, there is an 
interim region where
$|t| < m_V^2$, yet
$|t|$ remains large enough to apply perturbative techniques. Here, 
the transverse
momenta along the gluon ladder remain ordered and DGLAP
dynamics are expected to apply \cite{Gotsman:2001ne,Blok:2010ds}.
Studies have been made at HERA for 
light vector mesons \cite{Aktas:2006qs,Chekanov:2002rm} and
$J/\psi$ \cite{Aktas:2003zi,Chekanov:2009ab} mesons, 
as well as of
exclusively produced photons \cite{Aaron:2008ab}. 
Since it offers adequate statistics as well as a 
relatively clean theoretical interpretation, 
the $J/\psi$ channel has yielded the most precise tests. An example
analysis is shown in 
Fig.~\ref{Hight:Jpsi}.
The $t$ dependences generally follow the expected
approximate power law behaviour at large $|t|$ and 
the effective trajectory slope $\alphapom^\prime$ is 
much smaller than that describing soft diffraction, as expected
in a BFKL treatment. However, the overall description 
by specific BFKL predictions is not yet sufficiently good for strong
claims to be made. 

\section{Summary and Outlook}

Condensing into the previous 50 pages a total of about 200 publications from H1 and ZEUS on the hadronic final 
state has been a challenge.  Further reducing this to a 1-page summary is a task fraught with omissions and 
generalisations. With this caveat, the following represents an attempt to summarise the 
essential highlights of studies of the HERA hadronic final state as briefly as possible.

\begin{itemize}

\item Measurements of jet production among the most precise in the world, which have led to significant constraints on the 
parton density functions in the proton and precise determinations of $\alpha_s$.

\item Similarly, data on heavy-quark production which have led to 
important constraints on 
the charm contribution to the proton structure functions, unique to HERA, and on the heavy quark masses.

\item A demonstration of the power of QCD and in particular 
the applicability of the DGLAP equations for a multitude of processes
covering a wide kinematic range. 
Despite considerable attempts to isolate evidence for
BFKL or other parton evolution schemes and a handful of measurements which have
yet to be satisfactorily described, no unequivocal evidence has been found for the need for non-DGLAP dynamics.

\item Many measurements of light and heavy quark 
fragmentation processes and charged particle spectra which show similar behaviour to that observed in
$e^+e^-$ or other collisions. 
Despite some anomalies, the data are broadly 
consistent with the concept of universal fragmentation.

\item A wide variety of
searches for evidence of exotic QCD processes, such as pentaquarks and 
instantons.  
Of all searches performed, a glueball candidate decaying to a $K_S^0 K_S^0$ pair 
with a mass consistent with the $f_0(1710)$ state, was the only clear signal.

\item A detailed exploration of the quasi-elastic exclusive production of vector 
mesons ($\gamma^{(*)} p \rightarrow Vp$) and photons ($\gamma^{(*)} p \rightarrow  \gamma p$) over a 
wide range in scale, an appropriate choice for which is often $(Q^2 + M_V^2) / 4$. The data beautifully
illustrate and explore the transition with scale from a regime familiar from soft hadronic elastic and
diffractive scattering to a region governed by hard diffractive exchanges which can be 
interpreted in terms of partons.

\item Precise measurements of hard exclusive diffractive processes, for example $J/\psi$ photoproduction, 
and the development of successful methods for calculating related observables starting from a knowledge of
the parton densities of the proton. 

\item The precise measurement and interpretation of inclusive diffraction over a vast, usually 
three-dimensional, kinematic range, which has shown the process to be well modelled as the 
deep inelastic scattering of the electron from a factorisable soft object, not dissimilar from the 
`pomeron' of soft hadronic physics. 

\item The extraction of the partonic structure of the soft-pomeron-like exchange and its 
successful application in predicting diffractive final state observables using
standard NLO DGLAP-based tools.

\end{itemize}

Although these essential points are now unlikely to alter, the H1 and ZEUS collaborations are 
still regularly publishing new results at the time of writing, and in several important areas the most precise
measurements are still yet to come. Examples which have been discussed 
here include inclusive jet production in DIS and 
dijet photoproduction, for both of which only around 10\% of the available data has been used
in published measurements. Extending to the full dataset and, where possible, combining H1 and ZEUS
data will improve the precision and accessible kinematic range at 
high $E_T$ (equivalently large $x$). 
The precision on 
heavy flavour cross sections, particularly those in the beauty sector, is also often statistically limited
at present, with substantial power to improve if the full HERA data are exploited. 
In many areas, the understanding of the existing data 
can be improved considerably by the application of theoretical or phenomenological techniques which in
principle already exist. 
Most prominently, the development for $ep$ scattering of a Monte Carlo model which matches NLO QCD calculations to
parton showers would represent a major breakthrough in our ability to model and interpret jet and other
final state data. 
A major impact could also be achieved by
feeding into HERA analyses the improved understanding of the underlying event and of hadronisation which is currently
developing through model-tuning exercises 
at the LHC. The impact of these phenomenology improvements would be greatest in allowing 
low $E_T$ data to be 
exploited more fully, leading to better constraints on photon and low-$x$ proton structure.
New parametrisations of photon structure which include a wider range of HERA photoproduction
data would also be a significant step forward. 
Many of the above improvements would extend the list of HERA observables which could reliably be 
included in fits to extract information on the proton structure. 
Combining the final inclusive DIS HERA and fixed target data with carefully chosen HERA 
hadronic final state measurements has the potential to produce constraints on the proton PDFs, $\alpha_s$
and heavy quark masses which are unlikely to be surpassed for many years.  
Indeed, it may not be too ambitious to attempt a simultaneous extraction of proton and photon structure
from HERA and other data, a programme of work which may gain new impetus should a high energy 
$e^+ e^-$ linear collider be constructed.
 
Whilst there is still considerable room for improvement for the topics discussed above, there are also
areas of the HERA physics programme which are essentially complete. Given the current experimental 
precision and the limitations in our theoretical understanding, further substantial progress in areas such as the
search for novel low $x$ dynamics, the measurement of charged particle spectra and the unravelling of
soft and hard contributions to diffractive DIS appears unlikely. Particularly for 
unresolved issues in low $x$ physics such as parton saturation, we may have to wait for data from a future 
higher energy lepton--hadron collider before drawing firm conclusions. 

The impact of HERA hadronic final state data and the new techniques developed for its analysis  
has been felt strongly at the LHC. Apart from the obvious 
need for a precise knowledge of the proton structure 
as obtained from the well-matched HERA range of sensitivity, 
H1 and ZEUS input has contributed significantly in deciding 
how to make well-defined measurements of, for example, jet, heavy quark and diffractive processes.  
There are also less obvious fields which have benefited, for example the modelling of high energy
cosmic ray air showers and of the hadronic final states produced when neutrinos interact with 
hadronic matter in neutrino experiments. 
Whenever and wherever the next facility for high energy 
lepton--hadron scattering is built, the HERA 
results will give strong steers as to how and where to make precise 
measurements of familiar physics and to look for new phenomena, such as low $x$ parton saturation.  

In a sentence, the hadronic 
final state in electron--proton collisions at HERA has provided a rich source of data and deepened our understanding 
of strong interactions.

\section*{Acknowledgements}

PRN and MW are fortunate to have been members of the 
H1 and ZEUS collaborations, respectively, for around two decades. 
There are far too many colleagues with whom we have worked closely
to name in person, but 
we wish to record our thanks for many pleasant working relationships and
much intellectual stimulation. We would also like to thank 
J. Butterworth for providing input in the initial planning stages of this
review.

\bibliographystyle{apsrmp}
\bibliography{hera}

\end{document}